\documentclass[a4paper,11pt]{article}

\usepackage{ucs}
\usepackage[utf8x]{inputenc}
\usepackage{graphicx}
\usepackage{amsfonts}
\usepackage{amssymb}
\usepackage{amsmath}
\usepackage{amsthm}
\usepackage{dsfont}
\usepackage{enumerate}
\usepackage{fullpage}
\usepackage{ifthen}
\usepackage{epic}
\usepackage{authblk}
\usepackage{textcomp}
\usepackage[all]{xy}
\usepackage{mathrsfs}
\usepackage{subfigure}
\usepackage{titlesec}
\usepackage[small]{caption}
\usepackage{tikz,tkz-tab}
\usepackage{xcolor}
\usepackage{comment}

\usetikzlibrary{arrows,shapes,positioning}
\usetikzlibrary{decorations.markings}
\tikzstyle arrowstyle=[scale=1.5]
\tikzstyle directed=[postaction={decorate,decoration={markings,
    mark=at position .5 with {\arrow[arrowstyle]{stealth}}}}]
\tikzstyle reverse directed=[postaction={decorate,decoration={markings,
    mark=at position .5 with {\arrowreversed[arrowstyle]{stealth};}}}]

\usepackage[hypertexnames=false,colorlinks=true,linkcolor=blue,citecolor=blue]{hyperref}
\usepackage[numbers,comma,square,sort&compress]{natbib}
\usepackage[letterpaper,text={7in,9in},centering]{geometry}

\usepackage{titlesec}
\graphicspath{{eps/}{pdf/}}

\numberwithin{equation}{section}

\newtheorem{thm}{Theorem}

  {\begin{trivlist}\item[]{\bf Hypothesis #1 }\em}{\end{trivlist}}

\newcommand{\R}{\mathbb{R}}
\newcommand{\C}{\mathbb{C}}

\newcommand{\Z}{\mathbb{Z}}
\newcommand{\md}{\mathrm{d}}
\newcommand{\E}{\mathcal{E}}
\newcommand{\G}{\mathcal{G}}
\newcommand{\D}{\mathcal{D}}
\newcommand{\W}{\mathcal{W}}
\newcommand{\U}{\mathcal{U}}
\renewcommand{\S}{\mathcal{S}}

\renewcommand{\Re}{\mathrm{Re}}

\newcommand{\mbi}{\mathbf{i}}

\newcommand{\bqq}{\begin{equation}}
\newcommand{\eqq}{\end{equation}}
\newcommand{\bqs}{\begin{equation*}}
\newcommand{\eqs}{\end{equation*}}

 {\begin{trivlist}\item[]\textbf{Proof#1 }}%
 {\hspace*{\fill}$\rule{0.3\baselineskip}{0.35\baselineskip}$\end{trivlist}}

\title{Mathematical derivation of wave propagation properties \\in hierarchical neural networks \\with predictive coding feedback dynamics}
\author[1,3]{Gr\'egory Faye\footnote{ \texttt{gregory.faye@math.univ-toulouse.fr}}}
\author[1,2,3]{Guilhem Fouilh\'e}
\author[2,3]{Rufin VanRullen}
\affil[1]{Institut de Math\'ematiques de Toulouse ; UMR5219, Universit\'e de Toulouse ; UPS IMT, F-31062 Toulouse Cedex 9 France}
\affil[2]{CerCo, CNRS, 31052 Toulouse, France}
\affil[3]{ANITI, Universit\'e de Toulouse, 31062, Toulouse, France}
\begin{document}
\maketitle

\begin{abstract}
Sensory perception (e.g. vision) relies on a hierarchy of cortical areas, in which neural activity propagates in both directions, to convey information not only about sensory inputs but also about cognitive states, expectations and predictions. At the macroscopic scale, neurophysiological experiments have described the corresponding neural signals as both forward and backward-travelling waves, sometimes with characteristic oscillatory signatures. It remains unclear, however, how such activity patterns relate to specific functional properties of the perceptual apparatus. Here, we present a mathematical framework, inspired by neural network models of predictive coding, to systematically investigate neural dynamics in a hierarchical perceptual system. We show that stability of the system can be systematically derived from the values of hyper-parameters controlling the different signals (related to bottom-up inputs, top-down prediction and error correction). Similarly, it is possible to determine in which direction, and at what speed neural activity propagates in the system. Different neural assemblies (reflecting distinct eigenvectors of the connectivity matrices) can simultaneously and independently display different properties in terms of stability, propagation speed or direction. We also derive continuous-limit versions of the system, both in time and in neural space. Finally, we analyze the possible influence of transmission delays between layers, and reveal the emergence of oscillations at biologically plausible frequencies.
\end{abstract}

\newpage

\section{Introduction}
The brain's anatomy is characterized by a strongly hierarchical architecture, with a succession of brain regions that process increasingly complex information. This functional strategy is mirrored by the succession of processing layers found in modern deep neural networks (and for this reason, we use the term ``layer'' in this work to denote one particular brain region in this hierarchy, rather than the laminar organization of cortex that is well-known to neuroscientists). The hierarchical structure is especially obvious in the organization of the visual system~\cite{vanessen1991}, starting from the retina through primary visual cortex (V1) and various extra-striate regions, and culminating in temporal lobe regions for object recognition and in parietal regions for motion and location processing.\\\
In this hierarchy of brain regions, the flow of information is clearly bidirectional: there are comparable number of fibers sending neural signals down (from higher to lower levels of the hierachy) as there are going up~\cite{bullier2001}. While the bottom-up or ``feed-forward'' propagation of information is easily understood as integration of sensory input (and matches the functional structure found in artificial deep learning networks), the opposite feedback direction of propagation is more mysterious, and its functional role remains unknown.\\\
Predictive coding is one dominant theory to explain the function of cortical feedback~\cite{RB99}. Briefly, the theory states that each layer in the cortical hierarchy generates predictions about what caused their own activity; these predictions are sent to the immediately preceding layer, where a prediction error can be computed, and carried forward to the original layer, which can then iteratively update its prediction. Over time (and as long as the sensory input does not change), the system settles into a state where top-down predictions agree with bottom-up inputs, and no prediction error is transmitted. Like any large-scale theory of brain function, the predictive coding theory is heavily debated~\cite{millidge2021}. But macroscopic (EEG) experiments have revealed characteristic propagation signatures that could be hallmarks of predictive coding. For instance, Alamia and VanRullen~\cite{AVR19} showed evidence for alpha-band (7-15Hz) oscillatory travelling waves propagating in both directions (feed-forward and feedback); the oscillation frequency and dynamics were compatible with a simplistic hierarchical model that included a biologically plausible time delay for transmitting signals between layers, and was also confirmed by a rudimentary mathematical model. In another study, Bastos et al~\cite{bastos2012,bastos2015} found that beta (15-30Hz) and gamma-frequency (30-100Hz) oscillations could reflect, respectively, the predictions and prediction errors signals carried by backward and forward connections. \\\
More recently, predictive coding has been explored in the context of deep neural networks~\cite{Wen18,choski21,Pangetal}. For instance, Choksi et al~\cite{choski21} augmented existing deep convolutional networks with feedback connections and a mechanism for computing and minimizing prediction errors, and found that the augmented system displayed more robust perception, better aligned with human abilities. In another study, Pang et al~\cite{Pangetal} used a similar system and reported the emergence of illusory contour perception comparable to what humans (but not standard deep neural networks) would typically perceive. \\\
While the concept of predictive coding is potentially fundamental for understanding brain function, and its large-scale implementation in deep artificial neural networks provides empirical support for its potential functional relevance, there is a gap of theoretical knowledge about the type of brain activity that predictive coding could engender, and the potential conditions for its stability. Here, we propose a mathematical framework where a potentially infinite number of neuronal layers exchange signals in both directions according to predictive coding principles. The stable propagation of information in such a system can be explored analytically as a function of its initial state, its internal parameters (controlling the strength of inputs, predictions, and error signals) and its connectivity (e.g. convolution kernels). Our approach considers both a discrete approximation of the system, as well as continuous abstractions. We demonstrate the practical relevance of our findings by applying them to a ring model of orientation processing. Finally, we extend our analytical framework to a more biologically plausible situation with communication delays between successive layers. This gives rise to oscillatory signals resembling those observed in the brain.

\section{Model description}

Our initial model is inspired by the generic formulation of predictive coding proposed in the context of deep learning models by Choksi et al.~\cite{choski21}. This formulation considers different update terms at each time step: feed-forward inputs, memory term, feedback- and feed-forward prediction error corrections. By modulating the hyper-parameters controlling each of these terms, the model can be reconciled with different formulations of predictive coding (for instance, the Rao and Ballard model~\cite{RB99} by setting the feed-forward input term to zero) or other models of hierarchical brain function (e.g. similar to Heeger's model~\cite{heeger2017} by setting the feed-forward error correction to zero). Indeed, our objective is precisely to characterize the propagation dynamics inside the network as a function of the relative value of these hyper-parameters, which in turn alters the model's functionality.\\

We consider the following recurrence equation where $\E_j^n\in\R^d$ represents an encoder at step $n$ and layer $j$
\bqs
\E_j^{n+1}=\beta \W^f \E_{j-1}^{n+1} +(1-\beta) \E_j^{n}-\alpha \mathcal{F}_{j-1}^n -\lambda \mathcal{B}_{j}^n, \quad j=1,\cdots,J-1 \,,
\eqs
where $\W^f\in\mathscr{M}_d(\R)$ is a $d\times d$ square matrix representing the weights of feedforward connections which we assume to be the same for each layer such that $\W^f \E_{j-1}^{n+1}$ models an instantaneous feedforward drive from layer $j-1$ to layer $j$, controlled by hyper-parameter $\beta$. The term $\mathcal{F}_{j-1}^n$ encodes a feedforward error correction process, controlled by hyper-parameter $\alpha$, where the reconstruction error $\mathcal{R}_{j-1}^n$ at layer $j-1$, defined as the square error between the representation $\E_{j-1}^n$ and the predicted reconstruction $\W^b\E_j^n$, that is
\bqs
\mathcal{R}_{j-1}^n:= \frac{1}{2} \|\E_{j-1}^n-\W^b\E_j^n\|^2,
\eqs
propagates to the layer $j$ to update its representation. Here, $\W^b\in\mathscr{M}_d(\R)$ is a $d\times d$ square matrix representing the weights of feedback connections which we assume to be the same for each layer. Following \cite{RB99,choski21,Wen18,AVR19}, the contribution $\mathcal{F}_{j-1}^n$ is then taken to be the gradient of $\mathcal{R}_{j-1}^n$ with respect to $\E_j^n$, that is
\bqs
\mathcal{F}_{j-1}^n = \nabla \mathcal{R}_{j-1}^n =  - (\W^b) ^\mathbf{t}\E_{j-1}^n + (\W^b )^\mathbf{t} \W^b \E_{j}^n.
\eqs
On the other hand, $\mathcal{B}_{j}^n$ incorporates a top-down prediction to update the representation at layer $j$. This term thus reflects a feedback error correction process, controlled by hyper-parameter $\lambda$. Similar to the feedforward process, $\mathcal{B}_{j}^n$ is defined as the the gradient of $\mathcal{R}_{j}^n$ with respect to $\E_j^n$, that is
\bqs
\mathcal{B}_{j}^n = \nabla \mathcal{R}_{j}^n =  -\W^b \E_{j+1}^n+\E_j^n.
\eqs
As a consequence, our model reads
\bqq
\E_j^{n+1}=\beta \W^f \E_{j-1}^{n+1} +\alpha (\W^b) ^\mathbf{t}\E_{j-1}^n+\left[(1-\beta-\lambda)\mathbf{I}_d-\alpha (\W^b )^\mathbf{t} \W^b \right] \E_j^{n} + \lambda  \W^b \E_{j+1}^{n},
\label{model}
\eqq
for each $j=1,\cdots,J-1$ and $n\geq0$ where we denoted $\mathbf{I}_d$ the identity matrix of $\mathscr{M}_d(\R)$. We supplement the recurrence equation \eqref{model} with the following boundary conditions at layer $j=0$ and layer $j=J$. First, at layer $j=0$, we impose
\bqq
\E_0^n=\S_0^n, \quad n\geq0,
\label{layer0}
\eqq 
where $\S_0^n \in\R^d$ is a given source term, which can be understood as the network's constant visual input. At the final layer $j=J$, there is no possibility of incoming top-down signal, and thus one gets
\bqq
\E_J^{n+1}=\beta \W^f \E_{J-1}^{n+1} +\alpha (\W^b) ^\mathbf{t}\E_{J-1}^n+\left[(1-\beta)\mathbf{I}_d-\alpha (\W^b )^\mathbf{t} \W^b \right] \E_J^{n}, \quad n\geq0.
\label{layerJ}
\eqq
Finally, at the initial step $n=0$, we set
\bqq
\E^0_j=\mathcal{H}_j, \quad j=0,\cdots,J,
\label{IC}
\eqq 
for some given initial sequence $(\mathcal{H}_j)_{0,\cdots,J}$. For instance, in Choksi et al~\cite{choski21}, $\mathcal{H}_j$ was initialized by a first feedforward pass through the system, i.e. $\beta>0$ and $\alpha=\lambda=0$. Throughout we assume the natural following compatibility condition between the source terms and the initial condition, namely
\bqq
\S_0^0=\mathcal{H}_0.
\eqq
Regarding the hyper-parameters of the problem we assume that
\bqq
0\leq \beta <1,  \quad \text{ with } \quad 0 \leq \alpha+\lambda \leq 1.
\label{param}
\eqq

Our key objective is to characterize the behavior of the solutions of the above recurrence equation \eqref{model} as a function of the hyper-parameters and the feedforward and feedback connections matrices $\W^f$ and $\W^b$. We would like to stay as general as possible to encompass as many situations as possible, keeping in mind that we already made strong assumptions by imposing that the weight matrices of feedforward and feedback connections are identical from one layer to another. Motivated by concrete applications, we will mainly consider matrices $\W^f$ and $\W^b$ which act as convolutions on $\R^d$. 

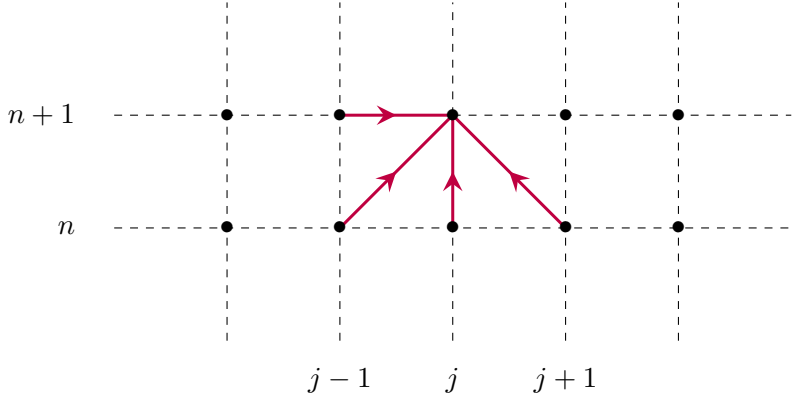
\begin{figure}
\begin{center}
\begin{tikzpicture}[scale=1.5]
\coordinate (A0) at (-1,0);
    \coordinate (A1) at (0,0);
    \coordinate (A2) at (1,0);
    \coordinate (A3) at (2,0);
    \coordinate (A4) at (3,0);
    \coordinate (A5) at (4,0);
 \coordinate (A6) at (5,0);

\coordinate (B0) at (-1,1);
   \coordinate (B1) at (0,1);
    \coordinate (B2) at (1,1);
    \coordinate (B3) at (2,1);
    \coordinate (B4) at (3,1);
    \coordinate (B5) at (4,1);
        \coordinate (B6) at (5,1);
    
        \coordinate (C2) at (1,-1.15);
    \coordinate (C3) at (2,-1.15);
 \coordinate (C4) at (3,-1.15);

    \coordinate (D1) at (0,2);
    \coordinate (D2) at (1,2);
    \coordinate (D3) at (2,2);
    \coordinate (D4) at (3,2);
    \coordinate (D5) at (4,2);

    \coordinate (E1) at (0,-1);
    \coordinate (E2) at (1,-1);
    \coordinate (E3) at (2,-1);
    \coordinate (E4) at (3,-1);
    \coordinate (E5) at (4,-1);

% \coordinate (D1) at (1.5,1);

\node[left] at (B0) {$n+1\quad$};
\node[left] at (A0) {$n\quad$};

    \node[below] at (C2) {$j-1$};
    \node[below] at (C3) {$j$};
    \node[below] at (C4) {$j+1$};
    
  %  \node[above] at (D1) {$\W^f$};

\draw[line width=0.4mm,color=purple,directed] (B2) -- (B3) ;
\draw[line width=0.4mm,color=purple,directed] (A3) -- (B3) ;
\draw[line width=0.4mm,color=purple,directed] (A2) -- (B3) ;
\draw[line width=0.4mm,color=purple,directed] (A4) -- (B3) ;

    \node at (A1) {$\bullet$};
    \node at (A2) {$\bullet$};
    \node at (A3) {$\bullet$};
    \node at (A4) {$\bullet$};
    \node at (A5) {$\bullet$};
    
    \node at (B1) {$\bullet$};
    \node at (B2) {$\bullet$};
    \node at (B3) {$\bullet$};
    \node at (B4) {$\bullet$};
    \node at (B5) {$\bullet$};

    \draw[dashed] (A0) -- (A6);
    \draw[dashed] (B0) -- (B2);
    \draw[dashed] (B3) -- (B6);
     \draw[dashed] (E1) -- (D1);
 \draw[dashed] (E2) -- (D2);
  \draw[dashed] (E4) -- (D4);
   \draw[dashed] (E5) -- (D5);
\draw[dashed] (B3) -- (D3);
\draw[dashed] (E3) -- (A3);

\end{tikzpicture}
  \end{center}
  \caption{Schematic illustration of the network structure of model \eqref{model} where each point represents a given neuronal layer index $j$ (x-axis) at a particular time step $n$ (y-axis), and the red arrows indicate the contributions leading to the update of $\E_j^{n+1}$. }
\end{figure}

\section{The identity case}

It turns out that we will gain much information by first treating the simplified case where $\W^f$ and $\W^b$ are both identity. That is, from now on, and throughout this section we assume that
\bqs
\W^f=\W^b=\mathbf{I}_d.
\eqs
That is, each neuron in a layer is only connected to the corresponding neuron in the immediately preceding and following layer, with unit weight in each direction. Under such a setting, the recurrence equation \eqref{model} reduces to a scalar equation, that is
\bqq
e_j^{n+1}=\beta  e_{j-1}^{n+1} +\alpha e_{j-1}^n+(1-\beta-\lambda- \alpha )e_j^{n} + \lambda  e_{j+1}^{n}, \quad j=1,\cdots,J-1,
\label{model1d}
\eqq
with this time the unknown $e_j^n\in\R$, together with 
\bqq
e_0^n=s_0^n, \quad n\geq0,
\label{layer0d1}
\eqq 
and
\bqq
e_J^{n+1}=\beta e_{J-1}^{n+1} +\alpha e_{J-1}^n+(1-\beta-\alpha )e_J^{n}, \quad n\geq0.
\label{layerJd1}
\eqq
with
\bqq
e^0_j=h_j, \quad j=0,\cdots,J.
\label{ICd1}
\eqq 

\subsection{Wave propagation on an infinite depth network}

It will be first useful to consider the above problem set on an infinite domain and look at 
\bqq
e_j^{n+1}=\beta  e_{j-1}^{n+1} +\alpha e_{j-1}^n+(1-\beta-\lambda- \alpha )e_j^{n} + \lambda  e_{j+1}^{n}, \quad j\in\Z,
\label{modelZ}
\eqq
given some initial sequence
\bqs
e^0_j=h_j, \quad j\in\Z.
\eqs
This situation has no direct equivalent in the brain, where the number of hierarchically connected layers is necessarily finite; but it is a useful mathematical construct. Indeed, such recurrence equations set on the integers $\Z$ are relatively well understood from the mathematical numerical analysis community. The behavior of the solution sequence $(e_j^n)_{j\in\Z}$ can be read out from the so-called amplification factor function defined as
\bqq
\rho(\theta):=\frac{\alpha \left(e^{-\mbi \theta}-1\right) +1-\beta +\lambda\left( e^{\mbi \theta}-1\right)}{1-\beta e^{-\mbi \theta}}, \quad \theta\in[-\pi,\pi],
\label{defrho}
\eqq
and which relates spatial and temporal modes. Indeed, formally, the sequence $(\rho(\theta)^n e^{\mbi j\theta})_{j\in\Z}$ is an explicit solution to \eqref{modelZ} for each $\theta\in[-\pi,\pi]$. Actually one can be much precise and almost explicit in the sense that one can relate the expression of the solutions to \eqref{modelZ} starting from some initial sequence $(h_j)_{j\in\Z}$ to the properties of $\rho$ in a systematic way that we now briefly explain.

Let us first denote by $\mathcal{G}^n=(\mathcal{G}^n_j)_{j\in\Z}$ the sequence which is the fundamental solution of \eqref{modelZ} in the special case where $(\mathcal{H}_j)_{j\in\Z}$ is the Dirac delta sequence $\boldsymbol{\delta}$. The Dirac delta sequence $\boldsymbol{\delta}$ is defined as $\boldsymbol{\delta}_0=1$ and $\boldsymbol{\delta}_j=0$ for all $j\in\Z\backslash\{0\}$. As a consequence, we have $\mathcal{G}^0=\boldsymbol{\delta}$ and for each $n\geq0$
\bqs
\G_j^{n+1}-\beta\G_{j-1}^{n+1}= \alpha \G_{j-1}^{n} +(1-\beta-\lambda-\alpha)\G_j^{n}+\lambda \G_{j+1}^{n}, \quad j\in\Z.
\eqs
 The starting point of the analysis is the following representation formula, obtained via inverse Fourier transform, which reads
\bqq
\G^n_j=\frac{1}{2\pi} \int_{-\pi}^\pi e^{\mbi j \theta} \rho(\theta)^n \md \theta, \quad n\geq1, \quad j\in\Z.
\label{formulaG}
\eqq
Then, given any initial sequence $(h_j)_{j\in\Z}$, the solution $(e_j^n)_{j\in\Z}$ to \eqref{modelZ} can be represented as the convolution product between the initial sequence and the fundamental solution, namely
\bqq
e_j^n=\sum_{\ell\in\Z} \G_{j-\ell}^nh_\ell, \quad j\in\Z, \quad n\geq1.
\label{solution_formula}
\eqq
That is, having characterized the fundamental solution for a simple input pattern ($\boldsymbol{\delta}$), with a unitary impulse provided to a single layer, we can now easily generalize to any arbitrary input pattern, by applying the (translated) fundamental solution to each layer.

Our aim is to understand under which conditions on the hyper-parameters we can ensure that the solutions of \eqref{modelZ} given through \eqref{solution_formula} remain bounded for all $n\geq1$ independently of the choice of the initial sequence $(h_j)_{j\in\Z}$. More precisely, we introduce the following terminology. We say that the recurrence equation is {\em stable} if for each bounded initial sequence $(h_j)_{j\in\Z}\in\ell^{\infty}(\Z)$, the corresponding solution $(e_j^n)_{j\in\Z}$ given by \eqref{solution_formula} satisfies 
\bqs
\underset{j\in\Z}{\sup}|e_j^n|\underset{n\rightarrow\infty}{\longrightarrow}0.
\eqs
On the other hand, we say that the recurrence equation is {\em unstable} if one can find a bounded initial sequence 
$(h_j)_{j\in\Z}\in\ell^{\infty}(\Z)$ such that the corresponding solution $(e_j^n)_{j\in\Z}$ given by \eqref{solution_formula} satisfies 
\bqs
\underset{j\in\Z}{\sup}|e_j^n|\underset{n\rightarrow\infty}{\longrightarrow}+\infty.
\eqs
Finally, we say that the recurrence equation is {\em marginally stable} if there exists a universal constant $C>0$ such that for each bounded initial sequence $(h_j)_{j\in\Z}\in\ell^{\infty}(\Z)$, the corresponding solution $(e_j^n)_{j\in\Z}$ given by \eqref{solution_formula} satisfies 
\bqs
\underset{j\in\Z}{\sup}|e_j^n|\leq C ~ \underset{j\in\Z}{\sup}|h_j|, \quad n\geq1.
\eqs
It turns out that one can determine the stability properties of the recurrence equation by solely looking at the amplification factor function. Indeed, from \cite{RSN}, we know that
\bqs
\underset{n\rightarrow\infty}{\lim} \| \mathcal{G}^n \|_{\ell^1(\Z)}^{1/n} = \underset{\theta\in[-\pi,\pi]}{\max}|\rho(\theta)|,
\eqs
where we have set
\bqs
\| \mathcal{G}^n \|_{\ell^1(\Z)}:=\sum_{j\in\Z}|\G_j^n|.
\eqs
As a consequence, we directly deduce that the recurrence equation is stable when $|\rho(\theta)|<1$ for all $\theta\in[-\pi,\pi]$, whereas it is unstable if there exists $\theta_0\in[-\pi,\pi]$ such that $|\rho(\theta_0)|> 1$. The limiting case occurs precisely when $\underset{\theta\in[-\pi,\pi]}{\max}|\rho(\theta)|=1$ and there is actually a long history of works \cite{Thomee,DSC14,RSC15,CF20,Coeuret22} that have studied the marginal stability of the recurrence equation in that case. All such results rely on a very precise understanding of the amplification factor function and lead to the following statement. 
\begin{thm}[\cite{Thomee,DSC14,RSC15,CF20,Coeuret22}]Suppose that there exist finitely many $\theta_1,\cdots,\theta_K\in[-\pi,\pi]$ such that  for all $\theta\in[-\pi,\pi]\backslash\left\{\theta_1,\cdots,\theta_k\right\}$ one has $|\rho(\theta)|<1$ and $|\rho(\theta_k)|=1$ for each $k=1,\cdots,K$. Furthermore, assume that there exist $c_k\in\R$, $\sigma_k\in\C$ with $\Re(\sigma_k)>0$ and an integer $\mu_k\geq1$  such that
\bqs
\frac{\rho(\theta_k+\theta)}{\rho(\theta_k)}=\exp\left(-\mbi c_k\theta  -\sigma_k \theta^{2\mu_k}+\mathcal{O}(|\theta|^{2\mu_k+1})\right), \text{ as } \theta \rightarrow0.
\eqs
Then the recurrence equation is marginally stable.
 \end{thm}
Based on the above notions of stability/instability, we see that the only interesting situation is when the recurrence equation is marginally stable, and thus when the amplification function is contained in the unit disk with finitely many tangent points to the unit circle with prescribed asymptotic expansions. This is also the only interesting situation from a biological standpoint, as it ensures that the network remains active, yet without runaway activations.

\subsubsection{Study of the amplification factor function}

Since we assumed that $0\leq \beta<1$, the denominator in \eqref{defrho} never vanishes and is well-defined. Next, we crucially remark that we always have 
\bqs
\rho(0)=1.
\eqs
We will now check under which conditions $|\rho(\theta)|\leq 1$ for all $\theta\in[-\pi,\pi]$ to guarantee marginal stability of the recurrence equation. 
%Indeed, for the solutions of \eqref{model1d} to be bounded for each $n\geq1$ it is necessary to ensure that $|\rho(\theta)|\leq 1$ for all $\theta\in[-\pi,\pi]$. This claim can be guessed from the explicit formula \eqref{formulaG} and we refer to \cite{Thomee} for a proof. 

To assess stability, we compute
\begin{align*}
\left|\rho(\theta)\right|^2&=\frac{\left((\lambda+\alpha)(\cos(\theta)-1)+1-\beta\right)^2+(\lambda-\alpha)^2\sin(\theta)^2}{1-2\beta \cos(\theta)+\beta^2}\\
&=\frac{(\lambda+\alpha)^2(\cos(\theta)-1)^2+2(1-\beta)(\lambda+\alpha)(\cos(\theta)-1)+\left(1-\beta\right)^2+(\lambda-\alpha)^2(1-\cos(\theta)^2)}{(1-\beta)^2+2\beta(1-\cos(\theta))}\\
&=\frac{(1-\cos(\theta))\left((\lambda+\alpha)^2(1-\cos(\theta))-2(1-\beta)(\lambda+\alpha)+(\lambda-\alpha)^2(1+\cos(\theta))\right)+\left(1-\beta\right)^2}{(1-\beta)^2+2\beta(1-\cos(\theta))}\\
&=\frac{(1-\cos(\theta))\left(-4\alpha\lambda \cos(\theta)-2(1-\beta)(\lambda+\alpha)+2(\lambda^2+\alpha^2)\right)+\left(1-\beta\right)^2}{(1-\beta)^2+2\beta(1-\cos(\theta))}
\end{align*}
such that $\left|\rho(\theta)\right|^2\leq1$ is equivalent to
\bqs
(1-\cos(\theta))\left(2\beta+4\alpha\lambda \cos(\theta) +2(1-\beta)(\lambda+\alpha)-2(\lambda^2+\alpha^2)\right)\geq 0, \quad \theta\in[-\pi,\pi],
\eqs
and since $1-\cos(\theta)\geq0$ we need to ensure 
\bqs
\beta+2\alpha\lambda \cos(\theta) +(1-\beta)(\lambda+\alpha)-\lambda^2-\alpha^2\geq 0, \quad \theta\in[-\pi,\pi],
\eqs
and evaluating at $\pm\pi$ the above inequality we get
\bqs
\beta +(1-\beta)(\lambda+\alpha)-(\lambda+\alpha)^2\geq 0.
\eqs
But we remark that the above expression can be factored as
\bqs
\left(\beta+\lambda+\alpha\right)\left(1-\lambda-\alpha\right)\geq 0.
\eqs
As a consequence, $\left|\rho(\theta)\right|^2\leq1$ if and only if $\lambda+\alpha \leq 1$. This is precisely the condition that we made in \eqref{param}. We can actually track cases of equality which are those values of $\theta\in[-\pi,\pi]$ for which we have
\bqs
(1-\cos(\theta))\left(2\beta+4\alpha\lambda \cos(\theta) +2(1-\beta)(\lambda+\alpha)-2(\lambda^2+\alpha^2)\right)= 0.
\eqs
We readily recover that at $\theta=0$ we have $|\rho(0)|=1$. So, now assuming that $\theta\neq0$, we need to solve
\bqs
\beta+2\alpha\lambda \cos(\theta) +(1-\beta)(\lambda+\alpha)-\lambda^2-\alpha^2 = 0,
\eqs
which we write as
\bqs
\beta-2\alpha\lambda +(1-\beta)(\lambda+\alpha)-\lambda^2-\alpha^2 +2\alpha\lambda\left( \cos(\theta)+1)\right) = 0,
\eqs
and using the previous factorization we get
\bqs
\left(\beta+\lambda+\alpha\right)\left(1-\lambda-\alpha\right)+2\alpha\lambda\left( \cos(\theta)+1)\right) = 0,
\eqs
and we necessarily get that both $1+\cos(\theta)=0$ and $1-\lambda-\alpha=0$ must be satisfied. As consequence, $|\rho(\pm \pi)|=1$ if and only if $1=\lambda+\alpha$.

\begin{figure}[t!]
\centering
\subfigure[$\beta=0$ and $\alpha+\lambda<1$.]{\includegraphics[width=.24\textwidth]{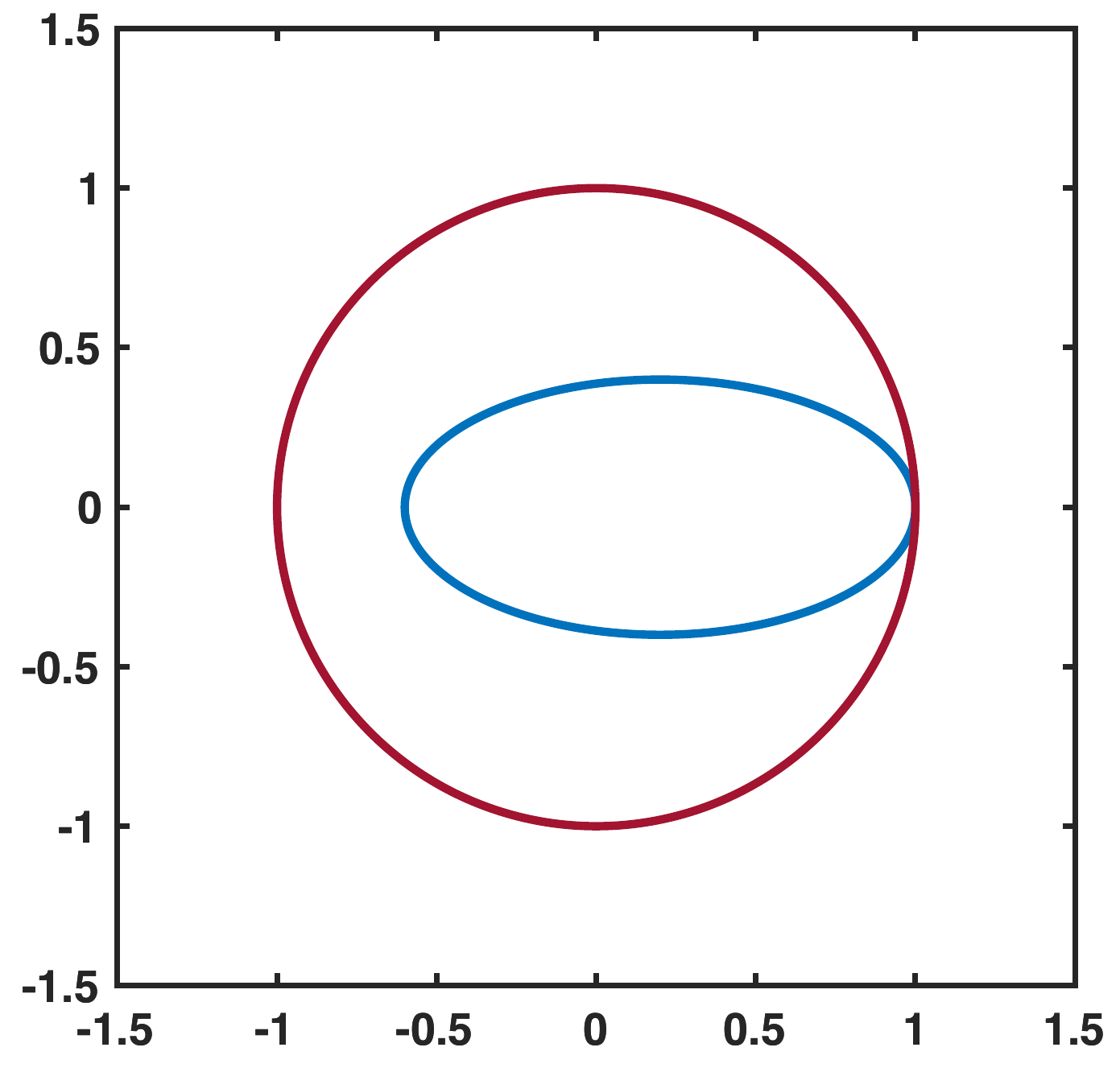}}
\subfigure[$\beta>0$ and $\alpha+\lambda<1$.]{\includegraphics[width=.24\textwidth]{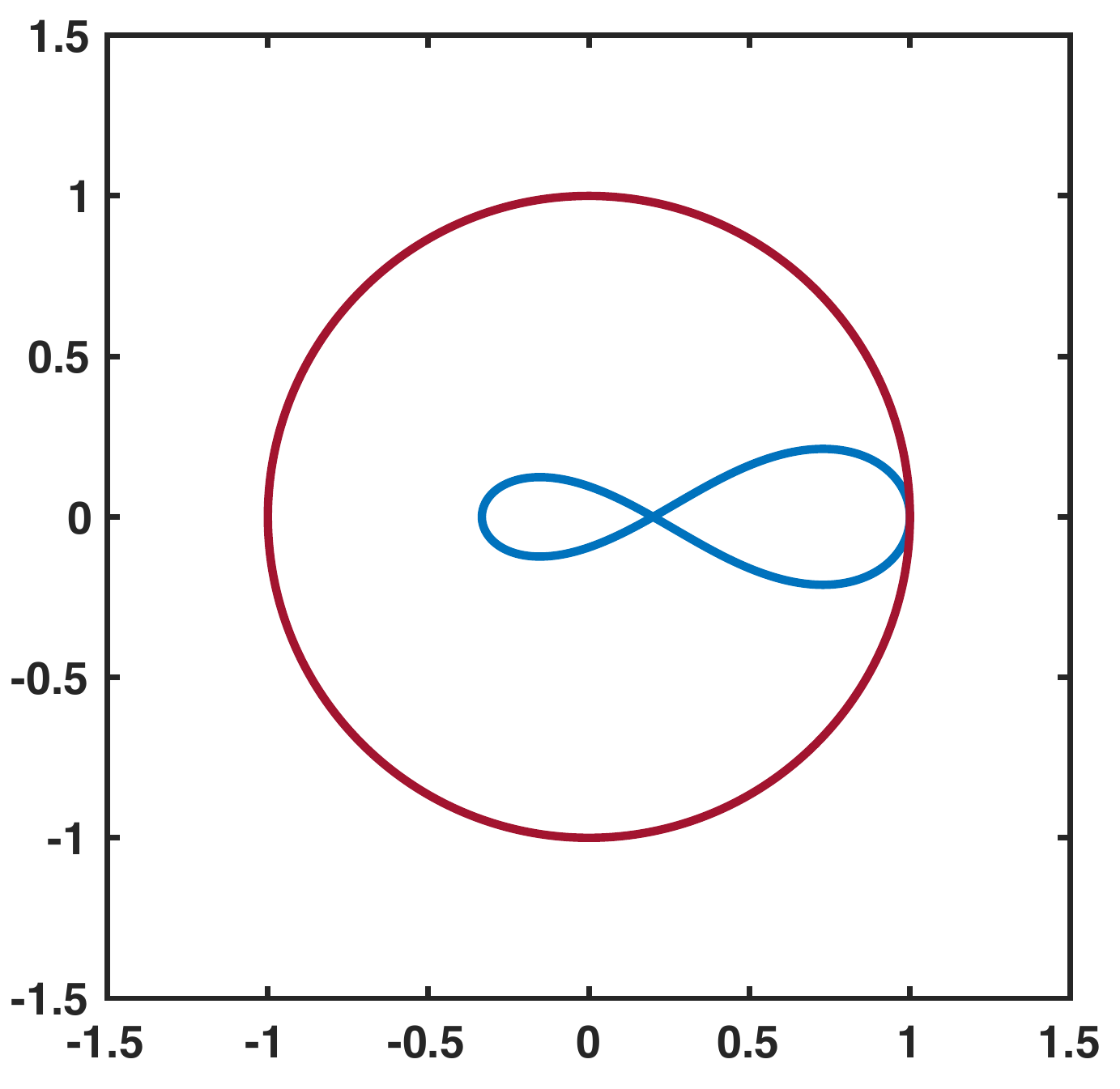}}
\subfigure[$\beta=0$ and $\alpha+\lambda=1$.]{\includegraphics[width=.24\textwidth]{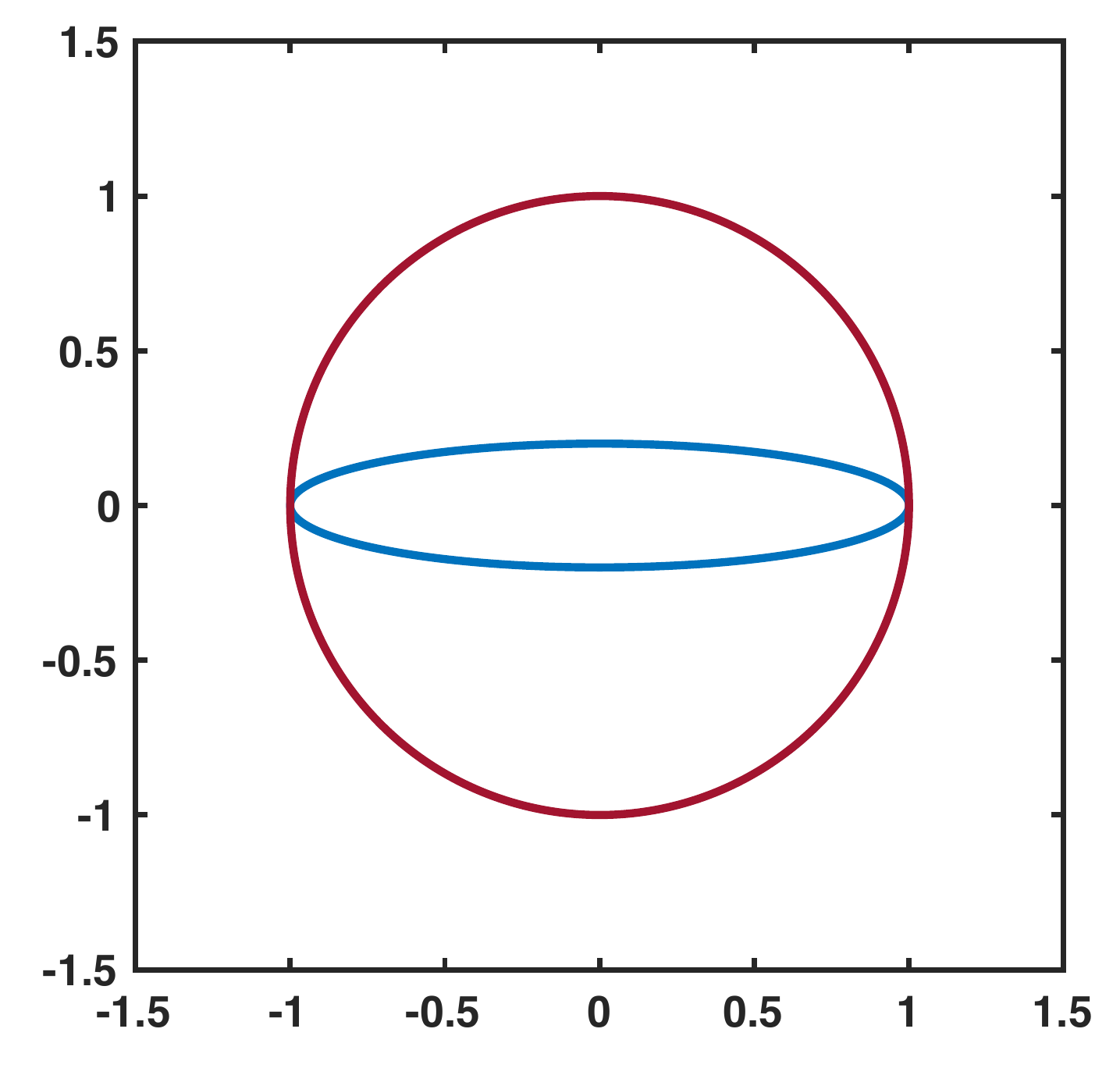}}
\subfigure[$\beta>0$ and $\alpha+\lambda=1$.]{\includegraphics[width=.24\textwidth]{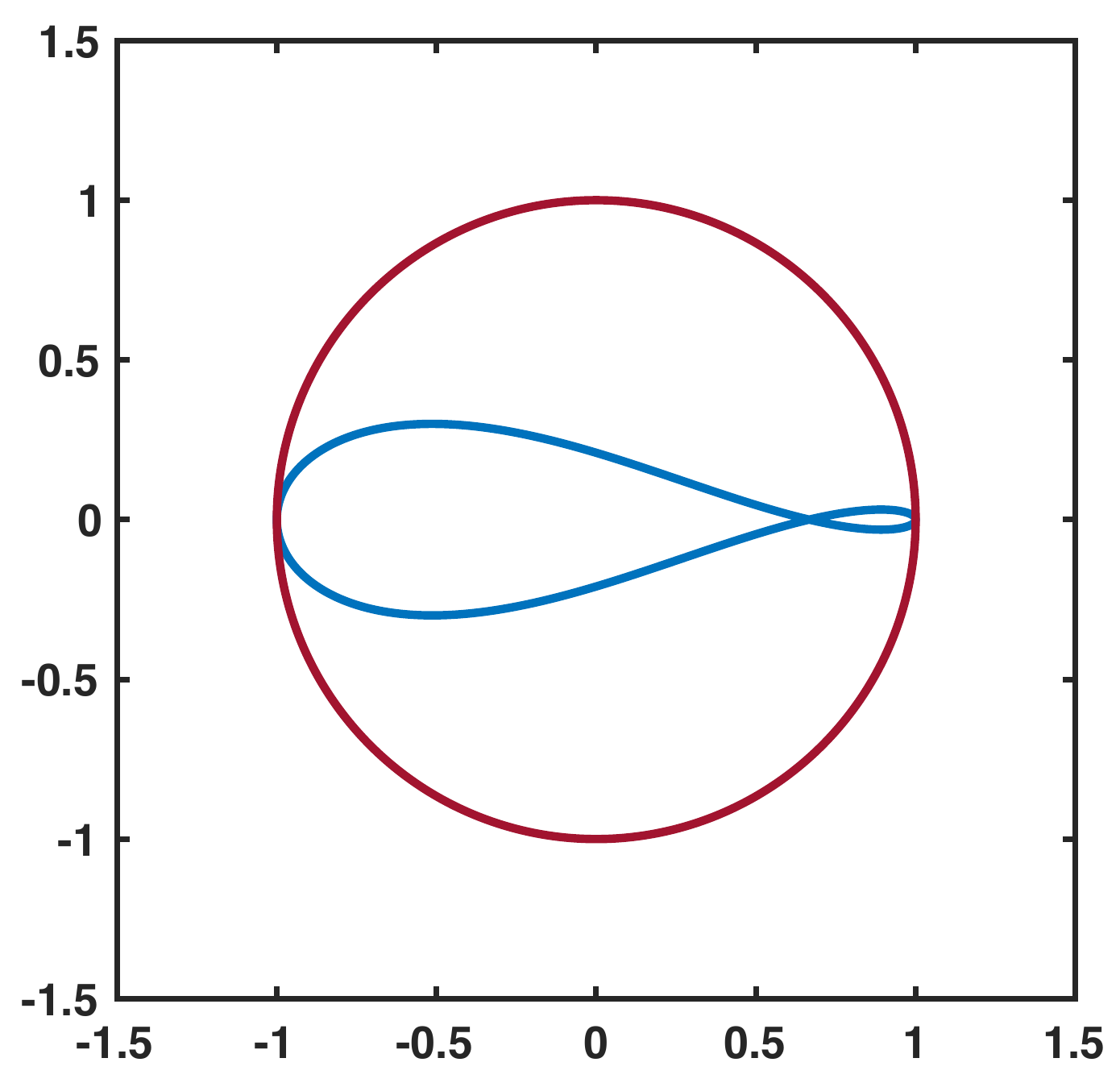}}
\caption{Several representative illustration of the curve $\theta\mapsto \rho(\theta)$ for $\theta\in[-\pi,\pi]$ in the case where $\beta=0$ and $\beta\neq0$. (a)-(b) Amplification factor function $\rho(\theta)$ (blue curve) with a unique tangency point on the unit circle at $z=1$ corresponding $\theta=0$. (c)-(f) When $\alpha+\lambda=1$ the function $\rho(\theta)$ (blue curve) has two tangency points on the unit circle at $z=1$ corresponding $\theta=0$ and $z=-1$ corresponding to $\theta=\pm\pi$.}
  \label{fig:spectrum}
\end{figure}

As a summary we have obtained that:
\begin{itemize}
\item if $0\leq \lambda+\alpha<1$ and $0\leq \beta <1$, then $|\rho(\theta)|<1$ for all $\theta\in[-\pi,\pi]\backslash \{0\}$ with $\rho(0)=1$;
\item if $\lambda+\alpha=1$ and $0\leq \beta <1$, then $|\rho(\theta)|<1$ for all $\theta\in(-\pi,\pi)\backslash \{0\}$ with $\rho(0)=1$ and $\rho(\pm\pi)=-1$.
\end{itemize}
We present in Figure~\ref{fig:spectrum} several representative illustrations of the spectral curves $\rho(\theta)$ for various values of the hyper-parameters recovering the results explained above.

Furthermore, near $\theta \sim 0$, we get that $\rho$ admits the following asymptotic expansion
\bqs
\rho(\theta) = \exp\left(- \mbi \frac{\beta+\alpha-\lambda }{1-\beta} \theta -\frac{\beta(1-\alpha-\lambda)+\alpha+\lambda-(\lambda-\alpha)^2}{2(1-\beta)^2}\theta^2+\mathcal{O}(|\theta|^3)\right), \text{ as } \theta\rightarrow 0,
\eqs
provided that
\bqs
\beta(1-\alpha-\lambda)+\alpha+\lambda-(\lambda-\alpha)^2 \neq0.
\eqs
In fact, since $-(\lambda-\alpha)^2\geq -(\lambda+\alpha)^2$ as both $\alpha$ and $\lambda$ are positive, we remark that
\bqs
\beta(1-\alpha-\lambda)+\alpha+\lambda-(\lambda-\alpha)^2 \geq \beta(1-\alpha-\lambda)+\alpha+\lambda-(\lambda+\alpha)^2=\left(\beta+\lambda+\alpha\right)\left(1-\lambda-\alpha\right)\geq0.
\eqs
Finally, we remark that when $\alpha+\lambda=1$ we have
\bqs
\rho(\theta+\pi)=-\exp\left(-\mbi \frac{-\beta+\alpha-\lambda}{1+\beta}\theta -\frac{1-(\alpha-\lambda)^2}{2(1+\beta)^2}\theta^2+\mathcal{O}(|\theta|^3) \right), \text{ as } \theta\rightarrow 0.
\eqs

From now on, we denote
\begin{align*}
c_0:= \frac{\beta+\alpha-\lambda }{1-\beta},& \quad \sigma_0:=\frac{\beta(1-\alpha-\lambda)+\alpha+\lambda-(\lambda-\alpha)^2}{2(1-\beta)^2},\\
c_\pi:=\frac{-\beta+\alpha-\lambda}{1+\beta},& \quad \sigma_\pi:=\frac{1-(\alpha-\lambda)^2}{2(1+\beta)^2},
\end{align*}
and we always assume that
\bqs
\sigma_0>0, \text{ and } \sigma_\pi>0,
\eqs
which is equivalent to assume that $0<\alpha<1$ and $0<\lambda<1$.

Here, $(c_0,\sigma_0)$ and $(c_\pi,\sigma_\pi)$ are derived, respectively, from the asymptotic expansions of the amplification factor function $\rho(\theta)$ near $\theta=0$ and $\theta=\pi$, as defined above. On the one hand $c_0$ reflects the propagation speed of the solution associated with $\rho(0)$, while $\sigma_0$ can be understood as its spatio-temporal spread (and similarly for the solution potentially associated with $\rho(\pi)$). In the following, we explore the fundamental solutions of this system for various values of its hyper-parameters.
%Then we denote by $\mathcal{G}^n=(\mathcal{G}^n_j)_{j\in\Z}$ the sequence which is the fundamental solution starting from the Dirac delta sequence $\boldsymbol{\delta}$. The Dirac delta sequence $\boldsymbol{\delta}$ is defined as $\boldsymbol{\delta}_0=1$ and $\boldsymbol{\delta}_j=0$ for all $j\in\Z\backslash\{0\}$. As a consequence, we have $\mathcal{G}^0=\boldsymbol{\delta}$ and for each $n\geq0$
%\bqs
%\G_j^{n+1}-\beta \G_{j-1}^{n+1}= \alpha \G_{j-1}^{n} +(1-\beta-\lambda-\alpha)\G_j^{n}+\lambda \G_{j+1}^{n}, \quad j\in\Z.
%\eqs
%Combining the results of \cite{Coeuret22,CF20}, for each $n\geq1$ and each $j\in\Z$, the element $\G_j^{n}$ can be decomposed as follows. 

\subsubsection{Turning off the instantaneous feedforward connections: case $\beta=0$}

We first investigate the case where there is no instantaneous feedforward connections in the network, that is we set $\beta=0$. This case, although less generic, is compatible with the prominent Rao-Ballard formulation of predictive coding~\cite{RB99}, in which feedforward connections---after contributing to setting the initial network activity---only convey prediction errors, as captured by the hyper-parameter $\alpha$. In that case, the model is fully explicit: the update at time step $n+1$ only depends on the internal states at the previous step $n$ since we simply have
\bqs
e_j^{n+1}= \alpha e_{j-1}^{n} +(1-\lambda-\alpha)e_j^{n}+\lambda e_{j+1}^{n}, \quad j\in\Z.
\eqs
As we assumed that $\alpha+\lambda\leq1$, the right-hand side of the recurrence equation is a positive linear combination of elements of the sequence $(e_j^{n})$ such that we have positivity principle of the solution, namely
\bqs
\forall j\in\Z, \quad  0 \leq h_j \quad  \Longrightarrow \quad \forall  j\in\Z, \quad n\geq1, \quad 0\leq e_j^{n}.
\eqs
Furthermore, since the recurrence equation is explicit, we have finite speed propagation, in the following sense. Recall that when $\beta=0$, the fundamental solution $\G^n$ is solution to
\bqs
\G_j^{n+1}= \alpha \G_{j-1}^{n} +(1-\lambda-\alpha)\G_j^{n}+\lambda \G_{j+1}^{n}, \quad n\geq1, \quad \quad j\in\Z,
\eqs
starting from $\mathcal{G}^0=\boldsymbol{\delta}$. Finite speed of propagation then refers to the property that 
\bqs
\G_j^{n}=0, \quad |j|>n.
\eqs
This in turn implies that necessarily $c_0\in(-1,1)$ which is readily seen from the explicit formula $c_0=\alpha-\lambda$ in that case. Actually, it is possible to be more precise and to give a general expression for the fundamental solution. Roughly speaking, each $\G_j^n$ ressembles a discrete Gaussian distribution centered at $j=c_0n$ and we refer to the recent theoretical results of \cite{DSC14,RSC15,Coeuret22,CF20} for a rigorous justification.  

Essentially, the results can be divided into two cases depending on whether or not $\alpha+\lambda=1$. As can be seen above, the special case $\alpha+\lambda=1$ results in a cancellation of the ``memory'' term, such that a neuronal layer $j$'s activity does not depend on its own activity at the previous time step, but only on the activity of its immediate neighbors $j-1$ and $j+1$. More precisely, we have the following:
\begin{figure}[t!]
\centering
\subfigure[$\beta=0$, $\lambda<\alpha$ and $\alpha+\lambda<1$.]{\includegraphics[width=.35\textwidth]{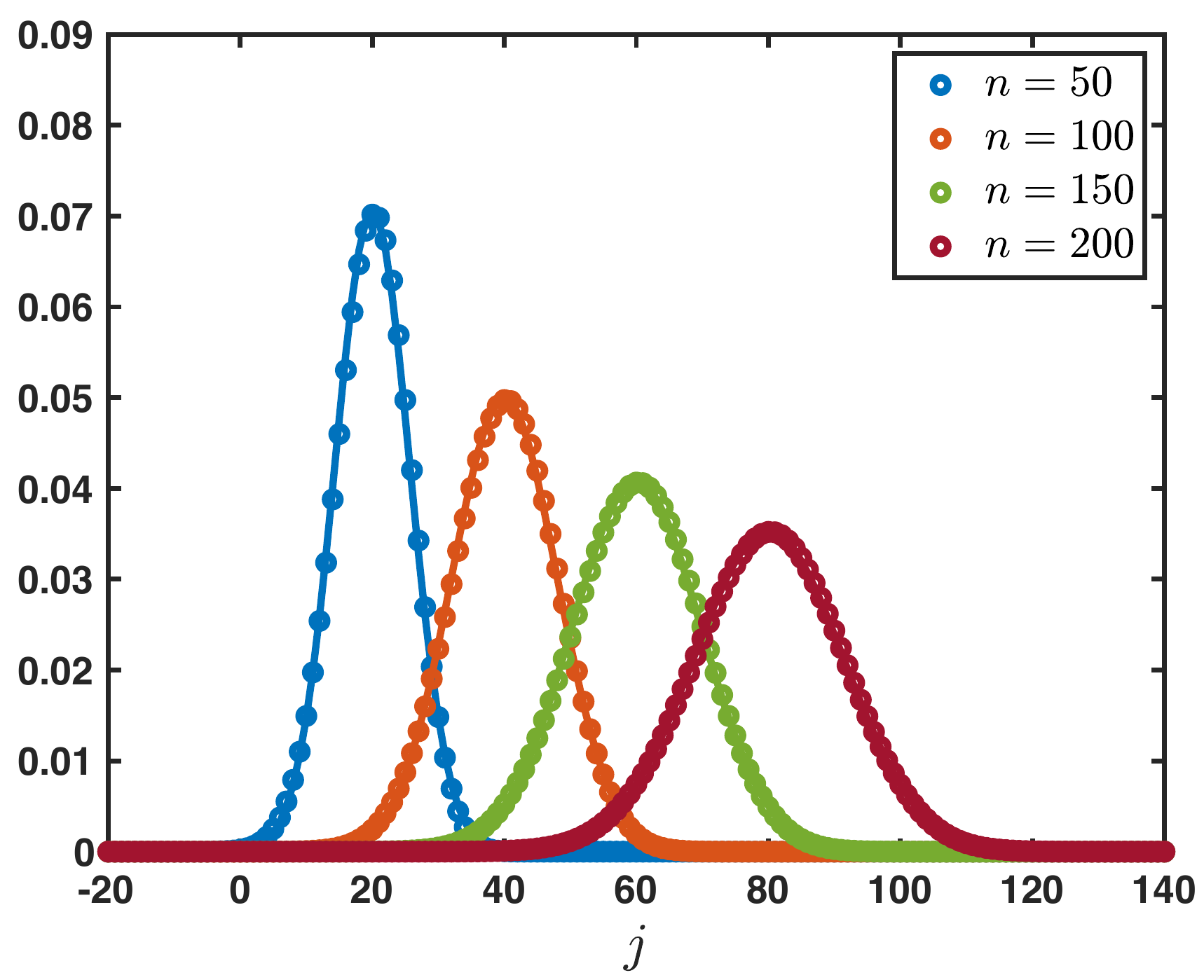}}\hspace{1cm}
\subfigure[$\beta=0$, $\alpha<\lambda$ and $\alpha+\lambda=1$.]{\includegraphics[width=.35\textwidth]{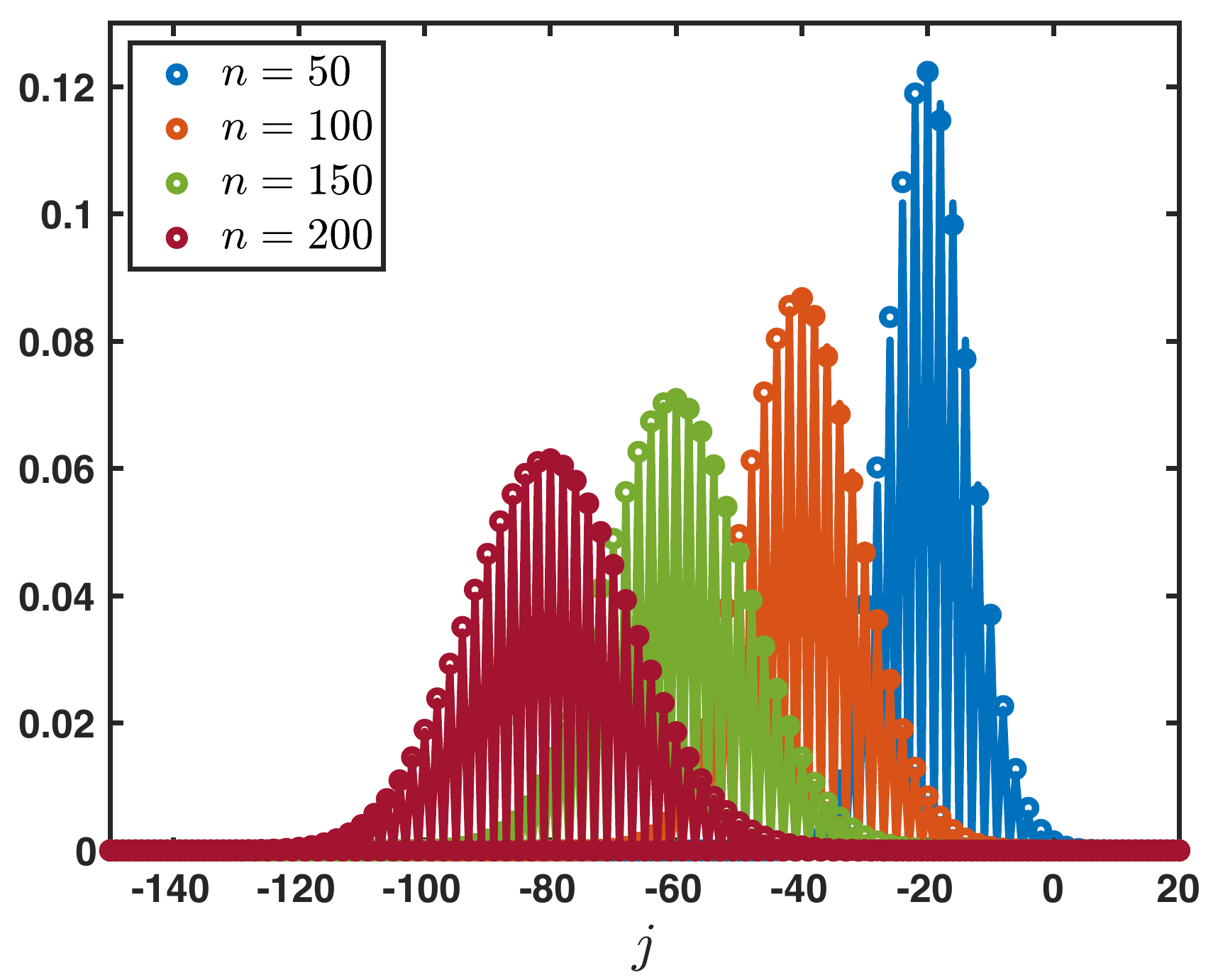}}
\caption{Illustration of the evolution of the fundamental solution $\G^n$ starting from the Dirac delta sequence at $j=0$ in the case $\beta=0$ at several time iterations. The circles represent the numerically computed solution while the plain lines represent the Gaussian approximation. (a) When $\lambda<\alpha$ there is a rightward propagation along a Gaussian profile whose leading profile is given by $\frac{1}{\sqrt{4\pi \sigma_0 n}}\exp\left( -\frac{|j-c_0n|^2}{4\sigma_0 n}\right)$. (b) When $\alpha<\lambda$ there is a leftward propagation along a Gaussian profile whose leading profile is given by $\frac{1+(-1)^{n+j}}{\sqrt{4\pi \sigma_0 n}}\exp\left( -\frac{|j-c_0n|^2}{4\sigma_0 n}\right)$ which vanishes whenever $n+j$ is odd.}
  \label{fig:profiles}
\end{figure}

\begin{figure}[t!]
\centering
\subfigure[$\beta=0$, $\lambda<\alpha$ and $\alpha+\lambda<1$.]{\includegraphics[width=.32\textwidth]{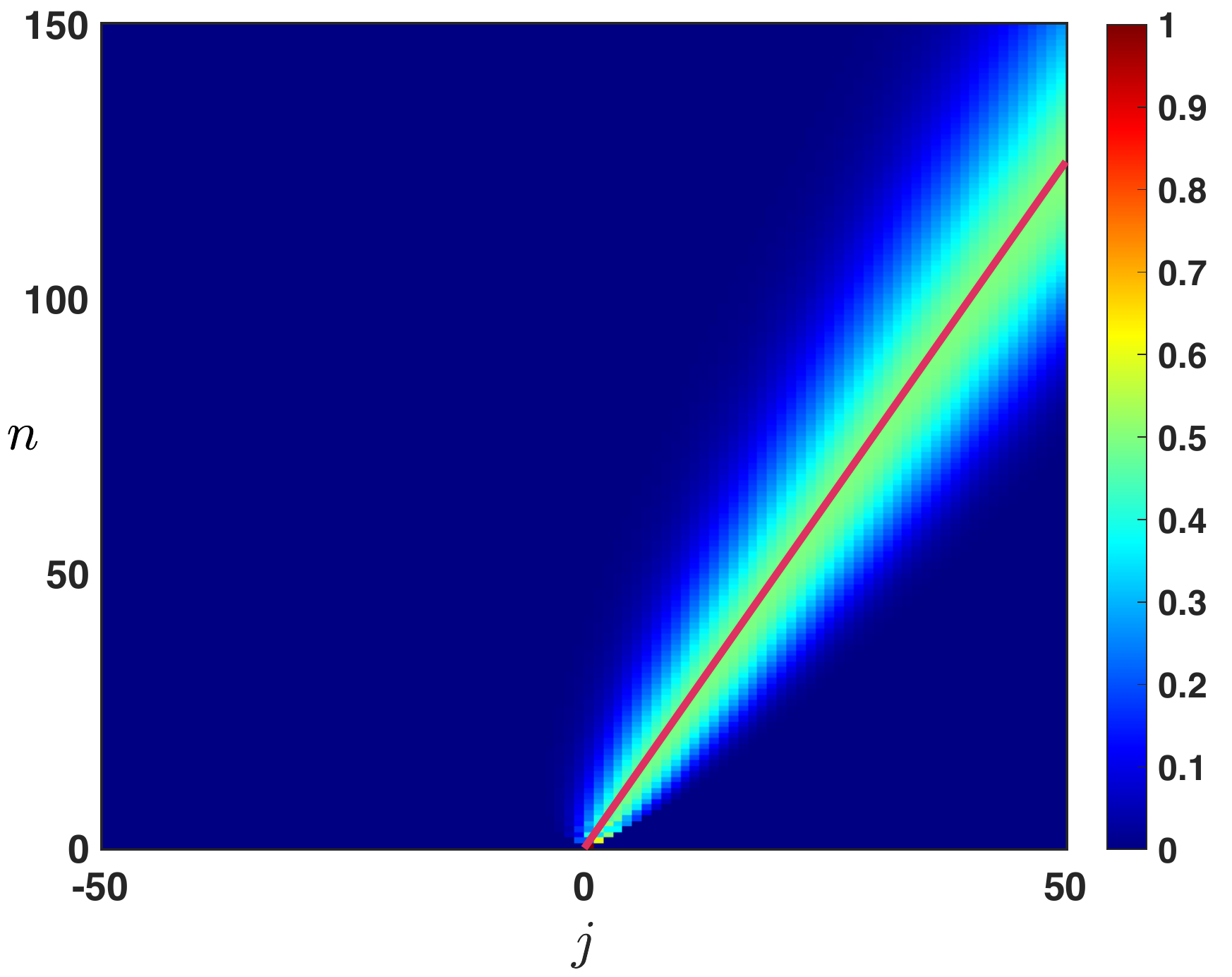}}
\subfigure[$\beta=0$, $\lambda=\alpha$ and $\alpha+\lambda<1$.]{\includegraphics[width=.32\textwidth]{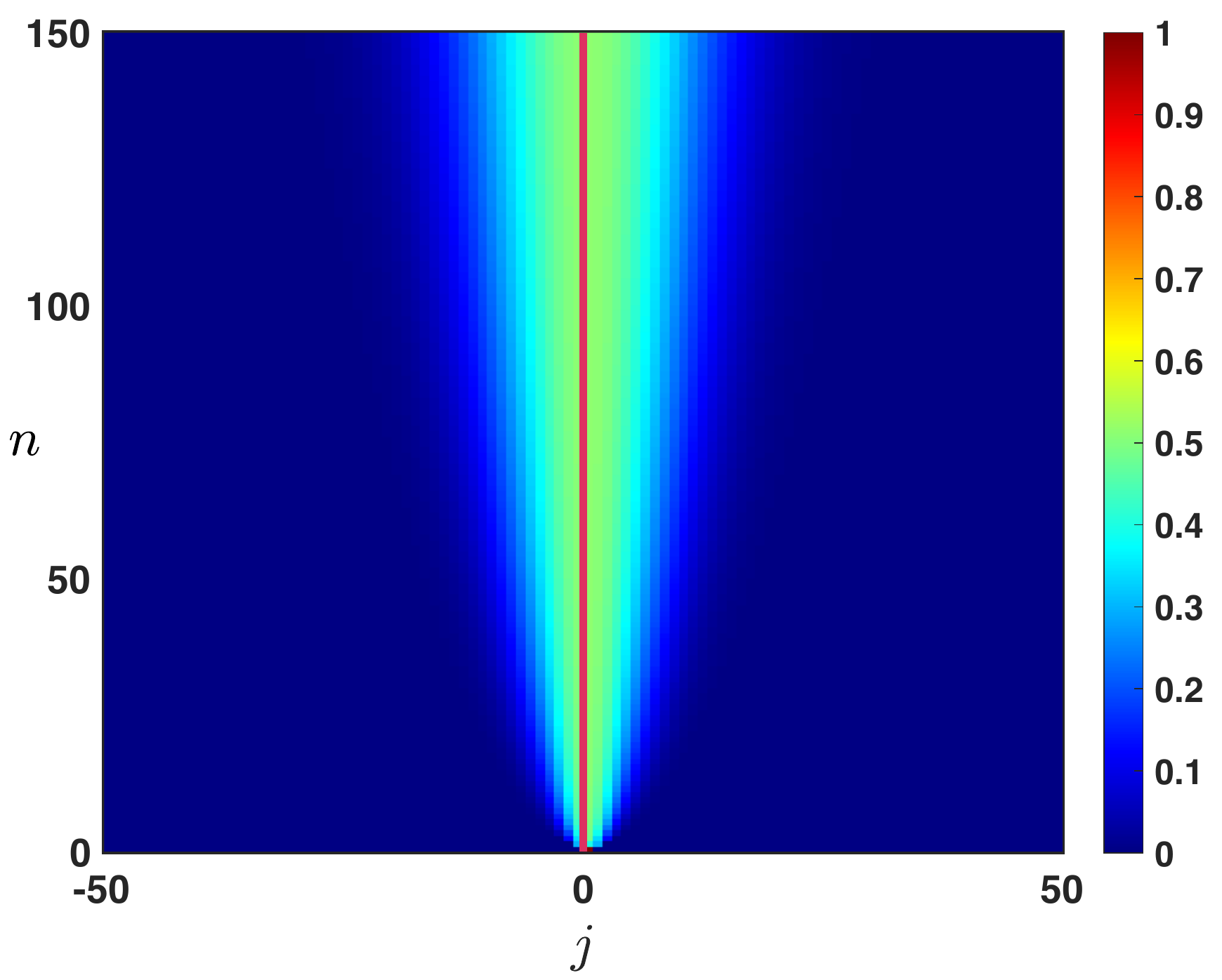}}
\subfigure[$\beta=0$, $\alpha<\lambda$ and $\alpha+\lambda<1$.]{\includegraphics[width=.32\textwidth]{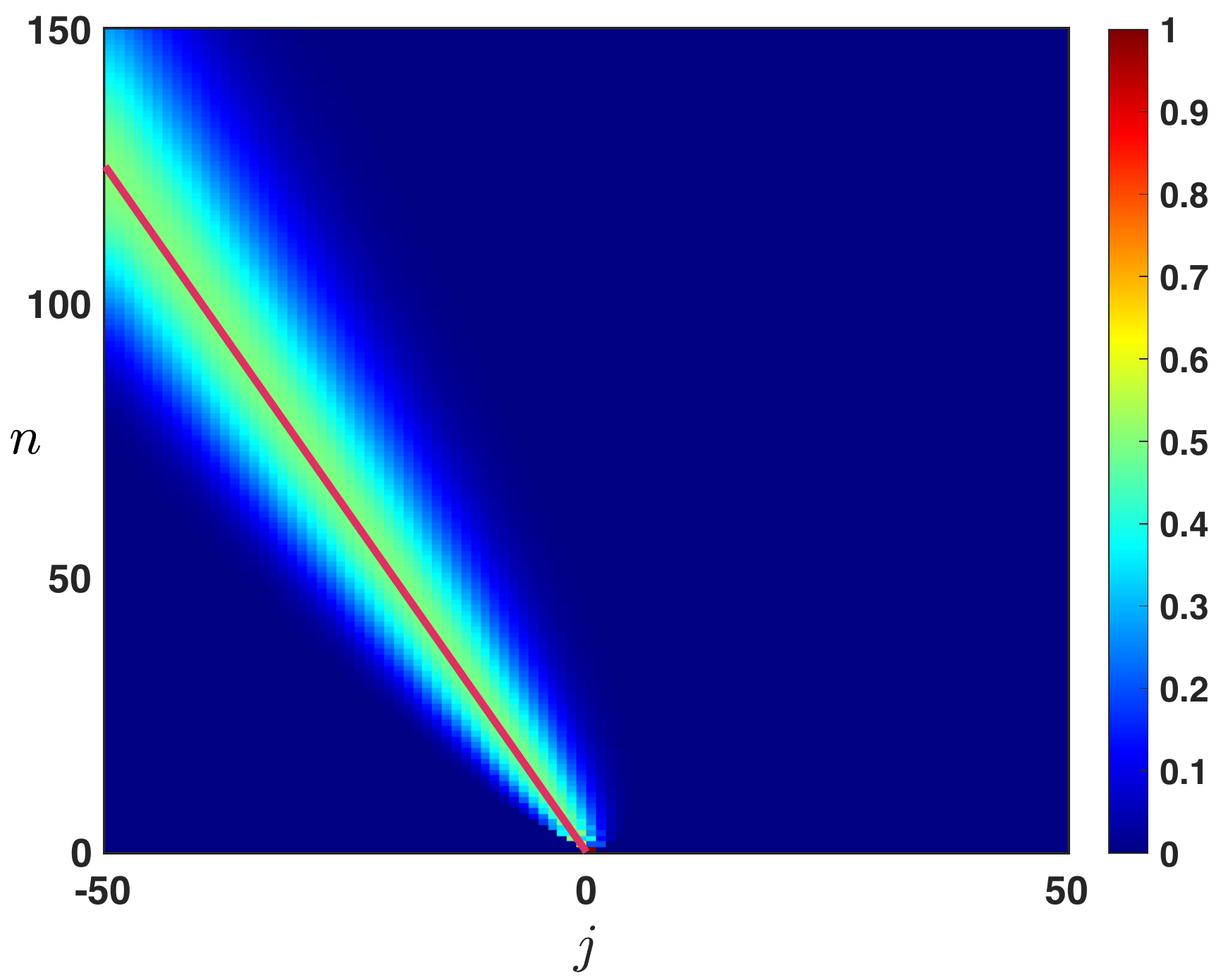}}
\subfigure[$\beta=0$, $\lambda<\alpha$ and $\alpha+\lambda=1$.]{\includegraphics[width=.32\textwidth]{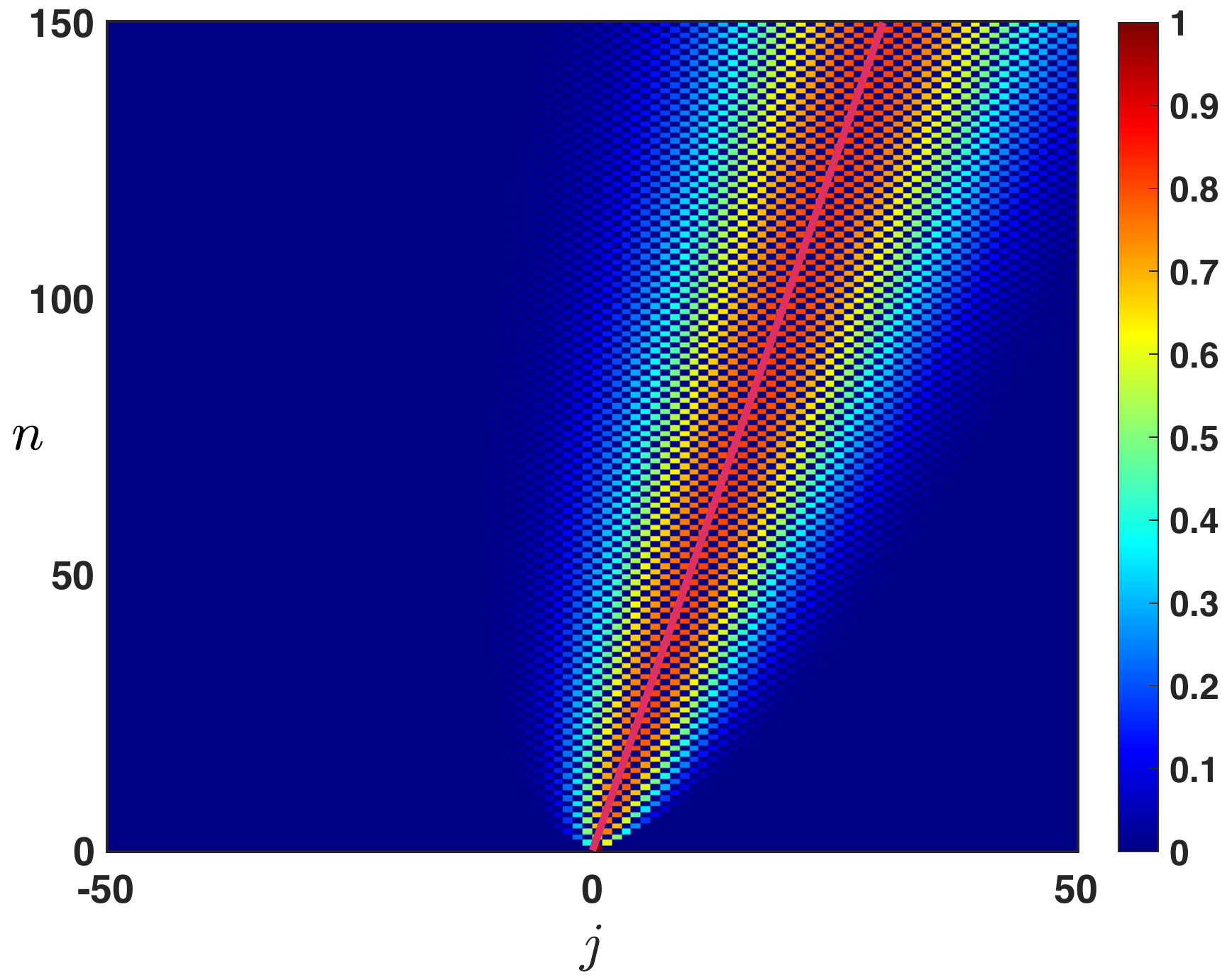}}
\subfigure[$\beta=0$, $\lambda=\alpha$ and $\alpha+\lambda=1$.]{\includegraphics[width=.32\textwidth]{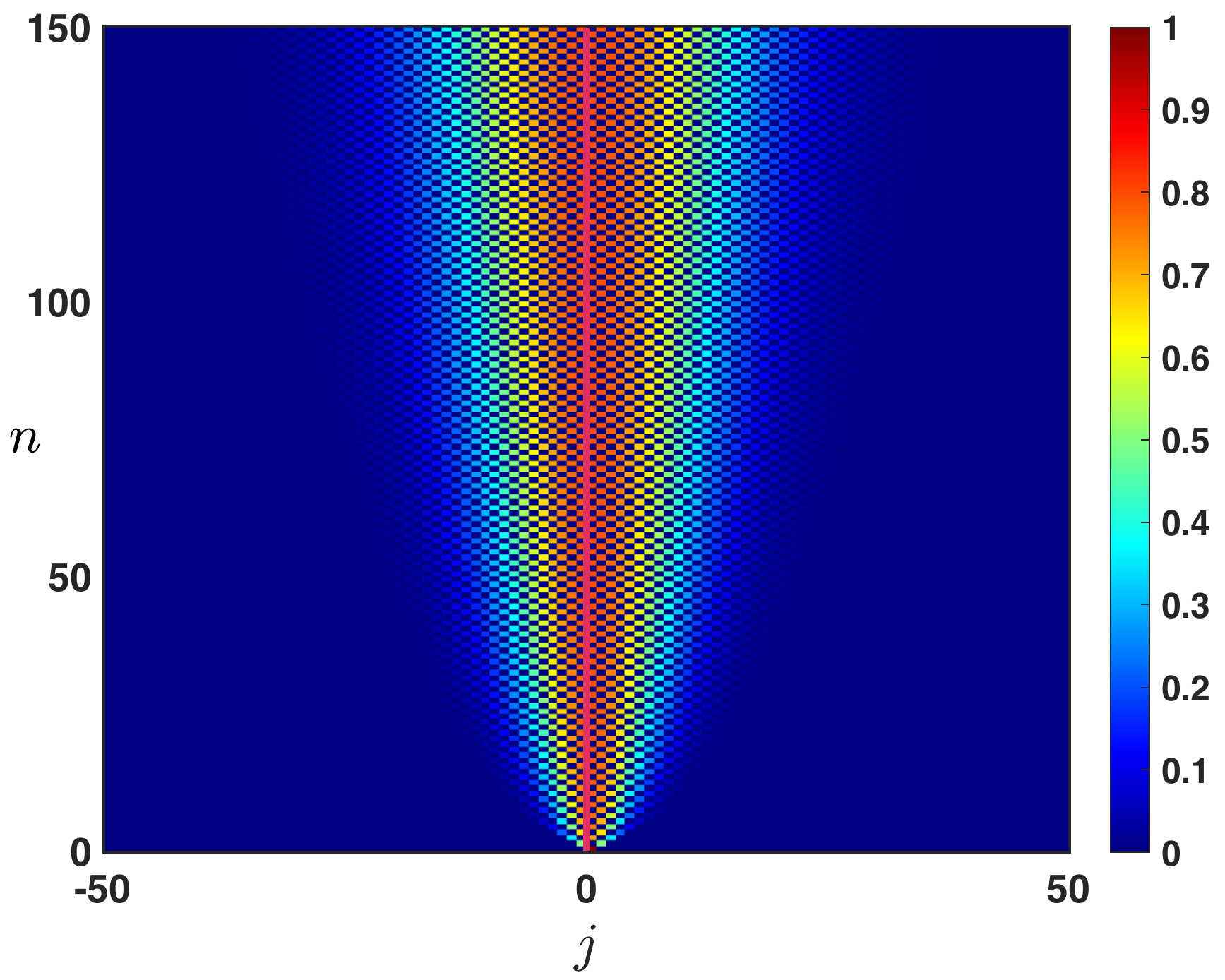}}
\subfigure[$\beta=0$, $\alpha<\lambda$ and $\alpha+\lambda=1$.]{\includegraphics[width=.32\textwidth]{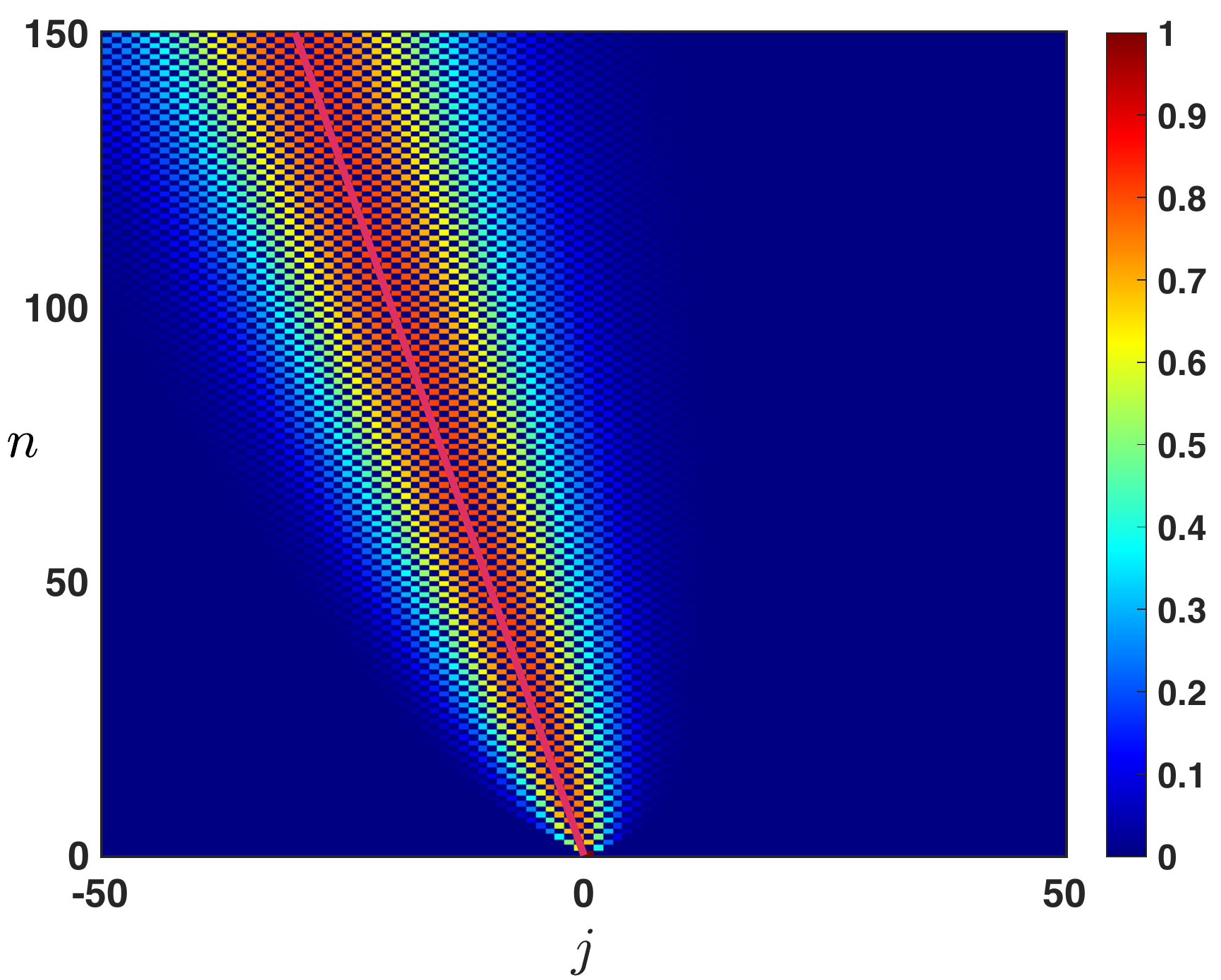}}
\caption{Illustration of the evolution of the rescaled solution sequence $(\sqrt{n}\,\G_j^n)_{j\in\Z}$ starting from the Dirac delta sequence at $j=0$ in the case $\beta=0$. First row: $\alpha+\lambda<1$ and second row: $\alpha+\lambda=1$. When $\lambda\lessgtr\alpha$, we observe a rightward/leftward propagation while when $\alpha=\lambda$ no propagation occurs. In all panels, the pink curve is given by $j=n c_0$, clearly illustrating the fact that $c_0$ measures the propagation speed of the solution. Note that in the case $\beta=0$ and $\alpha+\lambda=1$, we have $c_0=c_\pi$ which results in the tiled patterns observed in panels (d)-(e)-(f).}
  \label{fig:casebeta0}
\end{figure}

\begin{itemize}
\item Case: $0\leq \lambda+\alpha<1$. The fundamental solution can be decomposed as
\bqs
\G_j^{n}= \frac{1}{\sqrt{4\pi \sigma_0 n}}\exp\left( -\frac{|j-c_0n|^2}{4\sigma_0 n}\right)+\mathcal{N}_j^{n}, \quad |j|\leq n,
\eqs
where the remainder term satisfies an estimate
\bqs
\left|\mathcal{N}_j^{n}\right| \leq \frac{C}{n}\exp\left( -\kappa\frac{|j-c_0n|^2}{n}\right), \quad |j|\leq n,
\eqs
for some universal constants $C,\kappa>0$ which only depend on the hyper-parameters and not $n$ and $j$. In Figure~\ref{fig:profiles}(a), we represented the fundamental solution $\G_j^n$ at different time iterations (circles) in the case $\lambda<\alpha$ where there is rightward propagation with $c_0>0$ and compared it with the leading order fixed Gaussian profile centered at $j=c_0n$ (plain line). On the other hand, in Figure~\ref{fig:casebeta0}, panels (a)-(b)-(c), we illustrate the above results by presenting a space-time color plot of the fundamental solution rescaled by a factor $\sqrt{n}$. We observe rightward (respectively leftward) propagation with $c_0>0$ (respectively $c_0<0$) when $\lambda<\alpha$ (respectively $\alpha<\lambda$), while when $\alpha=\lambda$ we have $c_0=0$ and no propagation occurs.

\item Case: $\lambda+\alpha=1$. In this case, we first note that we have $c_0=c_\pi$ together with $\sigma_0=\sigma_\pi$ and
\bqs
\G_j^{n}= \frac{1+(-1)^{n+j}}{\sqrt{4\pi \sigma_0 n}}\exp\left( -\frac{|j-c_0n|^2}{4\sigma_0 n}\right)+\mathcal{N}_j^{n}, \quad |j|\leq n,
\eqs
where the remainder term satisfies an estimate
\bqs
\left|\mathcal{N}_j^{n}\right| \leq \frac{C}{n}\exp\left( -\kappa\frac{|j-c_0n|^2}{n}\right), \quad |j|\leq n,
\eqs
for some universal constants $C,\kappa>0$. In Figure~\ref{fig:profiles}(b), we represented the fundamental solution $\G_j^n$ at different time iterations (circles) in the case $\alpha<\lambda$ where there is leftward propagation with $c_0<0$ and compared it with the leading order fixed Gaussian profile centered at $j=c_0n$ (plain line). Similarly as in the previous case, in Figure~\ref{fig:casebeta0}, panels (d)-(e)-(f), we illustrate the above results by presenting a space-time color plot of the fundamental solution rescaled by a factor $\sqrt{n}$. The direction of propagation still depends on the sign of $c_0$ and whether or not $\lambda\lessgtr\alpha$. Unlike the case $\alpha+\lambda<1$, we observe a tiled pattern where $\G_j^{n}=0$ for even or odd integers alternatively for each time step.
\end{itemize}

As a partial intermediate summary, we note that the sign of $c_0$ (directly related to the sign of $\alpha-\lambda$) always indicates in which direction the associated Gaussian profile propagates. Namely if $\alpha>\lambda$ and $c_{0}>0$ (resp. $\alpha<\lambda$ and $c_{0}<0$) there is rightward (resp. leftward) propagation. Intuitively, this behavior reflects the functional role of each hyper-parameter, with $\alpha$ and $\lambda$ controlling feed-forward and feed-back prediction error correction, respectively. When $\alpha=\lambda$, the two terms are equally strong, and there is no dominant direction of propagation. In addition, when $\lambda+\alpha=1$, the Gaussian profile is \emph{oscillating} because of the presence of $(-1)^{n+j}$. As will be seen later when considering continuous versions of our model, this oscillatory pattern might not be truly related to neural oscillations observed in the brain, but could instead arise here as a consequence of discrete updating. 

Finally, we note that the fundamental solution sequence $(\mathcal{G}^n_j)_{j\in\Z}$ is uniformly integrable for all values of the parameters, that is there exists some universal constant $C>0$, depending only on the hyper-parameters such that
\bqs
\|\G^n\|_{\ell^1(\Z)}:=\sum_{j\in\Z}|\G_j^n|\leq C, \quad n\geq 1.
\eqs
As a consequence, since given any bounded initial sequence $(h_j)_{j\in\Z}\in \ell^\infty(\Z)$, the solution $(e_j^n)_{j\in\Z}$ to \eqref{modelZ} can be represented as the convolution product between the initial sequence and the fundamental solution, namely
\bqs
e_j^n=\sum_{\ell\in\Z} \G_{j-\ell}^nh_\ell, \quad j\in\Z, \quad n\geq1,
\eqs
we readily deduce that the solution $(e_j^n)_{j\in\Z}$ is uniformly bounded with respect to $n$, that is there exists some universal constant denoted $C>0$, such that
\bqs
\underset{j\in\Z}{\sup} \left|e_j^n\right| \leq C~ \underset{j\in\Z}{\sup} \left|h_j\right|, \quad n\geq 1.
\eqs

This is exactly our definition of marginal stability.

\subsubsection{Turning on the instantaneous feedforward connections: case $\beta>0$}

We now turn to the general case where $\beta>0$. That is, the feed-forward connections continue to convey sensory inputs at each time step following the network initializing, and $\beta$ controls the strength of these signals. In that case, the recurrence equation is no longer explicit but implicit and the positivity property together with the finite speed propagation no longer hold true in general. Indeed, upon introducing the shift operator 
\bqs
\mathbf{S}:(u_j)_{j\in\Z}\mapsto (u_{j+1})_{j\in\Z},
\eqs
we remark that equation \eqref{modelZ} can be written
\bqs
\left(\mathrm{Id}-\beta \mathbf{S}^{-1}\right)e^{n+1}=\alpha \mathbf{S}^{-1}e^n+(1-\beta-\lambda-\alpha)e^n+\lambda \mathbf{S}e^n, \quad n\geq0,
\eqs
with $e^n=(e_j^n)_{j\in\Z}$. Since $0<\beta<1$ and $\||\mathbf{S}^{-1}\||_{\ell^q(\Z)\to\ell^q(\Z)}=1$ for any $q\in[1,+\infty]$, the operator $\mathrm{Id}-\beta \mathbf{S}^{-1}$ is invertible on $\ell^q(\Z)$ for any $q\in[1,+\infty]$  with inverse
\bqs
\left(\mathrm{Id}-\beta \mathbf{S}^{-1}\right)^{-1}=\sum_{\ell=0}^\infty \beta^\ell \mathbf{S}^{-\ell}.
\eqs
As a consequence, the recurrence equation can be recast as a convolution operator across the network layers with infinite support, namely
\bqs
e^{n+1}_j=\alpha \sum_{\ell=0}^\infty \beta^\ell e^n_{j-\ell-1}+(1-\beta-\lambda-\alpha) \sum_{\ell=0}^\infty \beta^\ell e^n_{j-\ell}+\lambda\sum_{\ell=0}^\infty \beta^\ell e^n_{j-\ell+1}, \quad j\in\Z, \quad n\geq0.
\eqs
From the above expression, we readily deduce that the positivity of the solution is preserved whenever $0<\beta<1-\lambda-\alpha$. Furthermore, for the fundamental solution starting from the Dirac delta solution which solves
\bqs
\G_j^{n+1}-\beta \G_{j-1}^{n+1}= \alpha \G_{j-1}^{n} +(1-\beta-\lambda-\alpha)\G_j^{n}+\lambda \G_{j+1}^{n}, \quad j\in\Z, \quad n\geq0,
\eqs
we only have that
\bqs
\G_j^n=0, \quad j<-n,
\eqs
which implies that $-1<c_0,c_\pi<+\infty$. Indeed, from the formula of $c_0$ we get that
\bqs
c_0\sim \frac{1-\alpha-\beta}{1-\beta}\longrightarrow+\infty, \quad \text{ as } \beta \rightarrow 1^{-}.
\eqs
Once again, as in the case with $\beta=0$, we can precise the behavior of the fundamental solution by using the combined results of \cite{Coeuret22,CF20}.

\begin{comment}
\begin{figure}[t!]
\centering
\subfigure[$0<\beta<\lambda-\alpha$.]{\includegraphics[width=.32\textwidth]{betag0less1pos.pdf}}
\subfigure[$\beta=\lambda-\alpha$.]{\includegraphics[width=.32\textwidth]{betag0less1zero.pdf}}
\subfigure[$\lambda-\alpha<\beta$.]{\includegraphics[width=.32\textwidth]{betag0less1neg.pdf}}
\caption{Effects of turning on $\beta>0$ when $\alpha<\lambda$ and $\alpha+\lambda<1$. When $0<\beta<\lambda-\alpha$ we still have $c_0<0$ and observe backward propagation in panel (a) while when $\lambda-\alpha<\beta$ we have $c_0>0$ and have forward propagation as seen in panel (c). At the transition $\beta=\lambda-\alpha$, the wave speed vanishes $c_0=0$ and there is no propagation as illustrated in panel (b).  Intuitively, $\alpha$ and $\beta$, both propagating signals in the forward (rightward) direction, compete with $\lambda$ carrying the feedback (leftward) prediction signals; this competition determines the main direction of propagation of neural activity in the system.}
  \label{fig:caseneq0}
\end{figure}
\end{comment}

\begin{figure}[t!]
\centering
\subfigure[$\beta+\alpha<\lambda$.]{\includegraphics[width=.32\textwidth]{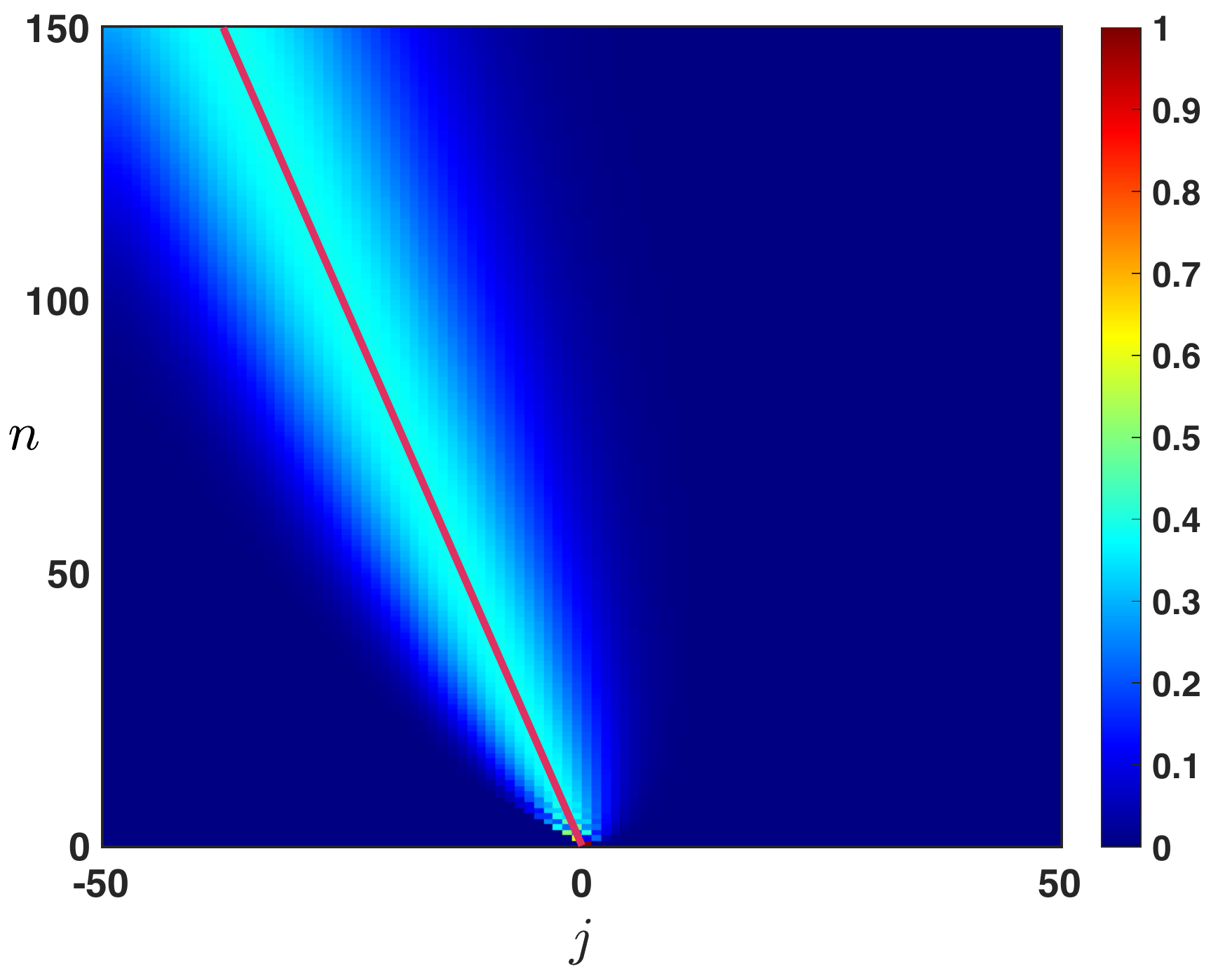}}
\subfigure[$\beta+\alpha=\lambda$.]{\includegraphics[width=.32\textwidth]{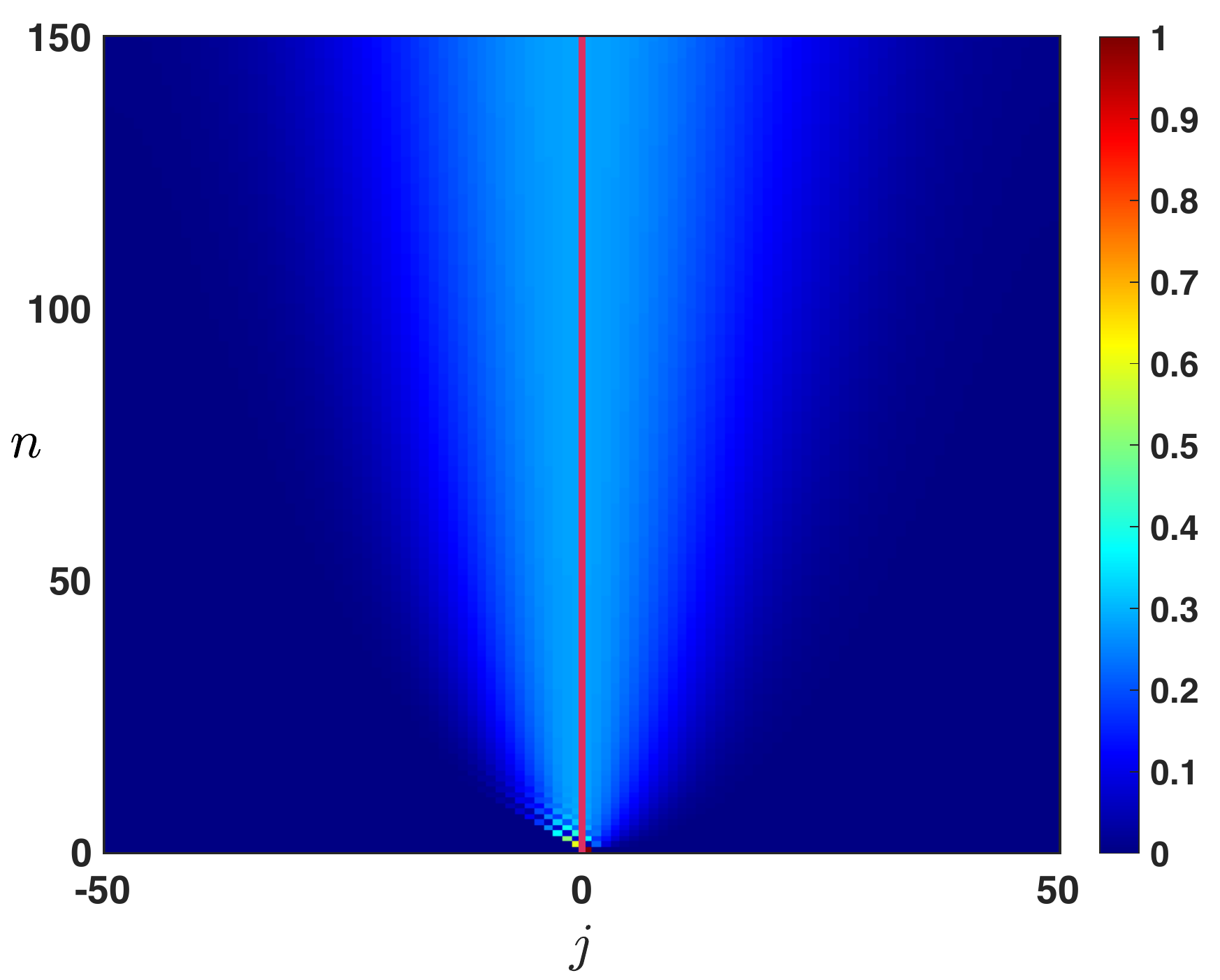}}
\subfigure[$\lambda<\beta+\alpha$.]{\includegraphics[width=.32\textwidth]{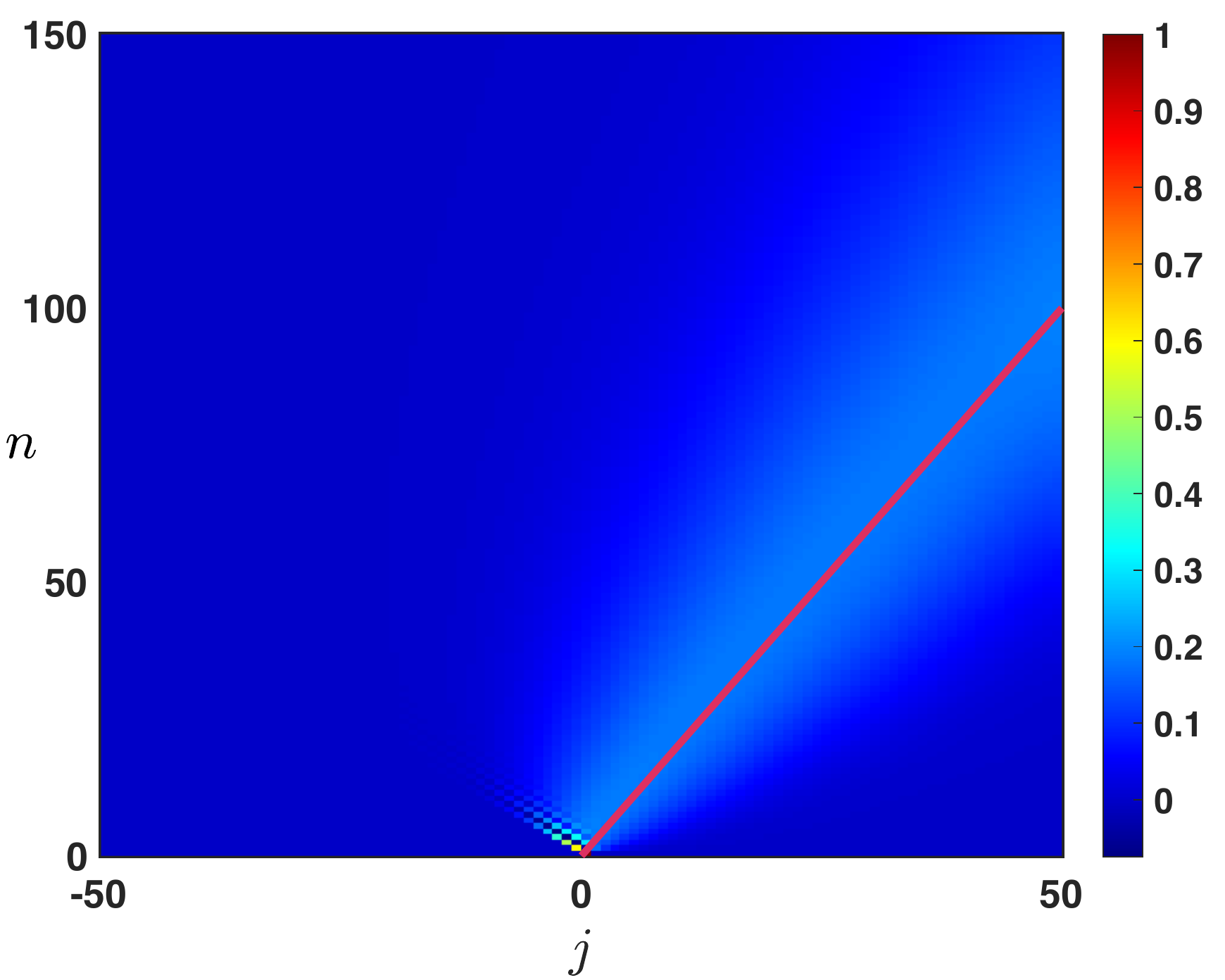}}
\caption{Effects of turning on $\beta>0$ when $\alpha+\lambda<1$. When $\beta+\alpha<\lambda$ we  have $c_0<0$ and observe backward propagation in panel (a) while when $\lambda<\beta+\alpha$ we have $c_0>0$ and have forward propagation as seen in panel (c). At the transition $\beta+\alpha=\lambda$, the wave speed vanishes $c_0=0$ and there is no propagation as illustrated in panel (b).Intuitively, $\alpha$ and $\beta$, both propagating signals in the forward (rightward) direction, compete with $\lambda$ carrying the feedback (leftward) prediction signals; this competition determines the main direction of propagation of neural activity in the system.}
  \label{fig:caseneq0}
\end{figure}

\begin{itemize}
\item Case: $0\leq \lambda+\alpha<1$. There exist some universal constants $C,\kappa>0$ and $L>0$ such that
\bqs
\G_j^{n}= \frac{1}{\sqrt{4\pi \sigma_0 n}}\exp\left( -\frac{|j-c_0n|^2}{4\sigma_0 n}\right)+\mathcal{N}_j^{n}, \quad -n\leq j \leq Ln,
\eqs
where the remainder term satisfies a Gaussian estimate
\bqs
\left|\mathcal{N}_j^{n}\right| \leq \frac{C}{n}\exp\left( -\kappa\frac{|j-c_0n|^2}{n}\right), \quad -n\leq j \leq Ln.
\eqs
While for $j>nL$ we simply get a pure exponential bound
\bqs
\left|\mathcal{G}_j^{n}\right| \leq C e^{-\kappa n-\kappa j }, \quad nL< j.
\eqs
Inspecting the formula for $c_0$, we notice that when $\alpha+\beta \lessgtr \lambda$ we have $c_0\lessgtr0$ and the wave speed vanishes precisely when $\alpha+\beta = \lambda$. This is illustrated in Figure~\ref{fig:caseneq0}, where we see that $\alpha$ and $\beta$, both propagating signals in the forward (rightward) direction, compete with $\lambda$ carrying the feedback (leftward) prediction signals; this competition determines the main direction of propagation of neural activity in the system.

\begin{figure}[t!]
\centering
\subfigure[$\alpha<\beta+\lambda$ and $\beta+\alpha<\lambda$.]{\includegraphics[width=.32\textwidth]{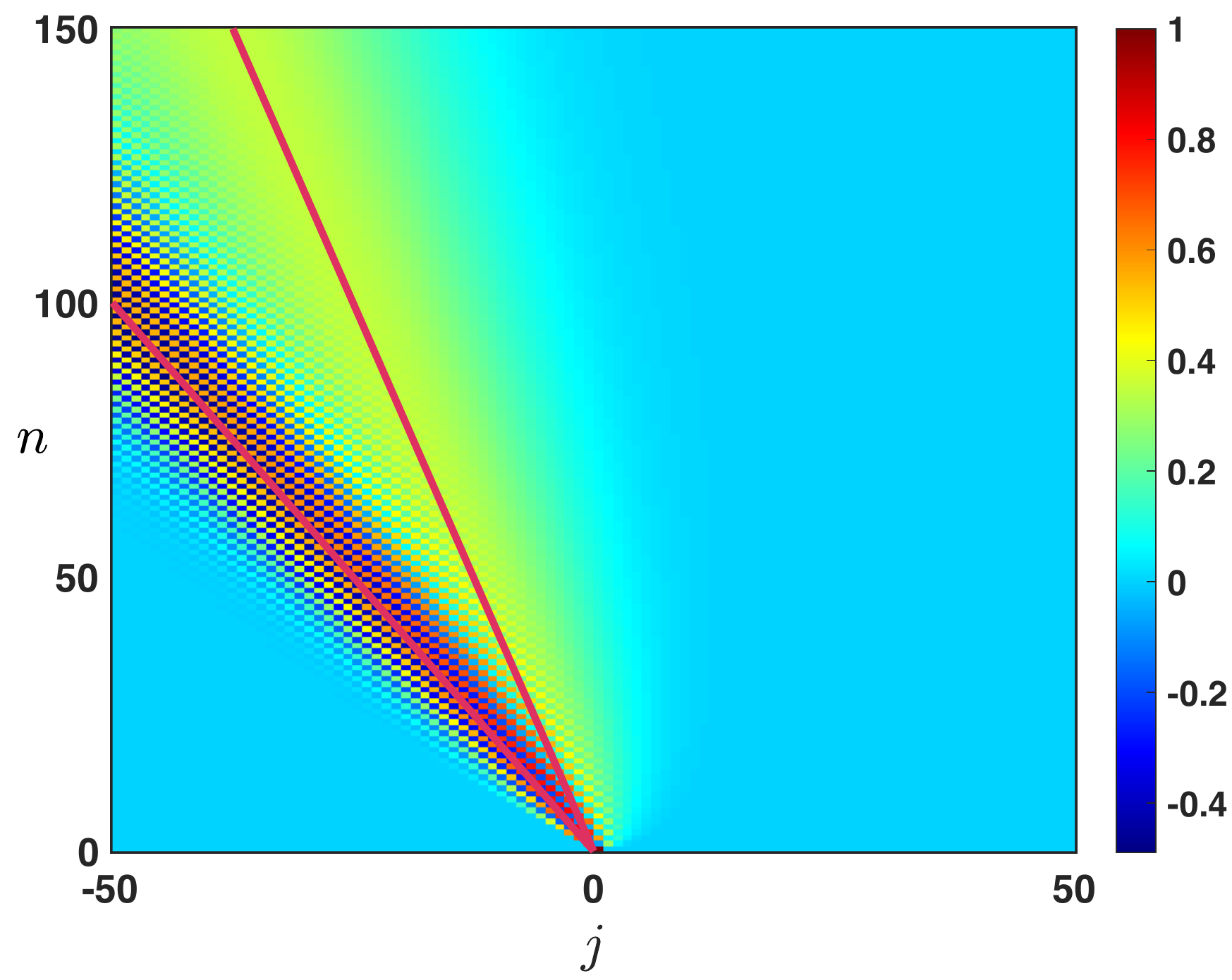}}
\subfigure[$\alpha<\beta+\lambda$ and $\beta+\alpha=\lambda$.]{\includegraphics[width=.32\textwidth]{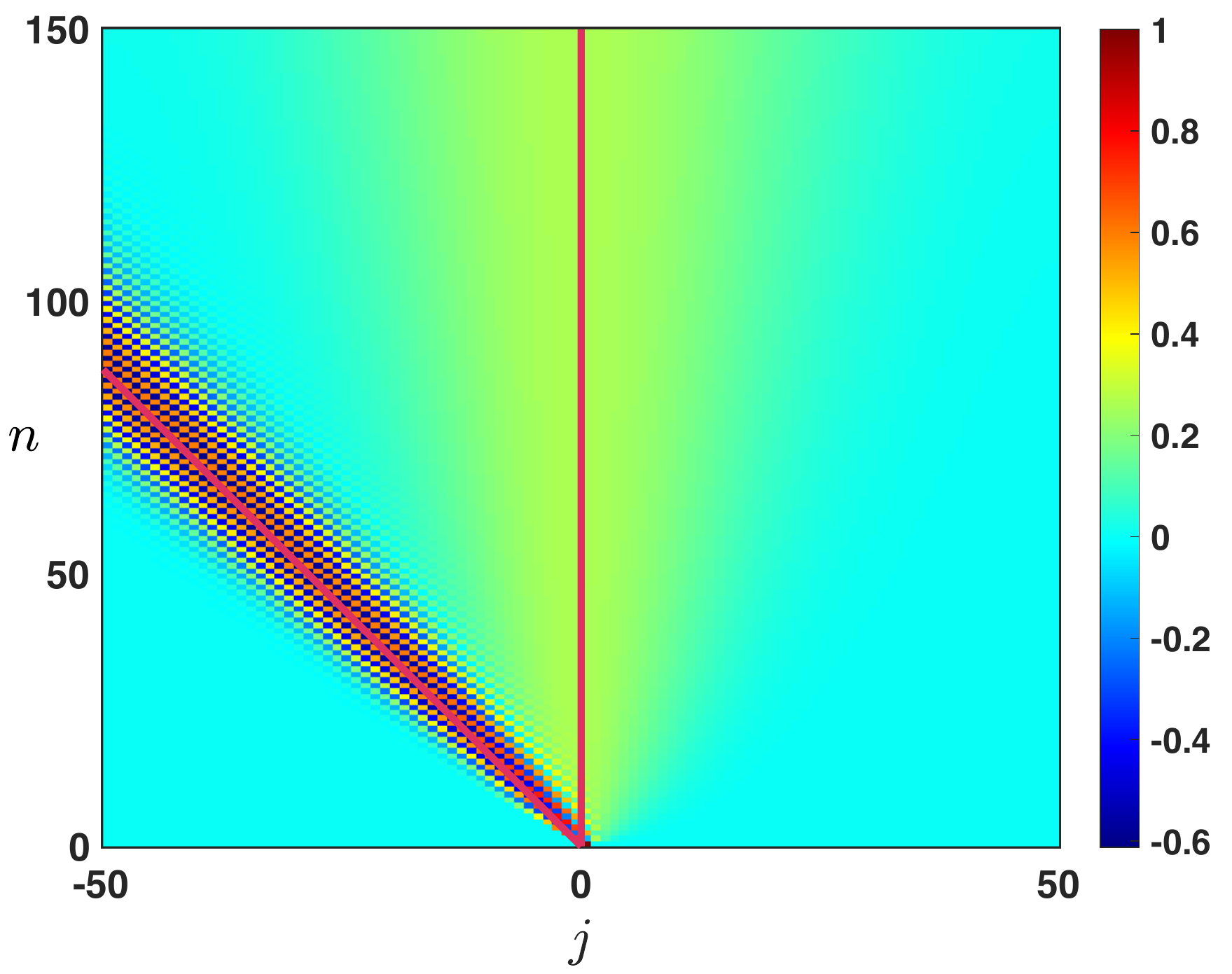}}
\subfigure[$\alpha<\beta+\lambda$ and $\lambda<\beta+\alpha$.]{\includegraphics[width=.32\textwidth]{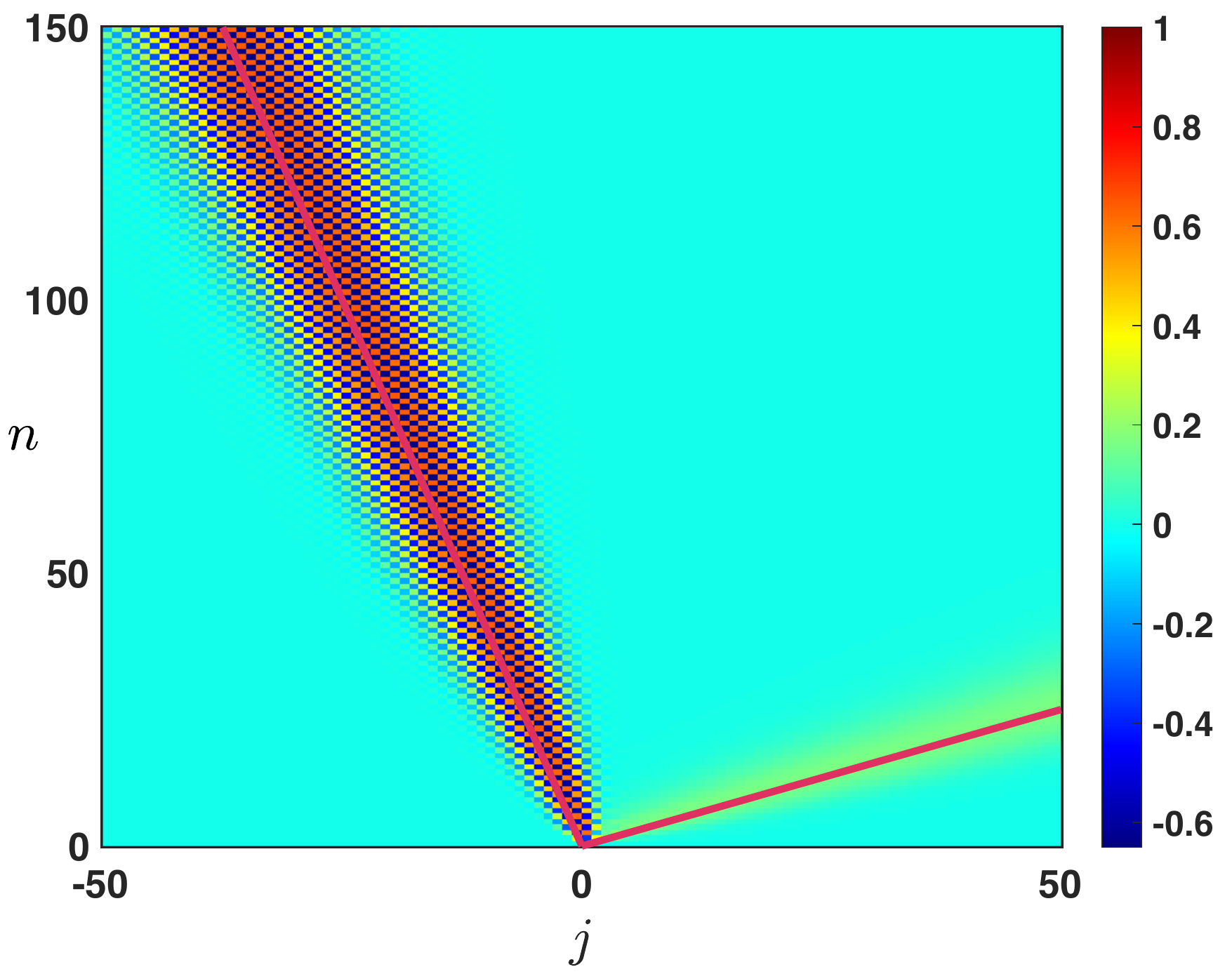}}
\subfigure[$\beta+\lambda<\alpha$.]{\includegraphics[width=.32\textwidth]{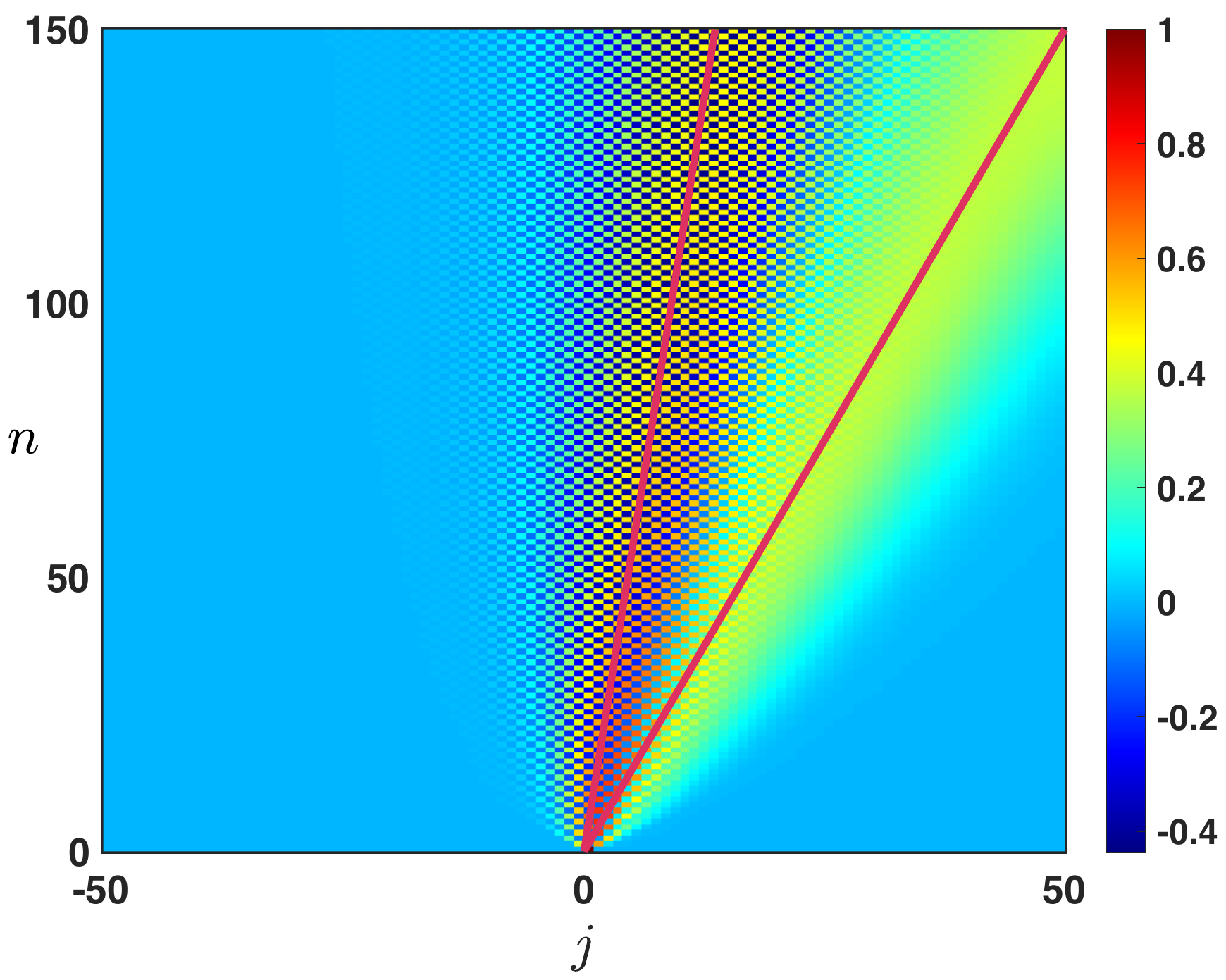}}
\subfigure[$\beta+\lambda=\alpha$.]{\includegraphics[width=.32\textwidth]{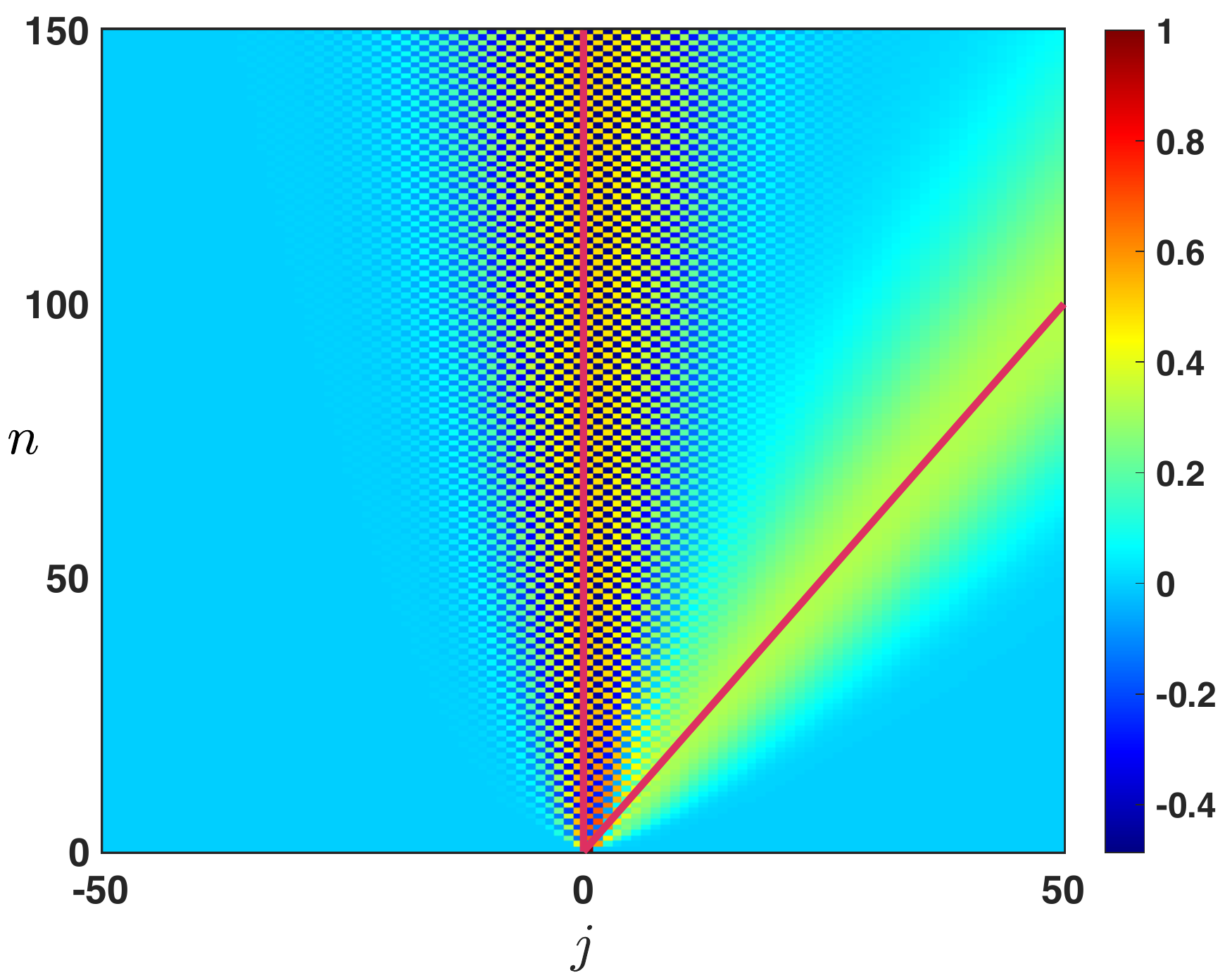}}
\subfigure[]{\includegraphics[width=.27\textwidth]{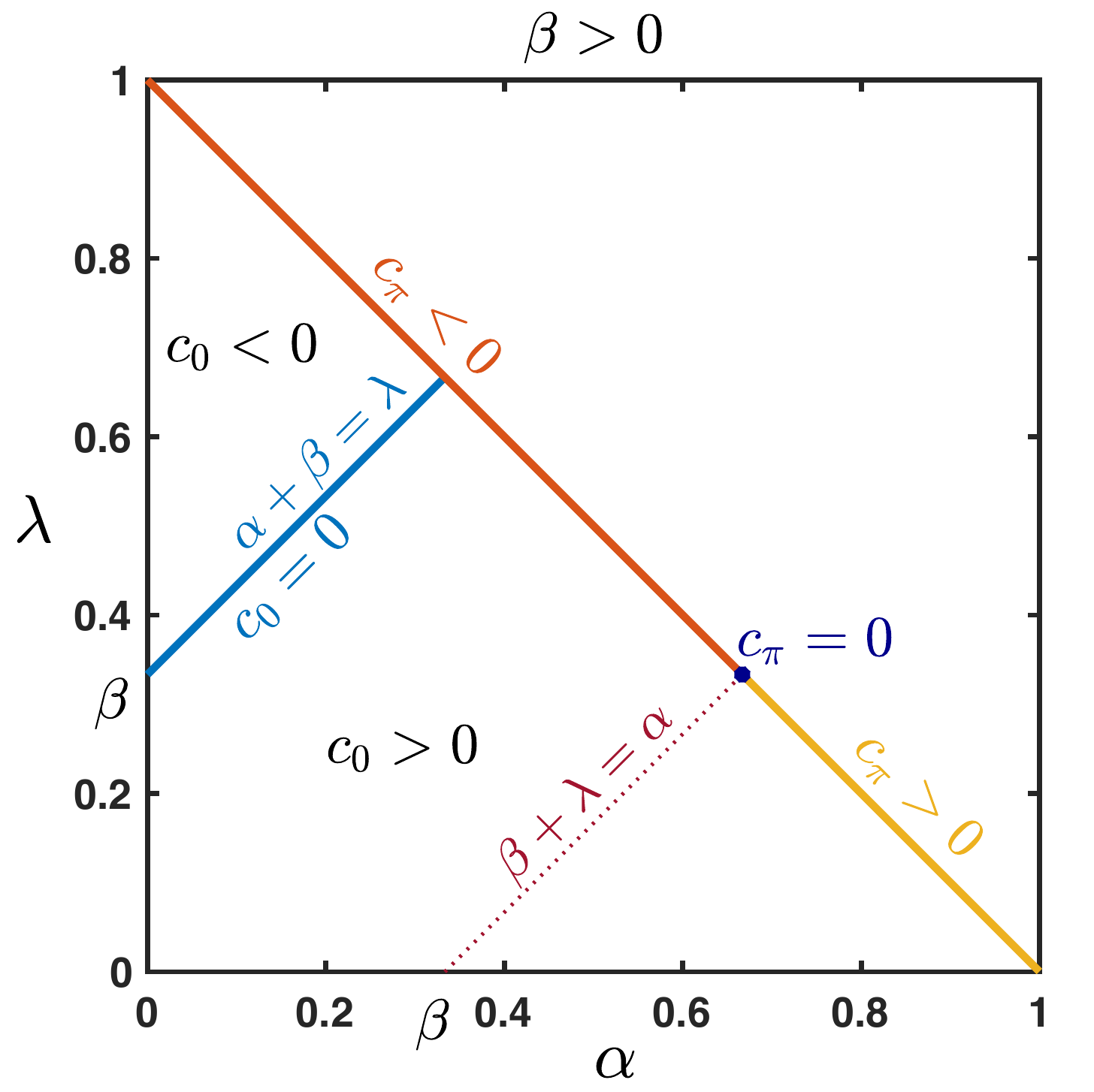}}
\caption{Effects of turning on $\beta>0$ when $\alpha+\lambda=1$. We now observe a secondary wave with associated wave speed $c_\pi$ whose sign depends on the competition between $\alpha$ and $\beta+\lambda$. (a)-(b)-(c) When $\alpha<\beta+\lambda$, the wave speed of the secondary wave always verifies $c_\pi<0$, and the competition between $\lambda$ and $\beta+\alpha$ gives the direction of the primary wave as previously reported in Figure~\ref{fig:caseneq0}. (d) When $\beta+\lambda<\alpha$ which always implies that $\alpha+\beta>\lambda$, we have $0<c_\pi<c_0$ traducing forward propagation for both waves. We remark that the secondary wave is slower. (e) When $\beta+\lambda=\alpha$ which also implies that $\alpha+\beta>\lambda$, we get $0=c_\pi<c_0$ such that the secondary wave is blocked. (f) Summary of the sign of the wave speeds $c_0$ and $c_\pi$ when $\beta>0$ when $\alpha+\lambda\leq 1$.}
  \label{fig:caseneq0eq1}
\end{figure}

\item Case: $\lambda+\alpha=1$. What changes in that case is the existence of a secondary wave with associated wave speed $c_\pi$ whose sign depends on the competition between $\alpha$ and $\beta+\lambda$. When $\alpha<\beta+\lambda$ then we have $c_\pi<0$, and the competition between $\lambda$ and $\beta+\alpha$ will determine the sign of $c_0$, as illustrated in panels (a)-(b)-(c) of Figure~\ref{fig:caseneq0eq1}. On the other hand, when $\beta+\lambda<\alpha$ implying that $c_\pi>0$, we note that $\alpha+\beta>\lambda$ and thus $c_0>0$. In that case, the explicit formula for $c_\pi$ and $c_0$ shows that $0<c_\pi<c_0$ and the secondary wave associated to $c_\pi$ is slower to propagate into the network, see Figure~\ref{fig:caseneq0eq1}(d). Finally, when $\beta+\lambda=\alpha$ we have $0=c_\pi<c_0$ and the secondary wave is blocked, see Figure~\ref{fig:caseneq0eq1}(e).
\end{itemize}

We have summarized in the diagram of Figure~\ref{fig:caseneq0eq1}(f) all possible configurations for the sign of the wave speeds $c_0$ and $c_\pi$ when $\beta\in(0,1)$ when $\alpha+\lambda\leq 1$. We notably observe that when $\beta$ is increased the region of parameter space where $c_0<0$ diminishes while the region of parameter space where $c_\pi<0$ increases, indicating that for high values of $\beta$ the primary wave is most likely to be forward while the secondary wave is most likely to be backward.

\subsection{Wave propagation on a semi-infinite network with a forcing source term}

Now that we have understood the intrinsic underlying mechanisms of wave propagation for our model \eqref{model1d} set on an infinite domain, we turn to the case where the network is semi-infinite. That is, the network admits an \emph{input} layer that is only connected to the layer above. The problem now reads
\bqq
\left\{
\begin{split}
e_j^{n+1}-\beta  e_{j-1}^{n+1} &=\alpha e_{j-1}^n+(1-\beta-\lambda- \alpha )e_j^{n} + \lambda  e_{j+1}^{n}, \quad j\geq1, \quad n\geq0,\\
e_0^n&=s_0^n, \quad n\geq0,\\
e_j^0&=h_j, \quad j\geq1.
\end{split}
\right.
\label{modelN}
\eqq
We see that the system depends on the source term $s_0^n$ applied to its input layer at each time step, also called a \emph{boundary value}, and on the starting activation value $(h_j)$ applied to each layer at the initial time point, also called the \emph{initial value}. In fact, the linearity principle tells us that the solutions of the above problem can be obtained as the linear superposition of the solutions to the following two problems, the boundary value problem, where all layers except the input layer are initialized at zero:
\bqq
\left\{
\begin{split}
g_j^{n+1}-\beta  g_{j-1}^{n+1} &=\alpha g_{j-1}^n+(1-\beta-\lambda- \alpha )g_j^{n} + \lambda  g_{j+1}^{n}, \quad j\geq1, \quad n\geq0,\\
g_0^n&=s_0^n, \quad n\geq0,\\
g_j^0&=0, \quad j\geq1,
\end{split}
\right.
\label{modelNbord}
\eqq
and the initial value problem, where the input layer source term is set to zero for all time steps:
\bqq
\left\{
\begin{split}
f_j^{n+1}-\beta  f_{j-1}^{n+1} &=\alpha f_{j-1}^n+(1-\beta-\lambda- \alpha )f_j^{n} + \lambda  f_{j+1}^{n}, \quad j\geq1, \quad n\geq0,\\
f_0^n&=0, \quad n\geq0,\\
f_j^0&=h_j, \quad j\geq1.
\end{split}
\right.
\label{modelNivp}
\eqq
Subsequently, the generic solution sequence $(e_j^n)_{j\geq1}$ can be obtained as
\bqs
e_j^n=f_j^n+g_j^n, \quad j\geq1, \quad n\geq1.
\eqs

\subsubsection{The initial value problem \eqref{modelNivp}}

\begin{figure}[t!]
\centering
\subfigure[]{\includegraphics[width=.32\textwidth]{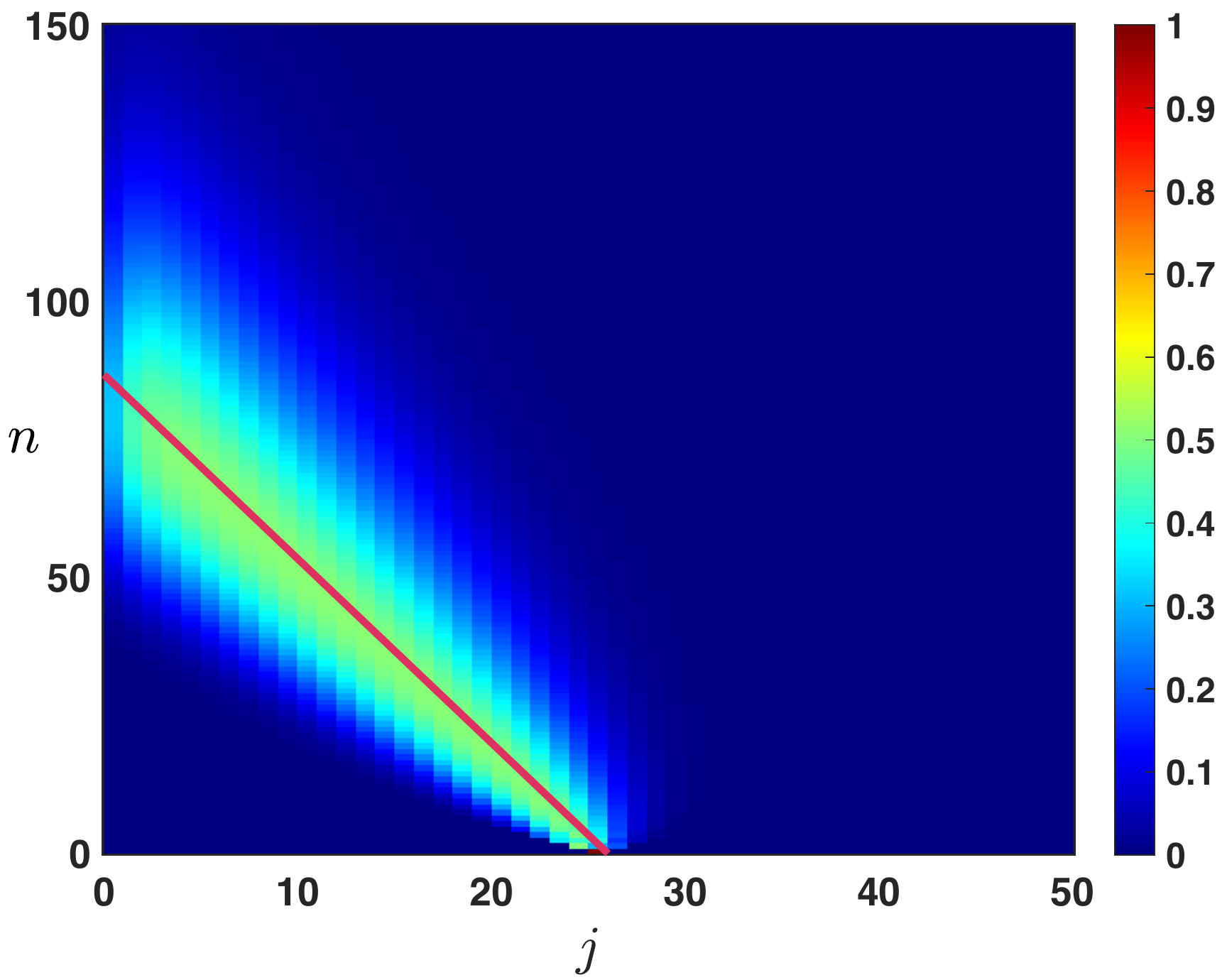}}
\subfigure[]{\includegraphics[width=.32\textwidth]{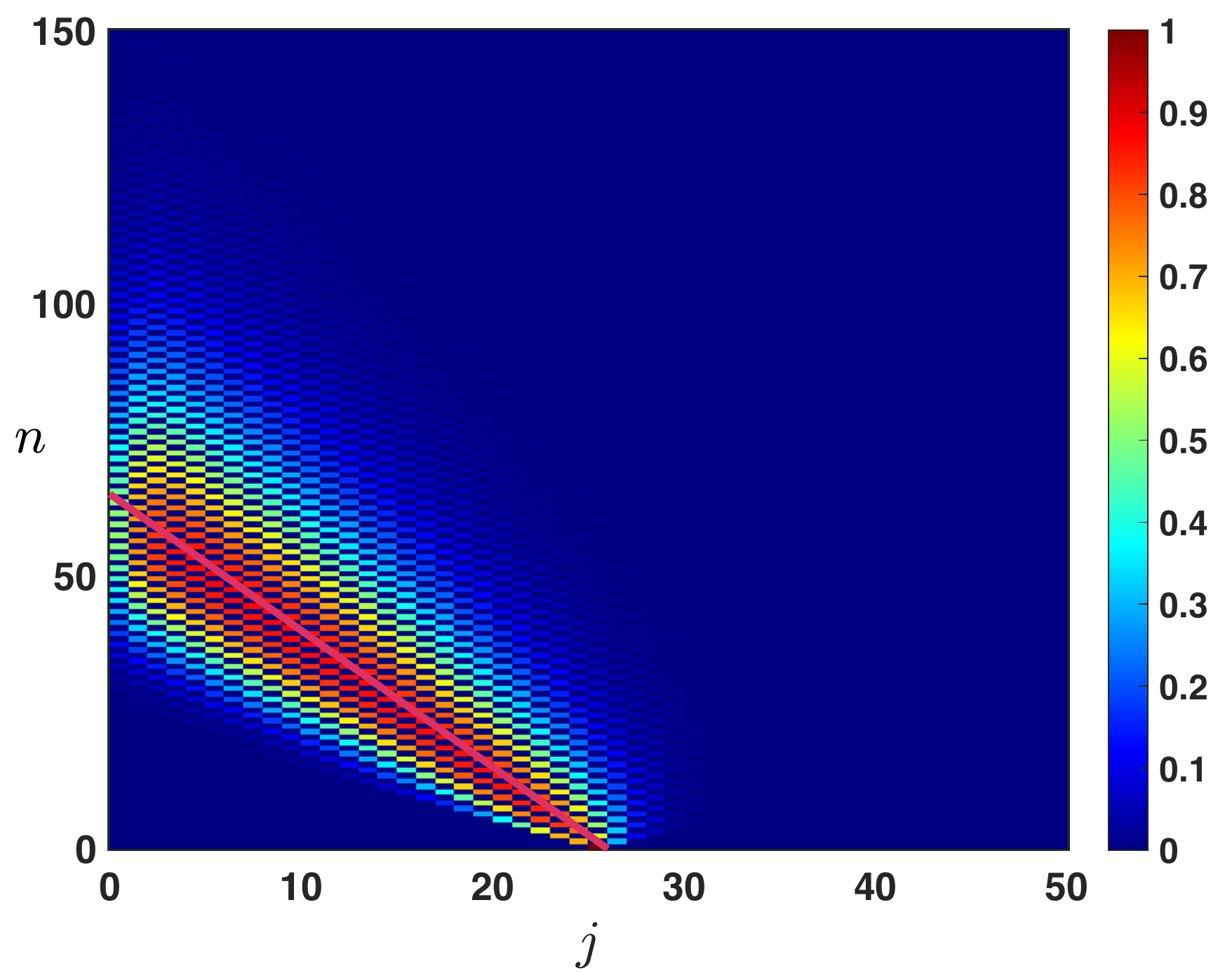}}
\subfigure[]{\includegraphics[width=.32\textwidth]{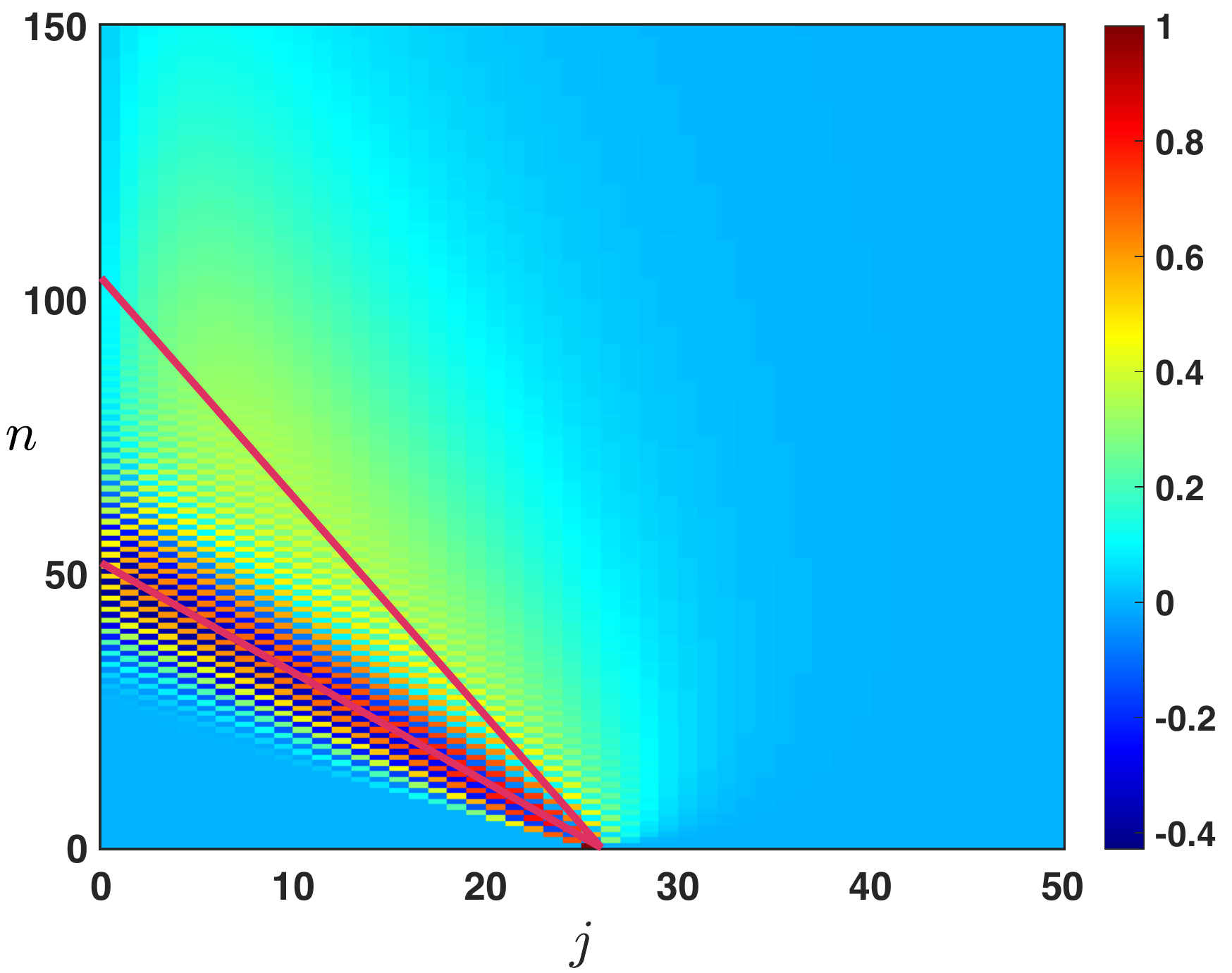}}
\caption{Space-time evolution of the rescaled solution sequence $(\sqrt{n} \, \G^n_{\mathrm{ivp}}(j,j_0))_{j\geq1}$ to \eqref{modelNivp} starting with a Dirac delta sequence at $j_0=25$ in different cases with leftward propagation. (a) $\beta=0$ \& $\alpha+\lambda<1$ with $c_0<0$. (b) $\beta=0$ \& $\alpha+\lambda=1$ with $c_0<0$. (c) $0<\beta<1$ \& $\alpha+\lambda=1$ with $-1<c_\pi<c_0<0$.}
  \label{fig:bordST}
\end{figure}

It is first natural to investigate the initial value problem \eqref{modelNivp} since it is really close to the infinite network case of the previous section. Here, we consider the effect of the initial value assigned to each layer $j>0$ at the first time step ($n=0$), except the input layer ($j=0$) which is set to zero. The dynamics of \eqref{modelNivp} is still read out from the amplification factor function $\rho$ defined in \eqref{defrho} and once again the solutions to \eqref{modelNivp} can be obtained as the convolution of the initial sequence with the fundamental solution associated to the problem. For $j_0\geq1$, we denote by $\boldsymbol{\delta}^{j_0}$ the Dirac delta sequence defined as $\boldsymbol{\delta}_{j_0}^{j_0}=1$ and $\boldsymbol{\delta}_j^{j_0}=0$ for all $j\geq1$ and $j\neq j_0$. Correspondingly, we denote by $\G^n_{\mathrm{ivp}}(\cdot,j_0)=(\G^n_{\mathrm{ivp}}(j,j_0))_{j\geq1}$ the solution to \eqref{modelNivp} starting from $\boldsymbol{\delta}^{j_0}$, and let us remark that the solutions to \eqref{modelNivp} starting from any initial condition $(h_j)_{j\geq1}$ can be represented as
\bqs
f_j^n=\sum_{j_0=1}^{+\infty}\G^n_{\mathrm{ivp}}(j,j_0)h_{j_0}, \quad j\geq1, \quad n\geq1.
\eqs
Combining the results of \cite{CF20,Coeuret22} together with those of \cite{CF21,GT1,GT2} which precisely deal with recurrence equations with boundary conditions, one can obtain very similar results as in the previous case. The very first obvious remark that we can make is that for all $j,j_0\geq1$ and $1\leq n<j_0$ we have
\bqs
\G^n_{\mathrm{ivp}}(j,j_0)=\G^n_{j-j_0},
\eqs
meaning that it takes $n=j_0$ iterations before the solution arrives at the boundary $j=0$ and for $1\leq n<j_0$ the problem is similar to the one set on the infinite network. This behavior is illustrated in Figure~\ref{fig:bordST} for several values of the hyper-parameters where we represent the spatio-temporal evolution of the rescaled solution sequence $(\sqrt{n} \, \G^n_{\mathrm{ivp}}(j,j_0))_{j\geq1}$. We clearly observe a Gaussian behavior before the solution reaches the boundary. And for all $n\geq j_0$, we can write
\bqs
\G^n_{\mathrm{ivp}}(j,j_0)=\G^n_{j-j_0}+\G_{\mathrm{bl}}^n(j,j_0),
\eqs
where $\G_{\mathrm{bl}}^n(j,j_0)$ is a remainder term generated by the boundary condition at $j=0$. It is actually possible to bound $\G_{\mathrm{bl}}^n(j,j_0)$ in each of the cases treated above. 

\begin{figure}[t!]
\centering
\subfigure[Gaussian approximation.]{\includegraphics[width=.32\textwidth]{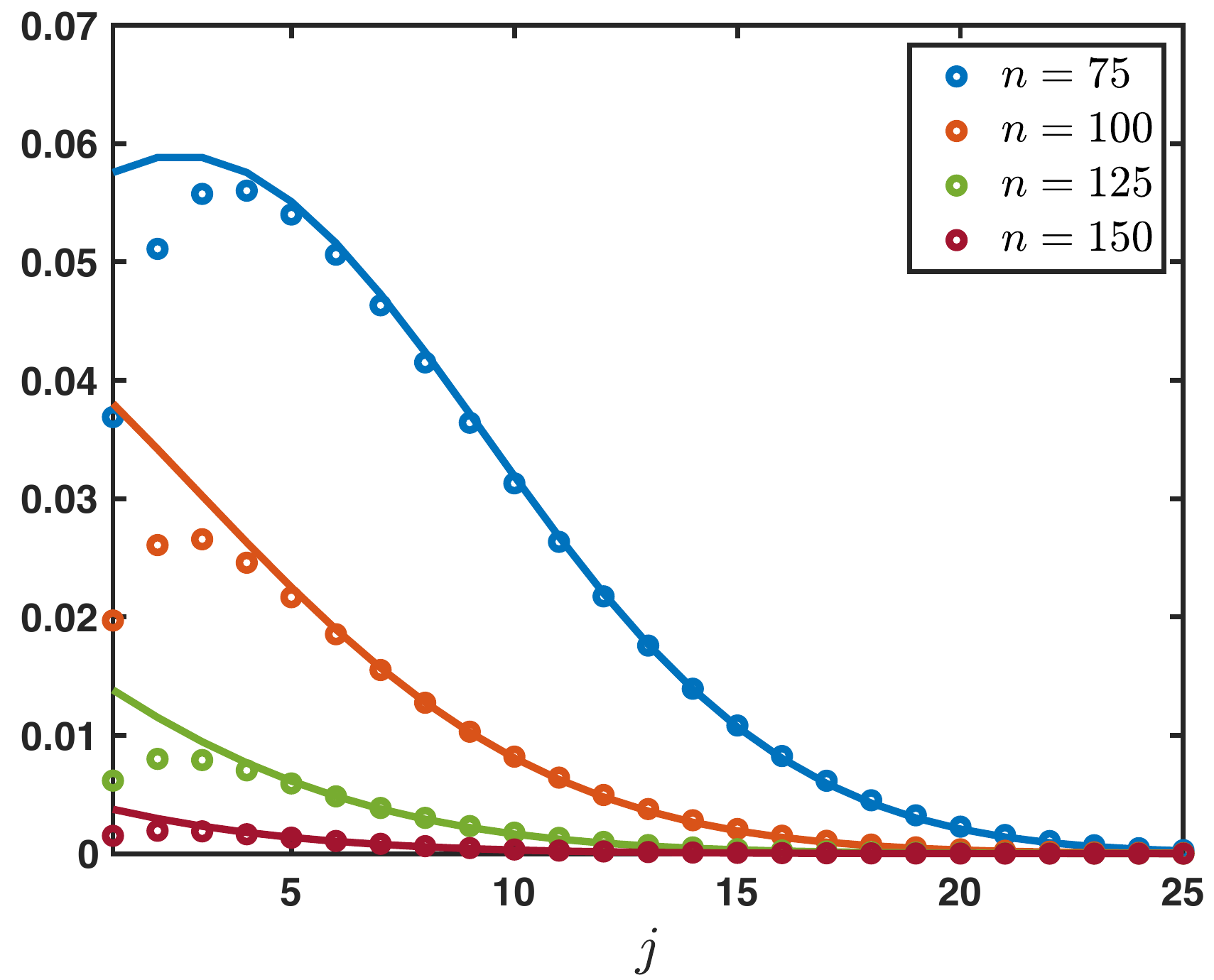}}
\subfigure[Visualization of the boundary layer.]{\includegraphics[width=.32\textwidth]{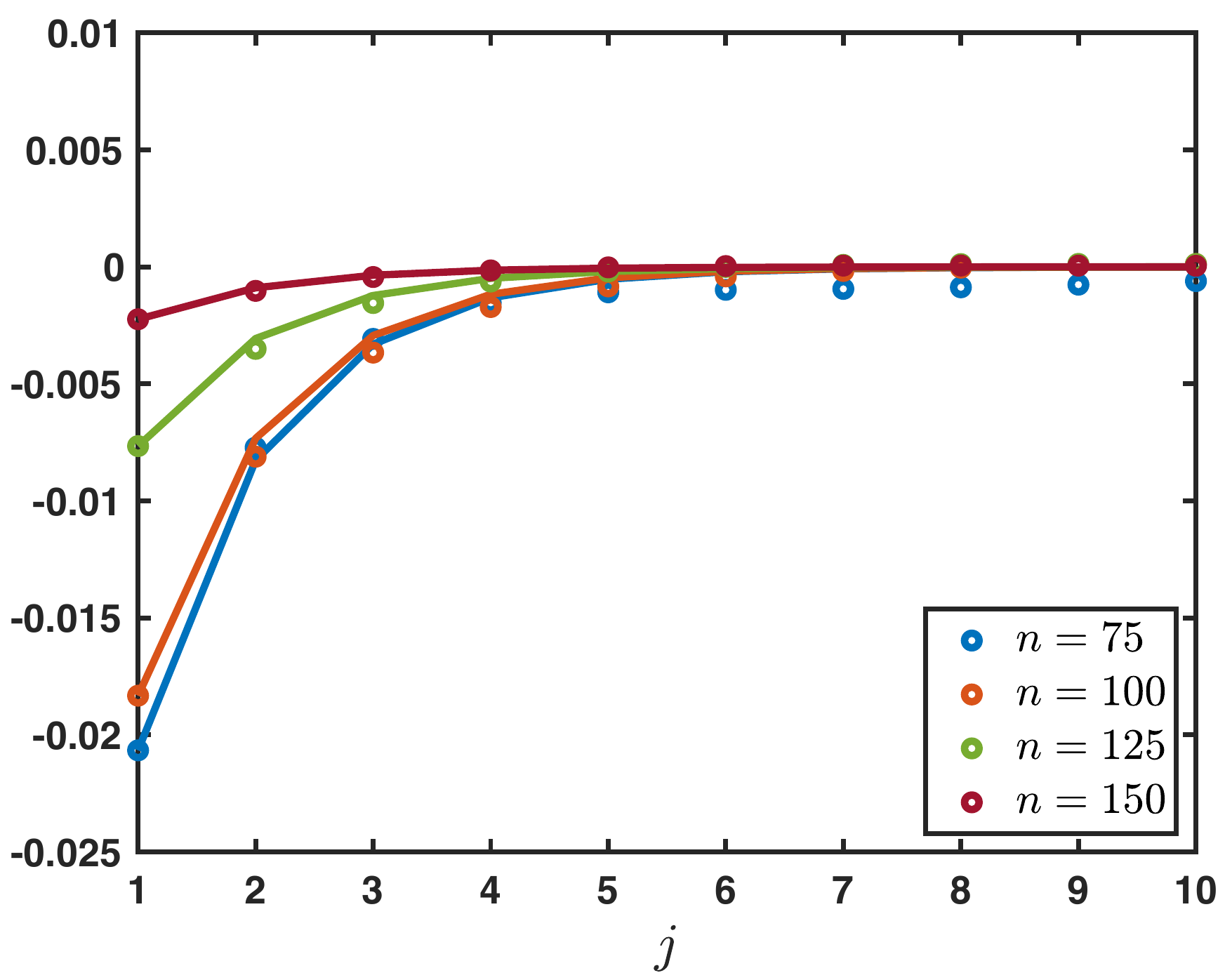}}
\subfigure[Gaussian \& boundary layer approximation.]{\includegraphics[width=.32\textwidth]{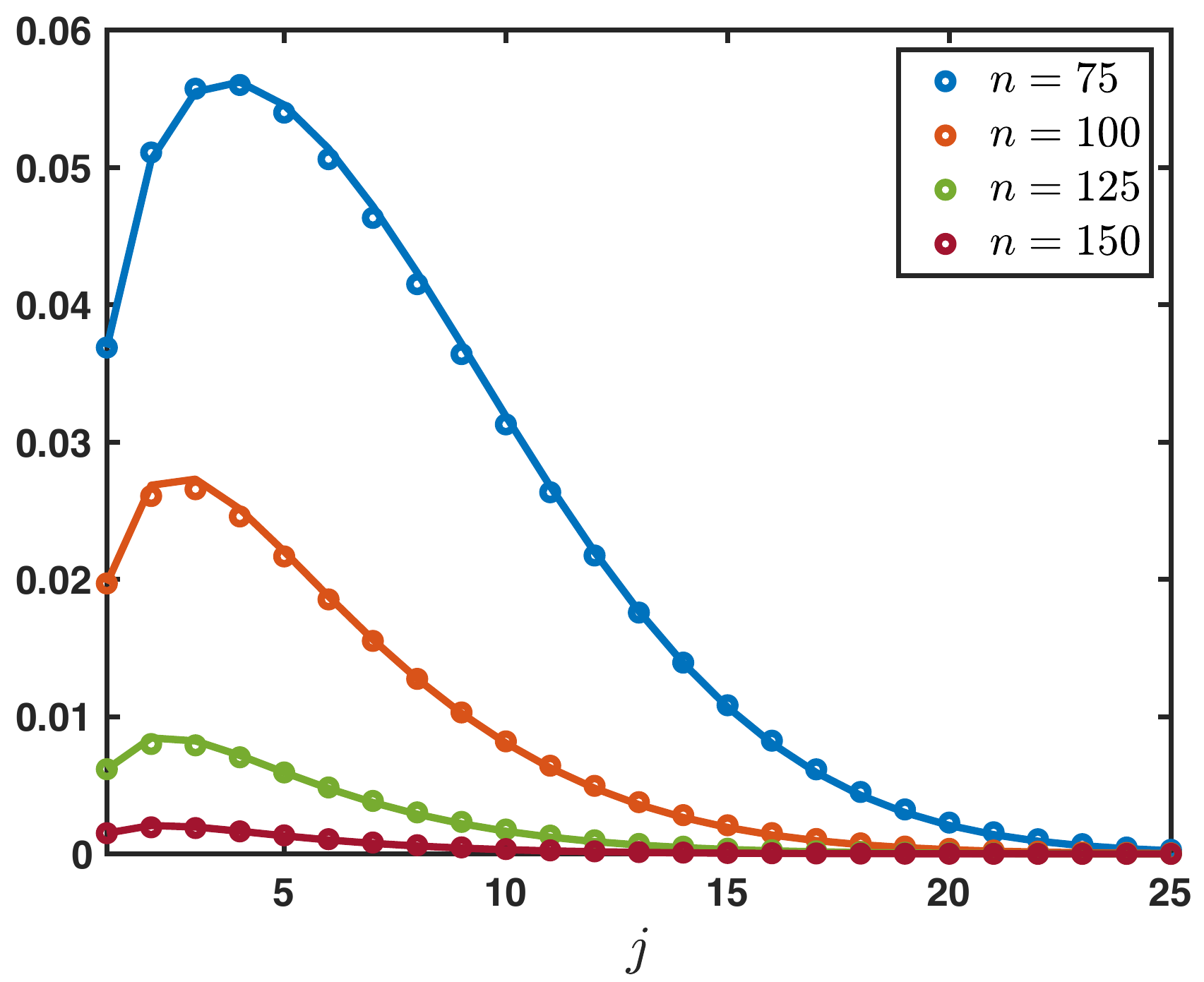}}
\caption{Illustration of the solution $\G^n_{\mathrm{ivp}}(j,j_0)$ to \eqref{modelNivp} in the case where $\beta=0$ and $\alpha+\lambda<1$ with $\alpha<\beta$. (a) Visualizations of the solution $\G^n_{\mathrm{ivp}}(j,j_0)$ (circles) at different time iterations. The plain lines correspond to the Gaussian approximation $\frac{1}{\sqrt{4\pi \sigma_0 n}}\exp\left( -\frac{|j-j_0-c_0n|^2}{4\sigma_0 n}\right)$ and remark the presence of a boundary layer (seen as a mismatch between the circles and the Gaussian lines approximation). (b) We represent the boundary layer by plotting (circles) $\G^n_{\mathrm{ivp}}(j,j_0)-\frac{1}{\sqrt{4\pi \sigma_0 n}}\exp\left( -\frac{|j-j_0-c_0n|^2}{4\sigma_0 n}\right)$ and we compare it to our boundary layer approximation $-\frac{1}{\sqrt{4\pi \sigma_0 n}}\exp\left( -\frac{|-j_0-c_0n|^2}{4\sigma_0 n}\right)\left(\frac{\alpha}{\lambda}\right)^j$ (plain lines). (c) Finally we compare the solution $\G^n_{\mathrm{ivp}}(j,j_0)$ (circles) to its first order approximation $\frac{1}{\sqrt{4\pi \sigma_0 n}}\exp\left( -\frac{|j-j_0-c_0n|^2}{4\sigma_0 n}\right)-\frac{1}{\sqrt{4\pi \sigma_0 n}}\exp\left( -\frac{|-j_0-c_0n|^2}{4\sigma_0 n}\right)\left(\frac{\alpha}{\lambda}\right)^j$ (plain lines).}
  \label{fig:bordprofile}
\end{figure}

When $\beta=0$ and $\alpha+\lambda<1$ with $\alpha<\lambda$ such that $c_0<0$, then $\G_{\mathrm{bl}}^n(j,j_0)$ is well approximated by
\bqs
\G_{\mathrm{bl}}^n(j,j_0)\approx
\left\{
\begin{array}{lc}
-\frac{1}{\sqrt{4\pi \sigma_0 n}}\exp\left( -\frac{|-j_0-c_0n|^2}{4\sigma_0 n}\right)\left(\frac{\alpha}{\lambda}\right)^j,& \quad 1\leq j \leq j_0,\\
e^{-\kappa n -\kappa (j-j_0)}, & \quad j>j_0,
\end{array}
\right.
\eqs
while when $\lambda<\alpha$ with $c_0>0$, then $\G_{\mathrm{bl}}^n(j,j_0)$ is well approximated by
\bqs
\G_{\mathrm{bl}}^n(j,j_0)\approx
\left\{
\begin{array}{lc}
e^{-\kappa n -\kappa (j_0-j)}, & \quad 1\leq j \leq j_0,\\
-\frac{1}{\sqrt{4\pi \sigma_0 n}}\exp\left( -\frac{|j-c_0n|^2}{4\sigma_0 n}\right)\left(\frac{\lambda}{\alpha}\right)^{j_0},& \quad j_0< j,
\end{array}
\right.
\eqs
this is illustrated in Figure~\ref{fig:bordprofile} in the case $c_0<0$. 

On the other hand for $\alpha+\lambda=1$ with $\alpha<\lambda$ such that $c_0<0$, then $\G_{\mathrm{bl}}^n(j,j_0)$ is well approximated by
\bqs
\G_{\mathrm{bl}}^n(j,j,_0)\approx
\left\{
\begin{array}{lc}
-\frac{1+(-1)^n}{\sqrt{4\pi \sigma_0 n}}\exp\left( -\frac{|-j_0-c_0n|^2}{4\sigma_0 n}\right)\left(\frac{\alpha}{\lambda}\right)^j,& \quad 1\leq j \leq j_0,\\
e^{-\kappa n -\kappa (j-j_0)}, & \quad j>j_0,
\end{array}
\right.
\eqs
while when $\lambda<\alpha$ with $c_0>0$, then $\G_{\mathrm{bl}}^n(j,j_0)$ is well approximated by
\bqs
\G_{\mathrm{bl}}^n(j,j_0)\approx
\left\{
\begin{array}{lc}
e^{-\kappa n -\kappa (j_0-j)}, & \quad 1\leq j \leq j_0,\\
-\frac{1+(-1)^n}{\sqrt{4\pi \sigma_0 n}}\exp\left( -\frac{|j-c_0n|^2}{4\sigma_0 n}\right)\left(\frac{\lambda}{\alpha}\right)^{j_0},& \quad j_0< j.
\end{array}
\right.
\eqs

When $0<\beta<1$ and $\alpha+\lambda<1$ the approximations are similar as for the case with $\beta=0$. We thus need to discuss three cases.

\begin{itemize}
\item Case $-1<c_\pi<c_0<0$. In that case, we have for $1\leq j \leq j_0$ that
\bqs
\G_{\mathrm{bl}}^n(j,j_0)\approx -\frac{1}{\sqrt{4\pi \sigma_0 n}}\exp\left( -\frac{|-j_0-c_0n|^2}{4\sigma_0 n}\right)\left(\frac{\alpha+\beta}{\lambda}\right)^j -\frac{(-1)^n}{\sqrt{4\pi \sigma_\pi n}}\exp\left( -\frac{|-j_0-c_\pi n|^2}{4\sigma_\pi n}\right)\left(\frac{\alpha-\beta}{\lambda}\right)^j,
\eqs
with an exponential bound for $j>j_0$. This situation is presented in Figure~\ref{fig:bordST}(c)
\item Case $-1<c_\pi<0<c_0$. In this case we have
\bqs
\G_{\mathrm{bl}}^n(j,j_0)\approx
\left\{
\begin{array}{lc}
-\frac{(-1)^n}{\sqrt{4\pi \sigma_\pi n}}\exp\left( -\frac{|-j_0-c_\pi n|^2}{4\sigma_\pi n}\right)\left(\frac{\alpha-\beta}{\lambda}\right)^j,& \quad 1\leq j \leq j_0,\\
-\frac{1}{\sqrt{4\pi \sigma_0 n}}\exp\left( -\frac{|j-c_0n|^2}{4\sigma_0 n}\right)\left(\frac{\lambda}{\alpha+\beta}\right)^{j_0}, & \quad j_0<j<Ln.
\end{array}
\right.
\eqs
\item Case $-1<0<c_\pi<c_0$. In this case we have
\bqs
\G_{\mathrm{bl}}^n(j,j_0)\approx -\frac{1}{\sqrt{4\pi \sigma_0 n}}\exp\left( -\frac{|j-c_0n|^2}{4\sigma_0 n}\right)\left(\frac{\lambda}{\alpha+\beta}\right)^{j_0}-\frac{(-1)^n}{\sqrt{4\pi \sigma_\pi n}}\exp\left( -\frac{|j-c_\pi n|^2}{4\sigma_\pi n}\right)\left(\frac{\lambda}{\alpha-\beta}\right)^{j_0}
\eqs
for $j_0<j<Ln$.
\end{itemize}

\subsubsection{The boundary value problem \eqref{modelNbord}}

We now turn our attention to the boundary value problem \eqref{modelNbord} where the network is initialized with zero activity, for all layers except the input. Motivated by applications, we will only focus on the case where $s_0^n=s_0\in\R$ for all $n\geq0$ (i.e., a constant sensory input) and thus study:
\bqq
\left\{
\begin{split}
g_j^{n+1}-\beta  g_{j-1}^{n+1} &=\alpha g_{j-1}^n+(1-\beta-\lambda- \alpha )g_j^{n} + \lambda  g_{j+1}^{n}, \quad j\geq1, \quad n\geq0,\\
g_0^n&=s_0, \quad n\geq0,\\
g_j^0&=0, \quad j\geq1.
\end{split}
\right.
\label{modelNbords0const}
\eqq

\begin{figure}[t!]
\centering
\subfigure[$\alpha<\lambda$.]{\includegraphics[width=.32\textwidth]{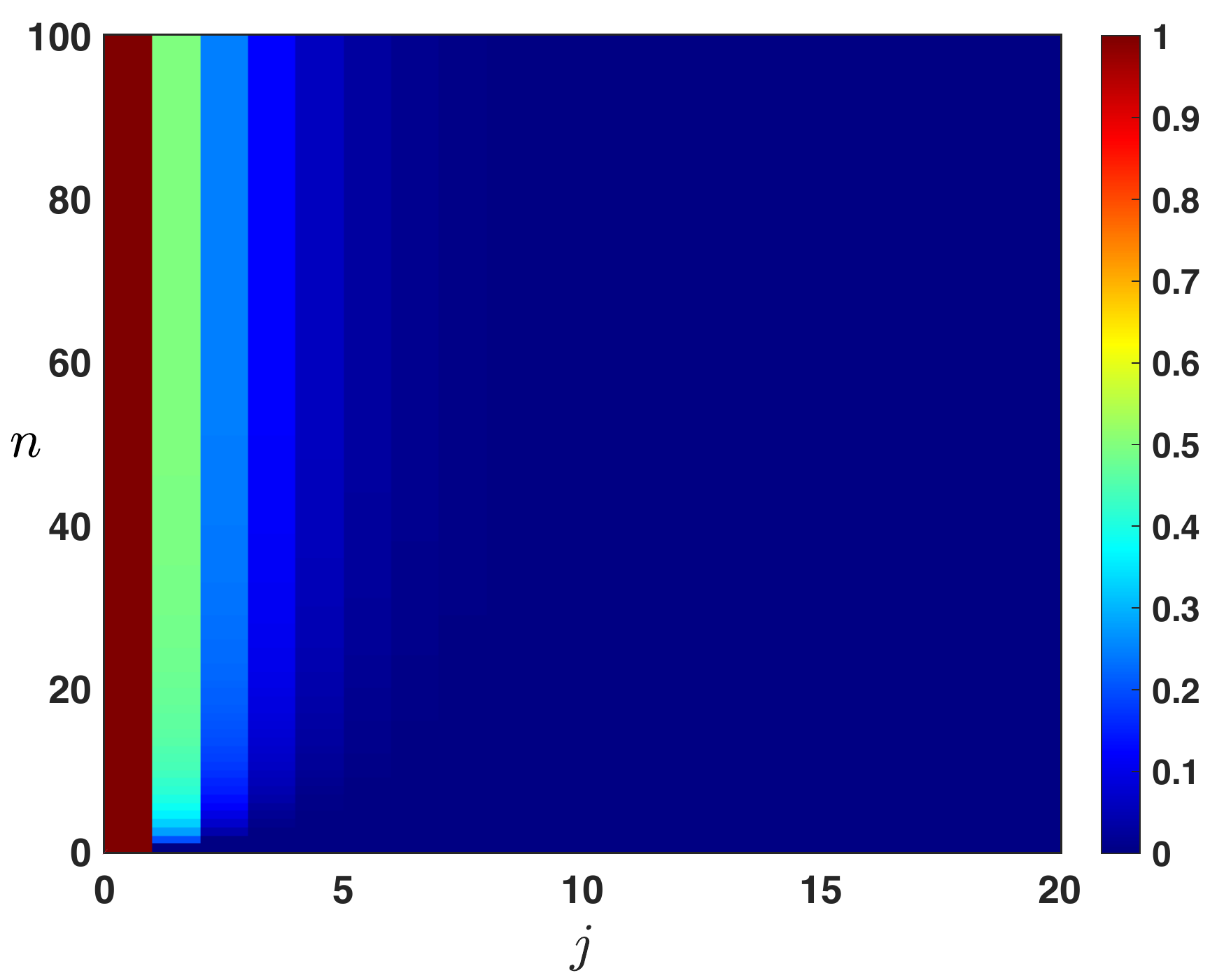}}
\subfigure[$\alpha=\lambda$.]{\includegraphics[width=.32\textwidth]{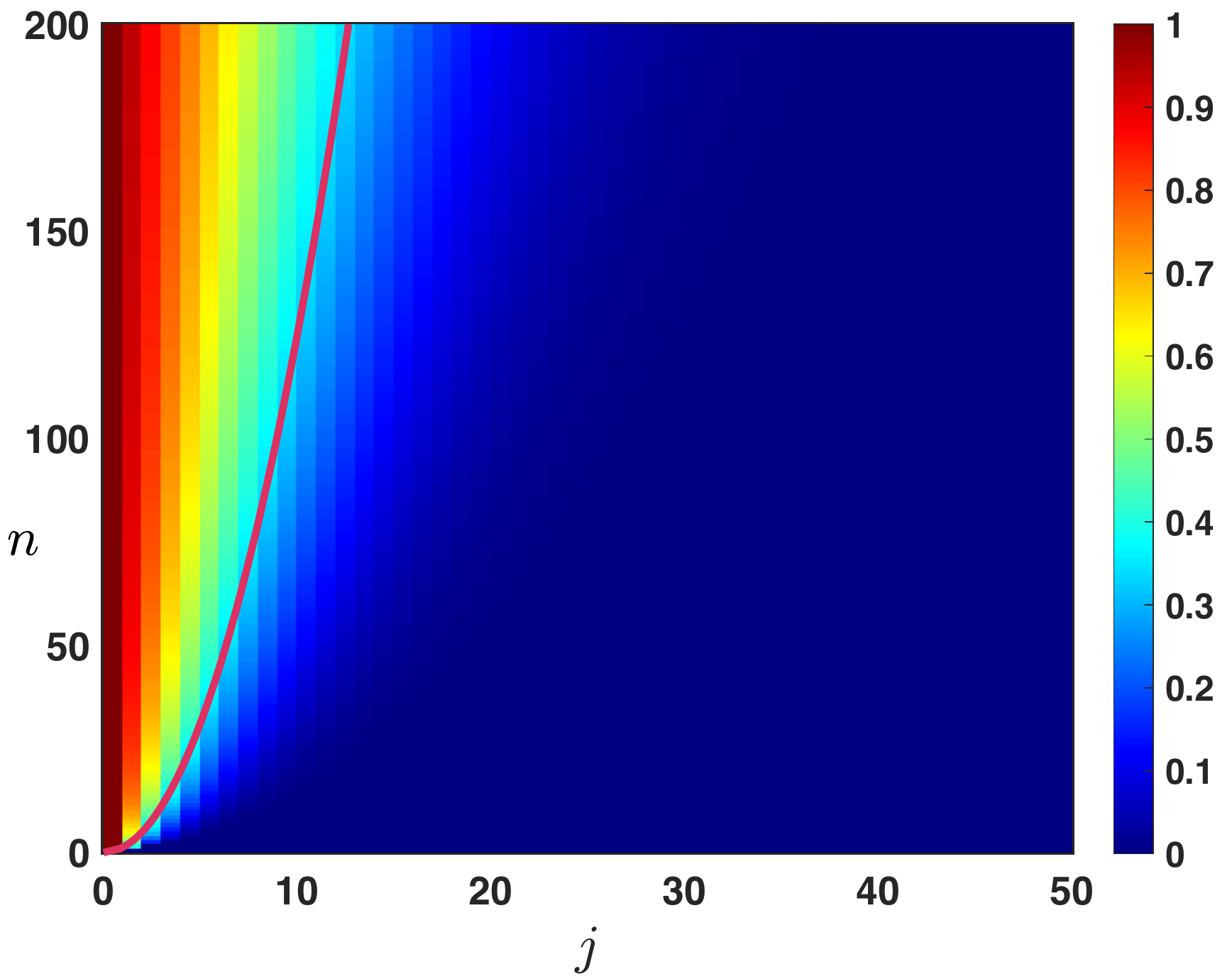}}
\subfigure[$\lambda<\alpha$.]{\includegraphics[width=.32\textwidth]{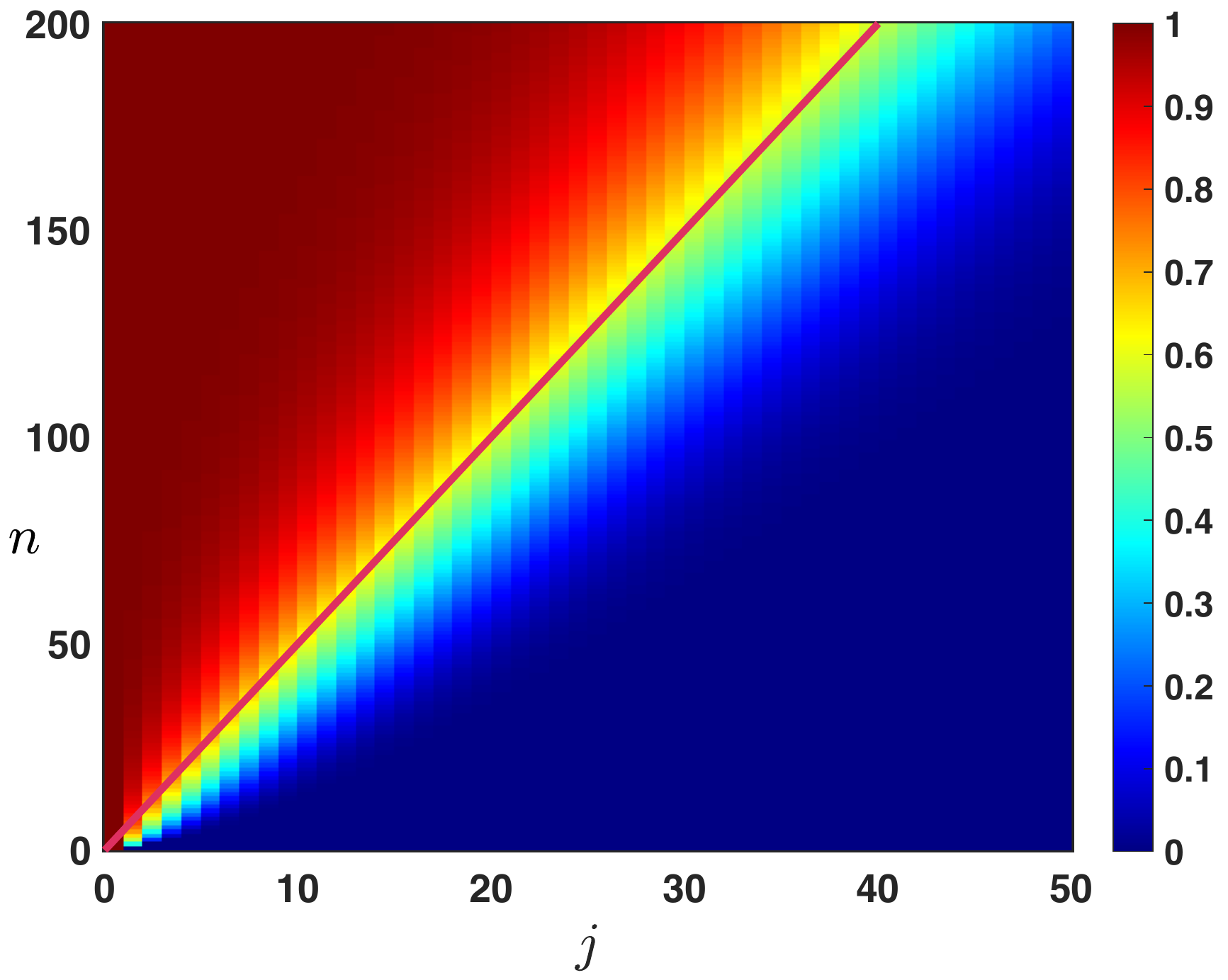}}
\subfigure[$\alpha<\lambda$.]{\includegraphics[width=.32\textwidth]{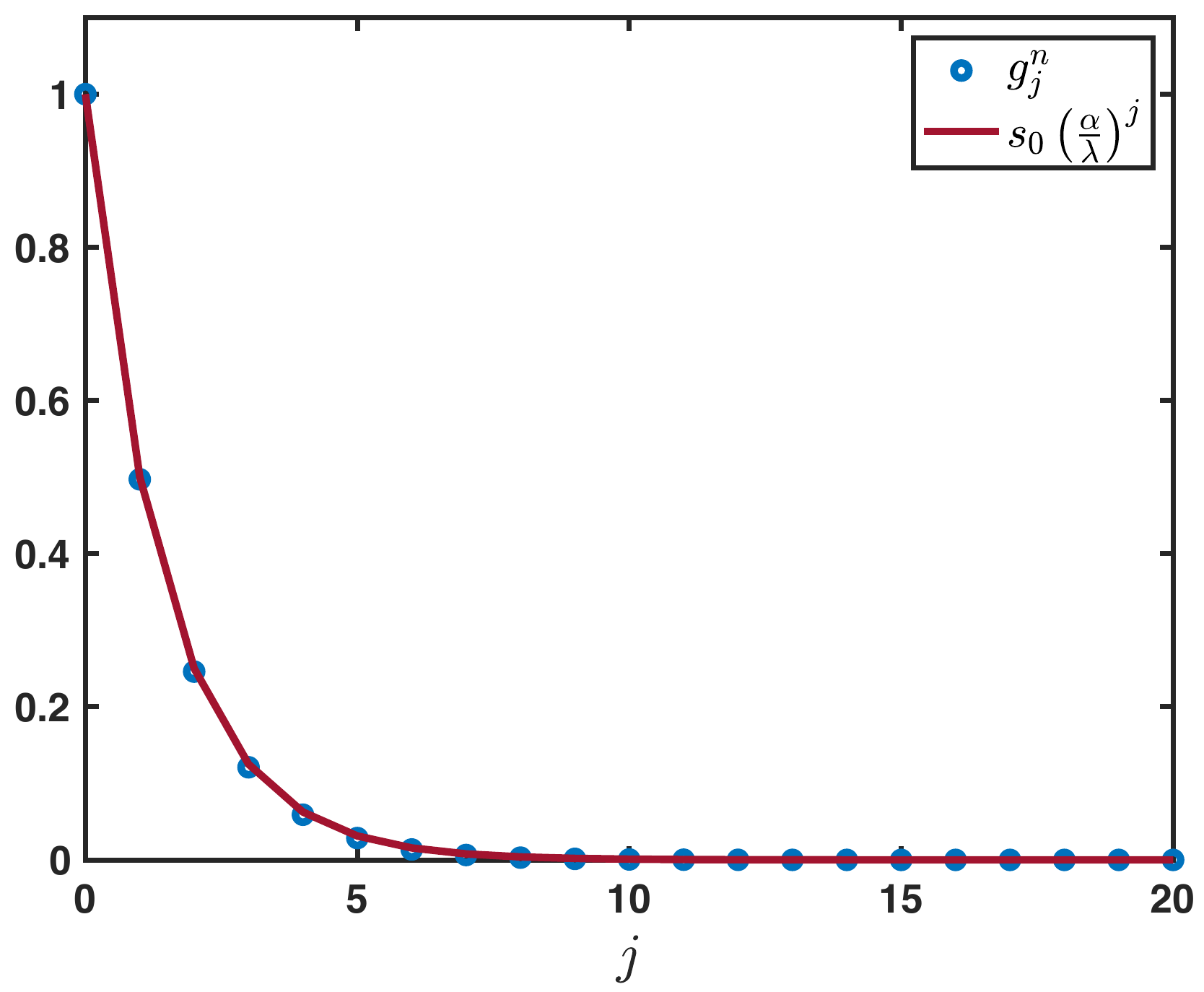}}
\subfigure[$\alpha=\lambda$.]{\includegraphics[width=.32\textwidth]{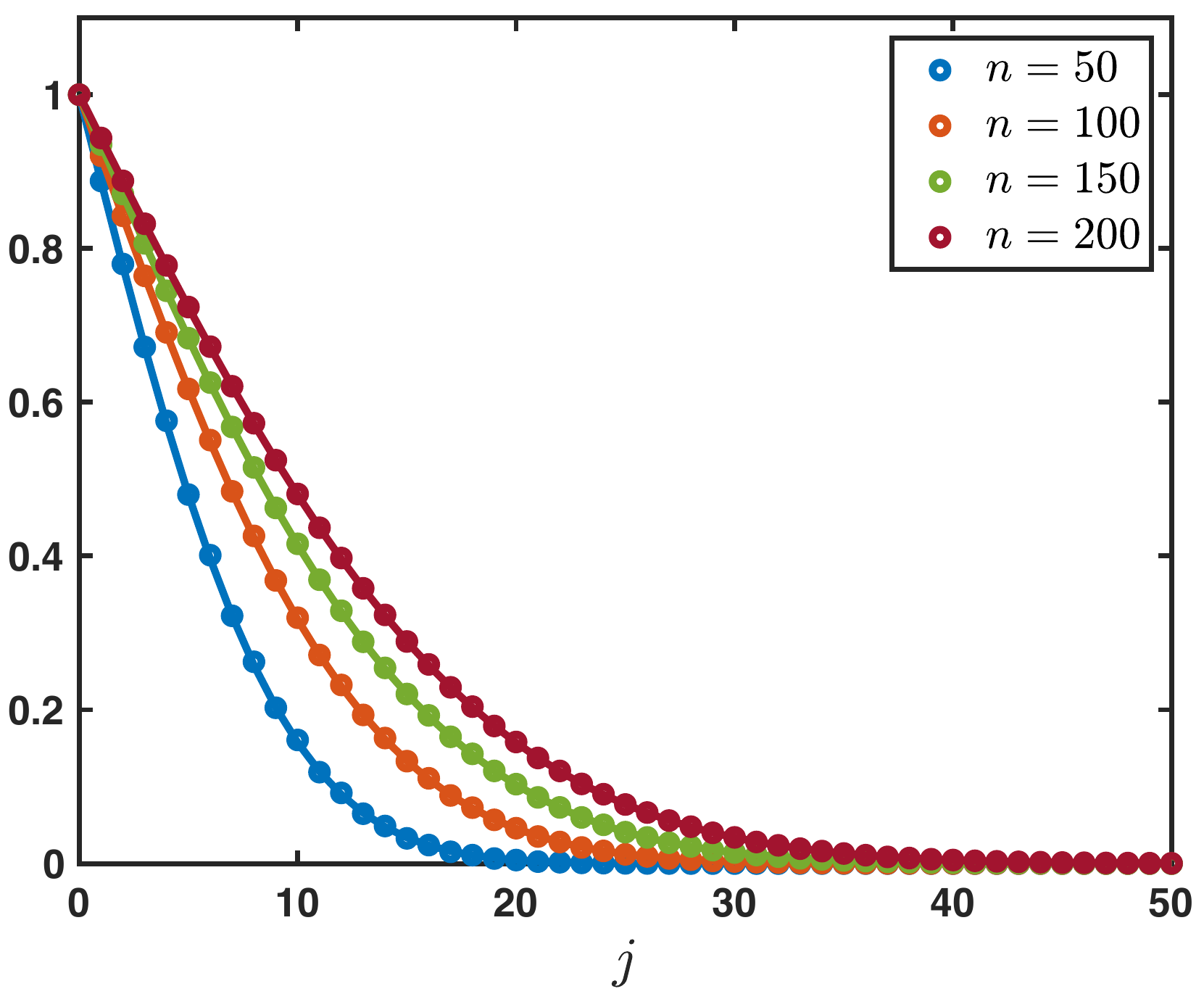}}
\subfigure[$\lambda<\alpha$.]{\includegraphics[width=.32\textwidth]{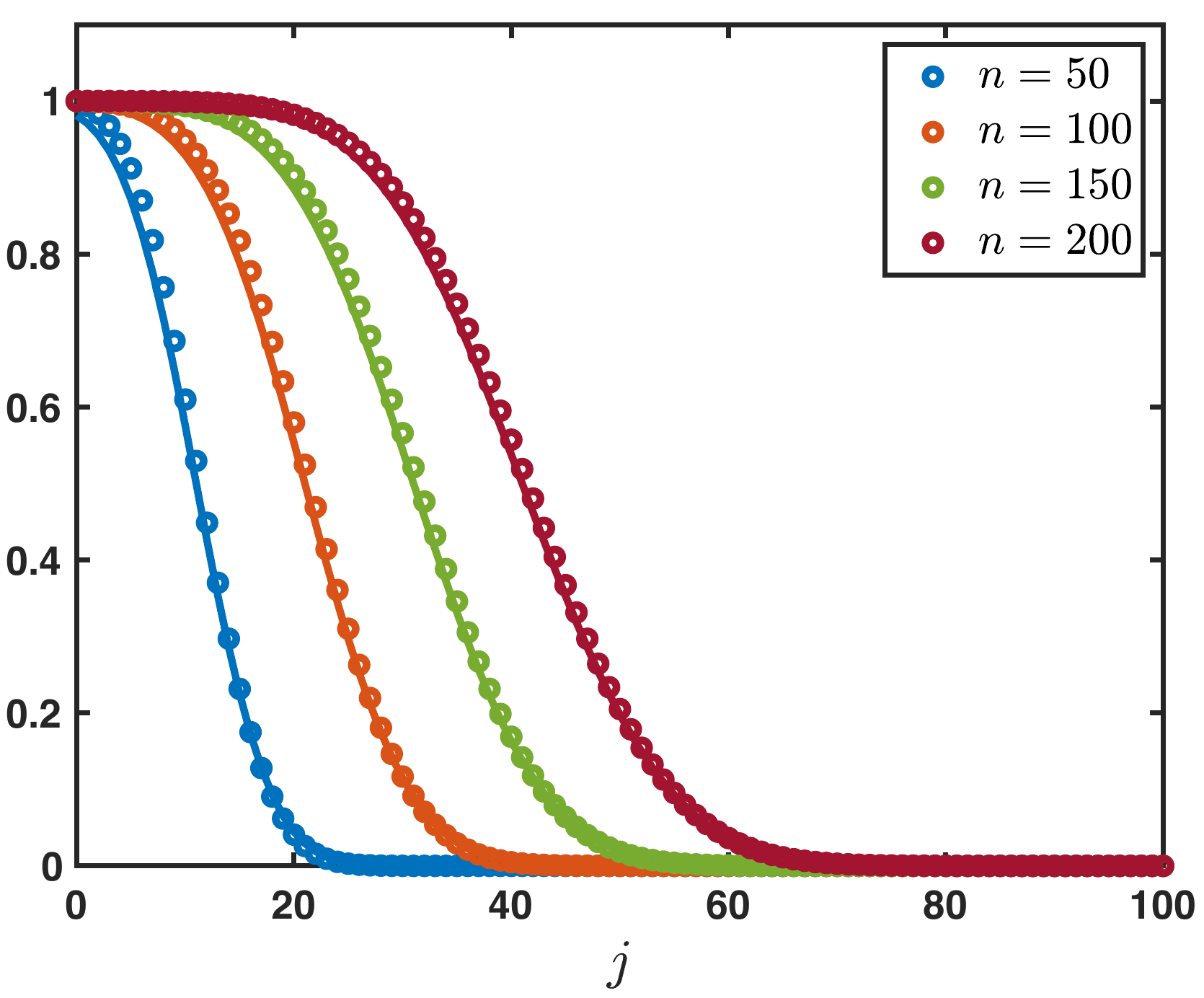}}
\caption{Case $\beta=0$ and $\alpha+\lambda\leq 1$. Visualization of the solution $(g_j^n)_{j\geq1}$ of \eqref{modelNbords0const} when $s_0^n=1$ for all $n\geq0$. Top row: space-time plots of the solution depending on $\alpha$ and $\lambda$. Bottom row: solution profiles at different time iterations (circles) compared with the leading order approximations. In the case $\alpha<\lambda$, we observe uniform convergence (in space) at exponential rate (in time) toward the  stationary solution $s_0 \left(\left(\frac{\alpha}{\lambda}\right)^j\right)_{j\geq1}$ while when $\lambda<\alpha$ we observe the propagation of a wavefront where the uniform steady state $s_0$ is propagating at speed $c_0>0$. In the intermediate case where $\alpha=\lambda$ we get a diffusive invasion as illustrated by the curve $j=\sqrt{4\sigma_0n}$.}
  \label{fig:BLbeta0}
\end{figure}

\paragraph{Case $\beta=0$.} Here, the stimulus information $s_o$ does not directly propagate through the network via its feedforward connections (since $\beta=0$), but may still propagate towards higher layers $j>0$ via the feedforward prediction error correction mechanism, governed by parameter $\alpha$. When $\alpha+\lambda\leq1$, we distinguish between three cases. Here and throughout, we denote by $\mathrm{erf}$ the error function defined by
\bqs
\mathrm{erf}(x):=\frac{2}{\sqrt{\pi}}\int_0^x e^{-z^2}\md z, \quad x\in\R.
\eqs
\begin{itemize}
\item Case $\alpha<\lambda$. In this case we have
\bqs
g_j^n =s_0 \left(\frac{\alpha}{\lambda}\right)^j\left( 1+\omega_j^n \right), \quad \text{ with } \quad  \left| \omega_j^n \right| \leq C e^{-\kappa n -\kappa j}, \quad j\geq 1, \quad n \geq 1.
\eqs
It is interesting to note that the sequence $\left( s_0 \left(\frac{\alpha}{\lambda}\right)^j \right)_{j\leq 1}$ is a stationary solution to \eqref{modelNbords0const} and we have uniform convergence at exponential rate toward this stationary solution, that is
\bqs
\underset{j\geq1}{\sup}\left| g_j^n -s_0 \left(\frac{\alpha}{\lambda}\right)^j\right| \leq C e^{-\kappa n} \underset{n\rightarrow+\infty}{\longrightarrow}0.
\eqs
We illustrate this uniform convergence in Figure~\ref{fig:BLbeta0}~(a)-(d).
\item Case $\alpha=\lambda$. We have
\bqs
\left| g_j^n -s_0\left(1-\mathrm{erf}\left(\frac{j}{\sqrt{4\sigma_0 n}} \right) \right)\right| \leq \frac{C}{n}\exp\left(-\kappa \frac{j^2}{n}\right), \quad j\geq 1, \quad n \geq 1.
\eqs
In this case, we observe a slow convergence to the steady state $s_0$. Indeed, for each $\delta\in(0,1/2)$ we have
\bqs
\underset{n\rightarrow+\infty}{\lim}\, \underset{1\leq j \leq n^{\delta}}{\sup}\,\left| g_j^n -s_0\right| =0,
\eqs
while for any $\delta>1/2$ we get
\bqs
\underset{n\rightarrow+\infty}{\lim}\, \underset{j \geq n^{\delta}}{\sup}\,\left| g_j^n \right| =0.
\eqs
The propagation is thus diffusive along $j \sim \sqrt{n}$. This can be seen in Figure~\ref{fig:BLbeta0}~(b)-(e).
\item Case $\lambda<\alpha$. In this case we have
\bqs
\left| g_j^n -\frac{s_0}{2}\left(1-\mathrm{erf}\left(\frac{j-c_0n}{\sqrt{4\sigma_0 n}} \right) \right)\right| \leq \frac{C}{\sqrt{n}}\exp\left(-\kappa \frac{(j-c_0n)^2}{n}\right), \quad j\geq 1, \quad n \geq 1.
\eqs
In this case, we deduce that we have local uniform convergence towards the steady state $s_0$, actually we have spreading at speed $c_0$. More precisely, for any $c\in(0,c_0)$ we have
\bqs
\underset{n\rightarrow+\infty}{\lim}\, \underset{1\leq j \leq cn}{\sup}\,\left| g_j^n -s_0\right| =0,
\eqs
while for any $c>c_0$, we get
\bqs
\underset{n\rightarrow+\infty}{\lim}\, \underset{ j \geq cn }{\sup}\,\left| g_j^n\right| =0.
\eqs
We refer to Figure~\ref{fig:BLbeta0}~(c)-(f) for an illustration. The figure clearly shows the competition between hyperparameters $\alpha$ and $\lambda$, with forward propagation of the sensory input only when $\alpha\geq\lambda$.
\end{itemize}

\paragraph{Case $0<\beta<1$.} Here, the stimulus information $s_o$ propagates through the network not only via its feedforward connections (governed by $\beta>0$) but also via the feedforward prediction error correction mechanism, governed by parameter $\alpha$.  In the case where $\alpha+\lambda\leq1$, the results from the case $\beta=0$ remain valid, the only differences coming from the fact that the above approximations in the case $\lambda\leq \alpha$ are only valid for $1\leq j \leq Ln$ for some large constant $L>0$ with exponential localized bounds for $j\geq Ln$ and that the steady state is now $\left(s_0 \left(\frac{\alpha+\beta}{\lambda}\right)^j\right)_{j\geq1}$ whenever $\alpha+\beta<\lambda$. This confirms that the feedforward propagation of the input $s_0$ is now dependent on both terms $\alpha$ and $\beta$, jointly competing against the feedback term $\lambda$.

\begin{figure}[t!]
\centering
\subfigure[Space-time plot.]{\includegraphics[width=.32\textwidth]{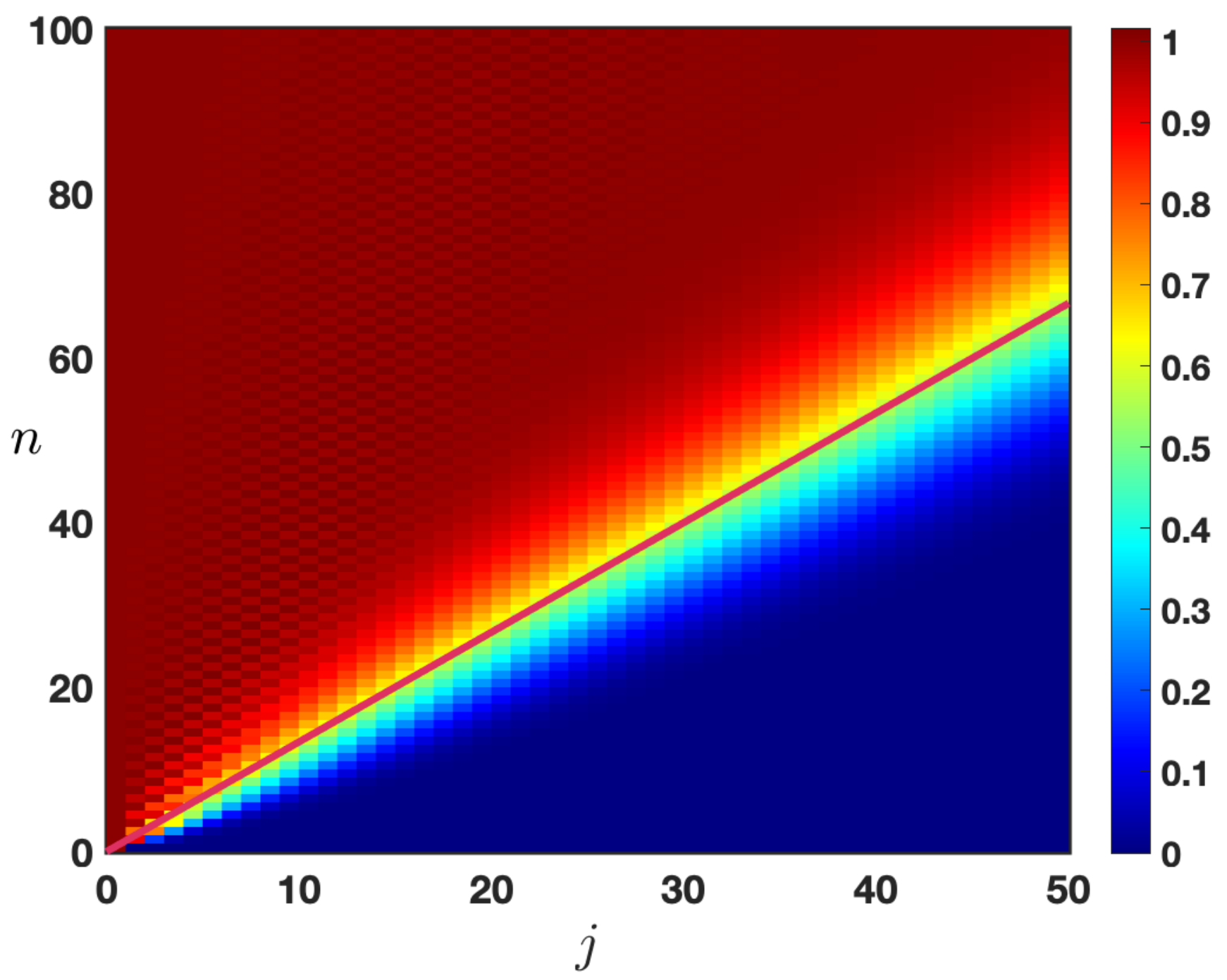}}
\subfigure[Profile of the solution at $n=70$.]{\includegraphics[width=.32\textwidth]{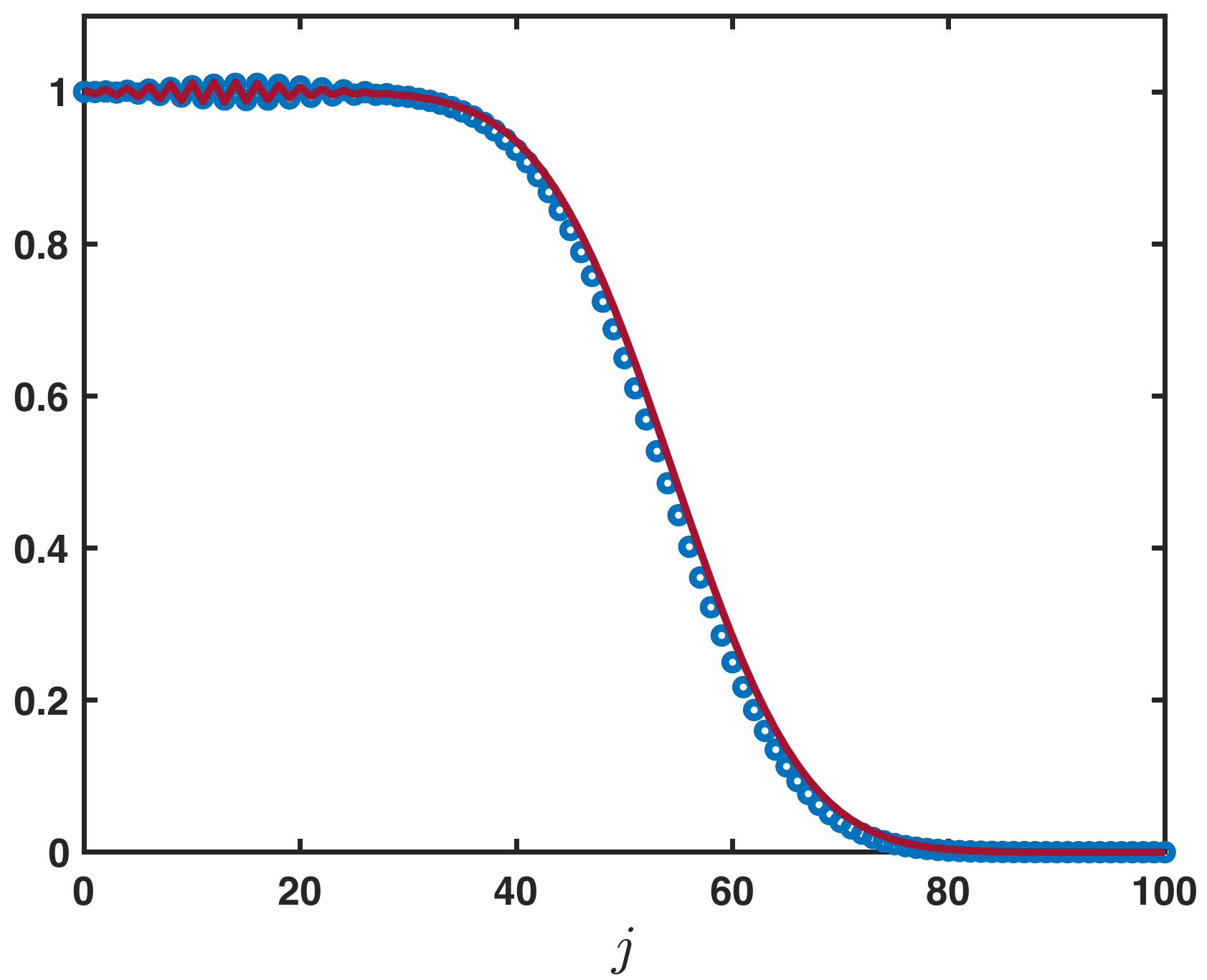}}
\subfigure[Zoom of the profile near the boundary.]{\includegraphics[width=.32\textwidth]{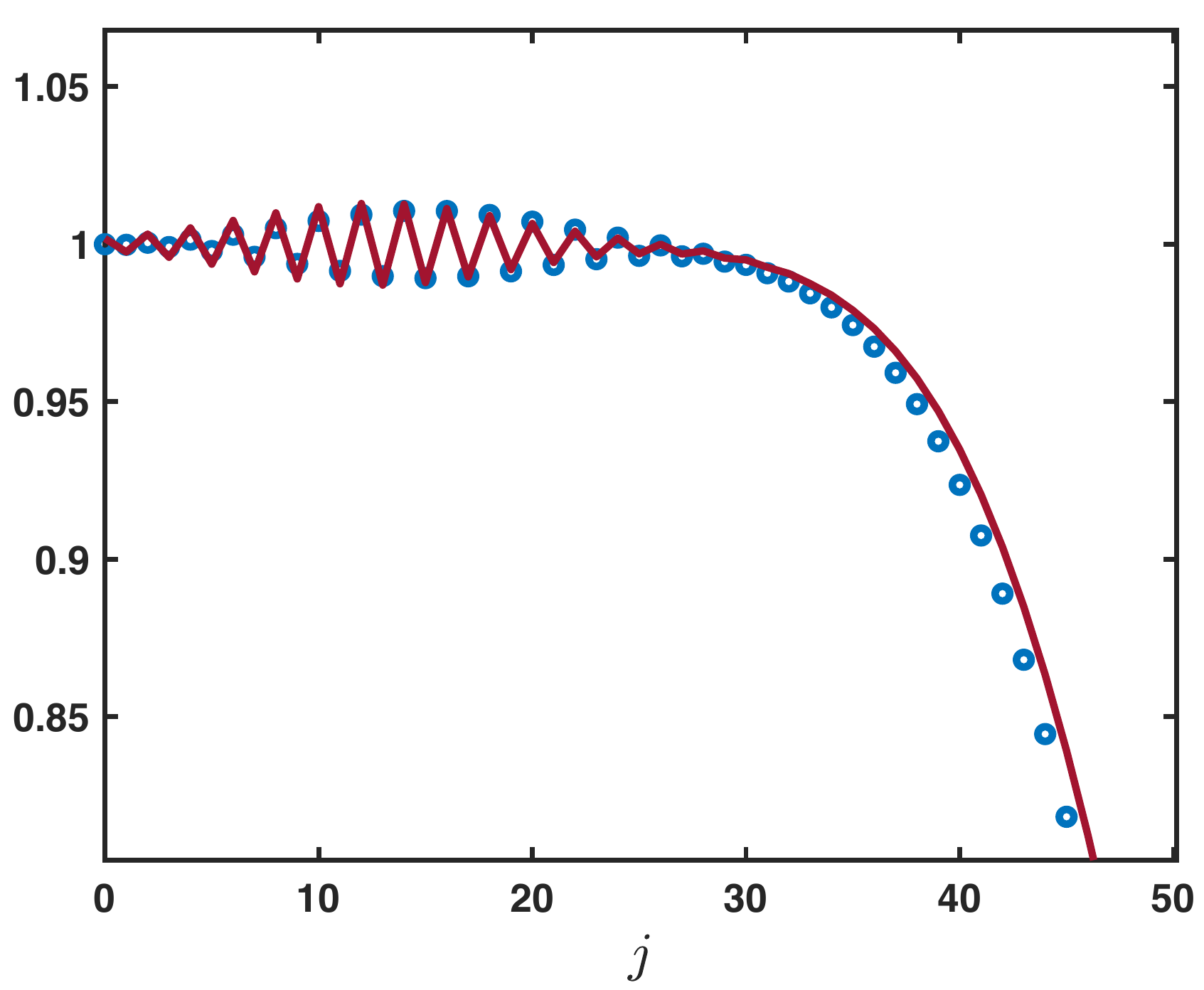}}
\caption{Case $0<\beta<1$ and $\alpha+\lambda=1$ with $0<\beta<\alpha-\lambda$. Visualization of the solution $(g_j^n)_{j\geq1}$ of \eqref{modelNbords0const} when $s_0^n=1$ for all $n\geq0$. The solution is the super-position of a leading rightward front spreading at speed $c_0>0$ and an oscillatory Gaussian profile propagating at speed $c_\pi>0$ with $c_\pi<c_0$.}
  \label{fig:BLbetageq0}
\end{figure}

Let us remark that when $0<\beta<\alpha-\lambda$ and in the special case $\alpha+\lambda=1$, where a second stable point exists for the amplification factor function at $\rho(\pi)$, we can get a slightly more accurate description of the solution in the form  
\bqs
g_j^n =\frac{s_0}{2}\left(1-\mathrm{erf}\left(\frac{j-c_0n}{\sqrt{4\sigma_0 n}} \right) \right)-\frac{s_0}{2(1+\beta)}\frac{(-1)^j}{\sqrt{4\pi \sigma_\pi n}}\exp\left(- \frac{(j-c_\pi n)^2}{4\sigma_\pi n}\right)+r_j^n, \quad n\geq1, \quad 1\leq j \leq Ln,
\eqs
where the remainder term satisfies an estimate of the form
\bqs
\left|r_j^n\right| \leq \frac{C}{\sqrt{n}}\exp\left(-\kappa \frac{(j-c_0n)^2}{n}\right)+\frac{C}{n}\exp\left(-\kappa \frac{(j-c_\pi n)^2}{n}\right).
\eqs
This is illustrated in Figure~\ref{fig:BLbetageq0}. It should be noted here that, while the main wavefront reflecting solution $c_0$ is a generic property of our network in the entire range of validity of parameters $0\leq\beta<1$ and $\alpha+\lambda\leq1$, the second oscillatory pattern reflecting $c_{\pi}$ only appears in the special case of $\beta\neq0$ and $\alpha+\lambda=1$. This oscillation is, in fact, an artifact from the discrete formulation of our problem, as will become evident in the next section, where we investigate continuous formulations of the problem.

\subsection{Towards continuous predictive models}
Starting from a discrete approximation of our system made sense, not only for mathematical convenience but also because artificial neural networks and deep learning systems implementing similar predictive coding principles are intrinsically discrete. Nonetheless, it can be useful to discard this discrete approximation and investigate our system in the continuous limit. Note that in the following, we will explore continuous extensions of our model in both time \textit{and} space. Biological neural networks, like any physical system, operate in continuous time and thus it is more biologically accurate to relax the temporal discretization assumption. This is what we do in the first part of this section. In the spatial domain, however, the discretization of our system into successive processing layers was not just an approximation, but also a reflection of the hierarchical anatomy of the brain. Nonetheless, we can still represent neuronal network depth continuously, even if only as a mathematical abstraction. This is what we will do in the subsequent part of this section. Understanding such continuous limits can allow us to test the robustness of our framework, and to relate it to canonical models whose dynamics have been more exhaustively characterized.

\subsubsection{Continuous in time interpretation} 

As a first step, we present a continuous in time interpretation of the model \eqref{modelN}.  We let $\Delta t>0$ be some parameter which will represent a time step and reformulate the recurrence equation as
\bqs
(1-\beta)\frac{e_j^{n+1}-e_j^n}{\Delta t}=\frac{\beta}{\Delta t} \left( e_{j-1}^{n+1} -e_j^{n+1}\right)+ \frac{\lambda}{\Delta t} \left( e_{j+1}^{n}-e_j^{n}\right)-\frac{\alpha}{\Delta t} \left(e_{j}^{n}-e_{j-1}^{n}\right).
\eqs
We now interpret $e_j^n$ as the approximation of some smooth function of time ${\bf e}_j(t)$ evaluated at $t_n:=n \Delta t$, that is $e_j^n \sim {\bf e}_j(t_n)$. As a consequence, we get that
\bqs
e_j^{n+1} \sim {\bf e}_j(t_{n+1})={\bf e}_j(t_{n}+\Delta t)={\bf e}_j(t_{n})+\Delta t \frac{\md }{\md t}{\bf e}_j(t_{n})+\mathcal{O}(|\Delta t|^2), \text{ as } \Delta t \rightarrow 0,
\eqs
such that
\bqs
\frac{e_j^{n+1}-e_j^n}{\Delta t} \sim \frac{\md }{\md t}{\bf e}_j(t_{n}), \text{ as } \Delta t \rightarrow 0.
\eqs
Now, introducing the scaled parameters 
\bqs
\widetilde{\beta}:=\frac{\beta}{\Delta t}, \quad \widetilde{\lambda}:=\frac{\lambda}{\Delta t}, \text{ and } \widetilde{\alpha}:=\frac{\alpha}{\Delta t},
\eqs
we get at the limit $\Delta t\rightarrow0$ the following lattice ordinary differential equation 
\bqq
\frac{\md }{\md t}{\bf e}_j(t)=(\widetilde{\beta}+\widetilde{\alpha}){\bf e}_{j-1}(t)-(\widetilde{\beta}+\widetilde{\alpha}+\widetilde{\lambda}){\bf e}_j(t)+\widetilde{\lambda}{\bf e}_{j+1}(t), \quad t>0.
\label{latticeODE}
\eqq
When defined on the infinite lattice $\Z$, one can represent the solutions as
\bqs
{\bf e}_j(t)=\sum_{k\in\Z}\mathbf{G}_{j-k}(t)h_k, \quad j\in\Z, \quad t>0,
\eqs
starting from the initial sequence ${\bf e}_j(t=0)=(h_j)_{j\in\Z}$ where $(\mathbf{G}_{j}(t))_{j\in\Z}$ is the fundamental solution to \eqref{latticeODE} starting from the Dirac delta sequence $\boldsymbol{\delta}$. Once again, each  $\mathbf{G}_{j}(t)$ can be represented by the inverse Fourier transform and reads
\bqs
\mathbf{G}_{j}(t)=\frac{1}{2\pi}\int_{-\pi}^{\pi} e^{\nu(\theta)t}e^{\mbi j \theta}\md \theta, \quad t>0, \quad j\in\Z,
\eqs
where the function $\nu(\theta)$ is defined as
\bqs
\nu(\theta):=(\widetilde{\beta}+\widetilde{\alpha})e^{-\mbi \theta}-(\widetilde{\beta}+\widetilde{\alpha}+\widetilde{\lambda})+\widetilde{\lambda}e^{\mbi \theta}, \quad \theta\in[-\pi,\pi].
\eqs
The function $\nu(\theta)$ serves as an amplification factor function for the time continuous equation \eqref{latticeODE}. To ensure stability\footnote{Note that the notions of stability/unstability and marginal stability introduced in the fully discrete setting naturally extend to the semi-continuous setting.}, one needs to impose that $\Re(\nu(\theta))\leq 0$ for each $\theta\in[-\pi,\pi]$. From its formula, we obtain that
\bqs
\Re(\nu(\theta))=(\widetilde{\beta}+\widetilde{\alpha}+\widetilde{\lambda})(\cos(\theta)-1), \quad \theta\in[-\pi,\pi],
\eqs
such that we deduce that $\Re(\nu(0))=0$ and $\Re(\nu(\theta))<0$ for all $\theta\in[-\pi,\pi]\backslash\{0\}$. In particular, it is now evident that, contrary to the discrete case, $\nu(\pi)$ cannot be a stable solution for the continuous system (except in the trivial case where all hyperparameters $\widetilde{\alpha},\widetilde{\beta},\widetilde{\lambda}$ are zero). This confirms that the previously observed oscillations associated with $\rho(\pi)$ in specific cases were merely an artifact of the temporal discretization. 

We note that, near the tangency point at $\theta=0$, the function $\nu(\theta)$ has the following asymptotic expansion
\bqs
\nu(\theta)=-(\widetilde{\beta}+\widetilde{\alpha}+\widetilde{\lambda})\mbi\theta -\frac{\widetilde{\beta}+\widetilde{\alpha}+\widetilde{\lambda}}{2}\theta^2+\mathcal{O}(|\theta|^3), \text{ as } \theta\rightarrow 0.
\eqs
It is also possible to prove a Gaussian approximation in that case, and following for example \cite{BFRZ}, we have 
\bqs
\mathbf{G}_{j}(t)=\frac{1}{\sqrt{4\pi \widetilde{\sigma_0}t}}\exp\left(-\frac{|j-\widetilde{c_0}t|^2}{4\widetilde{\sigma_0}t}\right)+\mathbf{R}_{j}(t), \quad j\in\Z, \quad t>0,
\eqs
with
\bqs
\left|\mathbf{R}_{j}(t) \right| \leq \frac{C}{\sqrt{t}}\exp\left(-\kappa \frac{|j-\widetilde{c_0}t|^2}{t}\right), \quad j\in\Z, \quad t>0,
\eqs
for some universal constants $C>0$ and $\kappa>0$. Here, $\widetilde{c_0}$ and $\widetilde{\sigma_0}$ are given by
\bqs
\widetilde{c_0}=\widetilde{\beta}+\widetilde{\alpha}-\widetilde{\lambda}, \quad \text{ and } \quad \widetilde{\sigma_0}=\frac{\widetilde{\beta}+\widetilde{\alpha}+\widetilde{\lambda}}{2}>0.
\eqs
We remark that both $\widetilde{c_0}$ and $\widetilde{\sigma_0}$ are linked to $c_0$ and $\sigma_0$ (the propagation speed and spread of the solution in the case of the discrete model) in the following sense
\bqs
\frac{c_0}{\Delta t}\rightarrow \widetilde{c_0}, \quad \text{ and } \quad \frac{\sigma_0}{\Delta t}\rightarrow \widetilde{\sigma_0}, \quad \text{ as } \Delta t\rightarrow0.
\eqs
We also note that the spatially homogeneous solutions of \eqref{latticeODE} are trivial in the sense that if we assume that ${\bf e}_j(t)={\bf e}(t)$ for all $j\in\Z$ then the equation satisfied by ${\bf e}(t)$ is simply
\bqs
\frac{\md }{\md t}{\bf e}(t)=0.
\eqs
Finally, we conclude by noticing that in this continuous in time regime, there is no possible oscillations either in space or time, in the sense that the fundamental solution always resembles a fixed Gaussian profile advected at wave speed $\widetilde{c_0}$. The formula for $\widetilde{c_0}$ highlights the intuitive functional relation between the propagation (or advection) direction and the ``competition'' between the feedforward influences $\widetilde{\alpha}+\widetilde{\beta}$ and the feedback influence $\widetilde{\lambda}$.

\subsubsection{Fully continuous  interpretation: both in time and depth} \label{sec:continuoustimedepth}

In this section, we give a possible physical interpretation of the discrete model \eqref{modelN} via continuous transport equations, in which both time and space (i.e., neuronal network depth) are made continuous. Let us introduce $\Delta t>0$, $\Delta x>0$ and set $\nu:=\frac{\Delta x}{\Delta t}$. As before, we can view ${\Delta t}$ as a time step for our system; additionally, ${\Delta x}$ can be viewed as a spatial step in the (continuous) neuronal depth dimension, and thus $\nu$ becomes akin to a neural propagation speed or a conduction velocity. We then rewrite the recurrence equation as
\bqs
(1-\beta)\frac{e_j^{n+1}-e_j^n}{\Delta t}=\beta \nu \frac{e_{j-1}^{n+1} -e_j^{n+1}}{\Delta x}+ \lambda \nu \frac{ e_{j+1}^{n}-e_j^{n}}{\Delta x}-\alpha \nu \frac{ e_{j}^{n}-e_{j-1}^{n}}{\Delta x}.
\eqs
The key idea is to now assume that $e_j^n$ represents an approximation of some smooth function ${\bf e}(t,x)$ evaluated at $t_n:=n \Delta t$ and $x_j:=j \Delta x$, that is $e_j^n \sim {\bf e}(t_n,x_j)$. Then passing to the limit $\Delta t\to 0$, $\Delta x \to 0$ with $\frac{\Delta x}{\Delta t}=\nu>0$ fixed and assuming that $\beta+\alpha\neq \lambda$, one gets the partial differential equation
\bqq
\partial_t {\bf e}(t,x) + \frac{\nu (\beta+\alpha-\lambda)}{1-\beta } \partial_x {\bf e}(t,x) =0, \quad t>0, \quad x>0,
\label{transport}
\eqq
with boundary condition ${\bf e}(t,x=0)=s_0(t)$ and initial condition ${\bf e}(t=0,x)=h(x)$ satisfying the compatibility condition $s_0(0)=h(0)$ where $s_0(t)$ is a smooth function such that $s_0(t^n)=s_0^n$ and $h(x_j)=h_j$. The above partial differential equation is a transport equation with associated speed $\frac{\nu (\beta+\alpha-\lambda)}{1-\beta }=\nu c_0$. Depending on the sign of $c_0$, we have a different representation for the solutions of \eqref{transport}. 
\begin{itemize}
\item Case $c_0<0$. Solution is given by
\bqs
{\bf e} (t,x)=h\left( x- \nu c_0  t\right), \quad t>0, \quad x>0.
\eqs
Let us remark that when $c_0<0$ the trace of the solution at $x=0$ is entirely determined by the initial data $h(x)$ since
\bqs
{\bf e} (t,x=0)=h\left(- \nu c_0  t\right), \quad t>0.
\eqs
Intuitively, this reflects the dominance of backward (leftward) propagation in this network, with solutions determined entirely by the initial value $h(x)$, even for $x=0$ (the source term, $s_0(t)$, having no influence in this case).
\item Case $c_0>0$. Solution is given by
\bqs
{\bf e} (t,x)=\left\{
\begin{array}{lc}
s_0\left(t-\frac{x}{\nu c_0}\right), & x \leq \nu c_0 t, \\
h\left( x- \nu c_0  t\right), & x > \nu c_0  t,
\end{array}
\right. \quad t>0, \quad x>0.
\eqs
Intuitively, this reflects the dominance of forward (rightward) propagation in this network, with both the source term $s_0(t)$ and the initial values $h(x)$ transported at constant velocity $\nu c_0$.
\end{itemize}

Thanks to the explicit form of the solutions, we readily obtain many qualitative properties of the solution ${\bf e} (t,x)$. Boundedness and positivity of the solutions are inherited from the functions $s_0(t)$ and $h(x)$. In the case where $\beta+\alpha= \lambda$ (i.e., with balanced feed-forward and feedback influences), the limiting equation is slightly different. Indeed, in this case, introducing $\delta:=\frac{\Delta x^2}{\Delta t}$ and letting $\Delta t\to 0$, $\Delta x \to 0$ with $\delta>0$ fixed, on gets the partial differential equation
\bqq
\partial_t {\bf e}(t,x) = \frac{\delta (\beta+\alpha+\lambda)}{2(1-\beta) } \partial_x^2 {\bf e}(t,x) , \quad t>0, \quad x>0,
\label{heat}
\eqq
and we readily observe that when $\beta+\alpha= \lambda$, we have that
\bqs
\frac{ \beta+\alpha+\lambda}{2(1-\beta) } = \frac{\beta(1-\alpha-\lambda)+\alpha+\lambda-(\lambda-\alpha)^2}{2(1-\beta)^2} =\sigma_0>0.
\eqs
We obtain a heat equation with a boundary condition ${\bf e}(t,x=0)=s_0(t)$ and initial condition ${\bf e}(t=0,x)=h(x)$. Upon denoting 
\bqs
\mathcal{S}(t,x,y):=\frac{1}{\sqrt{4\pi \delta \sigma_0 t}}\left(e^{-\frac{(x-y)^2}{4\delta \sigma_0 t}}-e^{-\frac{(x+y)^2}{4\delta \sigma_0 t}}\right),
\eqs
the solution of the equation is given by
\bqs
{\bf e}(t,x)=s_0(t)+\int_0^{+\infty}\mathcal{S}(t,x,y)\left(h(y)-s_0(0)\right)\md y-\int_0^t \int_0^{+\infty}\mathcal{S}(t-s,x,y)s_0'(s)\md y\md s, \quad t>0, \quad x>0,
%&=\int_0^{+\infty}\mathcal{S}(t,x,y)h(y)\md y+\int_0^t \int_0^{+\infty}\partial_s\mathcal{S}(t-s,x,y)s_0(s)\md y\md s.
\eqs
Let us remark that when $s_0(t)=s_0\in\R$ is constant for all $t\geq0$, the above expression simplifies to
\bqs
{\bf e}(t,x)=s_0\left(1-\mathrm{erf}\left(\frac{x}{\sqrt{4\delta \sigma_0 t}} \right)\right)+\int_0^{+\infty}\mathcal{S}(t,x,y)h(y)\md y, \quad t>0, \quad x>0.
\eqs

In conclusion, this section extended our discrete model towards a continuous limit in both space and time. In the temporal domain, it allowed us to understand our stable solution as an advection behavior, and alerted us that the other apparently oscillatory solutions previously observed in specific cases were mainly due to our discretization approximation. In the spatio-temporal domain, the continuous limit \eqref{transport} allowed us to realize that our main equation \eqref{model1d} was merely a discrete version of a transport equation. \\\
In the following sections, we will systematically return to discrete implementations (with gradually increasing functionality), before considering, again, their continuous formulations.

\section{Beyond the identity case}

In the previous section we have studied in depth the case where $\W^f$ and $\W^b$ are both the identity matrix: each neuron in any given layer directly conveys its activation value to a single corresponding neuron in the next layer, and to a single neuron in the previous layer. Motivated by concrete implementations of the model in deep neural networks \cite{Wen18,choski21}, we aim to investigate more realistic situations with more complex connectivity matrices. While the generic unconstrained case (i.e. two unrelated and dense connection matrices $\W^f$ and $\W^b$) does not easily lend itself to analytical study, we will consider here two situations of practical interest: in the first one, the forward and backward connection matrices are symmetric and identical; in the second case, each matrix is symmetric, but the two are not necessarily identical.

\subsection{The symmetric Rao \& Ballard case}

Following the pioneering work of Rao \& Ballard \cite{RB99}, we will assume in this section that $\W^f=(\W^b)^{\bf t}$ and $\W^f$ is symmetric, which implies that
\bqs
\W^f=\W^b\in \mathscr{S}_d(\R),
\eqs
where we denoted $\mathscr{S}_d(\R)$ the set of symmetric matrices on $\R^d$. 

The underlying interpretation is that, if a strong synaptic connection exists from neuron $a$ to neuron $b$, then there is also a strong connection from $b$ to $a$. This assumption, which follows from Hebbian plasticity rules (``neurons that fire together wire together'') does not capture all of the diversity of brain connectivity patterns, but can be considered a good first approximation \cite{RB99}. 

\subsubsection{Neural basis change and neural assemblies}

The spectral theorem ensures that $\W^f$ (and thus $\W^b$) is diagonalizable in an orthonormal basis. Namely, there exists an orthogonal invertible matrix $P\in\mathscr{M}_d(\R)$ such that $PP^{\bf t}=P^{\bf t}P=\mathbf{I}_d$ and there exists a diagonal matrix denoted $\D\in\mathscr{M}_d(\R)$ such that
\bqs
\W^f=\W^b=P \D P^{\bf t}.
\eqs
We denote by $\gamma_p\in\R^d$, $p=1,\cdots,d$ the diagonal elements of $\D$ and without loss of generality we may assume that
\bqs
\gamma_1\leq \cdots \leq \gamma_d.
\eqs
Thanks to this diagonalization, we can now perform a change of basis for our neuronal space. We set $\U_j^n:= P^t\E_j^n$ as the new basis, with $P\U_j^n:= \E_j^n$. Each $\U_j^n$ can now be understood as a neural \textit{assembly}, reflecting one of the principal components of the weight matrix $\W^f=\W^b$. Importantly, although assemblies may overlap, activity updates induced by feedforward or feedback connections to one given assembly do not affect the other assemblies, since the matrix $P$ is orthogonal. Therefore, our problem is much simplified when considering activity update equations at the level of these neural assemblies $\U_j^n$ rather than across individual neurons $\E_j^n$. Our model \eqref{model} becomes
\bqs
\U_j^{n+1}=\beta \D \U_{j-1}^{n+1} +\alpha \D\U_{j-1}^n+\left[(1-\beta-\lambda)\mathbf{I}_d-\alpha \D^2 \right] \U_j^{n} + \lambda  \D \U_{j+1}^{n}. %\quad j=1,\cdots J-1, \quad n\geq0,
\eqs
Note that, because all matrices in the above equation are diagonal, we have totally decoupled the $d$ components of the vector $\U_j^n$. More precisely, by denoting $u_{j,p}^{n}$ the $p$th component of $\U_j^n$, that is $\U_j^n=(u_{j,1}^{n},\cdots,u_{j,d}^{n})^\mathbf{t}$, we obtain
\bqs
u_{j,p}^{n+1}-\beta \gamma_p u_{j-1,p}^{n+1} =\alpha \gamma_p u_{j-1,p}^n +(1-\beta-\lambda-\alpha \gamma_p^2) u_{j,p}^{n} +\lambda  \gamma_p u_{j+1,p}^{n}, \quad p=1,\cdots,d.
\eqs
This indicates that one needs to study 
\bqq
u_{j}^{n+1}-\beta \gamma u_{j-1}^{n+1} =\alpha \gamma u_{j-1}^n +(1-\beta-\lambda-\alpha \gamma^2) u_{j}^{n} +\lambda  \gamma u_{j+1}^{n}, %\quad j=1,\cdots J-1, \quad n\geq0,
\label{modelgam}
\eqq
where $\gamma\in\R$ is a given parameter. Here, $\gamma$ can be thought of as the connection strength across layers (both feedforward and feedback, since we assumed here symmetric connectivity) of the neural assembly under consideration. By construction, each assembly in a given layer is only connected to the corresponding assembly in the layer above, and similarly in the layer below. Note that when $\gamma=1$, we encounter again the exact situation that we studied in the previous section (\ref{modelN}), but now with neural assemblies in lieu of individual neurons. 

\subsubsection{Study of the amplification factor function}

Based on our previous analysis, the behavior of the solutions to \eqref{modelgam} are intrinsically linked to the properties of the amplification factor function:
\bqs
\rho_\gamma(\theta) = \frac{\alpha \gamma \left(e^{-\mbi \theta}-\gamma\right) +1-\beta +\lambda\left(\gamma e^{\mbi \theta}-1\right)}{1-\beta \gamma e^{-\mbi \theta}}, \quad \theta\in[-\pi,\pi],
\eqs
where one needs to ensure that $|\rho_\gamma(\theta)|\leq 1$ for all $\theta\in[-\pi,\pi]$. The very first condition is to ensure that $1\neq \beta |\gamma|$ (to avoid division by zero). Next, we investigate the behavior of $\rho_\gamma(\theta)$ at $\theta=0$ and check under which condition on $\gamma$ we can ensure that $-1\leq \rho_\gamma(0)\leq 1$. We have
\bqs
\rho_\gamma(0)=1+\frac{(1-\gamma)(\alpha\gamma-\lambda-\beta)}{1-\beta \gamma},
\eqs
which readily tells us that $\rho_\gamma(0)=1$ if and only if $\gamma=1$ or $\gamma=\frac{\lambda+\beta}{\alpha}$. And on the other hand $\rho_\gamma(0)=-1$ if and only if $\gamma=\gamma_\pm^0$ with 
\bqs
\gamma_\pm^0=\frac{\lambda+\alpha-\beta \pm \sqrt{(\lambda+\alpha-\beta)^2+4\alpha(2-\lambda-\beta)}}{2\alpha},
\eqs
with $\gamma_-^0<0<\gamma_+^0$ since $\lambda+\beta<2$ by assumption. One also notices that
\bqs
(\lambda+\alpha-\beta)^2+4\alpha(2-\lambda-\beta)=(\lambda+3\alpha-\beta)^2+8\alpha(1-\alpha-\lambda),
\eqs
such that either $\alpha+\lambda=1$ and in that case $\gamma_-^0=-1$ and $\gamma_+^0=1+\frac{1-\beta}{\alpha}>1$, or $\alpha+\lambda<1$ and in that case $\gamma_-^0<-1$ and $\gamma_+^0>1+\frac{1-\beta}{\alpha}>1$. Next, we remark that
\bqs
\rho_\gamma(\pi)=1-\frac{(1+\gamma)(\alpha\gamma+\lambda+\beta)}{1+\beta \gamma}=\rho_{-\gamma}(0),
\eqs
which then implies that $\rho_\gamma(\pi)=1$ if and only if $\gamma=-1$ or $\gamma=-\frac{\lambda+\beta}{\alpha}$ and $\rho_\gamma(\pi)=-1$ if and only if $\gamma=-\gamma_\pm^0$. 

As explained in the beginning, our aim is to completely characterize under which conditions on $\gamma\in\R$, $0\leq \beta<1$ and $0<\alpha,\lambda<1$ with $\alpha+\lambda\leq1$, one can ensure that $|\rho_\gamma(\theta)|\leq 1$ for all $\theta\in[-\pi,\pi]$.

Potential regions of marginal stability are thus given by those values of the parameters satisfying $\gamma=\pm1$, $\gamma=\pm\frac{\lambda+\beta}{\alpha}$, $\gamma=\pm\gamma_+^0$ and $\gamma=\pm\gamma_-^0$, and it is important to determine the intersections among the above regions. We have already proved that $\gamma_-^0=-1$ whenever $\alpha=1-\lambda$. Next, we compute that $\gamma_-^0=-\frac{\lambda+\beta}{\alpha}$ whenever $\alpha=\frac{\lambda(\lambda+\beta)}{1-\lambda-\beta}=:\Lambda$, while $\gamma_-^0=-\gamma_+^0$ whenever $\alpha=\beta-\lambda$ and $\gamma_+^0=\frac{\lambda+\beta}{\alpha}$ when $\alpha=\beta(\lambda+\beta)$. Let us already point out that $\Lambda$ is only defined if $\lambda+\beta<1$ and in that case $\Lambda>0$.

\begin{figure}[t!]
\centering
\subfigure[]{\includegraphics[width=.29\textwidth]{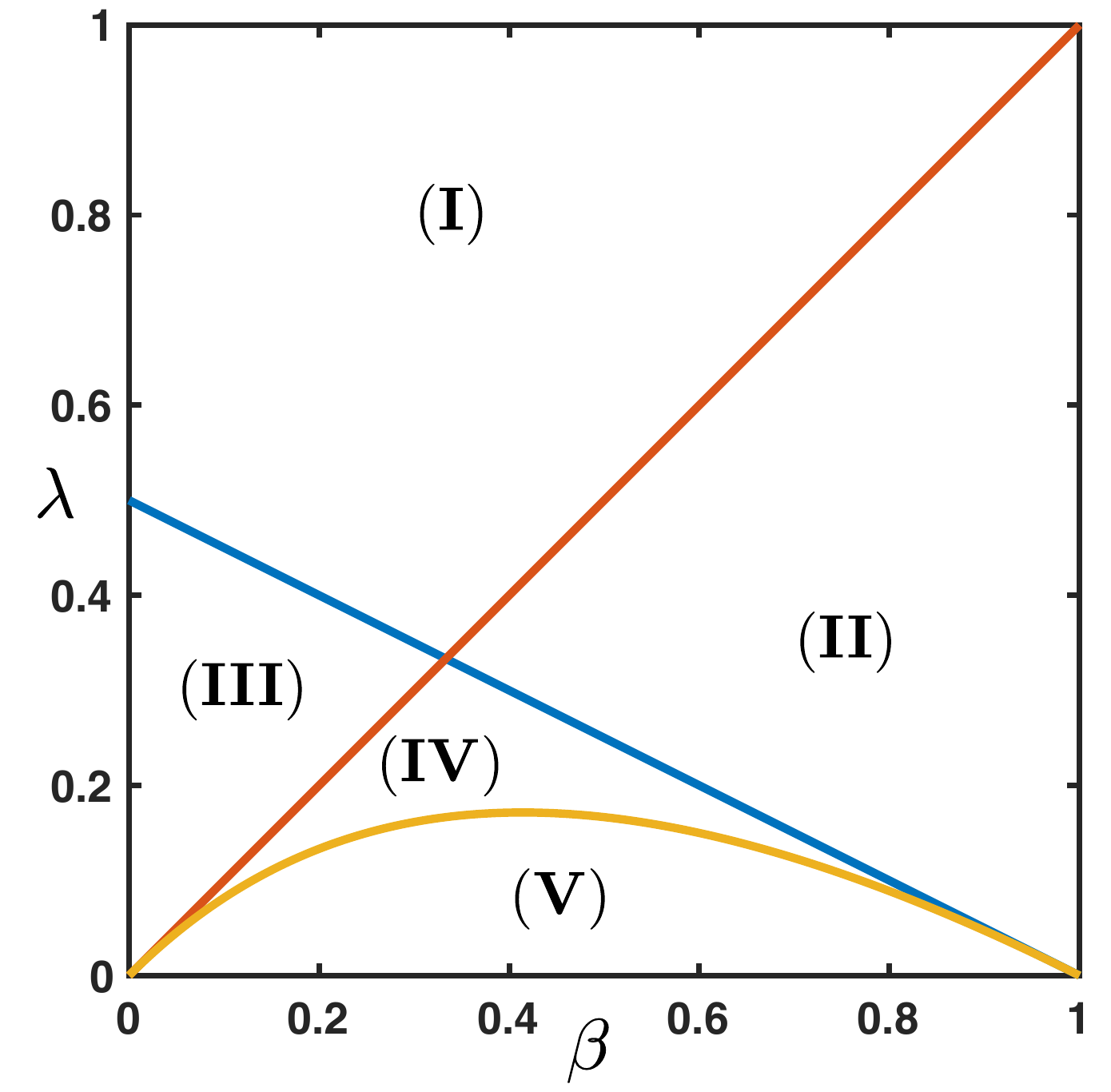}}
\subfigure[$(\lambda,\beta)$ in Region $\mathbf{(I)}$.]{\includegraphics[width=.3\textwidth]{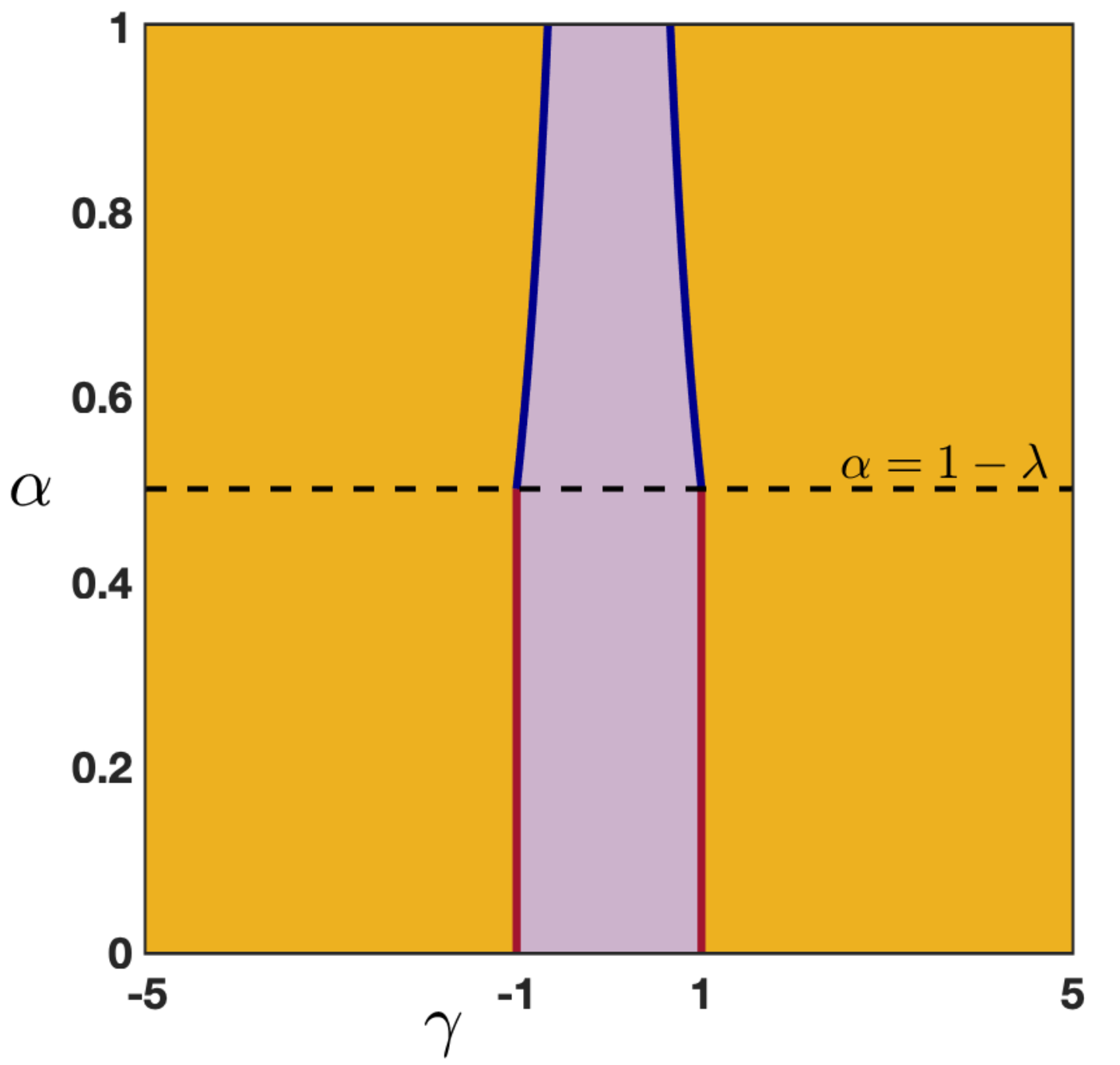}}
\subfigure[$(\lambda,\beta)$ in Region $\mathbf{(II)}$.]{\includegraphics[width=.3\textwidth]{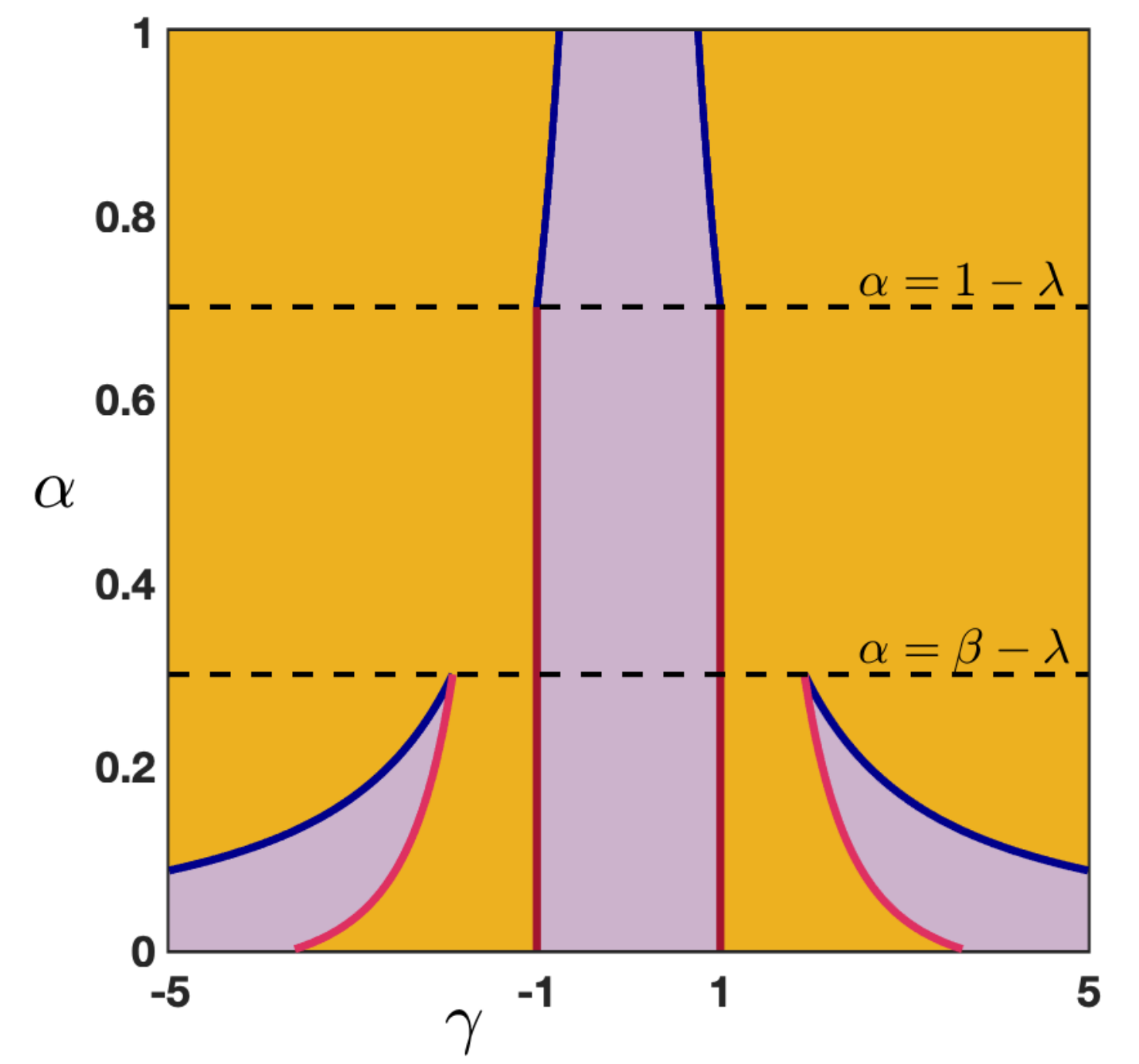}}
\subfigure[$(\lambda,\beta)$ in Region $\mathbf{(III)}$.]{\includegraphics[width=.3\textwidth]{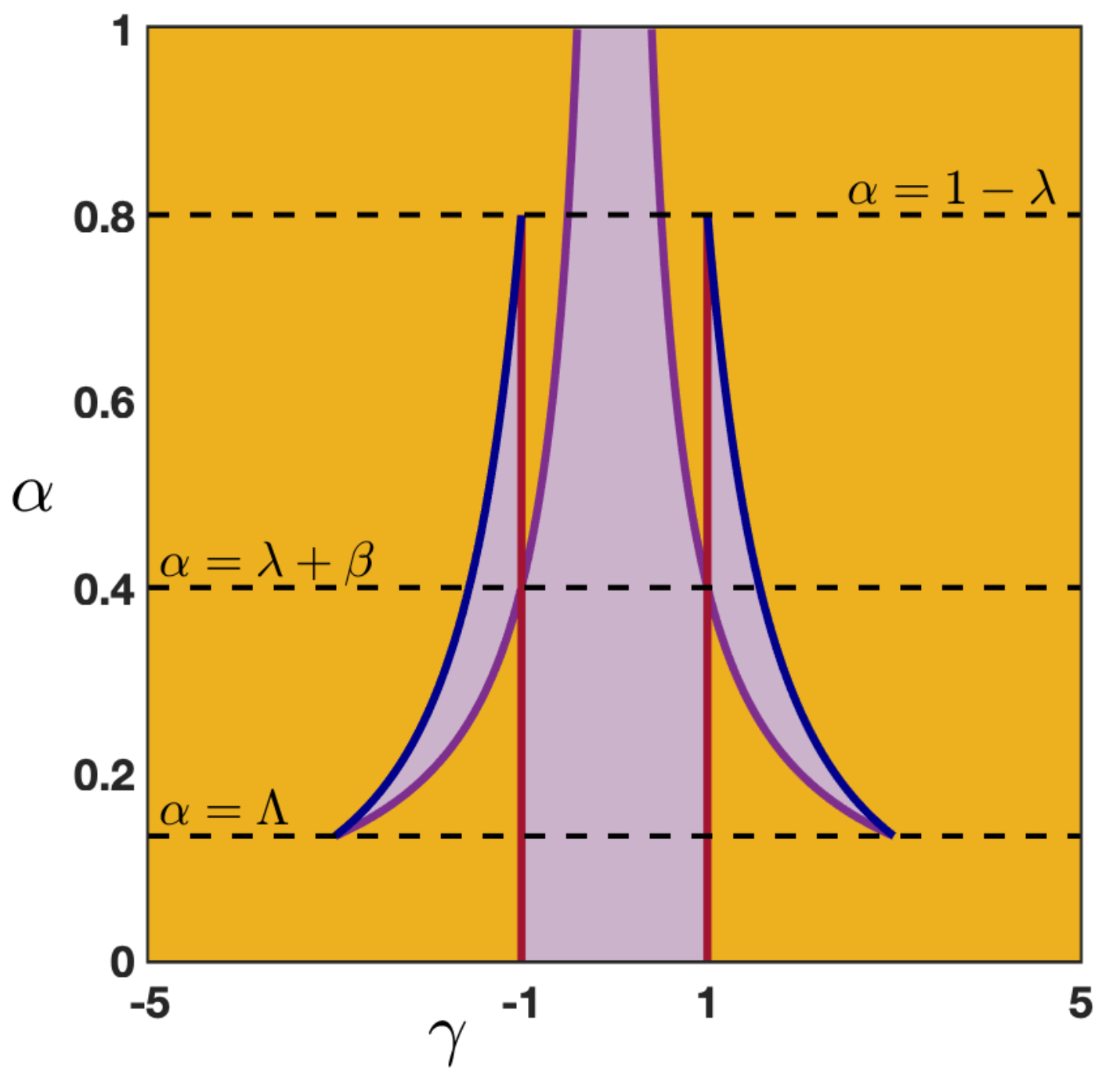}}
\subfigure[$(\lambda,\beta)$ in Region $\mathbf{(IV)}$.]{\includegraphics[width=.3\textwidth]{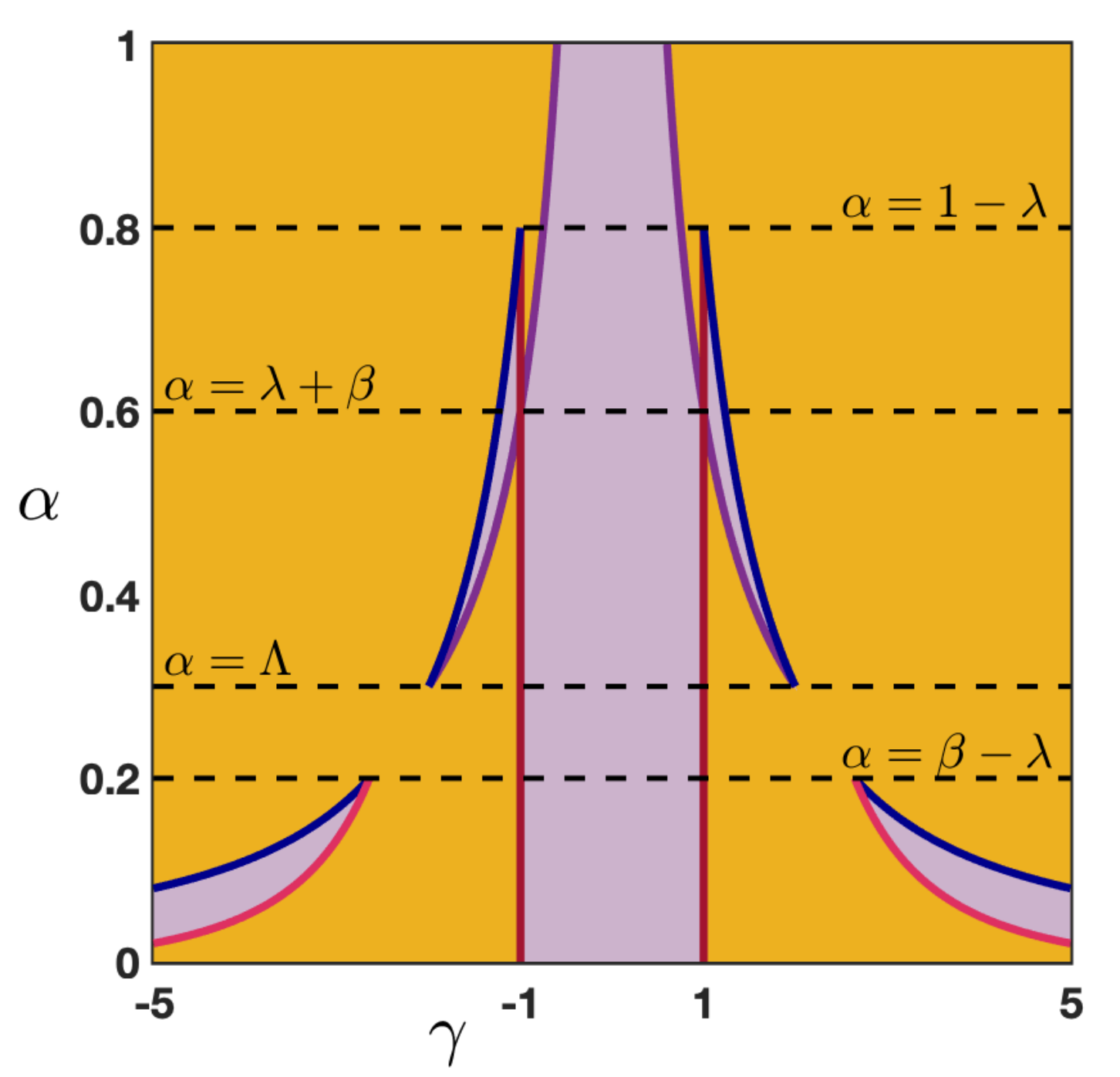}}
\subfigure[$(\lambda,\beta)$ in Region $\mathbf{(V)}$.]{\includegraphics[width=.3\textwidth]{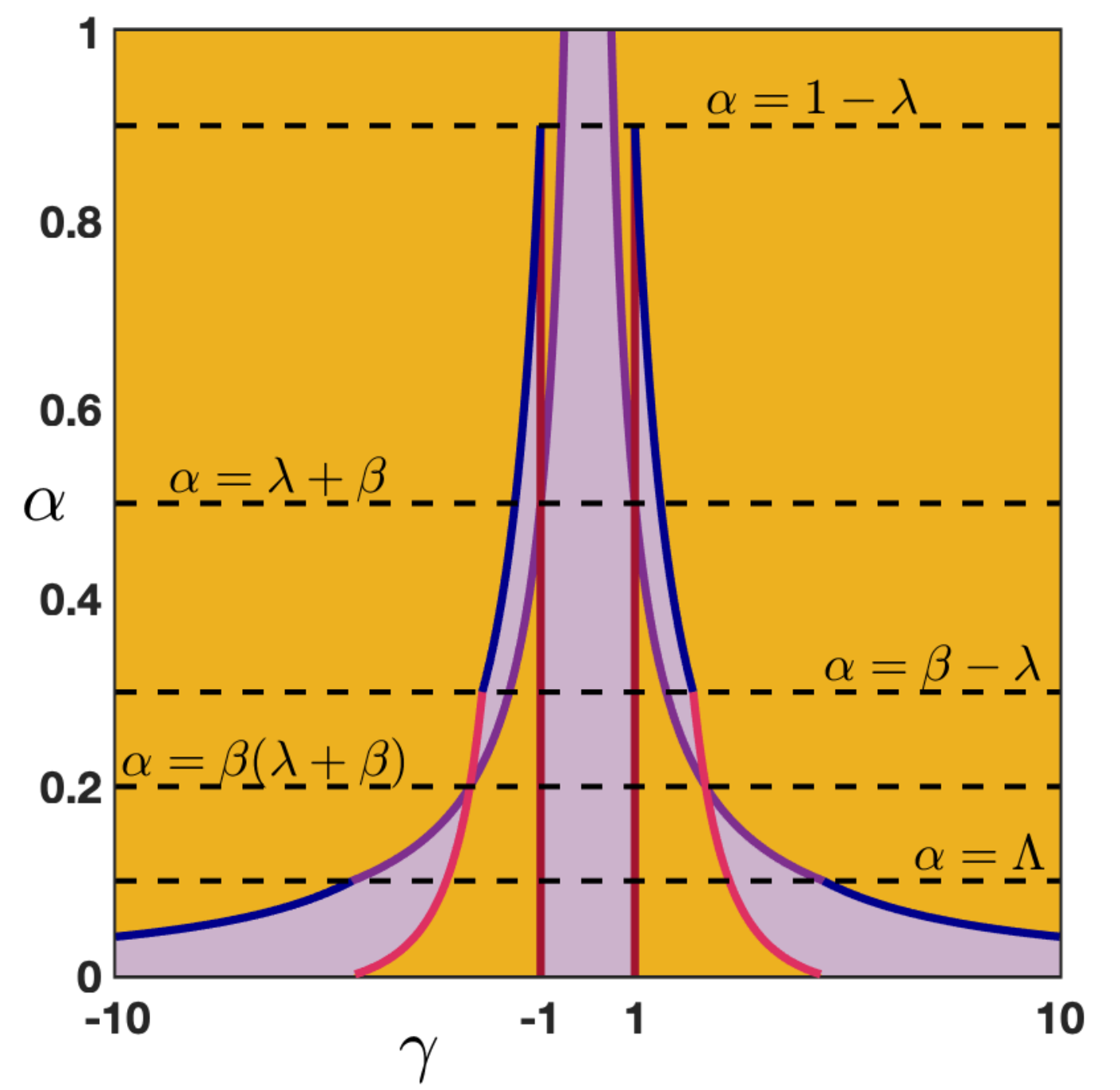}}
\caption{Stability/instability regions and their boundaries as a function of $(\alpha,\gamma)$ for \eqref{modelgam} while $(\lambda,\beta)$ being fixed in one the five regions given in panel (a). Shaded orange regions correspond to an instability for \eqref{modelgam}  while  purple regions correspond to a stability for \eqref{modelgam}. The boundaries of the stability/instability regions are given by the intersections of the parametrized curves $\gamma=\pm1$ (dark red curves), $\gamma=\pm \gamma_-^0$ (dark blue curves), $\gamma=\pm \gamma_+^0$ (pink curves) and $\gamma=\pm\frac{\lambda+\beta}{\alpha}$ (magenta curves). Note that the region of interest is $0<\alpha\leq 1-\lambda$. Along each such parametrized curves equation \eqref{modelgam} is marginally stable.}
\label{fig:StabRegionGam}
\end{figure}

We now introduce five regions in the quadrant $(\beta,\lambda)\in[0,1)\times(0,1)$ which are depicted in Figure~\ref{fig:StabRegionGam}(a). First, since $1-\lambda=\Lambda=\lambda+\beta$ if and only if $2\lambda+\beta=1$ (which corresponds to the blue line in Figure~\ref{fig:StabRegionGam}(a)), we deduce that when $2\lambda+\beta\geq1$ we have  $1-\lambda\leq \min(\Lambda,\lambda+\beta)$ which leads us to define the following two regions:
\begin{align*}
\mathbf{(I)}&:=\left\{ (\beta,\lambda)\in[0,1)\times(0,1) ~|~ 2\lambda+\beta\geq1 \text{ and } \beta\leq\lambda\right\}, \\
\mathbf{(II)}&:=\left\{ (\beta,\lambda)\in[0,1)\times(0,1) ~|~ 2\lambda+\beta\geq1 \text{ and } \beta>\lambda\right\}.
\end{align*}
Now, when $2\lambda+\beta<1$, we have the strict ordering $0<\Lambda<\lambda+\beta<1-\lambda$ and when $\beta>\lambda$ it is thus necessary to compare $\Lambda$ to $\beta-\lambda$. We remark that $\Lambda=\beta-\lambda=\beta(\lambda+\beta)$ if and only if $\beta^2-\beta(1-\lambda)+\lambda=0$, which corresponds to the yellow parabola in Figure~\ref{fig:StabRegionGam}(a). We thus define the following three regions:
\begin{align*}
\mathbf{(III)}&:=\left\{ (\beta,\lambda)\in[0,1)\times(0,1) ~|~ 2\lambda+\beta<1 \text{ and } \beta\leq\lambda\right\}, \\
\mathbf{(IV)}&:=\left\{ (\beta,\lambda)\in[0,1)\times(0,1) ~|~ 2\lambda+\beta<1, \beta>\lambda \text{ and } \beta^2-\beta(1-\lambda)+\lambda\geq0 \right\},\\
\mathbf{(V)}&:=\left\{ (\beta,\lambda)\in[0,1)\times(0,1) ~|~ 2\lambda+\beta<1, \beta>\lambda \text{ and } \beta^2-\beta(1-\lambda)+\lambda<0\right\}.
\end{align*}
Note that when $(\beta,\lambda)$ is in region $\mathbf{(IV)}$, we have the ordering
\bqs
0<\beta-\alpha\leq \Lambda<\lambda+\beta<1-\lambda,
\eqs
while for  $(\beta,\lambda)$ in region $\mathbf{(V)}$, we have
\bqs
0<\Lambda<\beta(\lambda+\beta)<\beta-\alpha<\lambda+\beta<1-\lambda.
\eqs
We can  characterize the stability of our equation separately for each of the five regions defined in Figure~\ref{fig:StabRegionGam}(a). Since the region already determines the value of the parameters $\beta$ and $\lambda$, the stability will be expressed as a function of the two remaining parameters $\alpha$ and $\gamma$ (Figure~\ref{fig:StabRegionGam}(b-f)). We refer to Figures~\ref{fig:StabRegionGam}(b-f) for a comprehensive representation of the stability regions. Note that the boundaries of the stability/instability regions are precisely given by the intersections of the parametrized curves $\gamma=\pm1$ (dark red curves), $\gamma=\pm \gamma_-^0$ (dark blue curves), $\gamma=\pm \gamma_+^0$ (pink curves) and $\gamma=\pm\frac{\lambda+\beta}{\alpha}$ (magenta curves). Along each such parametrized curves equation \eqref{modelgam} is marginally stable. We comment below the case $(\lambda,\beta)$ in Region $\mathbf{(III)}$. The other cases can be described in the same way, but we leave this out for conciseness.

\begin{table}
\begin{center}
\begin{tabular}{|c|c|c|c|c|c|c|}
\hline
 & $\gamma=1$ & $\gamma=-1$ & $\gamma=\frac{\lambda+\beta}{\alpha}$ & $\gamma=-\frac{\lambda+\beta}{\alpha}$ & $\gamma= \gamma_\pm^0$  & $\gamma=- \gamma_\pm^0$ \\
\hline
$c_0^\gamma$ & $\frac{\beta+\alpha-\lambda}{1-\beta}$ &  $\frac{\alpha-\lambda-\beta}{1+\beta}$ & $\frac{(\beta+\lambda)(\beta+\alpha-\lambda)}{\alpha-\beta(\lambda+\beta)}$ & &  $\frac{\gamma_\pm^0(\beta+\lambda-\alpha)}{1-\beta\gamma_\pm^0}$ &  \\
\hline
 $c_\pi^\gamma$ & $\frac{\alpha-\lambda-\beta}{1+\beta}$ & $\frac{\beta+\alpha-\lambda}{1-\beta}$ &  & $\frac{(\beta+\lambda)(\beta+\alpha-\lambda)}{\alpha-\beta(\lambda+\beta)}$ &  & $\frac{\gamma_\pm^0(\beta+\lambda-\alpha)}{1-\beta\gamma_\pm^0}$   \\
 \hline
\end{tabular}
\end{center}
\vspace{-5mm}
\caption{Expressions of the wave speed $c_0^\gamma$ and $c_\pi^\gamma$ for values of $\gamma$ corresponding the boundaries of the stability regions from Figure~\ref{fig:StabRegionGam}. Let us note that $c_\pi^1$ and $c_0^{-1}$ only exist when $\alpha+\lambda=1$.}
\label{table}
\end{table}

Suppose that $(\lambda,\beta)$ belongs to Region $\mathbf{(III)}$. We present the results of Figure~\ref{fig:StabRegionGam}(d) by letting $\alpha$ vary between $0$ and $1-\lambda$ and $\gamma\in\R$. More precisely, for each fixed $\alpha\in(0,1-\lambda)$ we investigate the stability properties as a function of $\gamma$. We have to distinguish between several subcases.
\begin{enumerate}
\item[(i)] If $0<\alpha<\Lambda$. Then, equation \eqref{modelgam} is stable for each $\gamma\in(-1,1)$, unstable for $|\gamma|>1$ and marginally stable at $\gamma=\pm1$ with $\rho_1(0)=1$ and $\rho_{-1}(\pm\pi)=1$.
\item[(ii)] If $\alpha=\Lambda$. Then  $\gamma_-^0=-\frac{\lambda+\beta}{\alpha}$ and equation \eqref{modelgam} is stable for each $\gamma\in(-1,1)$, unstable for $|\gamma|>|\gamma_-^0|$ and $|\gamma_-^0|>|\gamma|>1$, whereas it is marginally stable at $\gamma=\pm1$ and at $\gamma=\pm\gamma_-^0$ with $\rho_1(0)=1$,  $\rho_{-1}(\pm\pi)=1$, $\rho_{\gamma_-^0}(0)=-1$ and $\rho_{-\gamma_-^0}(\pm\pi)=-1$ together with $\rho_{-\gamma_-^0}(0)=1$,  $\rho_{\gamma_-^0}(\pm\pi)=1$.
\item[(iii)] If $\Lambda<\alpha<\lambda+\beta$. Then, equation \eqref{modelgam} is stable for each $\gamma\in(-1,1)$ and $\frac{\lambda+\beta}{\alpha}<|\gamma|<|\gamma_-^0|$, unstable for $|\gamma|>|\gamma_-^0|$ and $\frac{\lambda+\beta}{\alpha}>|\gamma|>1$ and marginally stable at $\gamma\in\left\{\pm1,\pm\frac{\lambda+\beta}{\alpha},\pm\gamma_-^0\right\}$ with $\rho_1(0)=1$,  $\rho_{-1}(\pm\pi)=1$, $\rho_{\frac{\lambda+\beta}{\alpha}}(0)=1$,  $\rho_{-\frac{\lambda+\beta}{\alpha}}(\pm\pi)=1$, $\rho_{\gamma_-^0}(0)=-1$ and $\rho_{-\gamma_-^0}(\pm\pi)=-1$.
\item[(iv)] If $\alpha=\lambda+\beta$. Then, equation \eqref{modelgam} is stable for each $\gamma\in(\gamma_-^0,-\gamma_-^0)\backslash\{\pm1\}$, unstable for $|\gamma|>|\gamma_-^0|$ and marginally stable at $\gamma\in\left\{\pm1,\pm\gamma_-^0\right\}$ with $\rho_1(0)=1$,  $\rho_{-1}(\pm\pi)=1$, $\rho_{\gamma_-^0}(0)=-1$ and $\rho_{-\gamma_-^0}(\pm\pi)=-1$. Remark that in this case we have $\frac{\lambda+\beta}{\alpha}=1$.
\item[(v)] If $\lambda+\beta<\alpha<1-\lambda$. Then, equation \eqref{modelgam} is stable for each $\gamma\in\left(-\frac{\lambda+\beta}{\alpha},\frac{\lambda+\beta}{\alpha}\right)$ and $1<|\gamma|<|\gamma_-^0|$, unstable for $|\gamma|>|\gamma_-^0|$ and $1>|\gamma|>\frac{\lambda+\beta}{\alpha}$ and marginally stable at $\gamma\in\left\{\pm1,\pm\frac{\lambda+\beta}{\alpha},\pm\gamma_-^0\right\}$ with $\rho_1(0)=1$,  $\rho_{-1}(\pm\pi)=1$, $\rho_{\frac{\lambda+\beta}{\alpha}}(0)=1$,  $\rho_{-\frac{\lambda+\beta}{\alpha}}(\pm\pi)=1$, $\rho_{\gamma_-^0}(0)=-1$ and $\rho_{-\gamma_-^0}(\pm\pi)=-1$.
\item[(vi)] If $\alpha=1-\lambda$. Then, equation \eqref{modelgam} is stable for each $\gamma\in\left(-\frac{\lambda+\beta}{\alpha},\frac{\lambda+\beta}{\alpha}\right)$, unstable for $|\gamma|>|\gamma_-^0|=1$ and $1>|\gamma|>\frac{\lambda+\beta}{\alpha}$ and marginally stable at $\gamma\in\left\{\pm1,\pm\frac{\lambda+\beta}{\alpha}\right\}$ with $\rho_1(0)=1$ and $\rho_{1}(\pm\pi)=-1$,  $\rho_{-1}(\pm\pi)=1$ and $\rho_{-1}(0)=-1$, with $\rho_{\frac{\lambda+\beta}{\alpha}}(0)=1$,  $\rho_{-\frac{\lambda+\beta}{\alpha}}(\pm\pi)=1$. Remark that in this case we have $\gamma_-^0=-1$.
\end{enumerate}

\paragraph{Summary.} In summary, we see that stability is nearly guaranteed whenever $-1<\gamma<1$, regardless of the values of other parameters (as long as $0<\alpha<1-\lambda$). This makes intuitive sense, as $\gamma$ represents the connection strength across layers of a particular neural assembly, a connection weight $|\gamma|<1$ implies that activity of this assembly will remain bounded across layers. Additionally, and perhaps more interestingly, in some but not all regions (e.g. Regions II and V) stability can be obtained for much larger values of $|\gamma|$; this, however, appears to coincide with low values of the $\alpha$ parameter. In other words, for high connection strengths $|\gamma|$, the feedforward error correction term $\alpha$ makes the system unstable.

\subsubsection{Wave speed characterization}
\begin{figure}[t!]
\centering
\includegraphics[width=.4\textwidth]{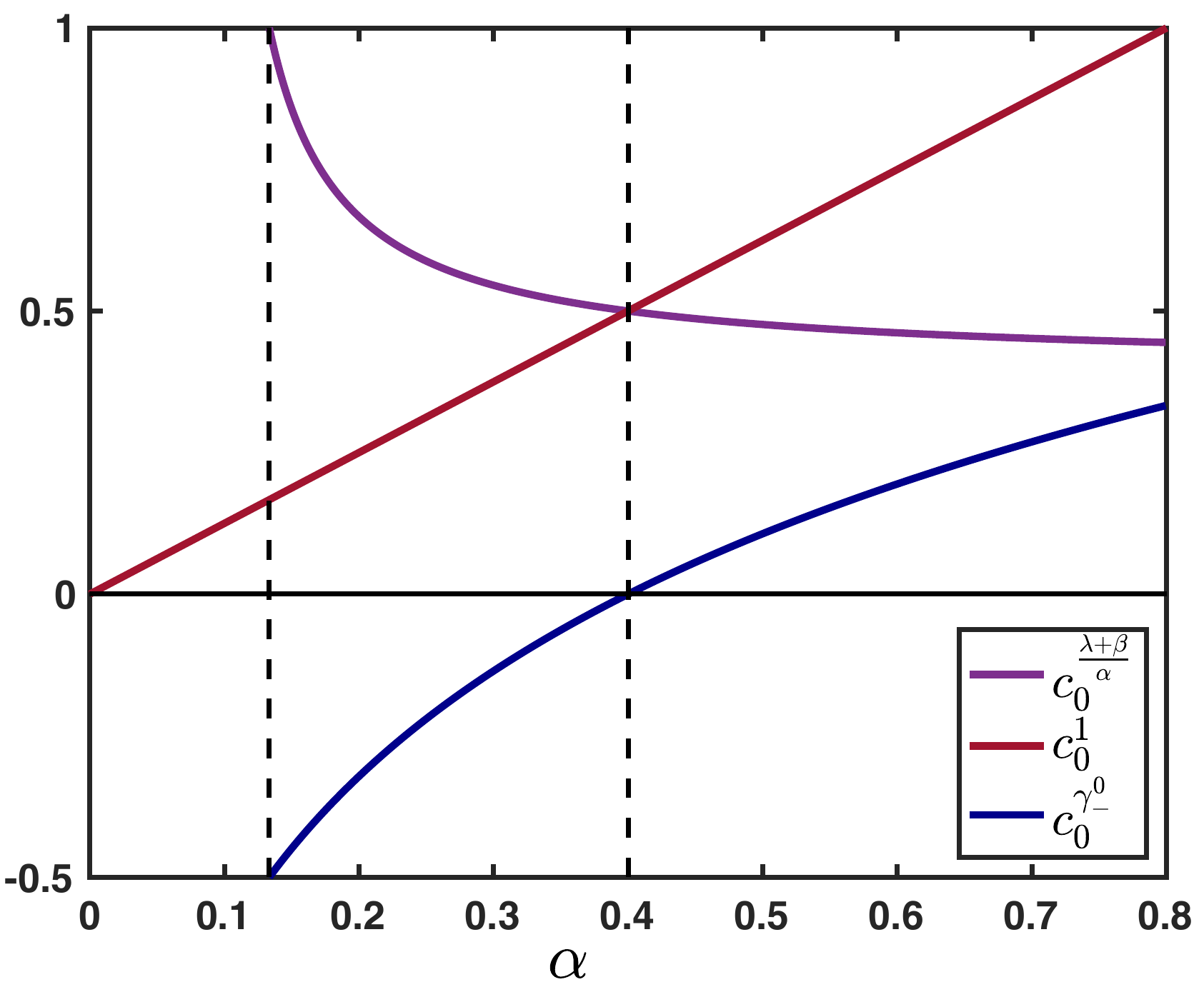}
\caption{Plot of the wave speeds $c_0^1$, $c_0^{\frac{\lambda+\beta}{\alpha}}$ and $c_0^{\gamma_-^0}$ as a function of $\alpha$ in the case $(\lambda,\beta)$ in Region $\mathbf{III}$ associated to Figure~\ref{fig:StabRegionGam}(d).}
\label{fig:WaveSpeedGam}
\end{figure}

In the previous section (The Identity Case), we have proved that the direction of propagation was given by the sign of $c_0$ and $c_\pi$ whenever they exist which could be read off from the behavior of $ \rho_\gamma(\theta)$ near $\theta=0$ or $\theta=\pm\pi$. We have reported the values of $c_0^\gamma$ and $c_{\pi}^\gamma$ for different values of $\gamma$ in Table~\ref{table}.  For example, in Figure~\ref{fig:WaveSpeedGam} we illustrate the changes in propagation speed and direction for $c_0^\gamma$ for the case $(\lambda,\beta)$ in Region $\mathbf{(III)}$ (as defined in Figure~\ref{fig:StabRegionGam}a), but the calculations remain valid for the other regions.

It is worth emphasizing that for fixed values of the hyper-parameters $\alpha$, $\beta$ and $\lambda$, we see here that varying $\gamma$ can give rise to different propagation speeds or even different directions. As each neuronal assembly $u_{j,p}$ in a given layer $j$ is associated with its own connection strength $\gamma_p$, it follows that different speeds and even different directions of propagation can concurrently be obtained in a single network, one for each assembly. For instance, in a given network with hyperparameters $\alpha=0.2$, $\beta=0.2$ and $\lambda=0.3$ (region $\mathbf{III}$), a neural assembly with a connection strength of $\gamma=1$ would propagate forward at a relatively slow speed, while another with $\gamma=2.5$ would propagate in the same direction at a much faster speed, and yet another assembly with $\gamma=\gamma^0_-\approx-2.09$ would simultaneously propagate in the opposite backward direction.

\subsubsection{Continuous in time interpretation}

We can repeat the ``continuous system'' analysis conducted in the previous section (The Identity Case), which has lead to \eqref{latticeODE}, but this time with Rao-Ballard connection matrices between layers. With the same scaling on the hyperparameters 
\bqs
\widetilde{\beta}:=\frac{\beta}{\Delta t}, \quad \widetilde{\lambda}:=\frac{\lambda}{\Delta t}, \text{ and } \widetilde{\alpha}:=\frac{\alpha}{\Delta t},
\eqs
we get that, at the limit $\Delta t\rightarrow0$, the equation \eqref{modelgam} becomes the following lattice ordinary differential equation 
\bqq
\frac{\md }{\md t}{\bf u}_j(t)=(\widetilde{\beta}+\widetilde{\alpha})\gamma{\bf u}_{j-1}(t)-(\widetilde{\beta}+\widetilde{\lambda}+\widetilde{\alpha}\gamma^2){\bf u}_j(t)+\widetilde{\lambda}\gamma{\bf u}_{j+1}(t), \quad t>0.
\label{latticeODEgam}
\eqq
Note that the neuronal layer activity is now expressed in terms of neural assemblies ${\bf u}_j$ rather than individual neurons ${\bf e}_j$.

\begin{figure}[t!]
\centering
\includegraphics[width=.3\textwidth]{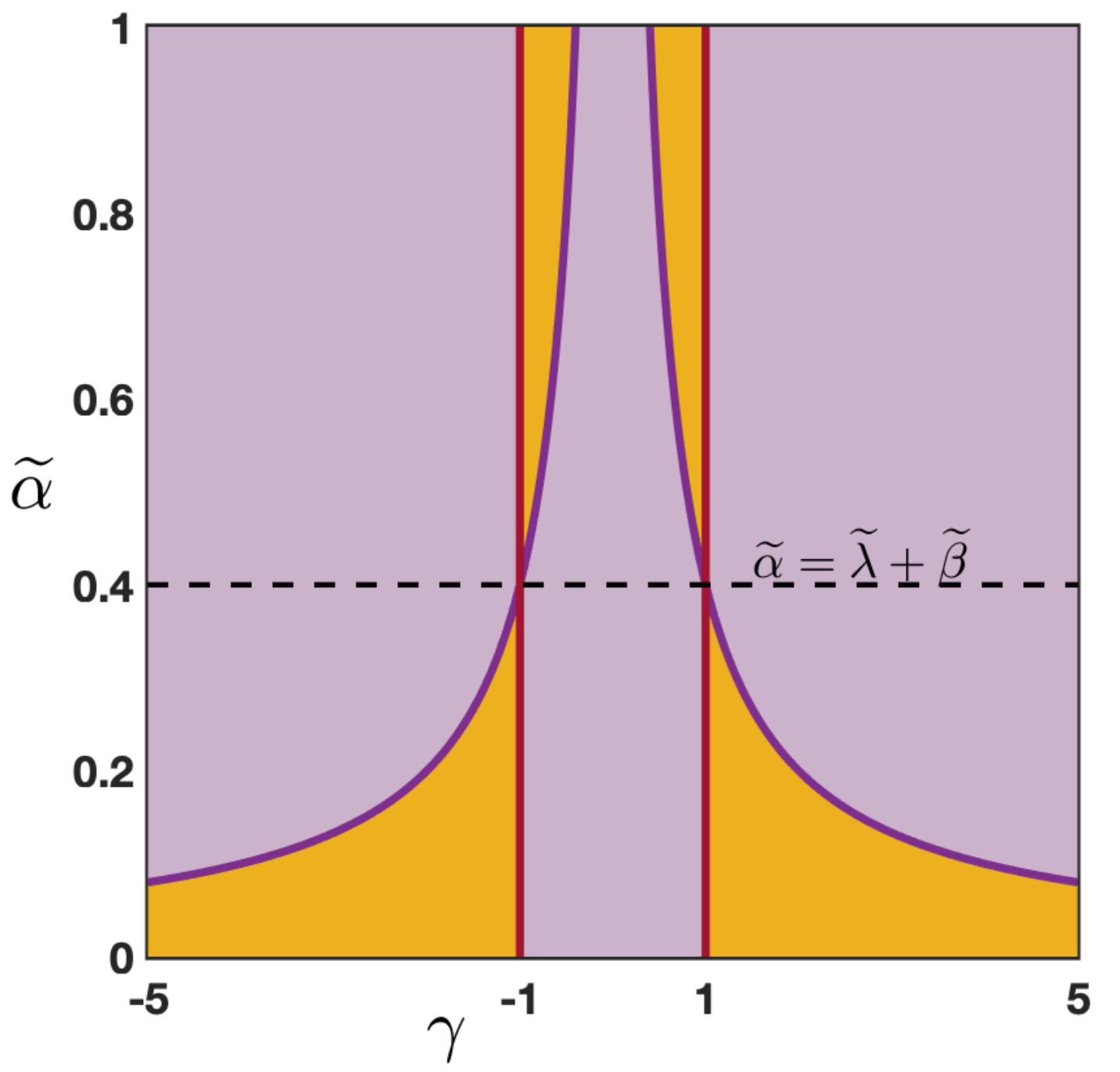}
\caption{Stability/instability regions and their boundaries as a function of $(\widetilde{\alpha},\gamma)$ for \eqref{latticeODEgam} for any $(\widetilde{\lambda},\widetilde{\beta})$ fixed. The shaded orange region corresponds to an instability for \eqref{latticeODEgam} while the purple region corresponds to a stability for \eqref{latticeODEgam}. The boundaries of the stability/instability regions are given by the intersections of the parametrized curves $\gamma=\pm1$ (dark red curves) and $\gamma=\pm\frac{\lambda+\beta}{\alpha}$ (magenta curves) where equation \eqref{latticeODEgam} is marginally stable.}
\label{fig:StabRegionGamContinuous}
\end{figure}

The amplification factor function in this case is given by
\bqs
\nu_\gamma(\theta)=(\widetilde{\beta}+\widetilde{\alpha})\gamma e^{-\mbi\theta}-(\widetilde{\beta}+\widetilde{\lambda}+\widetilde{\alpha}\gamma^2)+\widetilde{\lambda}\gamma e^{\mbi\theta}, \quad \theta\in[-\pi,\pi],
\eqs
whose real part is given by
\bqs
\Re(\nu_\gamma(\theta))=(\widetilde{\beta}+\widetilde{\alpha}+\widetilde{\lambda})\gamma \cos(\theta)-(\widetilde{\beta}+\widetilde{\lambda}+\widetilde{\alpha}\gamma^2), \quad \theta\in[-\pi,\pi].
\eqs
When $\gamma>0$, we observe that
\bqs
\underset{\theta\in[-\pi,\pi]}{\max}\Re(\nu_\gamma(\theta))=\Re(\nu_\gamma(0))=(\widetilde{\lambda}+\widetilde{\beta}-\widetilde{\alpha}\gamma)(\gamma-1),
\eqs
whereas when $\gamma<0$, we have
\bqs
\underset{\theta\in[-\pi,\pi]}{\max}\Re(\nu_\gamma(\theta))=\Re(\nu_\gamma(\pm\pi))=-(\widetilde{\lambda}+\widetilde{\beta}+\widetilde{\alpha}\gamma)(\gamma+1).
\eqs
As a consequence, the stability analysis in this case is very simple and depends only on the relative position of $\gamma$ with respect to $\pm1$ and $\pm\frac{\widetilde{\lambda}+\widetilde{\beta}}{\widetilde{\alpha}}$. It is summarized in Figure~\ref{fig:StabRegionGamContinuous}.

The simple behavior illustrated in Figure~\ref{fig:StabRegionGamContinuous} for our continuous system contrasts with the number and diversity of behaviors obtained for the discrete version of the same system (Figure~\ref{fig:StabRegionGam}). A number of points are worth highlighting. For instance, although the values of $\beta$ and $\lambda$ were critical for the discrete system (to define the region (I) to (V)), they do not affect the qualitative behavior of the continuous system.  Furthermore, some observations in the continuous system appear to contradict the conclusions made previously in the discrete case. We see that stability can still be obtained with high values of the connection weight $\gamma>>1$, but this time the stable regions coincide with high $\alpha$ values, whereas it was the opposite in Figure~\ref{fig:StabRegionGam} panels (b),(f). This qualitative difference in behavior can be taken as a point of caution, to remind us that a discrete approximation of the system can be associated with important errors in interpretation.\\
Finally we note that, while stability regions are qualitatively different in the continuous case compared to the discrete approximation, the speed and direction of propagation of neural signals (reflected in the variables $c_0$ and $c_{\pi}$ when they exist) remains comparable.

\subsubsection{A class of examples}

In this section, we provide a class of examples of $\W^f$ amenable to a complete analysis. Namely we consider $\W^f$ as the following linear combination
\bqq
\W^f=\zeta \mathbf{I}_d+\xi \mathbf{A},
\label{convmatrix}
\eqq
for some $\zeta,\xi\in\R$ where $\mathbf{A}\in\mathscr{M}_d(\R)$ is given by
\bqs
\mathbf{A}=\left(
\begin{matrix}
-2 & 1 & 0 &  \cdots & \cdots & 0  \\
1 & -2 & 1 &  \ddots & \ddots & \vdots \\
0 & \ddots & \ddots &  \ddots & \ddots & \vdots \\
\vdots & \ddots & \ddots & \ddots & \ddots & \vdots \\
\vdots & \ddots & \ddots & 1 & -2 & 1 \\
0 & \cdots & \cdots & 0 & 1 & -2
\end{matrix}
\right).
\eqs
The matrix $\mathbf{A}$ is nothing but the discrete laplacian and $\W^f$ acts as a convolution operator on $\R^d$. More precisely, $\W^f$ combines a convolution term with a residual connection term, as in the well-known ResNet architecture~\cite{he2016}. Let us also note that the spectrum of $\mathbf{A}$ is well known and given by
\bqs
\mathrm{Spec}(\mathbf{A})=\left\{-4\sin^2\left(\frac{p\pi}{2(d+1)}\right), \quad p=1,\cdots,d\right\}.
\eqs
As a consequence, the spectrum of $\W^f$ is simply given by
\bqs
\mathrm{Spec}(\mathbf{\W^f})=\left\{\zeta-4\xi\sin^2\left(\frac{p\pi}{2(d+1)}\right), \quad p=1,\cdots,d\right\}.
\eqs

One can for example set
\bqs
\zeta=\frac{\sin^2\left(\frac{d\pi}{2(d+1)}\right)+\sin^2\left(\frac{\pi}{2(d+1)}\right)}{\sin^2\left(\frac{d\pi}{2(d+1)}\right)-\sin^2\left(\frac{\pi}{2(d+1)}\right)}\quad \text{ and } \quad \xi=\frac{1}{2\left(\sin^2\left(\frac{d\pi}{2(d+1)}\right)-\sin^2\left(\frac{\pi}{2(d+1)}\right)\right)},
\eqs
such that
\bqs
\zeta-4\xi\sin^2\left(\frac{d\pi}{2(d+1)}\right)=-1, \quad \zeta-4\xi\sin^2\left(\frac{\pi}{2(d+1)}\right)=1,
\eqs
and for all $p=2,\cdots,d-1$
\bqs
\zeta-4\xi\sin^2\left(\frac{p\pi}{2(d+1)}\right) \in(-1,1).
\eqs
Next, for any $p=1,\cdots,d$ the eigenvector corresponding to the eigenvalue $-4\sin^2\left(\frac{p\pi}{2(d+1)}\right)$ is 
\bqs
U_p=\left(\cos\left(\frac{p\pi}{d+1}\right),\cdots,\cos\left(\frac{pk\pi}{d+1}\right),\cdots,\cos\left(\frac{pd\pi}{d+1}\right)\right)^{\bf t}\in\R^d.
\eqs
$U_p$ is the projection vector that corresponds to the $p^{th}$ neural assembly $u_{j,p}$ as defined above.\\
Along $U_1$, the recurrence equation reduces to \eqref{modelgam} with $\gamma=1$, while along $U_d$, the recurrence equation reduces to \eqref{modelgam} with $\gamma=-1$, and we can apply the results of the previous section (the Identity case). In between (for all $1\leq p\leq d$) we see that the eigenvalues of our connection matrix $\mathbf{\W^f}$ span the entire range between $-1$ and $1$, that they can be explicitly computed, and thus that the stability, propagation speed and direction of activity in the corresponding neural assembly can be determined.

\subsubsection{Fully continuous interpretation in time, depth and width.} 
For the same class of example (connection matrix composed of a convolution and residual terms), we now wish to provide a fully continuous interpretation for model \eqref{model} in the special case $\zeta=1$ and $\xi$ adjusted as follows. By fully continuous, we mean that we explore the limit of our model when not only time $t$, but also network depth $x$ \textit{and} neuronal layer width $y$ are considered as continuous variables. Although we already presented a model that was continuous in both \textit{time} and \textit{depth} in subsection~\ref{sec:continuoustimedepth}, the layers in that model only comprised a single neuron, and had no intrinsic spatial dimension. We now introduce this third continuous dimension. The starting point is to see $\E_{j,k}^n$, the $k$th element of  $\E_{j}^n$, as an approximations of some continuous function $\E(t,x,y)$ evaluated at $t_n=n\Delta t$, $x_j=j\Delta x$ and $y_k=k\Delta y$ for some $\Delta t>0$, $\Delta x>0$ and $\Delta y>0$. Let us first remark that the action of $\mathbf{A}$ on $\E_j^n$ is given by
\bqs
(\mathbf{A}\E_j^n)_k=\E_{j,k-1}^n-2\E_{j,k}^n+\E_{j,k+1}^n,
\eqs
which can be seen at a discrete approximation of $\partial_y^2 \E(t_n,x_j,y_k)$ up to a scaling factor of order $\Delta y^2$. Once again, setting $\nu=\frac{\Delta x}{\Delta t}$ and introducing $\kappa=\frac{\Delta y^2}{\Delta t}$, we may rewrite \eqref{model} with $\W^f=\W^b=\mathbf{I}_d+\xi \mathbf{A}$ as
\begin{align*}
(1-\beta)\frac{\E_j^{n+1}-\E_j^n}{\Delta t}&=\beta \nu \frac{\E_{j-1}^{n+1} -\E_j^{n+1}}{\Delta x}+ \lambda \nu \frac{ \E_{j+1}^{n}-\E_j^{n}}{\Delta x}-\alpha \nu \frac{ \E_{j}^{n}-\E_{j-1}^{n}}{\Delta x}\\
&~~~+\beta \xi \kappa \frac{\mathbf{A}}{\Delta y^2}\E_{j-1}^{n+1}+\alpha \xi \kappa \frac{\mathbf{A}}{\Delta y^2}\E_{j-1}^{n}-2\alpha \xi \kappa \frac{\mathbf{A}}{\Delta y^2}\E_{j}^{n}-\alpha \xi^2 \kappa \Delta y^2 \frac{\mathbf{A}}{\Delta y^2} \frac{\mathbf{A}}{\Delta y^2}\E_{j}^{n}\\
&~~~+\lambda \xi \kappa \frac{\mathbf{A}}{\Delta y^2}\E_{j+1}^{n}.
\end{align*}
Now letting $\Delta t\to0$, $\Delta x\to0$ and $\Delta y\to0$ with $\nu$ and $\kappa$ fixed, we obtain the following partial differential equation  
\bqs
\partial_t \E(t,x,y) + \frac{\nu (\beta+\alpha-\lambda)}{1-\beta } \partial_x \E(t,x,y) =\frac{\xi \kappa (\beta+\lambda-\alpha)}{1-\beta}\partial_y^2\E(t,x,y).
\eqs
This is a diffusion equation along the $y$ dimension while being a transport in the $x$ direction. As such, it is only well defined (or stable) when the sign of the diffusion coefficient in front of $\partial_y^2\E(t,x,y)$ is positive. This depends on the sign of $\xi$ and $\beta+\lambda-\alpha$, which need to verify $\xi(\beta+\lambda-\alpha)>0$. In that case, the system diffuses neural activity along the dimension $y$ such that the entire neuronal layer converges to a single, uniform activation value when $t\to\infty$.

\subsection{The general symmetric case}\label{subsec:gensymcase}

Finally, we now wish to relax some of the assumptions made in the previous Rao-Ballard case. Thus, the last case that we present is one where we assume that
\begin{itemize}
\item[(i)] $\W^f$ and $\W^b$ are symmetric matrices, that is $\W^f,\W^b\in\mathscr{S}_d(\R)$,
\item[(ii)] $\W^f$ and $\W^b$ commute, that is $\W^f\W^b=\W^b\W^f$.
\end{itemize}
But we do not necessarily impose that $\W^f=(\W^b)^{\bf t}$ as in the Rao \& Ballard's previous case. Let us already note that examples of matrices verifying the above conditions are residual convolution matrices introduced in \eqref{convmatrix}, that is $\W^f=\zeta_f \mathbf{I}_d+\xi_f \mathbf{A}$ and $\W^b=\zeta_b \mathbf{I}_d+\xi_b \mathbf{A}$ for some $\zeta_{b,f},\xi_{b,f}\in\R$. Under assumptions (i) and (ii), $\W^f$ and $\W^b$ can be diagonalized in the same orthonormal basis, meaning that there exist an invertible orthogonal matrix $P\in \mathscr{M}_d(\R)$ such that $PP^\mathbf{t}=P^\mathbf{t}P=I_d$, and two diagonal matrices $\D^f\in \mathscr{M}_d(\R)$ and $\D^b\in \mathscr{M}_d(\R)$ with the properties that
\bqs
P^\mathbf{t}\W^fP= \D^f, \quad \text{ and } \quad P^\mathbf{t}\W^bP= \D^b.
\eqs
For future reference, we denote by $\gamma_p^{f,b}$ for each $1\leq p \leq d$ the diagonal elements of $\D^{f,b}$. Once again, we can use the matrix $P$ to apply an orthonormal basis change and create neural \textit{asssemblies} $\U_j^n:=P^t\E_j^n$. With $P\U_j^n :=\E_j^n$, the recurrence equation becomes
\bqs
\U_j^{n+1}-\beta \D^f \U_{j-1}^{n+1} =\alpha \D^b \U_{j-1}^n+\left[(1-\beta-\lambda)\mathbf{I}_d-\alpha \D^b \D^b\right]  \U_j^{n} + \lambda  \D^b \U_{j+1}^{n}.
\eqs
Note that, because all matrices in the above equation are diagonal, we have also totally decoupled the $d$ components of the vector $\U_j^n$. More precisely, by denoting $u_{j,p}^{n}$ the $p$th component of $\U_j^n$, that is $\U_j^n=(u_{j,1}^{n},\cdots,u_{j,d}^{n})^\mathbf{t}$, we obtain
\bqs
u_{j,p}^{n+1}-\beta \gamma_p^f u_{j-1,p}^{n+1} =\alpha \gamma_p^b u_{j-1,p}^n +(1-\beta-\lambda-\alpha \left(\gamma_p^b\right)^2) u_{j,p}^{n} +\lambda  \gamma_p^b u_{j+1,p}^{n}, \quad p=1,\cdots,d.
\eqs
This indicates that one needs to study 
\bqq
u_{j}^{n+1}-\beta \gamma_1 u_{j-1}^{n+1} =\alpha \gamma_2 u_{j-1}^n +(1-\beta-\lambda-\alpha \gamma_2^2) u_{j}^{n} +\lambda  \gamma_2 u_{j+1}^{n}, 
\label{modelgambis}
\eqq
where $\gamma_{1,2}\in\R$ are now two given parameters. As before, $\gamma_{1,2}$ can be thought of as the connection strength across layers of the neural assembly under consideration. By construction, each assembly in a given layer is only connected to the corresponding assembly in the layer above, and similarly in the layer below, with $\gamma_1$ for the feedforward direction and $\gamma_2$ for the feedback direction. Note that $\gamma_1=\gamma_2$ would then correspond to the Rao-Ballard situation studied previously.

\subsubsection{Study of the amplification factor function}

\begin{figure}[t!]
\centering
\includegraphics[width=.35\textwidth]{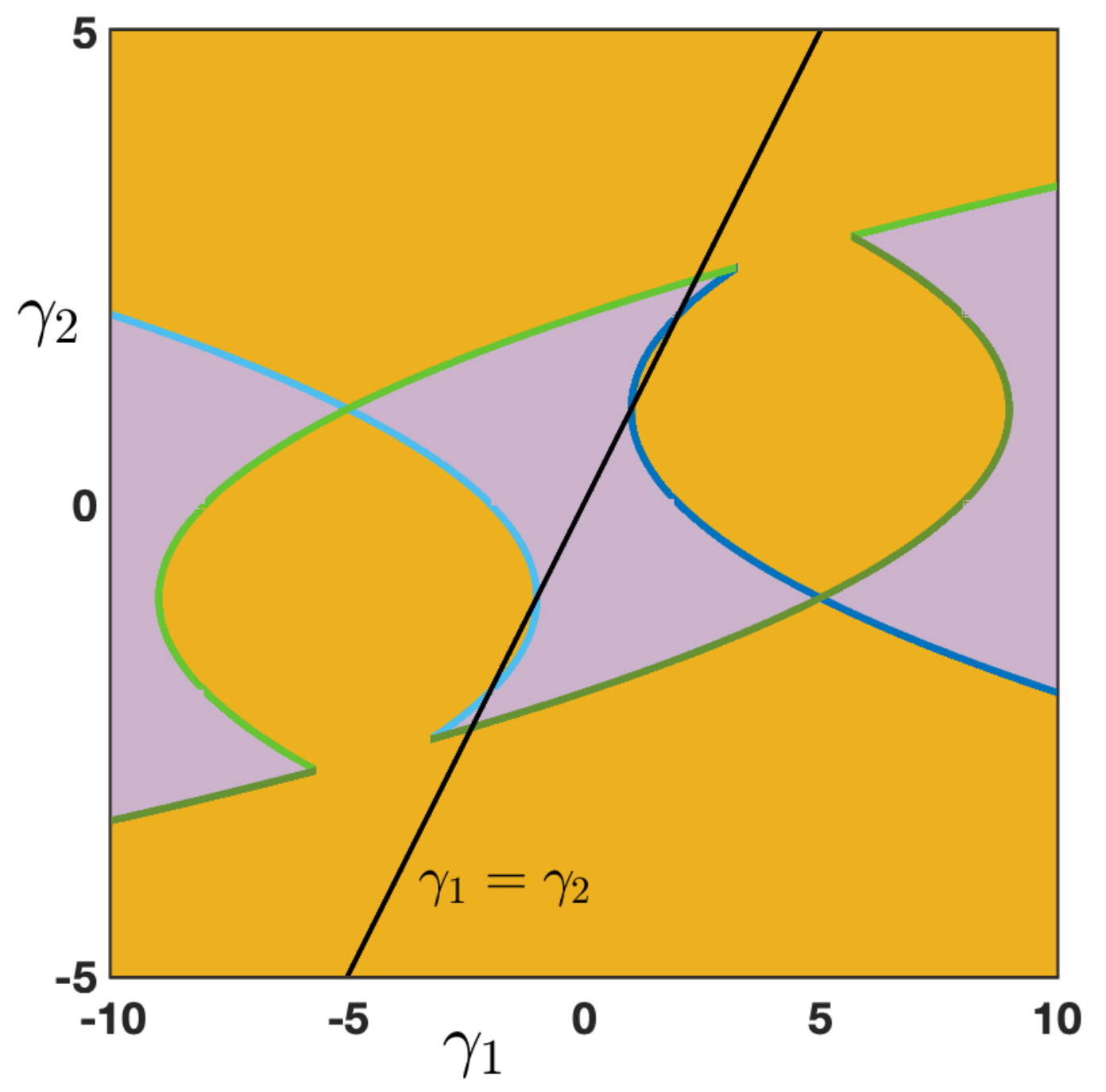}
\caption{Stability/instability regions and their boundaries as a function of $(\gamma_1,\gamma_2)$ for \eqref{modelgambis} for fixed values of the hyperparameters $(\alpha,\beta,\lambda)$. The shaded orange region corresponds to an instability for \eqref{modelgambis} while the purple region corresponds to a stability for \eqref{modelgambis}. The boundaries of the stability/instability regions are given by the intersections of the parametrized curves $\gamma_1=\chi(\gamma_2)$ (blue curve), $\gamma_1=-\chi(-\gamma_2)$ (light blue curve), $\gamma_1=\zeta(\gamma_2)$ (dark green curves) and $\gamma_1=-\zeta(-\gamma_2)$ (light green curves) where equation \eqref{modelgambis} is marginally stable. We represented the line $\gamma_1=\gamma_2$ (black curve) which corresponds to the case studied in Figure~\ref{fig:StabRegionGam}(d) with $(\beta,\lambda)$ in Region $\mathbf{(III)}$ and $\alpha$ fixed in $(\Lambda,\lambda+\beta)$.}
\label{fig:StabRegionGam12}
\end{figure}

Repeating the previous analysis, one needs to understand the amplification factor
\bqs
\rho_{\gamma_1,\gamma_2}(\theta) = \frac{\alpha\gamma_2 \left(e^{-\mbi \theta}-\gamma_2\right) +1-\beta +\lambda\left(\gamma_2 e^{\mbi \theta}-1\right)}{1-\beta \gamma_1 e^{-\mbi \theta}}, \quad \theta\in[-\pi,\pi].
\eqs
We already note a symmetry property of the amplification factor function which reads
\bqs
\rho_{\gamma_1,\gamma_2}(\theta)=\rho_{-\gamma_1,-\gamma_2}(\theta\pm\pi), \quad \theta\in[-\pi,\pi].
\eqs
As a consequence, whenever $\rho_{\gamma_1,\gamma_2}(0)=\pm1$ one has $\rho_{-\gamma_1,-\gamma_2}(\pm\pi)=\pm1$ for the same values of the parameters. Then, we note that
\bqs
\rho_{\gamma_1,\gamma_2}(0)=1 \Longleftrightarrow  \gamma_1=\chi(\gamma_2),
\eqs
where the function $\chi(x)$, depending only on the hyper-parameters, is given by
\bqq
\chi(x):=\frac{\alpha x^2-(\alpha+\lambda)x+\lambda+\beta}{\beta}, x\in\R.
\label{fctchi}
\eqq 
Thus, using the above symmetry, we readily deduce that
\bqs
\rho_{\gamma_1,\gamma_2}(\pm\pi)=1 \Longleftrightarrow  \gamma_1=-\chi(-\gamma_2).
\eqs
Finally, we compute that
\bqs
\rho_{\gamma_1,\gamma_2}(0)=-1 \Longleftrightarrow  \gamma_1=\zeta(\gamma_2),
\eqs
where the function $\zeta(x)$, depending only on the hyper-parameters, is given by
\bqq
\zeta(x):=\frac{-\alpha x^2+(\alpha+\lambda)x+2-\lambda-\beta}{\beta}, x\in\R.
\label{fctzeta}
\eqq 
Using the above symmetry, we readily deduce that
\bqs
\rho_{\gamma_1,\gamma_2}(\pm\pi)=-1 \Longleftrightarrow  \gamma_1=-\zeta(-\gamma_2).
\eqs
A complete and exhaustive characterization of all possible cases as a function of $\gamma_{1,2}$ and the hyper-parameters is beyond the scope of this paper. Nevertheless, we can make some few further remarks. The four above  curves $ \gamma_1=\chi(\gamma_2)$, $ \gamma_1=-\chi(-\gamma_2)$,  $\gamma_1=\zeta(\gamma_2)$ and $ \gamma_1=-\zeta(-\gamma_2)$ form parabolas in the plane $(\gamma_1,\gamma_2)$ that can intersect and provide the boundaries of the stability regions. For example, we can notice that $\gamma_1=\zeta(\gamma_2)$ and $ \gamma_1=-\zeta(-\gamma_2)$ intersect if and only if $\gamma_2=\pm\sqrt{\frac{2-\lambda-\beta}{\alpha}}$ whereas $ \gamma_1=\chi(\gamma_2)$ and $\gamma_1=-\chi(-\gamma_2)$ can never intersect. We refer to Figure~\ref{fig:StabRegionGam12} for an illustration of the stability regions and their boundaries in the case $(\alpha,\beta,\lambda)=(0.4,0.2,0.3)$. Here, we see that stability can be obtained with large values of the feedforward connection strength $\gamma_1$, but this requires the feedback connections strength $\gamma_2$ to remain low. Of course, different qualitative behaviors and stability regions may be obtained for different choices of the hyperparameters $(\alpha,\beta,\lambda)$; while it is beyond the scope of the present study to characterize them all, it is important to point out that such a characterization is feasible using the present method, for any choice of the hyperparameters.

More interestingly, we can investigate the dependence of the wave speed as a function of the parameters $\gamma_1$ and $\gamma_2$. For example, when $\gamma_1=\chi(\gamma_2)$, we have that
\bqs
\rho_{\chi(\gamma_2),\gamma_2}(\theta)=\exp\left(-\mbi \frac{(\alpha-\lambda)\gamma_2+\beta \chi(\gamma_2)}{1-\beta\chi(\gamma_2)}\theta+\mathcal{O}(|\theta|^2)\right), \text{ as } \theta \rightarrow 0,
\eqs
such that the associated wave speed is given by
\bqs
c_0^\chi=\frac{(\alpha-\lambda)\gamma_2+\beta \chi(\gamma_2)}{1-\beta\chi(\gamma_2)},
\eqs
whose sign may vary as $\gamma_2$ is varied. We refer to the forthcoming section~\ref{secRMO} below for a practical example (see Figure~\ref{fig:Speed}).

\subsubsection{Continuous in time interpretation}

\begin{figure}[t!]
\centering
\includegraphics[width=.3\textwidth]{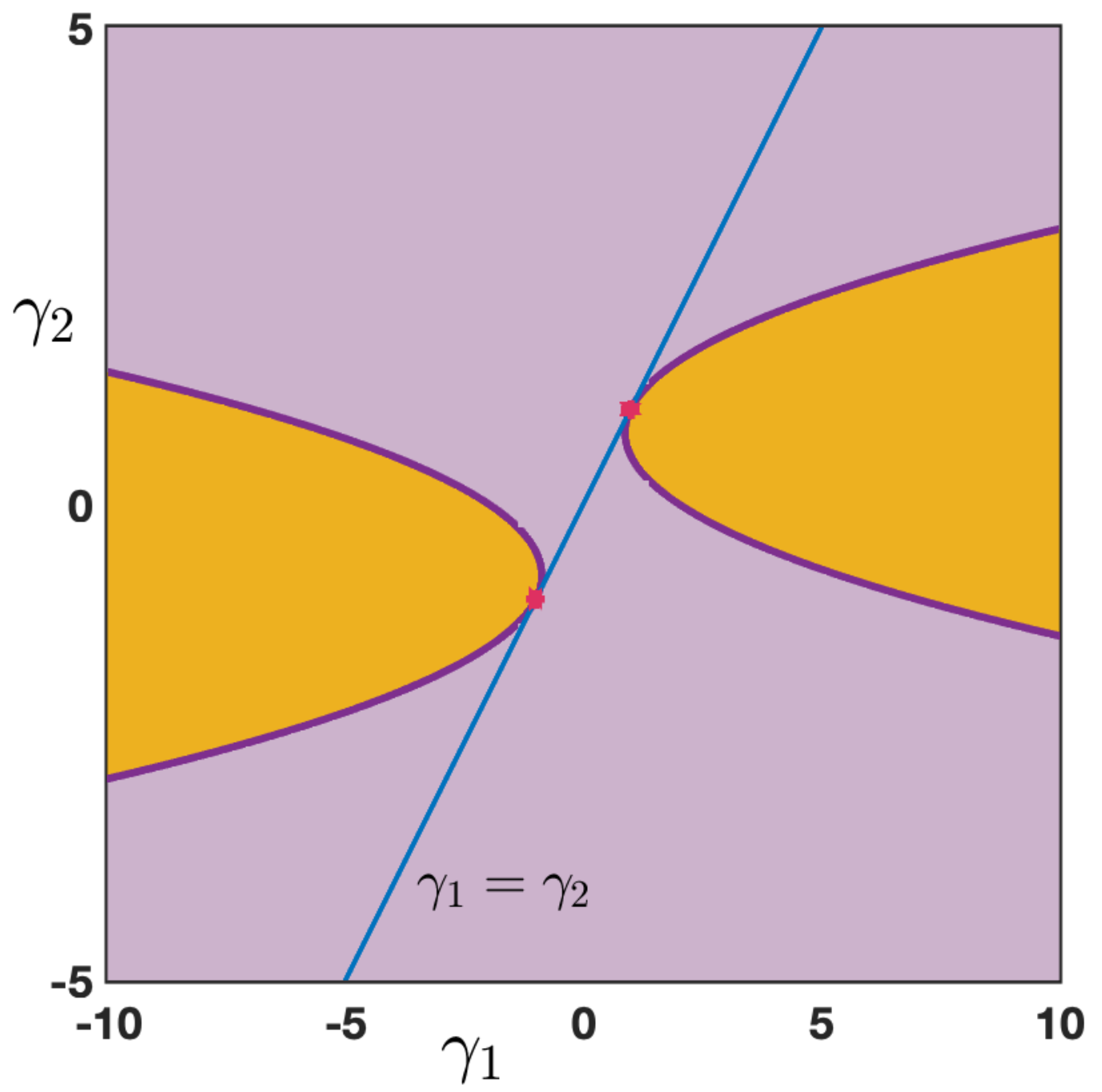}
\caption{Stability/instability regions and their boundaries as a function of $(\gamma_1,\gamma_2)$ for \eqref{latticeODEgambis} for any $(\widetilde{\alpha},\widetilde{\lambda},\widetilde{\beta})$ fixed with $\widetilde{\alpha}=\widetilde{\lambda}+\widetilde{\beta}$. The shaded orange region corresponds to an instability for \eqref{latticeODEgambis} while the purple region corresponds to a stability for \eqref{latticeODEgambis}. The boundaries of the stability/instability regions are given by the intersections of the parabolas $\gamma_1= \frac{\widetilde{\alpha}\gamma_2^2-(\widetilde{\alpha}+\widetilde{\lambda})\gamma_2+\widetilde{\beta}+\widetilde{\lambda}}{\widetilde{\beta}}$ and $\gamma_1= \frac{-\widetilde{\alpha}\gamma_2^2-(\widetilde{\alpha}+\widetilde{\lambda})\gamma_2-\widetilde{\beta}-\widetilde{\lambda}}{\widetilde{\beta}}$ (magenta curves) where equation \eqref{latticeODEgambis} is marginally stable. We represented the line $\gamma_1=\gamma_2$ (blue curve) which corresponds to the case studied in Figure~\ref{fig:StabRegionGamContinuous}.}
\label{fig:StabRegionGamfbContinuous}
\end{figure}

As done in previous sections, we now perform a continuous in time limit of the model \eqref{modelgambis}. With the same scaling on the hyperparameters 
\bqs
\widetilde{\beta}:=\frac{\beta}{\Delta t}, \quad \widetilde{\lambda}:=\frac{\lambda}{\Delta t}, \text{ and } \widetilde{\alpha}:=\frac{\alpha}{\Delta t},
\eqs
we get that, at the limit $\Delta t\rightarrow0$, the equation \eqref{modelgambis} becomes the following lattice ordinary differential equation 
\bqq
\frac{\md }{\md t}{\bf u}_j(t)=(\widetilde{\beta}\gamma_1+\widetilde{\alpha}\gamma_2){\bf u}_{j-1}(t)-(\widetilde{\beta}+\widetilde{\lambda}+\widetilde{\alpha}\gamma_2^2){\bf u}_j(t)+\widetilde{\lambda}\gamma_2{\bf u}_{j+1}(t), \quad t>0.
\label{latticeODEgambis}
\eqq
The amplification factor function in this case is given by
\bqs
\nu_{\gamma_1,\gamma_2}(\theta)=(\widetilde{\beta}\gamma_1+\widetilde{\alpha}\gamma_2) e^{-\mbi\theta}-(\widetilde{\beta}+\widetilde{\lambda}+\widetilde{\alpha}\gamma_2^2)+\widetilde{\lambda}\gamma_2 e^{\mbi\theta}, \quad \theta\in[-\pi,\pi],
\eqs
whose real part is given by
\bqs
\Re(\nu_{\gamma_1,\gamma_2}(\theta))=(\widetilde{\beta}\gamma_1+(\widetilde{\alpha}+\widetilde{\lambda})\gamma_2)\cos(\theta)-(\widetilde{\beta}+\widetilde{\lambda}+\widetilde{\alpha}\gamma_2^2), \quad \theta\in[-\pi,\pi].
\eqs
When $\widetilde{\beta}\gamma_1+(\widetilde{\alpha}+\widetilde{\lambda})\gamma_2>0$, we observe that
\bqs
\underset{\theta\in[-\pi,\pi]}{\max}\Re(\nu_{\gamma_1,\gamma_2}(\theta))=\Re(\nu_{\gamma_1,\gamma_2}(0))=\widetilde{\beta}\gamma_1+(\widetilde{\alpha}+\widetilde{\lambda})\gamma_2-(\widetilde{\beta}+\widetilde{\lambda}+\widetilde{\alpha}\gamma_2^2),
\eqs
such that
\bqs
\underset{\theta\in[-\pi,\pi]}{\max}\Re(\nu_{\gamma_1,\gamma_2}(\theta))=0 \Longleftrightarrow \gamma_1 = \frac{\widetilde{\alpha}\gamma_2^2-(\widetilde{\alpha}+\widetilde{\lambda})\gamma_2+\widetilde{\beta}+\widetilde{\lambda}}{\widetilde{\beta}}.
\eqs
Whereas, when $\widetilde{\beta}\gamma_1+(\widetilde{\alpha}+\widetilde{\lambda})\gamma_2<0$, we observe that
\bqs
\underset{\theta\in[-\pi,\pi]}{\max}\Re(\nu_{\gamma_1,\gamma_2}(\theta))=\Re(\nu_{\gamma_1,\gamma_2}(\pm\pi))=-\widetilde{\beta}\gamma_1-(\widetilde{\alpha}+\widetilde{\lambda})\gamma_2-(\widetilde{\beta}+\widetilde{\lambda}+\widetilde{\alpha}\gamma_2^2),
\eqs
such that
\bqs
\underset{\theta\in[-\pi,\pi]}{\max}\Re(\nu_{\gamma_1,\gamma_2}(\theta))=0 \Longleftrightarrow \gamma_1 = \frac{-\widetilde{\alpha}\gamma_2^2-(\widetilde{\alpha}+\widetilde{\lambda})\gamma_2-\widetilde{\beta}-\widetilde{\lambda}}{\widetilde{\beta}}.
\eqs
As a consequence, the stability regions are determined by the locations of the parabolas $\gamma_2\mapsto \frac{\widetilde{\alpha}\gamma_2^2-(\widetilde{\alpha}+\widetilde{\lambda})\gamma_2+\widetilde{\beta}+\widetilde{\lambda}}{\widetilde{\beta}}$ and $\gamma_2\mapsto \frac{-\widetilde{\alpha}\gamma_2^2-(\widetilde{\alpha}+\widetilde{\lambda})\gamma_2-\widetilde{\beta}-\widetilde{\lambda}}{\widetilde{\beta}}$ in the plane $(\gamma_1,\gamma_2)$. We observe that they never intersect and are oriented in the opposite directions and refer to Figure~\ref{fig:StabRegionGamfbContinuous} for a typical configuration. Here, we see that the system is stable for a very large range of values of both $\gamma_1$ and $\gamma_2$. In particular, for large enough values of the feedback connection weight (e.g. $|\gamma_2|>3$), stability is guaranteed regardless of the value of the feedforward connection weight $\gamma_1$ (within a reasonable range, e.g. $\gamma_1\in (-10,10)$). This is the opposite behavior as that obtained for the discrete system in Figure~\ref{fig:StabRegionGam12}, where stability was impossible under the same conditions for $\gamma_{1,2}$. This highlights again the errors of interpretation that can potentially be caused by discrete approximation of a continuous system.

\subsubsection{Fully continuous interpretation when $\W^f=\mathbf{I}_d+\xi_f \mathbf{A}$ and $\W^b=\mathbf{I}_d+\xi_b \mathbf{A}$.} When $\W^f=\mathbf{I}_d+\xi_f \mathbf{A}$ and $\W^b=\mathbf{I}_d+\xi_b \mathbf{A}$, one can once again identify $\E_{j,k}^t$ as the approximation of some smooth function $\E(t,x,y)$ at $t_n=n\Delta t$, $x_j=j\Delta x$ and $y_k = k \Delta y$, along the three dimensions of time, network depth and neuronal layer width. We may rewrite \eqref{model} in this case as
\begin{align*}
(1-\beta)\frac{\E_j^{n+1}-\E_j^n}{\Delta t}&=\beta \nu \frac{\E_{j-1}^{n+1} -\E_j^{n+1}}{\Delta x}+ \lambda \nu \frac{ \E_{j+1}^{n}-\E_j^{n}}{\Delta x}-\alpha \nu \frac{ \E_{j}^{n}-\E_{j-1}^{n}}{\Delta x}\\
&~~~+\beta \xi_f \kappa \frac{\mathbf{A}}{\Delta y^2}\E_{j-1}^{n+1}+\alpha \xi_b \kappa \frac{\mathbf{A}}{\Delta y^2}\E_{j-1}^{n}-2\alpha \xi_b \kappa \frac{\mathbf{A}}{\Delta y^2}\E_{j}^{n}-\alpha \xi_b^2 \kappa \Delta y^2 \frac{\mathbf{A}}{\Delta y^2} \frac{\mathbf{A}}{\Delta y^2}\E_{j}^{n}\\
&~~~+\lambda \xi_b \kappa \frac{\mathbf{A}}{\Delta y^2}\E_{j+1}^{n},
\end{align*}
such that in the limit $\Delta t\to0$, $\Delta x\to0$ and $\Delta y\to0$ with $\nu$ and $\kappa$ fixed, we obtain the following partial differential equation  
\bqs
\partial_t \E(t,x,y) + \frac{\nu (\beta+\alpha-\lambda)}{1-\beta } \partial_x \E(t,x,y) =\kappa \frac{\beta \xi_f+(\lambda-\alpha)\xi_b}{1-\beta}\partial_y^2\E(t,x,y).
\eqs
As before, this is a diffusion equation along the $y$ dimension, whose stability depends on the positivity of the diffusion coefficient, i.e. $\beta \xi_f+(\lambda-\alpha)\xi_b \geq 0$ .

\subsubsection{Application to a ring model of orientations}\label{secRMO}

Going back to our discrete system, in this section we consider the case where neurons within each layer encode for a given orientation in $[0,\pi]$. Here, we have in mind visual stimuli which are made of a fixed elongated black bar on a white background with a prescribed orientation. We introduce the following matrix $\mathbf{A}_{\mathrm{per}}\in\mathscr{M}_d(\R)$ given by
\bqs
\mathbf{A}_{\mathrm{per}}=\left(
\begin{matrix}
-2 & 1 & 0 &  \cdots & 0 & 1  \\
1 & -2 & 1 &  \ddots & \ddots & 0 \\
0 & \ddots & \ddots &  \ddots & \ddots & \vdots \\
\vdots & \ddots & \ddots & \ddots & \ddots & \vdots \\
0 & \ddots & \ddots & 1 & -2 & 1 \\
1 & 0 & \cdots & 0 & 1 & -2
\end{matrix}
\right),
\eqs
which is nothing but the discretizing of the Laplacian with boundary condition. Indeed, for each $\E_j^n\in\R^d$, we assume that neuron $\E_{j,k}^n$ encodes for orientation $\frac{k}{d}\pi$ for $k=1,\cdots,d$. We readily remark that $0\in \mathrm{Spec}(\mathbf{A}_{\mathrm{per}})$ with corresponding eigenvector $U_1=(1,\cdots,1)^{\bf t}\in\R^d$. Furthermore, we have:
\begin{itemize}
\item if $d=2m+1$ is odd, then $\lambda_p=-4\sin\left(\frac{p\pi}{d}\right)^2$ with $p=1,\cdots, m$ is an eigenvalue of $\mathbf{A}_{\mathrm{per}}$ of multiplicity $2$ with associated eigenvectors 
\begin{align*}
U_{2p}&=\left(\cos\left(\frac{2p\pi}{d}\right),\cdots,\cos\left(\frac{2kp\pi}{d}\right),\cdots,1\right)^{\bf t}\in\R^d,\\
U_{2p+1}&= \left(\sin\left(\frac{2p\pi}{d}\right),\cdots,\sin\left(\frac{2kp\pi}{d}\right),\cdots,0\right)^{\bf t}\in\R^d;
\end{align*}
\item if $d=2m$ is even, then $\lambda_p=-4\sin\left(\frac{p\pi}{d}\right)^2$ with $p=1,\cdots, m-1$ is an eigenvalue of $\mathbf{A}_{\mathrm{per}}$ of multiplicity $2$ with associated eigenvectors $U_{2p}$ and $U_{2p+1}$ as above. And $\lambda=-4$ is a simple eigenvalue of $\mathbf{A}_{\mathrm{per}}$ with associated eigenvector $U_d=(-1,1,-1,1,\cdots,-1,1)\in\R^d$.
\end{itemize}

\begin{figure}[t!]
  \centering
\subfigure[Eigenvectors $U_{2}$ and $U_{3}$.]{\includegraphics[width=.35\textwidth]{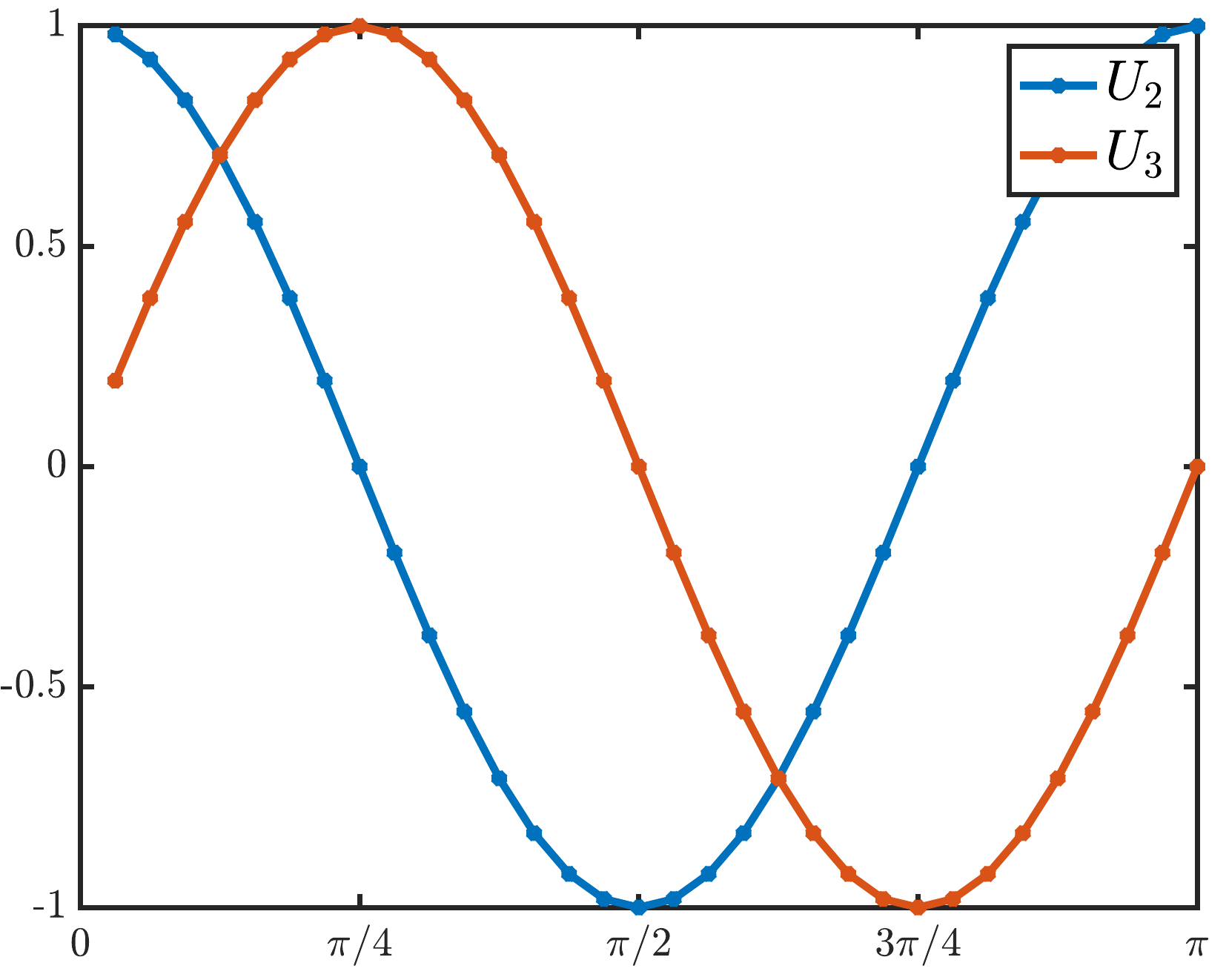}}\hspace{1cm}
\subfigure[Eigenvectors $U_{4}$ and $U_{5}$.]{\includegraphics[width=.35\textwidth]{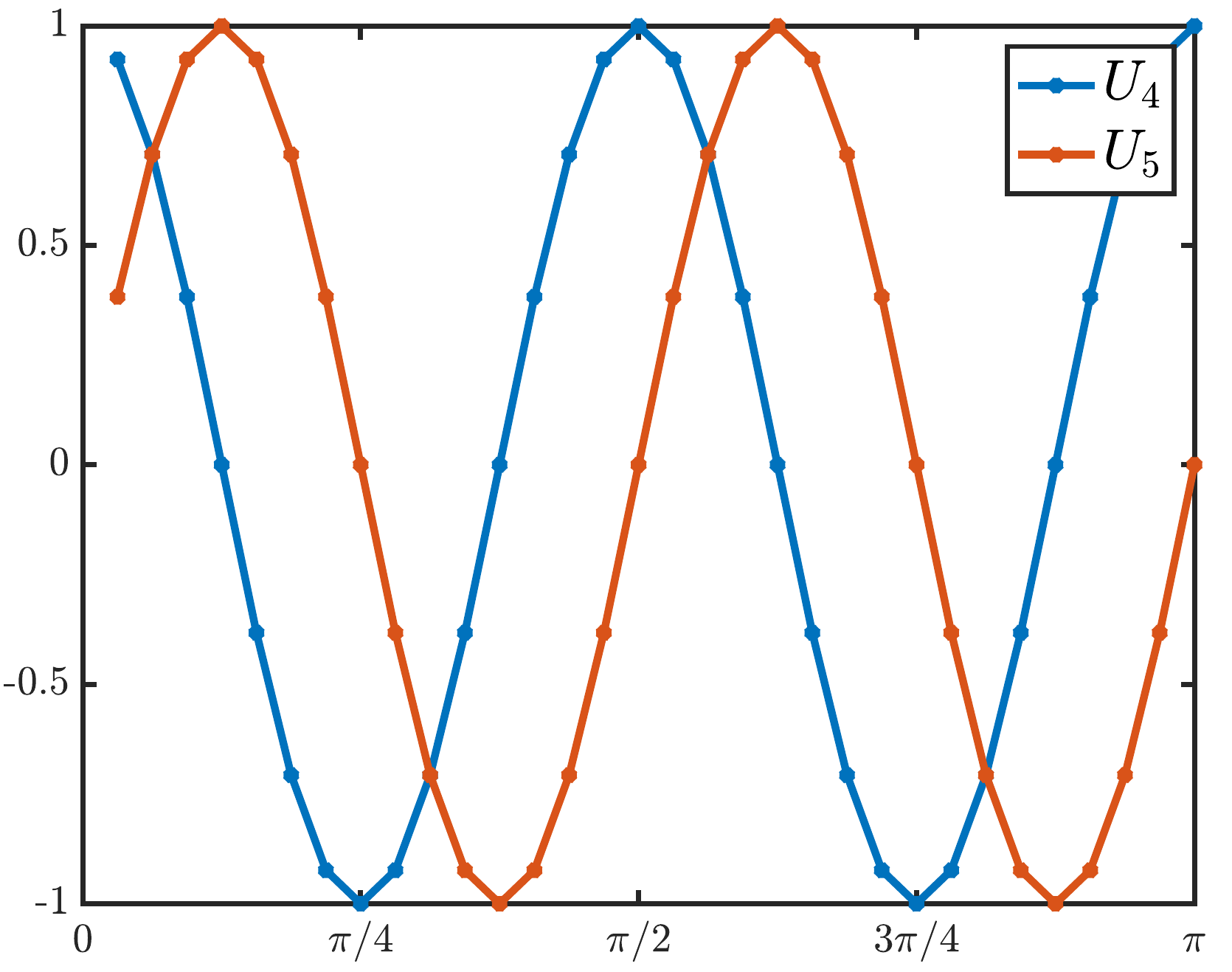}}
\caption{We plot the eigenvectors $U_{2p}$ and $U_{2p+1}$ for $p=1$ and $p=2$ as a function of $\frac{k}{d}\pi$ for $k=1,\cdots,d$. We note that $U_{2p}$ and $U_{2p+1}$ encode the first Fourier modes. Here we have set $d=2^5$.}
  \label{fig:Modes}
\end{figure}

It may be interesting to note that any linear combinations of $U_{2p}$ and $U_{2p+1}$ can always be written in the form
\bqs
a U_{2p}+bU_{2p+1} = A \left(\cos\left(\frac{2p\pi}{d}+\varphi\right),\cdots,\cos\left(\frac{2kp\pi}{d}+\varphi\right),\cdots,\cos(\varphi)\right)^{\bf t}\in\R^d,
\eqs
where $A=\sqrt{a^2+b^2}>0$ and $\varphi=-\mathrm{arctan}\left(\frac{b}{a}\right)\in(-\pi/2,\pi/2)$ whenever $a\neq0$ and $b\neq0$. This means that $U_{2p}$ and $U_{2p+1}$ span all possible translations modulo $[0,\pi]$ of a fixed profile. We refer to Figure~\ref{fig:Modes} for a visualization of the first eigenvectors. In short, these eigenvectors $U_i$ implement a Fourier transform of the matrix $\mathbf{A}_{\mathrm{per}}$.

\begin{figure}[t!]
  \centering
\subfigure[$\W^f$.]{\includegraphics[width=.35\textwidth]{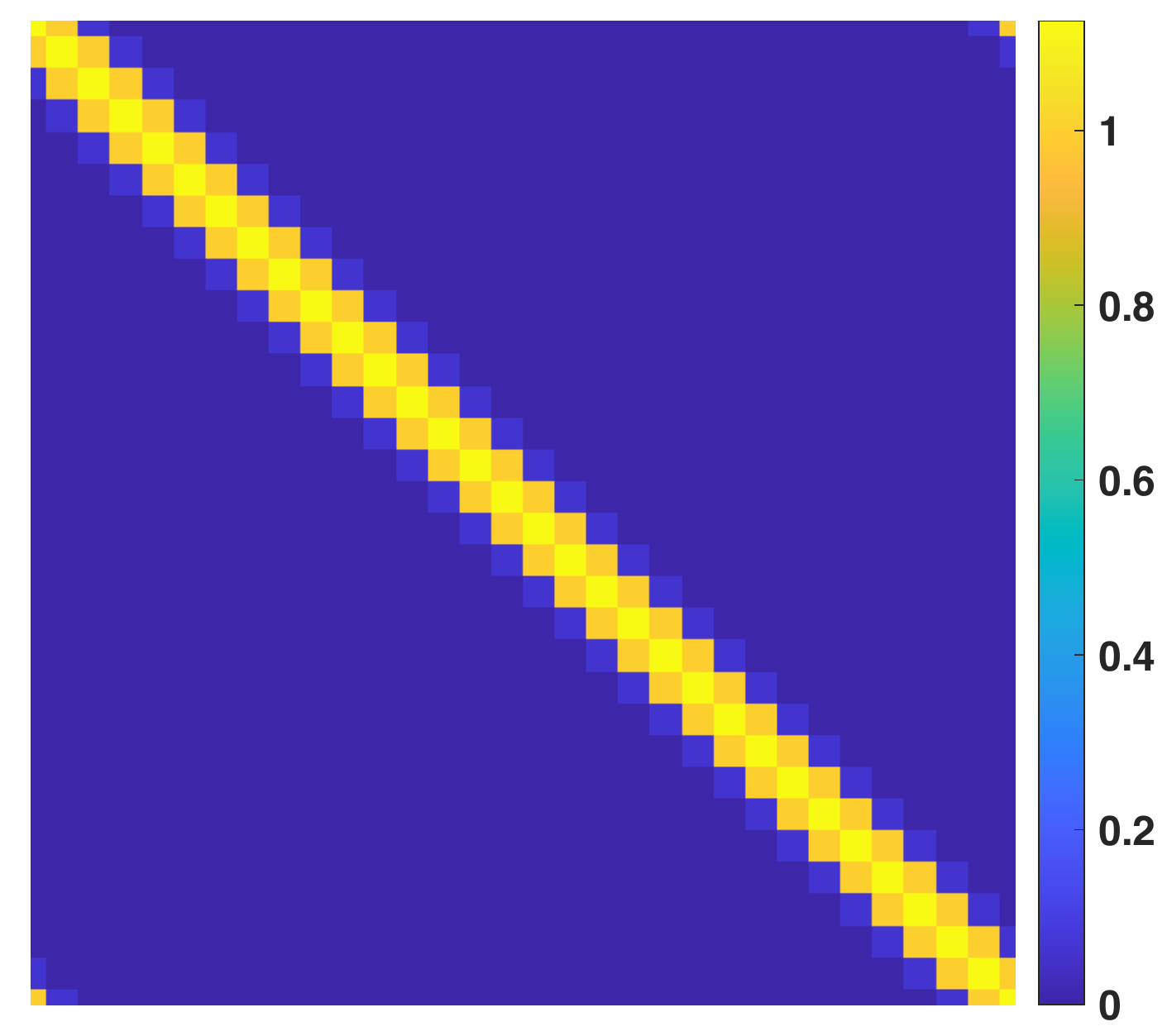}}\hspace{1cm}
\subfigure[$\W^b$.]{\includegraphics[width=.35\textwidth]{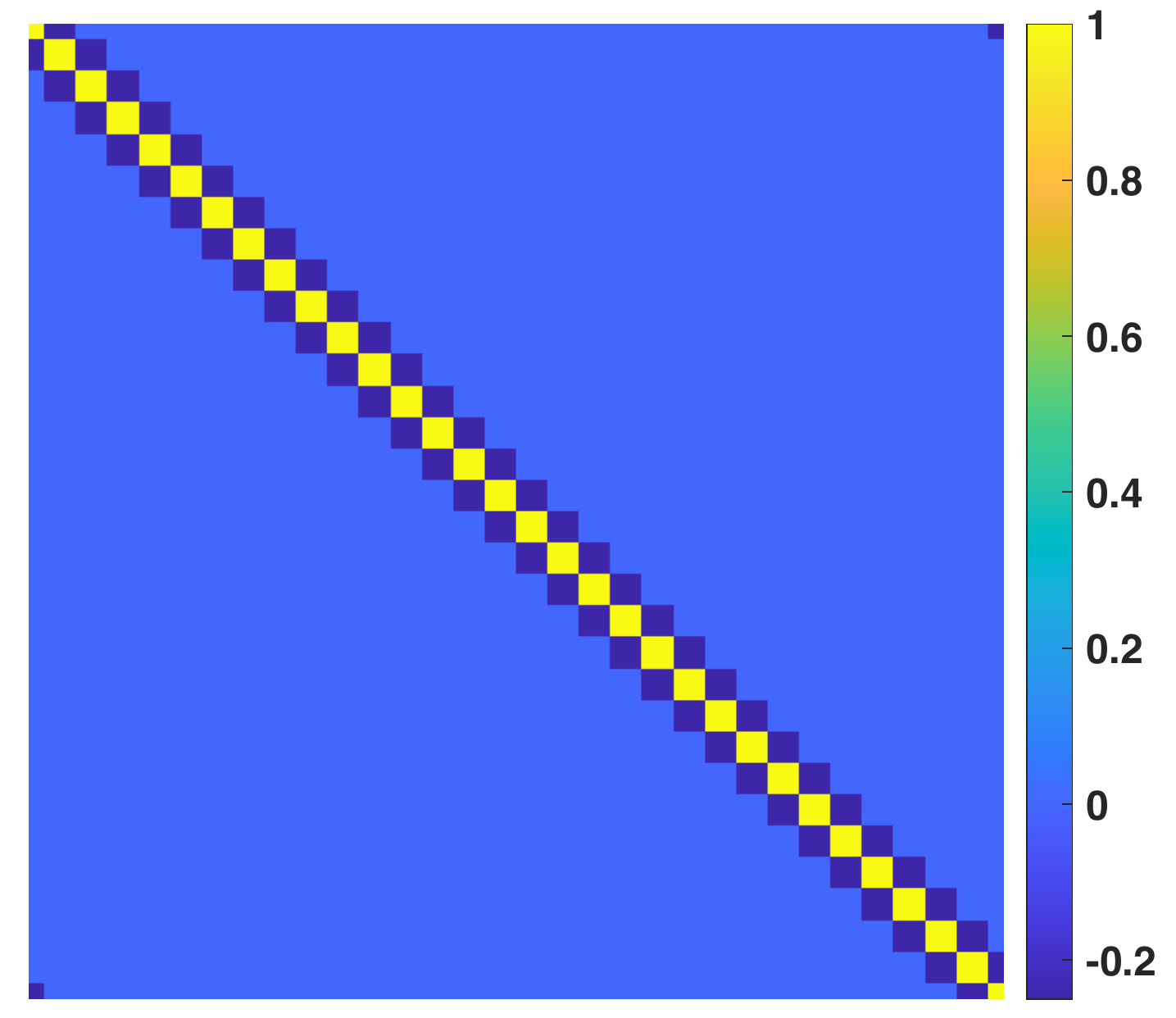}}
\caption{Illustration of the matrices $\W^f$ and $\W^b$ for $d=2^5$ neurons and values of the hyper-parameters fixed to $(\alpha,\beta,\lambda)=(0.1,0.1,0.5)$. Note the band structure of $\W^f$ with local excitation.}
  \label{fig:Matrix}
\end{figure}

We now set $\W^b$ to be
\bqs
\W^b=\frac{1}{2}\mathbf{I}_d-\frac{1}{4}\mathbf{A}_{\mathrm{per}}=\left(
\begin{matrix}
1 & -\frac{1}{4} & 0 &  \cdots & 0 & -\frac{1}{4}  \\
-\frac{1}{4} & 1 & -\frac{1}{4} &  \ddots & \ddots & 0 \\
0 & \ddots & \ddots &  \ddots & \ddots & \vdots \\
\vdots & \ddots & \ddots & \ddots & \ddots & \vdots \\
0 & \ddots & \ddots & -\frac{1}{4} & 1 & -\frac{1}{4} \\
-\frac{1}{4} & 0 & \cdots & 0 & -\frac{1}{4} & 1
\end{matrix}
\right),
\eqs
which means that $\W^b$ acts as a convolution with local excitation and lateral inhibition. From now on, to fix ideas, we will assume that $d=2m$ is even. We define the following matrix
\bqs
P=\left(U_1,U_2,\cdots,U_d\right)\in\mathscr{M}_d(\R).
\eqs
As a consequence, we have the decomposition
\bqs
P^\mathbf{t}\W^bP= \D^b, 
\eqs
with $\D^b=\mathrm{diag}\left(\frac{1}{2},\frac{1}{2}-\frac{1}{4}\lambda_1,\frac{1}{2}-\frac{1}{4}\lambda_1,\cdots,\frac{1}{2}-\frac{1}{4}\lambda_{m-1},\frac{1}{2}-\frac{1}{4}\lambda_{m-1},\frac{3}{2}\right)\in\mathscr{M}_d(\R)$. Now, for given values of the hyper-parameters $(\alpha,\beta,\lambda)$ with $\beta>0$,  we set $\D^f:=\chi(\D^b)$ where the map $\chi$, defined in \eqref{fctchi}, is applied to the diagonal elements of $\D^b$, that is
\bqs
\D^f=\mathrm{diag}\left(\chi\left(\frac{1}{2}\right),\chi\left(\frac{1}{2}-\frac{1}{4}\lambda_1\right),\cdots,\chi\left(\frac{1}{2}-\frac{1}{4}\lambda_{m-1}\right),\chi\left(\frac{3}{2}\right)\right)\in\mathscr{M}_d(\R).
\eqs
And then we set $\W^f:=P\D^fP^{\bf t}$. We refer to Figure~\ref{fig:Matrix} for an illustration of the structures of matrices $\W^f$ and $\W^b$. For large set of values of the hyper-parameters, $\W^f$ still present a band structure with positive elements on the diagonals indicating that $\W^f$ can also be interpreted as a convolution with local excitation. For the values of the hyper-parameters fixed in Figure~\ref{fig:Matrix}, the feedforward matrix $\W^f$ is purely excitatory.

\begin{figure}[t!]
  \centering
\includegraphics[width=.35\textwidth]{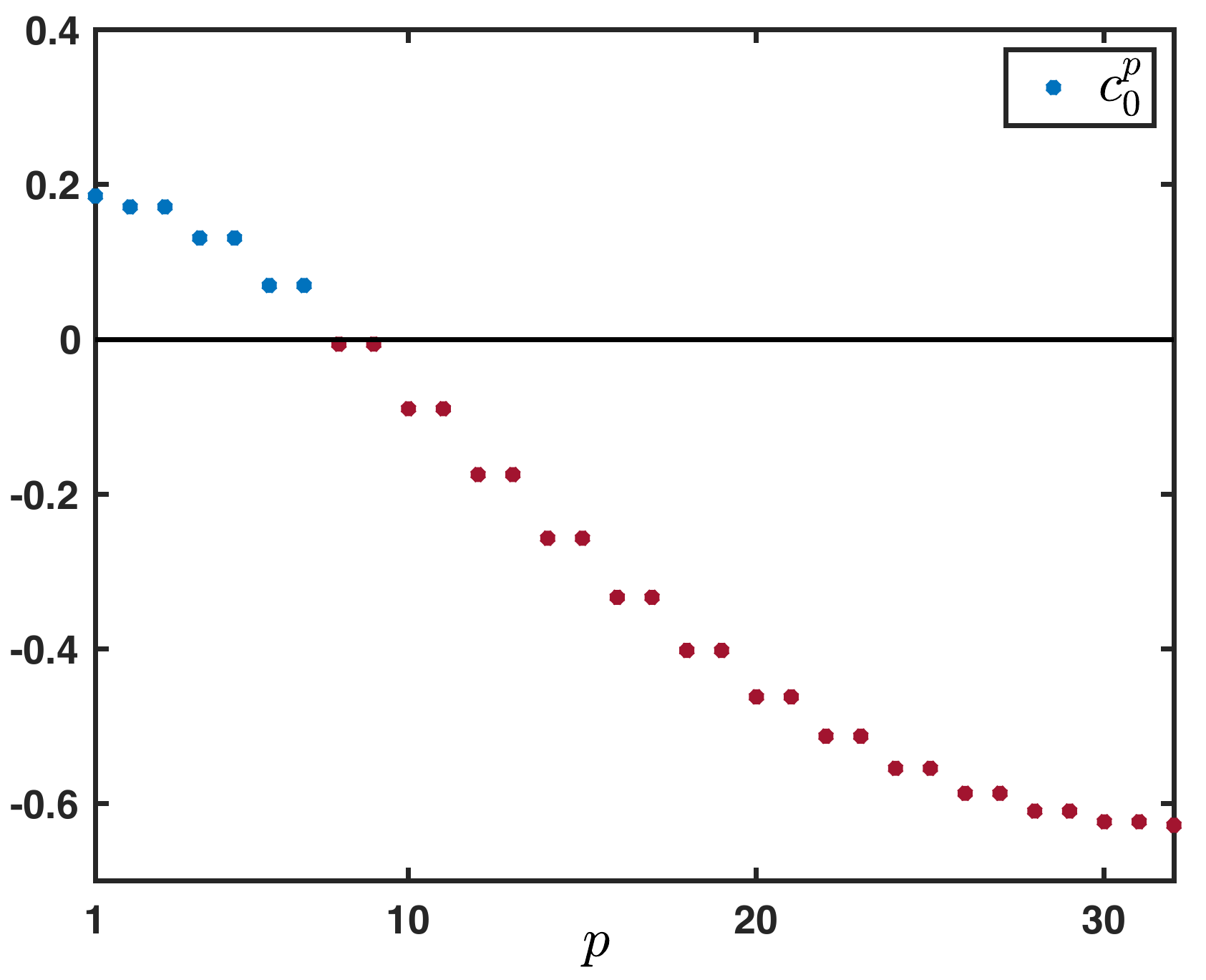}
\caption{Plot of the wave speed $c_0^p$ for $p=1,\cdots,d$ (colored dots). The color code (blue/red) refers to the sign of $c_0^p$: blue when positive and dark red when negative. Note that only the elements associated to the eigenvalues $\frac{1}{2}$, $\frac{1}{2}-\frac{1}{4}\lambda_1$, $\frac{1}{2}-\frac{1}{4}\lambda_2$ and $\frac{1}{2}-\frac{1}{4}\lambda_3$ are positive. We also remark that $c_0^p$ is a monotone decreasing function. Here $d=2^5$ and values of the hyper-parameters are fixed to $(\alpha,\beta,\lambda)=(0.1,0.1,0.5)$.}
  \label{fig:Speed}
\end{figure}

Reproducing the analysis developed in the previous Subsection~\ref{subsec:gensymcase}, we perform a change of orthonormal basis to express neural activities in terms of the relevant \textit{assemblies} $\U_j^n:=P^t\E_j^n$. With $P\U_j^n :=\E_j^n$, the recurrence equation becomes
\bqs
\U_j^{n+1}-\beta \D^f \U_{j-1}^{n+1} =\alpha \D^b \U_{j-1}^n+\left[(1-\beta-\lambda)\mathbf{I}_d-\alpha \D^b \D^b\right]  \U_j^{n} + \lambda  \D^b \U_{j+1}^{n}.
\eqs
Then, if we denote by $\gamma_p$ the $p$th diagonal element of $\D^b$, then for each $p=1,\cdots, d$ the above recurrence writes
\bqs
u_{j,p}^{n+1}-\beta \chi(\gamma_p) u_{j-1,p}^{n+1} =\alpha \gamma_p u_{j-1,p}^n +(1-\beta-\lambda-\alpha \gamma_p^2) u_{j,p}^{n} +\lambda  \gamma_p u_{j+1,p}^{n},
\eqs
where $u_{j,p}^n$ is the $p$th component (or neural assembly) of $\U_j^n$. For each $p=1,\cdots, d$, the associated amplification factor function reads
\bqs
\rho_p(\theta) = \frac{\alpha\gamma_p \left(e^{-\mbi \theta}-\gamma_p\right) +1-\beta +\lambda\left(\gamma_p e^{\mbi \theta}-1\right)}{1-\beta \chi(\gamma_p) e^{-\mbi \theta}}, \quad \theta\in[-\pi,\pi],
\eqs
and with our specific choice of function $\chi$, we have that $\rho_p(0)=1$ with 
\bqs
\rho_p(\theta)=\exp\left(-\mbi \frac{(\alpha-\lambda)\gamma_p+\beta \chi(\gamma_p)}{1-\beta\chi(\gamma_p)}\theta-\sigma_0^p\theta^2+\mathcal{O}(|\theta|^3)\right), \text{ as } \theta \rightarrow 0,
\eqs
such that the associated wave speed is given by
\bqs
c_0^p=\frac{(\alpha-\lambda)\gamma_p+\beta \chi(\gamma_p)}{1-\beta\chi(\gamma_p)},
\eqs
and where we have set
\bqs
\sigma_0^p=\frac{\alpha(1+4\lambda)\gamma_p^2+\beta+\lambda - (\alpha + \lambda)\gamma_p(\alpha\gamma_p^2 + \beta + \lambda)}{2(1-\beta \chi(\gamma_p))^2}.
%\frac{\alpha(\alpha +\lambda)\gamma_p^3 + (-4\alpha\lambda - \alpha)\gamma_p^2 + (\alpha + \lambda)(\beta + \lambda)\gamma_p - \beta - \lambda}{2(1-\beta \chi(\gamma_p))^2}.
\eqs
From now on, we assume that we have tuned the hyper-parameters such that $|\rho_p(\theta)|<1$ for all $\theta\in[-\pi,\pi]\backslash\left\{0\right\}$ and each $p=1,\cdots,d$. This can in fact be systematically checked numerically for a given set of hyper-parameters. We report in Figure~\ref{fig:Speed} the shape of $p\mapsto c_0^p$ for the same values of the hyper-parameters as the ones in Figure~\ref{fig:Matrix} and $d=2^5$. We first remark that $p\mapsto c_0^p$ is a monotone decreasing map, and in our specific case we have
\bqs
c^d_0<c^{d-1}_0=c^{d-2}_0<\cdots<c^9_0=c^8_0<0<c^7_0=c^6_0<c^5_0=c^4_0<c^3_0=c^2_0<c^1_0.
\eqs

Given a fixed input entry $\E_0\in\R^d$ presented at $j=0$ to the network continually at each time step, we can deduce which components of $\E_0\in\R^d$ will be able to propagate forward through the network. More precisely, we can decompose $\E_0$ along the basis $\left(U_1,\cdots U_d\right)$ of eigenvectors, that is
\bqs
\E_0= \sum_{p=1}^{d}a_p U_{p},
\eqs
for some real coefficients $a_p$ for $p=1,\cdots,d$. Assuming that the network was at rest initially, we get that the dynamics along each eigenvector (or neural assembly) is given by
\bqq
\left\{
\begin{split}
u_{j,p}^{n+1}-\beta \chi(\gamma_p) u_{j-1,p}^{n+1} &=\alpha \gamma_p u_{j-1,p}^n +(1-\beta-\lambda-\alpha \gamma_p^2) u_{j,p}^{n} +\lambda  \gamma_p u_{j+1,p}^{n}, \quad j\geq1, \quad n\geq0,\\
u_{0,p}^n&=a_p, \quad n\geq0,\\
u_{j,p}^0&=0, \quad j\geq1.
\end{split}
\right.
\label{bordp}
\eqq
Thus, we readily obtain that
\bqs
\E_j^n= \sum_{p=1}^{d}u_{j,p}^{n} U_{p}, \quad j\geq1, \quad n\geq 1,
\eqs
where $u_{j,p}^{n}$ is a solution to \eqref{bordp}.

\begin{figure}[t!]
  \centering
\subfigure[Case $p=2$ with $\gamma_2=\frac{1}{2}-\frac{1}{4}\lambda_1\sim0.5096$.]{\includegraphics[width=.35\textwidth]{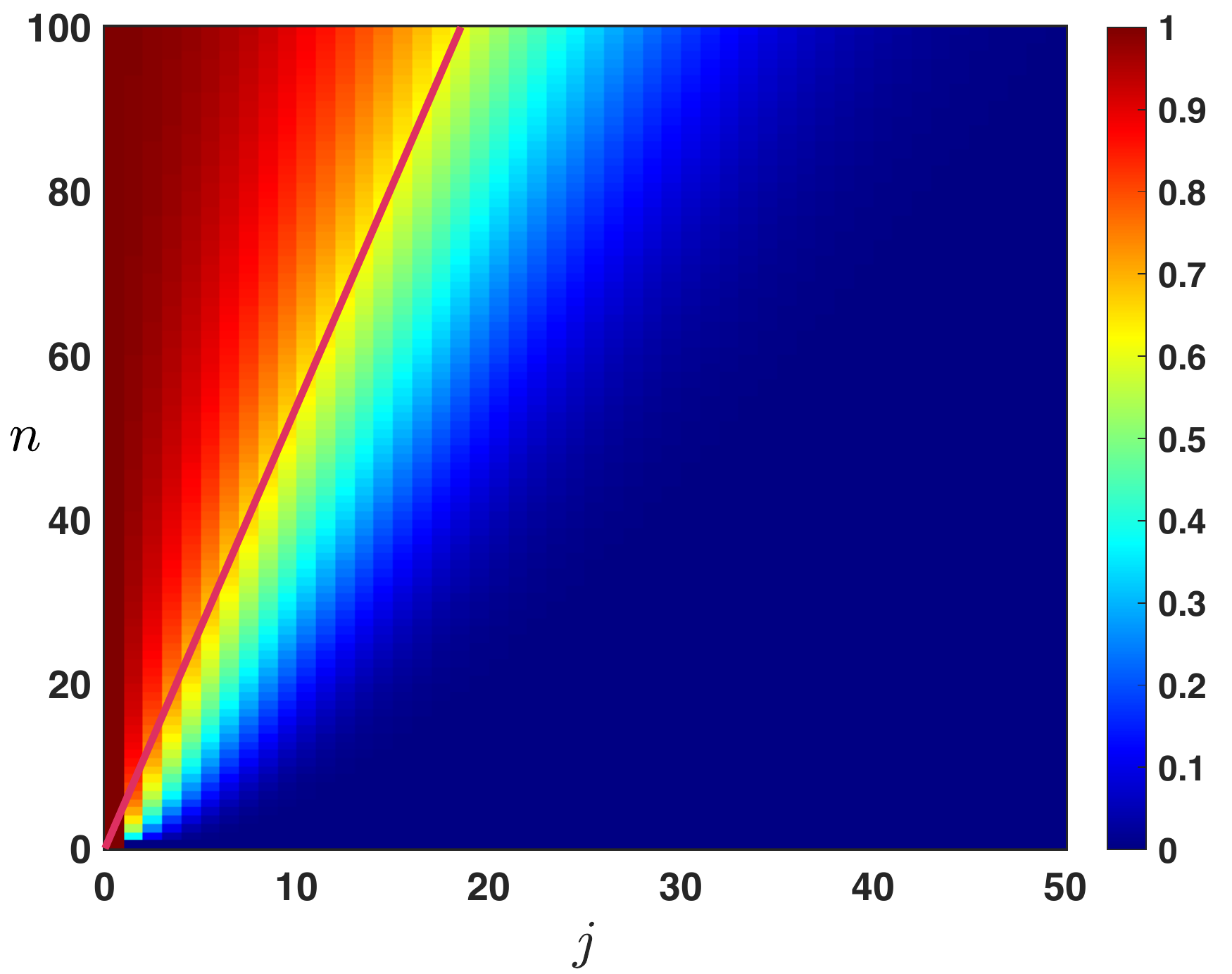}}\hspace{1cm}
\subfigure[Case $p=10$ with $\gamma_{10}=\frac{1}{2}-\frac{1}{4}\lambda_5\sim 0.7222$.]{\includegraphics[width=.35\textwidth]{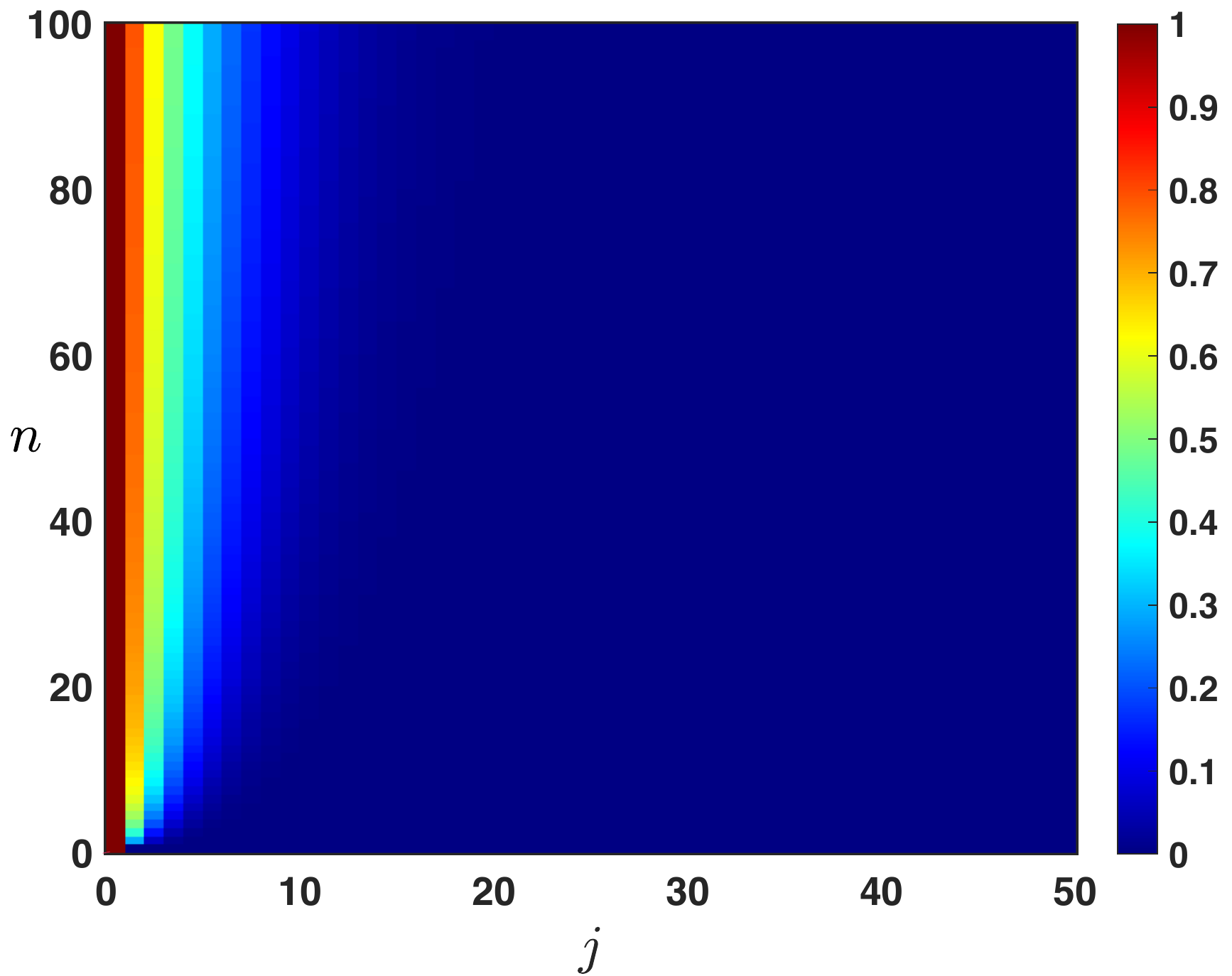}}
\caption{Space-time plot of the solution of the recurrence equation for \eqref{bordp} for $p=2$ and $p=10$ associated to respectively positive wave speed $c_0^2>0$ and negative wave speed $c_0^{10}<0$. The neural assembly associated with the 2nd eigenvector of the connectivity matrix propagates its input signal into the network at constant speed; but the neural assembly associated with the 10th eigenvector does not propagate the signals it receives on the input layer.}
  \label{fig:STrmo}
\end{figure}

As a consequence, the monotonicity property of the map $p\mapsto c_0^p$ indicates that the homogeneous constant mode $U_1$ is the fastest to propagate forward into the network with associated spreading speed $c^1_0$, it is then followed by the modes $(U_{2},U_{3})$ propagating at speed $c^2_0=c^3_0$. In our numerics, we have set the parameters such that $c^1_0 \approx c^2_0=c^3_0$ with a significant gap with the other wave speeds. Lets us remark, that all modes $U_p$ with $p\geq8$ are not able to propagate into the network (see Figure~\ref{fig:STrmo}). Thus our architecture acts as a mode filter.

Even more precisely, let us remark that the sequence $\left(a_p\left(\frac{\alpha \gamma_p+\beta \chi(\gamma_p)}{\lambda\gamma_p}\right)^j\right)_{j\geq0}$ is a stationary solutions of \eqref{bordp} which remains bounded whenever $p$ is such that the associated wave speed is negative, that is $c_0^p<0$, since in that case, one has $\alpha \gamma_p+\beta \chi(\gamma_p)<\lambda\gamma_p$.  The solution $\E_j^n$ can then be approximated as 
\bqs
\E_j^n\simeq \sum_{p~:~ c_0^p>0} \frac{a_p}{2}\left(1-\mathrm{erf}\left(\frac{j-c_0^pn}{\sqrt{4\sigma_0^pn}}\right) \right)U_{p}+\sum_{p~:~ c_0^p<0} a_p\left(\frac{\alpha \gamma_p+\beta \chi(\gamma_p)}{\lambda\gamma_p}\right)^jU_{p}, \quad j\geq1, \quad n\geq 1,
\eqs

\begin{figure}[t!]
  \centering
\subfigure[$j=0$.]{\includegraphics[width=.32\textwidth]{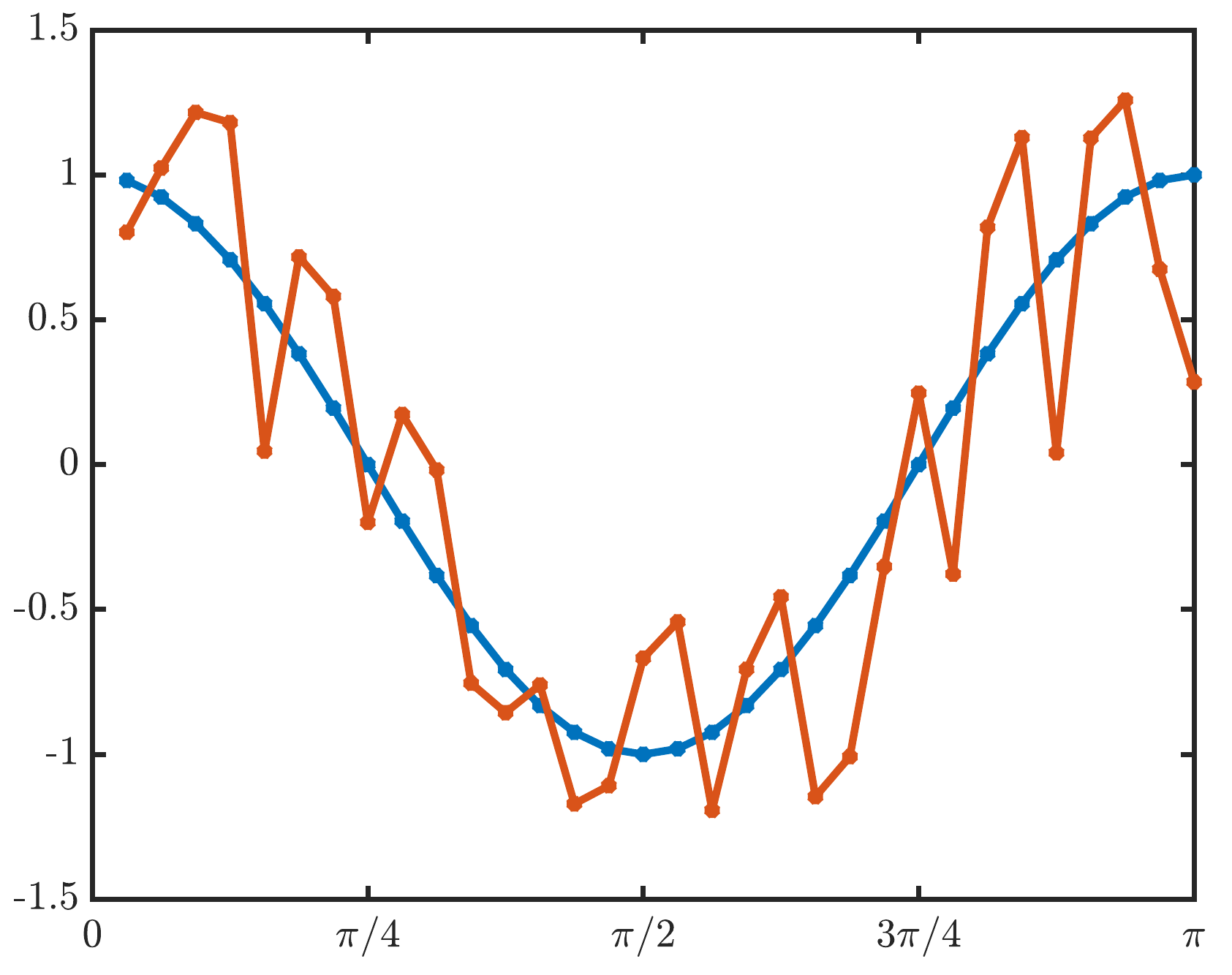}}
\subfigure[$j=1$.]{\includegraphics[width=.32\textwidth]{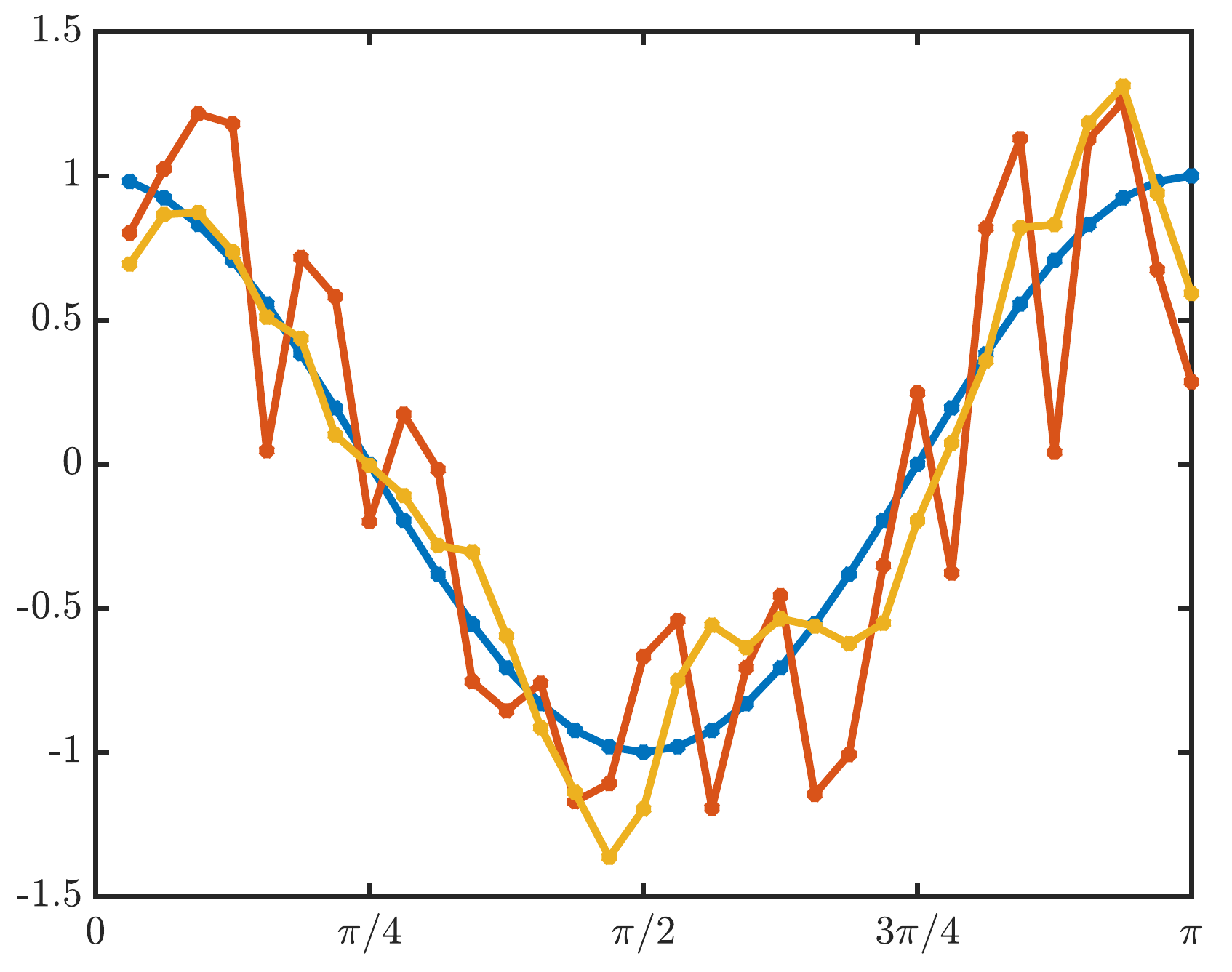}}
\subfigure[$j=2$.]{\includegraphics[width=.32\textwidth]{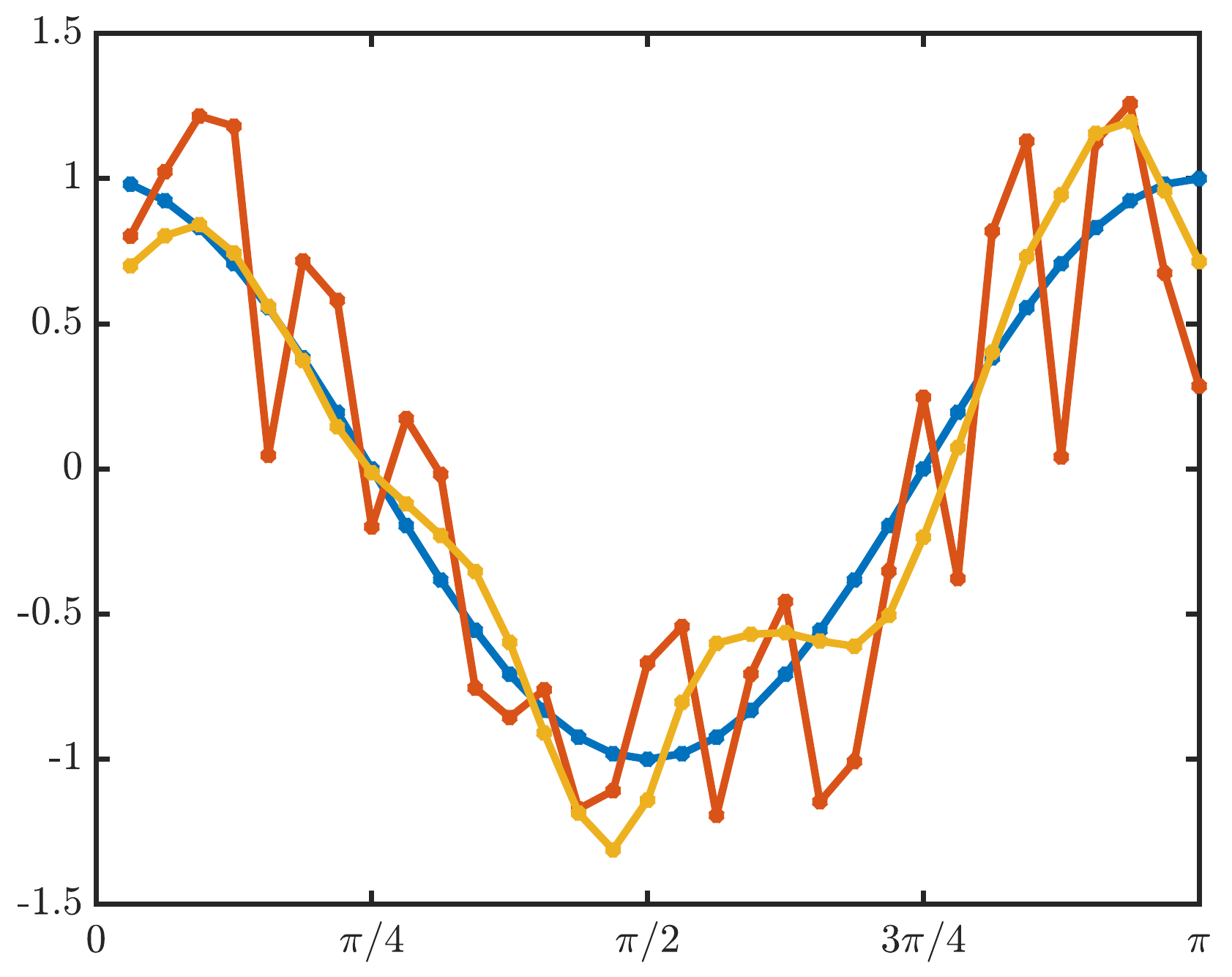}}
\subfigure[$j=3$.]{\includegraphics[width=.32\textwidth]{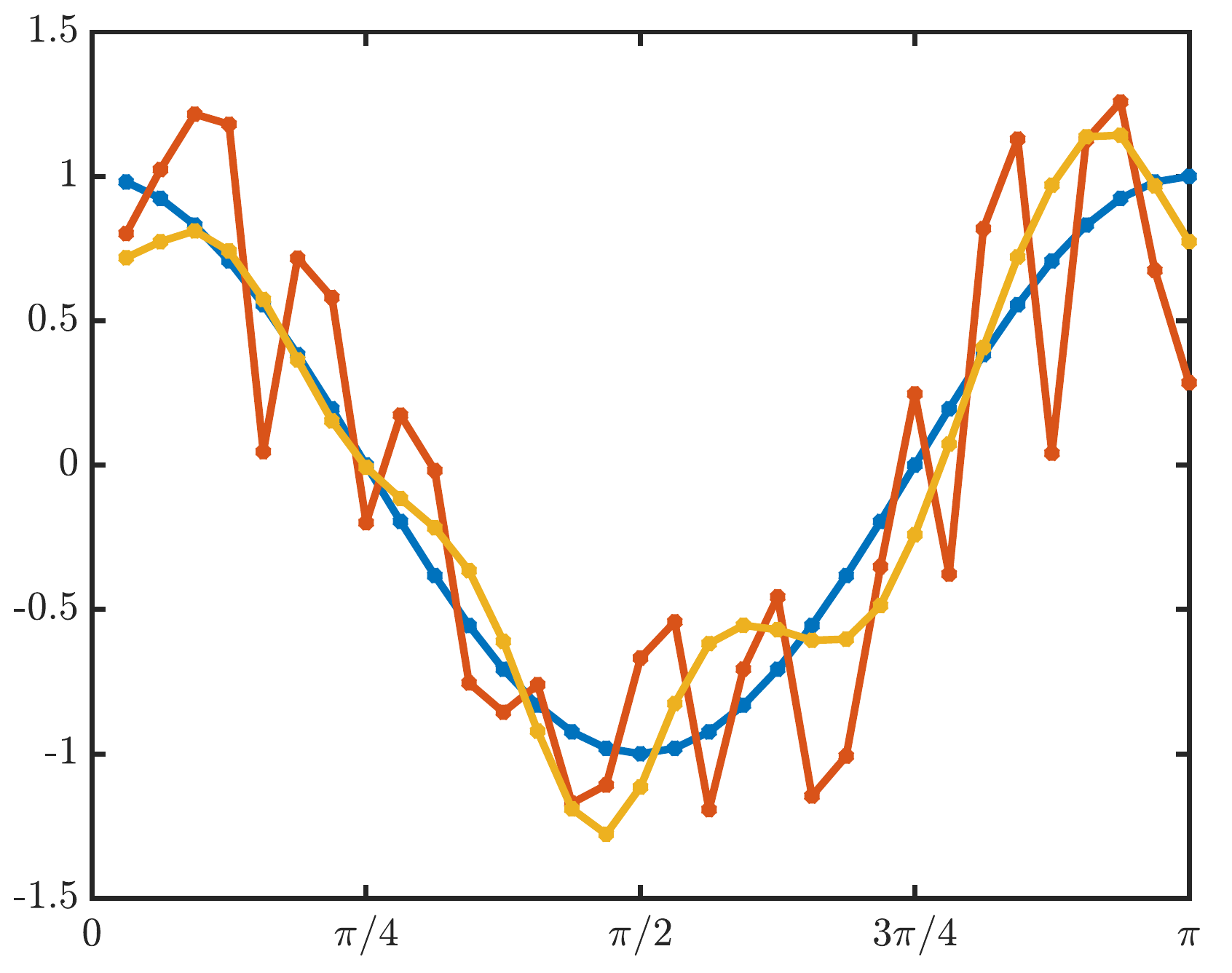}}
\subfigure[$j=4$.]{\includegraphics[width=.32\textwidth]{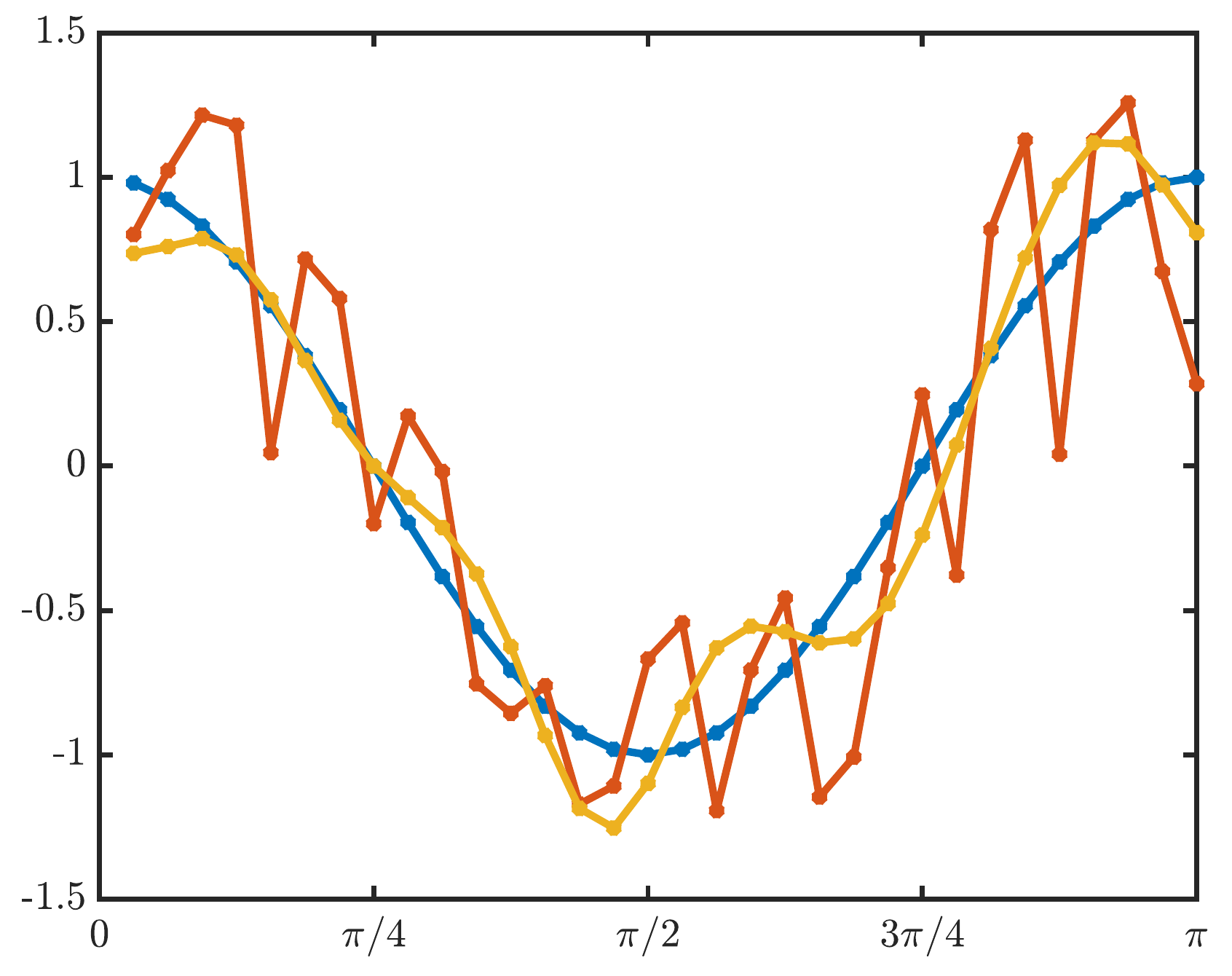}}
\subfigure[$j=5$.]{\includegraphics[width=.32\textwidth]{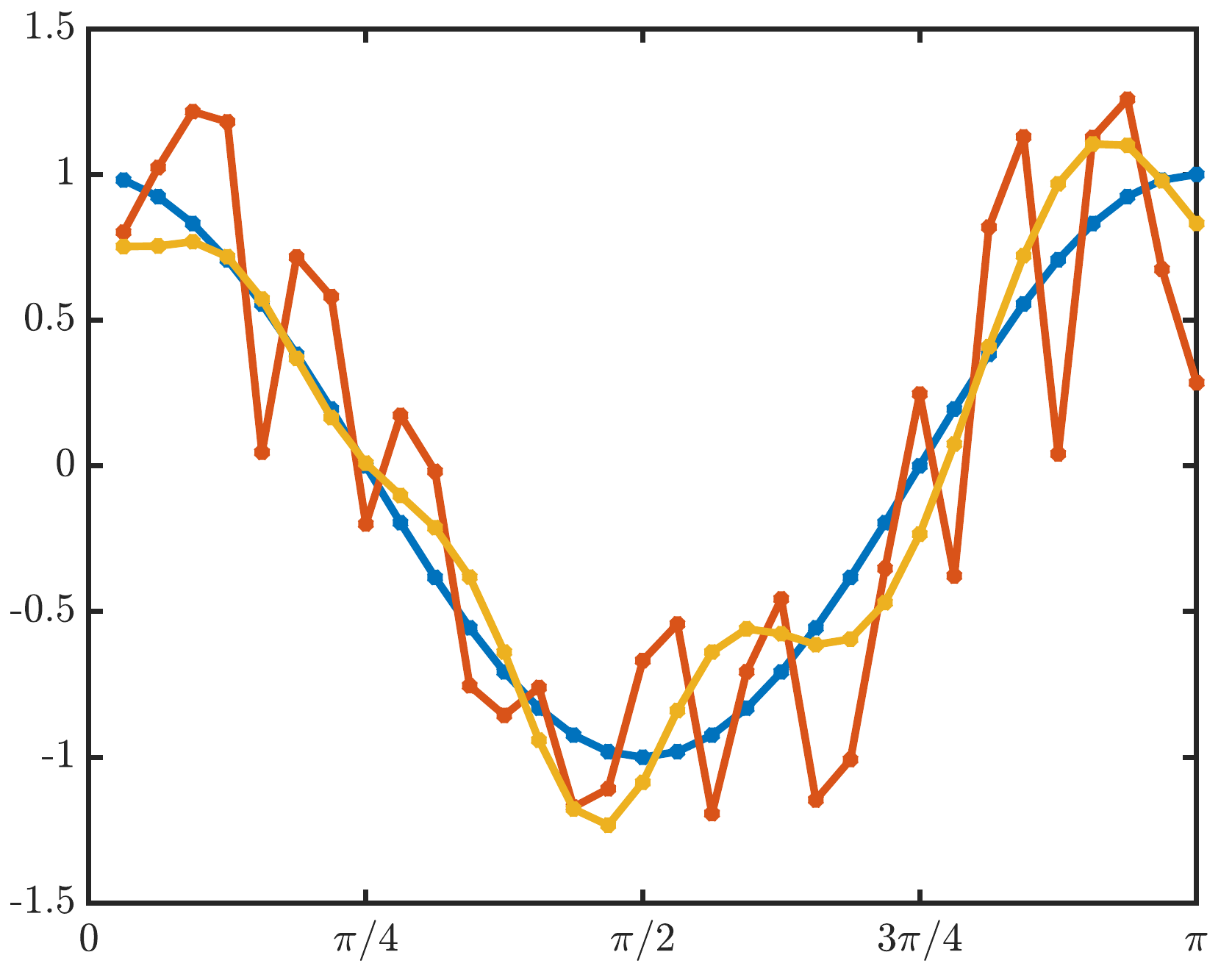}}
\caption{ A fixed input $\E_0$ (red) which is the superposition of a tuned curve at $\vartheta=0$ (blue) with some fixed random noise is presented at layer $j=0$. Profile (yellow) of $\E_j^n$ at time iteration $n=200$ along the first layers of the network $j\in\left\{1,2,3,4,5\right\}$. }
  \label{fig:PropagLayersrmo}
\end{figure}

\begin{figure}[t!]
  \centering
\subfigure[$n=0$.]{\includegraphics[width=.32\textwidth]{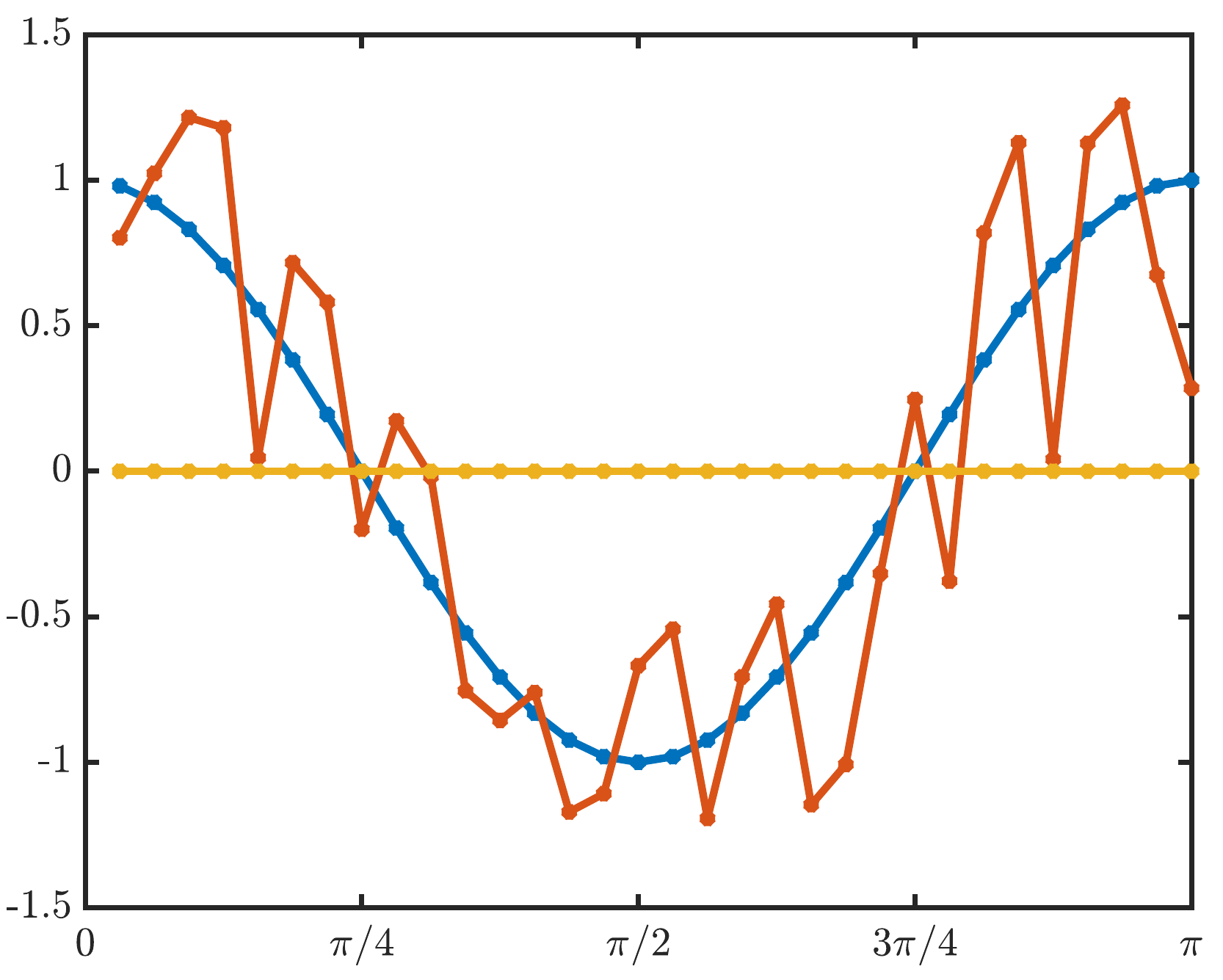}}
\subfigure[$n=20$.]{\includegraphics[width=.32\textwidth]{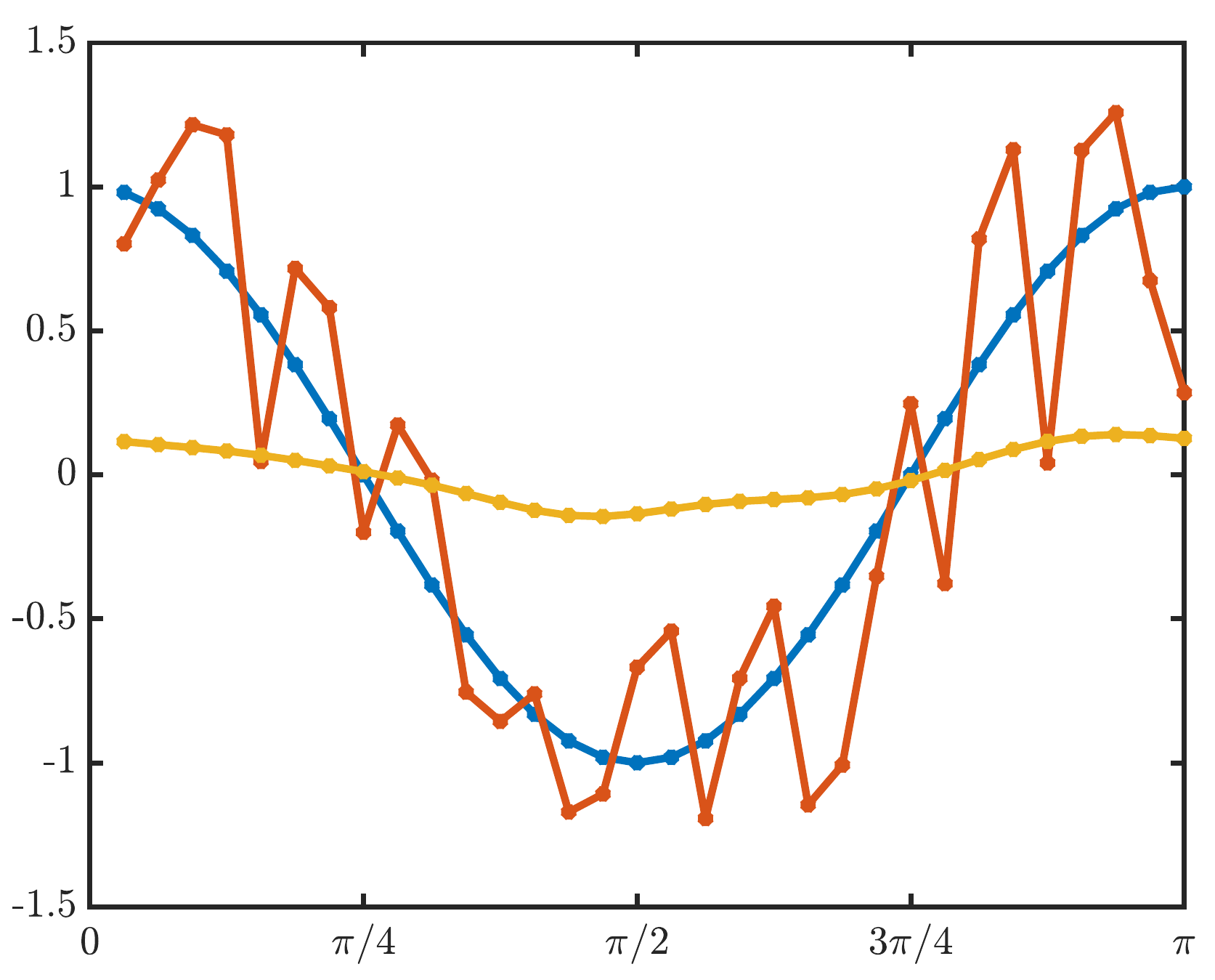}}
\subfigure[$n=40$.]{\includegraphics[width=.32\textwidth]{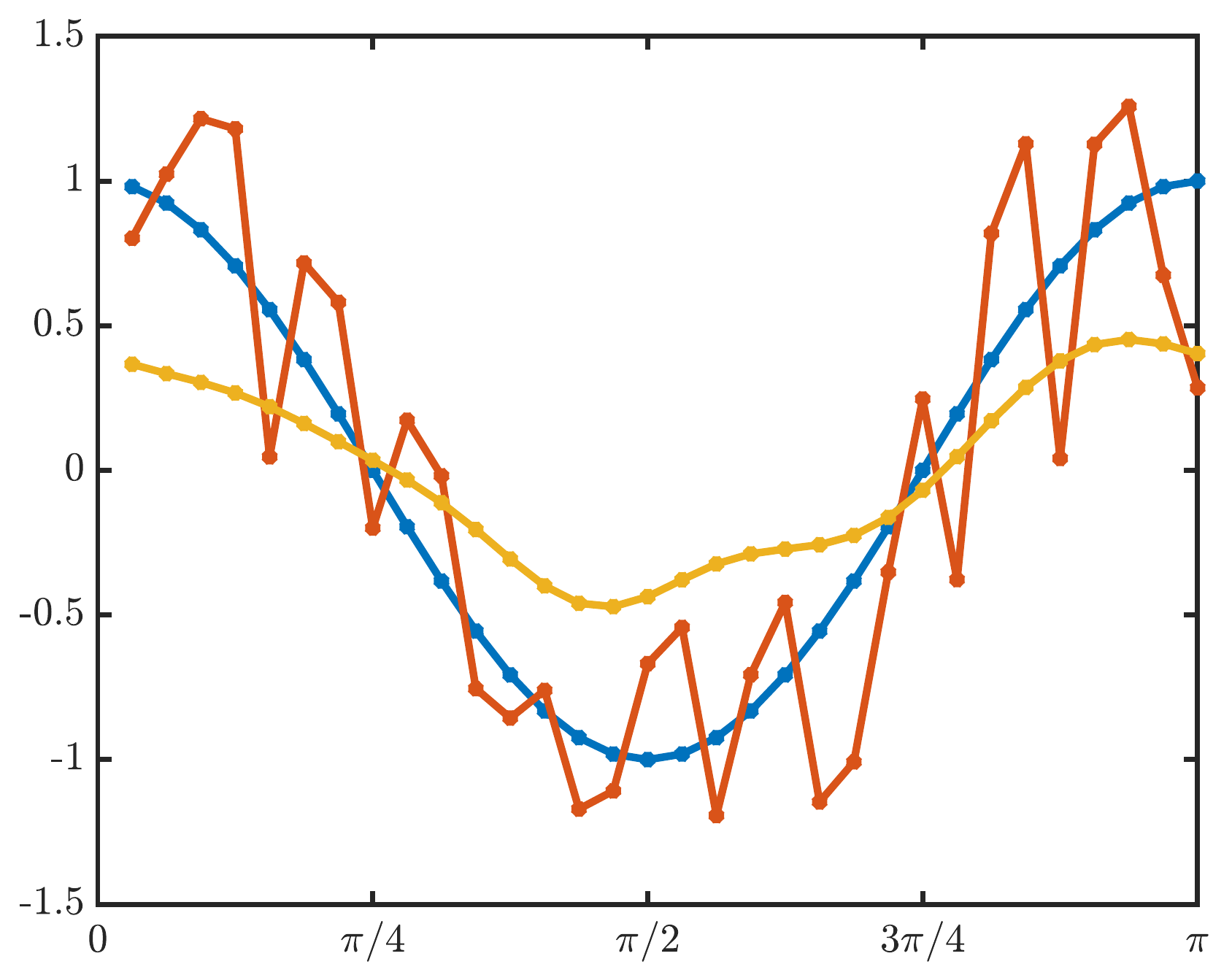}}
\subfigure[$n=60$.]{\includegraphics[width=.32\textwidth]{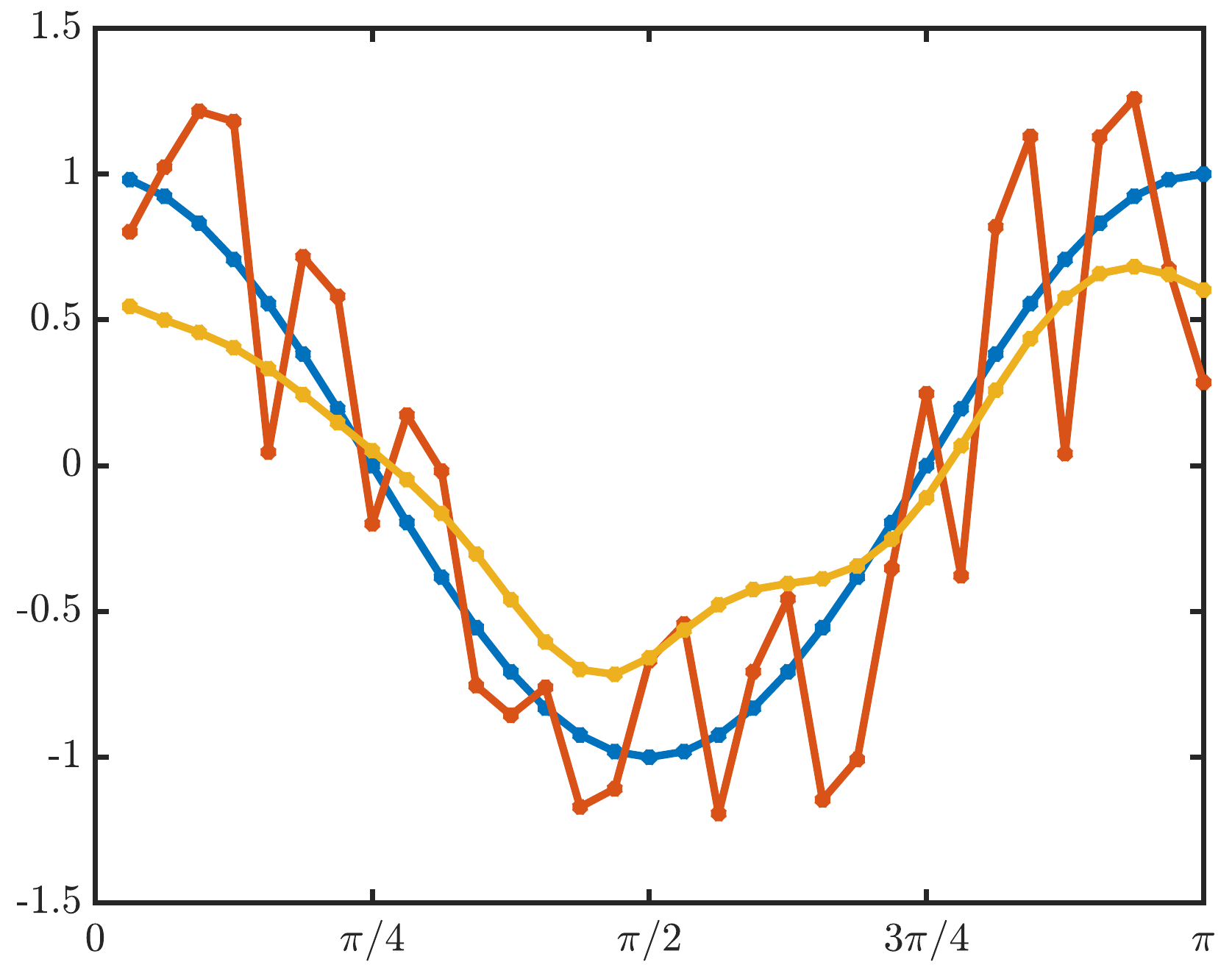}}
\subfigure[$n=80$.]{\includegraphics[width=.32\textwidth]{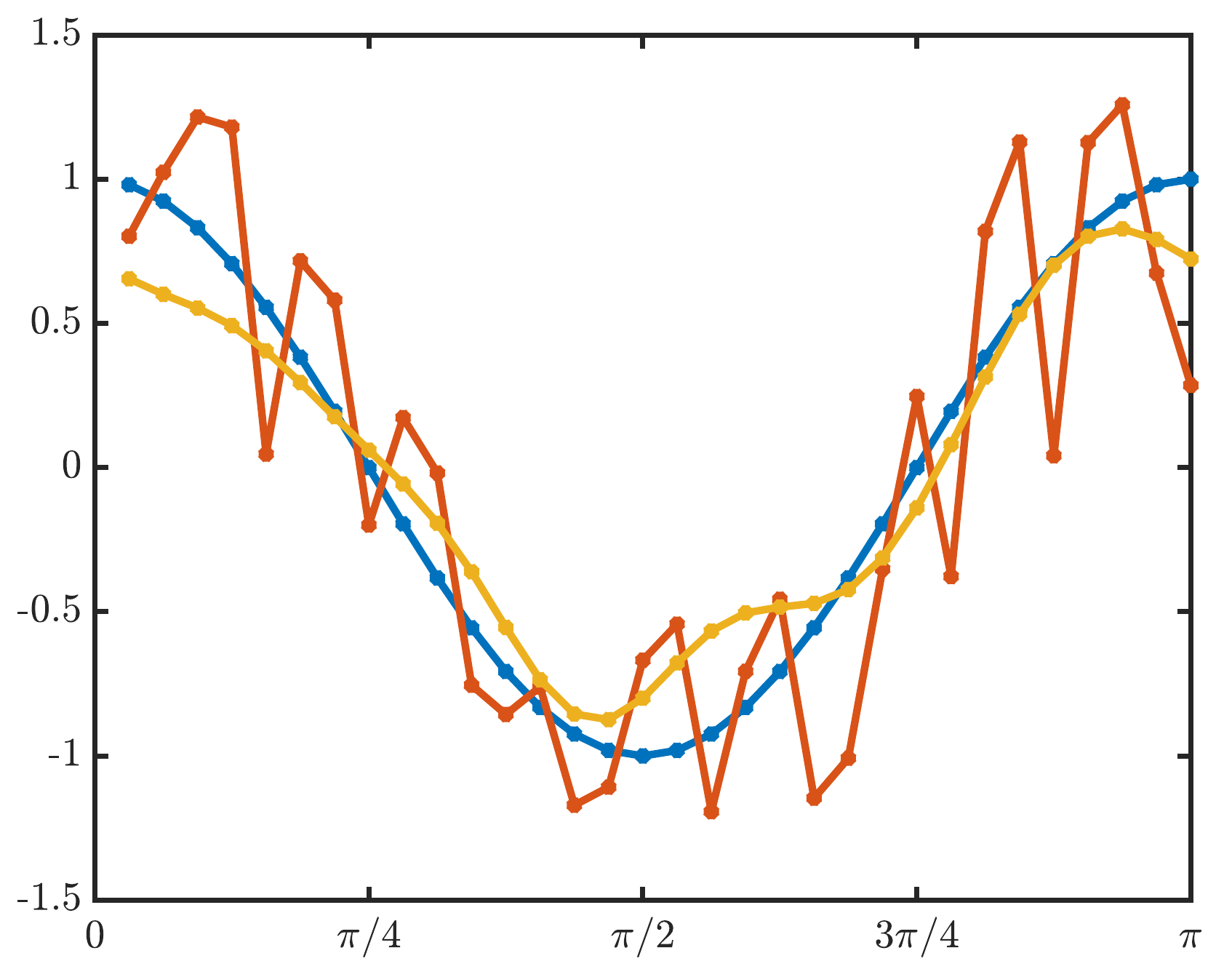}}
\subfigure[$n=100$.]{\includegraphics[width=.32\textwidth]{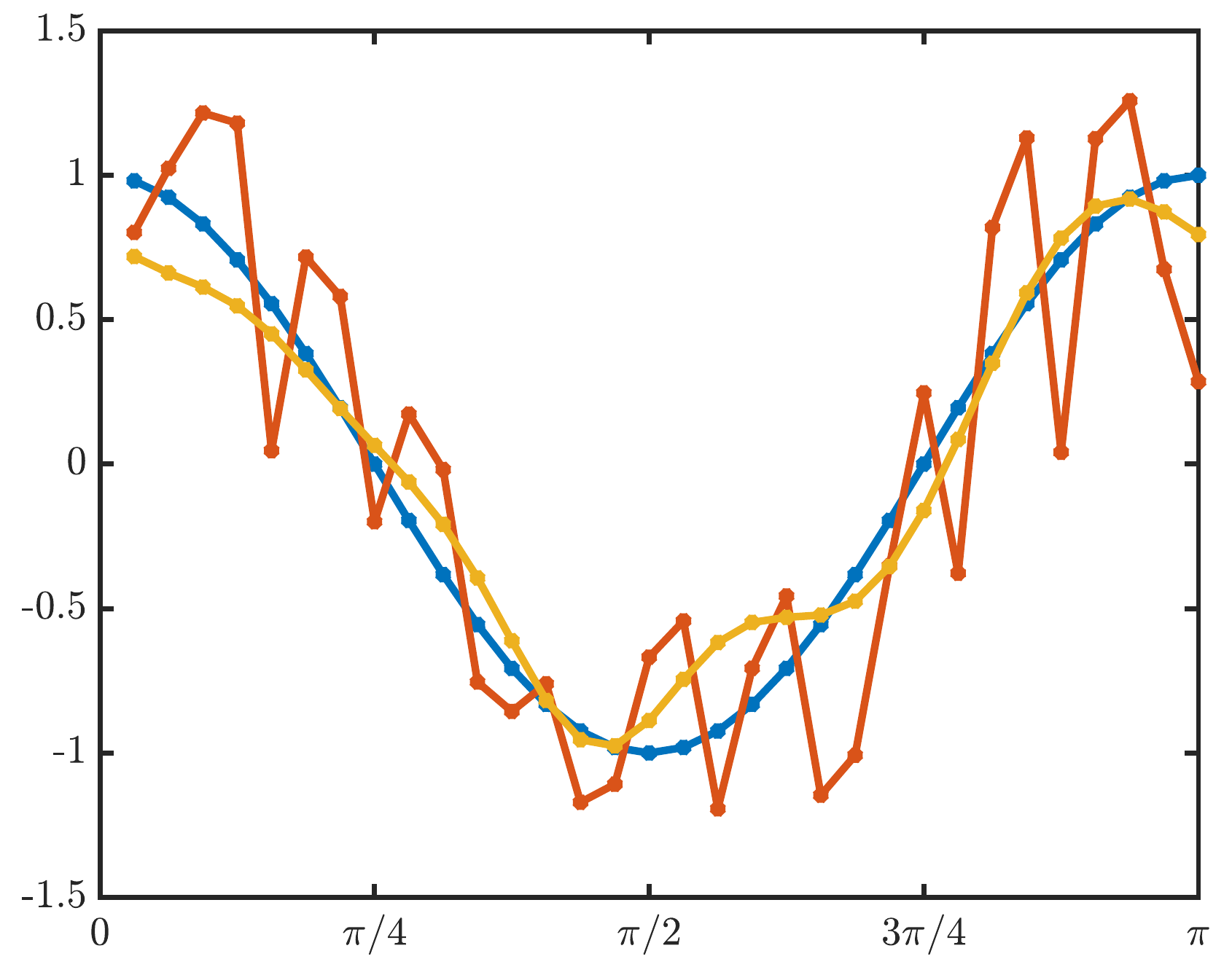}}
\caption{Time evolution of $\E_j^n$ at layer $j=10$ (yellow) for $n\in\left\{0,20,40,60,80,100\right\}$ with fixed input $\E_0$ at layer $j=0$ (red). The input $\E_0$ is the superposition of a tuned curve at $\vartheta=0$ (blue) with some fixed random noise.}
  \label{fig:TimeEvolJ10rmo}
\end{figure}

This is illustrated by a first example simulation in Figures~\ref{fig:PropagLayersrmo} and~\ref{fig:TimeEvolJ10rmo}. We present at $j=0$ a fixed input $\E_0$ which is generated as the superposition of a tuned curve at $\vartheta=0$ (blue) with some fixed random noise: namely we select $a_1=0$, $a_2=1$, $a_3=0$ and all other coefficients $a_p$ for $p=4,\cdots,d$ are drawn from a normal law with an amplitude pre-factor of magnitude $\varepsilon$ set to $\varepsilon=0.1$. The shape of the input $\E_0$ is shown in Figure~\ref{fig:PropagLayersrmo}(a). The profile of $\E_j^n$ at time iteration $n=200$ along the first layers of the network $j\in\left\{1,2,3,4,5\right\}$ is given in Figure~\ref{fig:PropagLayersrmo}(b)-(c)-(d)-(e)-(f) respectively. We first observe that the network indeed acts as a filter since across the layers of the network the solution profile $\E_j^n$ has been denoised and get closer to the tuned curve at $\vartheta=0$. Let us also remark that the filtering is more efficient for layers away from the boundary and is less efficient for those layers near the boundary. This is rather natural since the impact of the input $\E_0$ is stronger on the first layers. We see that already at layer $j=5$, we have almost fully recovered the tuned curve at $\vartheta=0$ (see Figure~\ref{fig:PropagLayersrmo}(f)). On the other hand, in Figure~\ref{fig:TimeEvolJ10rmo}, we show the time evolution of $\E_j^n$ at a fixed layer far away from the boundary, here $j=10$. Initially, at $n=0$, the layer is inactivated (see Figure~\ref{fig:TimeEvolJ10rmo}(a)), and we see that after several time iterations that the solution profile $\E_j^n$ start to be activated. It is first weakly tuned (see Figures~\ref{fig:TimeEvolJ10rmo}(b)-(c)-(d)) and then it becomes progressively fully tuned and converges to the tuned curve at $\vartheta=0$ (see Figures~\ref{fig:TimeEvolJ10rmo}(e)-(f)).

\begin{figure}[t!]
  \centering
\subfigure[$n=0$.]{\includegraphics[width=.32\textwidth]{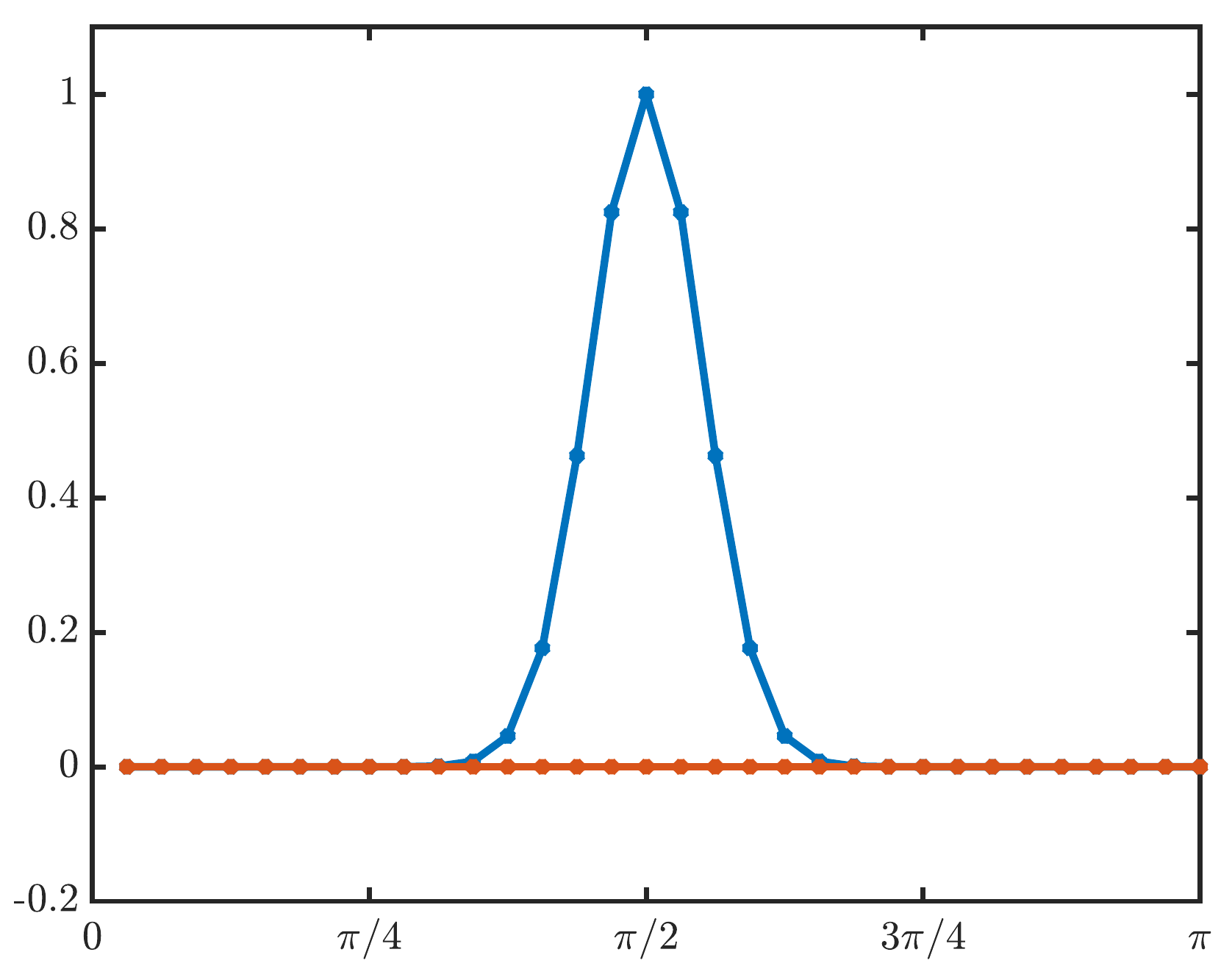}}
\subfigure[$n=20$.]{\includegraphics[width=.32\textwidth]{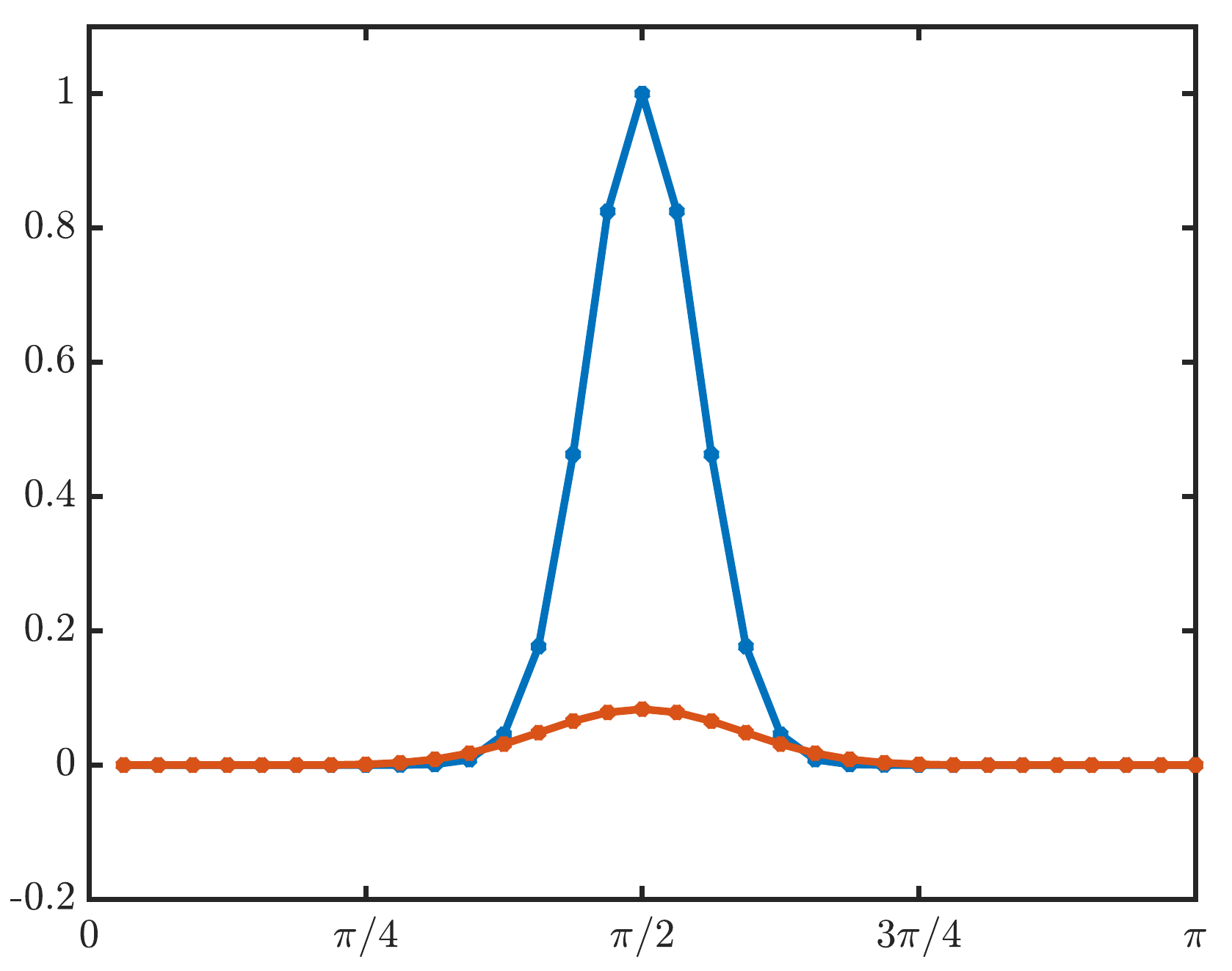}}
\subfigure[$n=40$.]{\includegraphics[width=.32\textwidth]{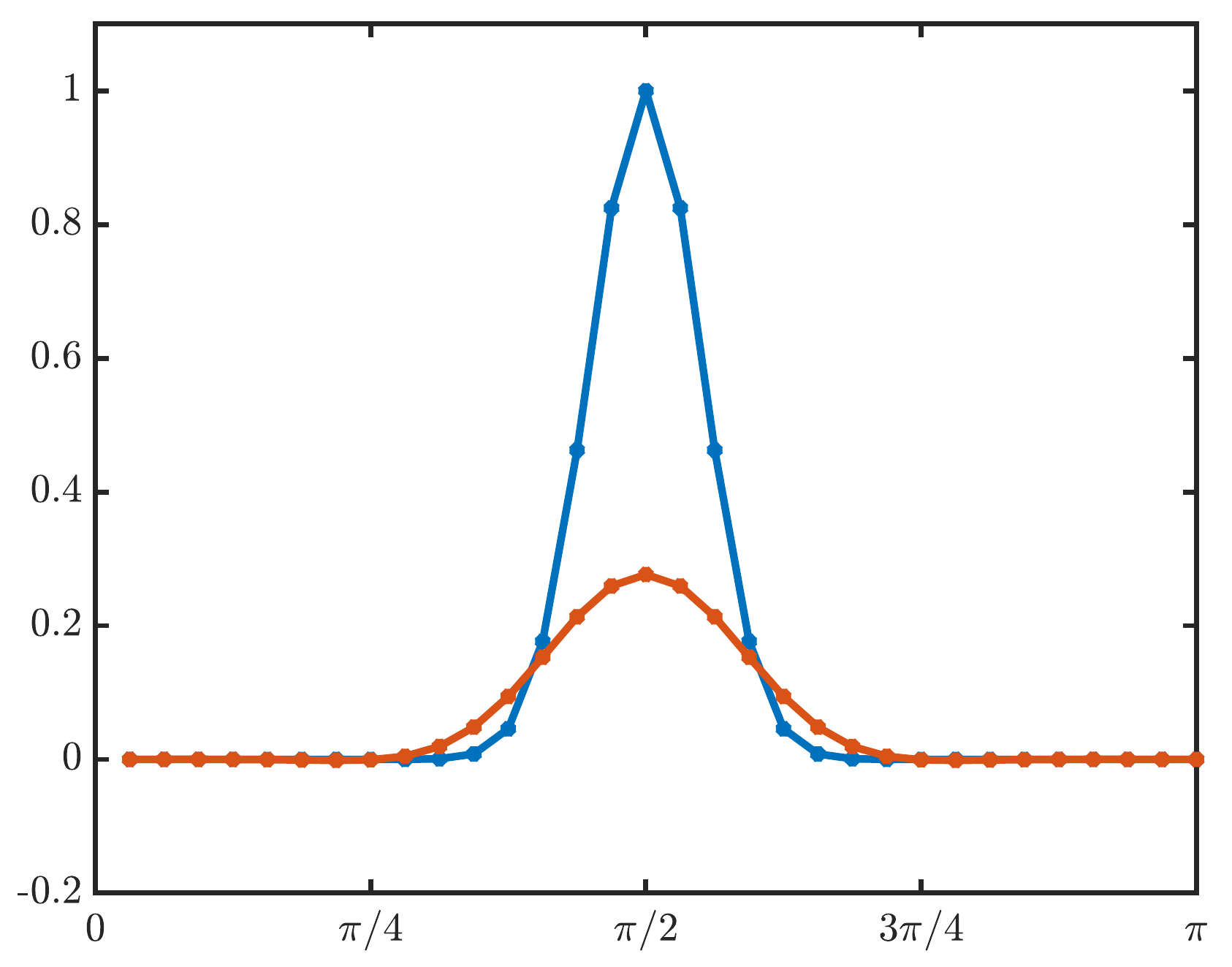}}
\subfigure[$n=60$.]{\includegraphics[width=.32\textwidth]{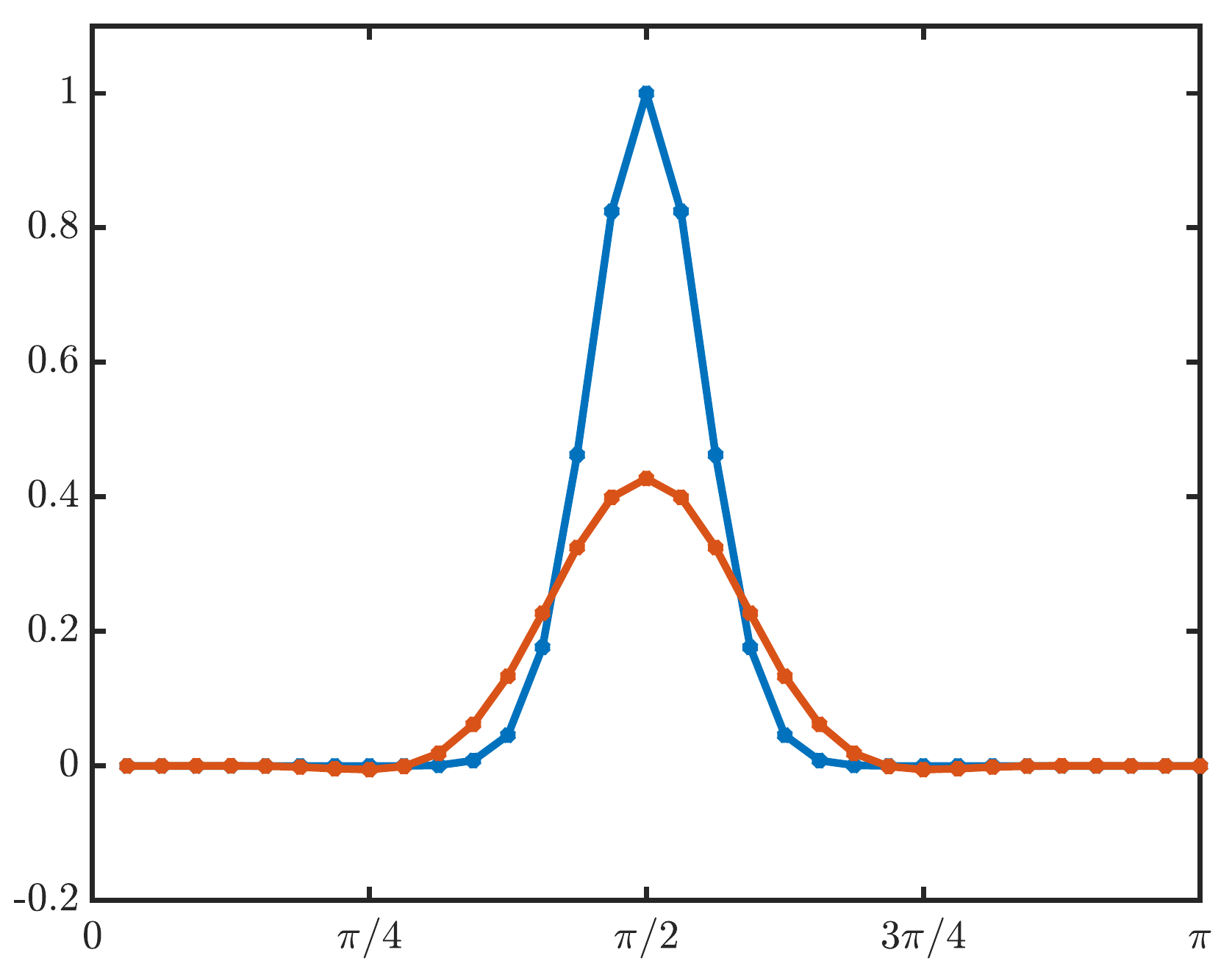}}
\subfigure[$n=80$.]{\includegraphics[width=.32\textwidth]{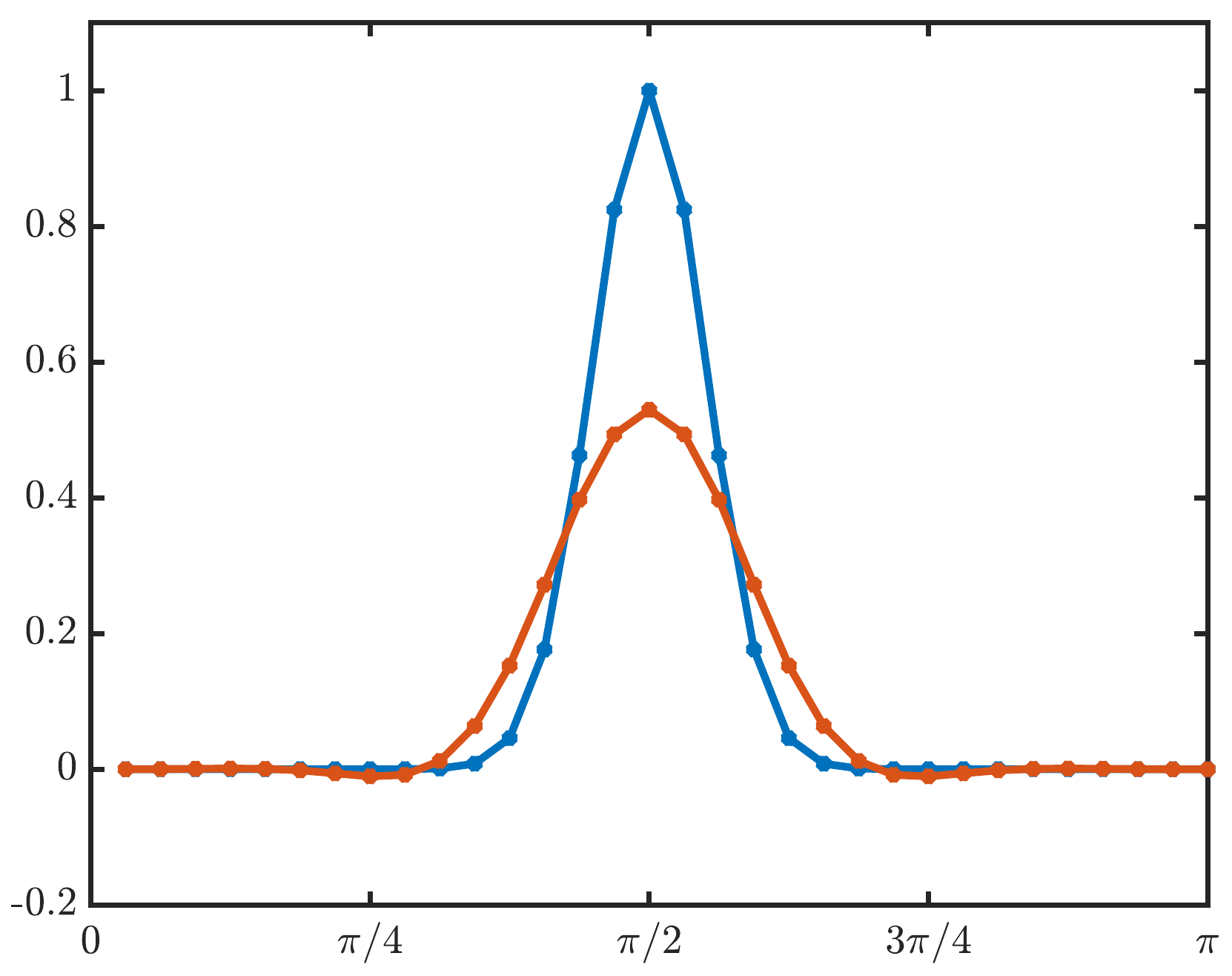}}
\subfigure[$n=100$.]{\includegraphics[width=.32\textwidth]{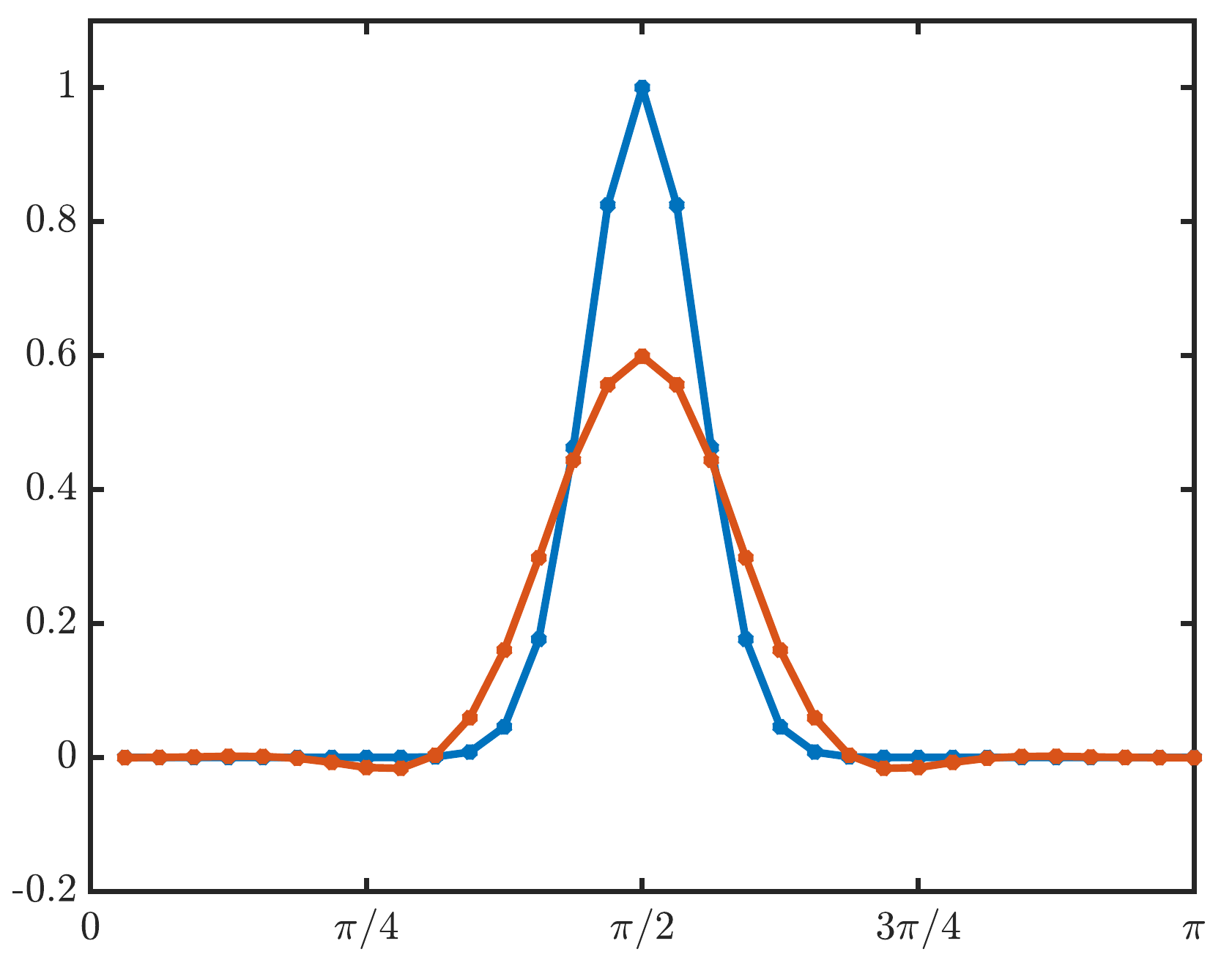}}
\subfigure[$n=200$.]{\includegraphics[width=.32\textwidth]{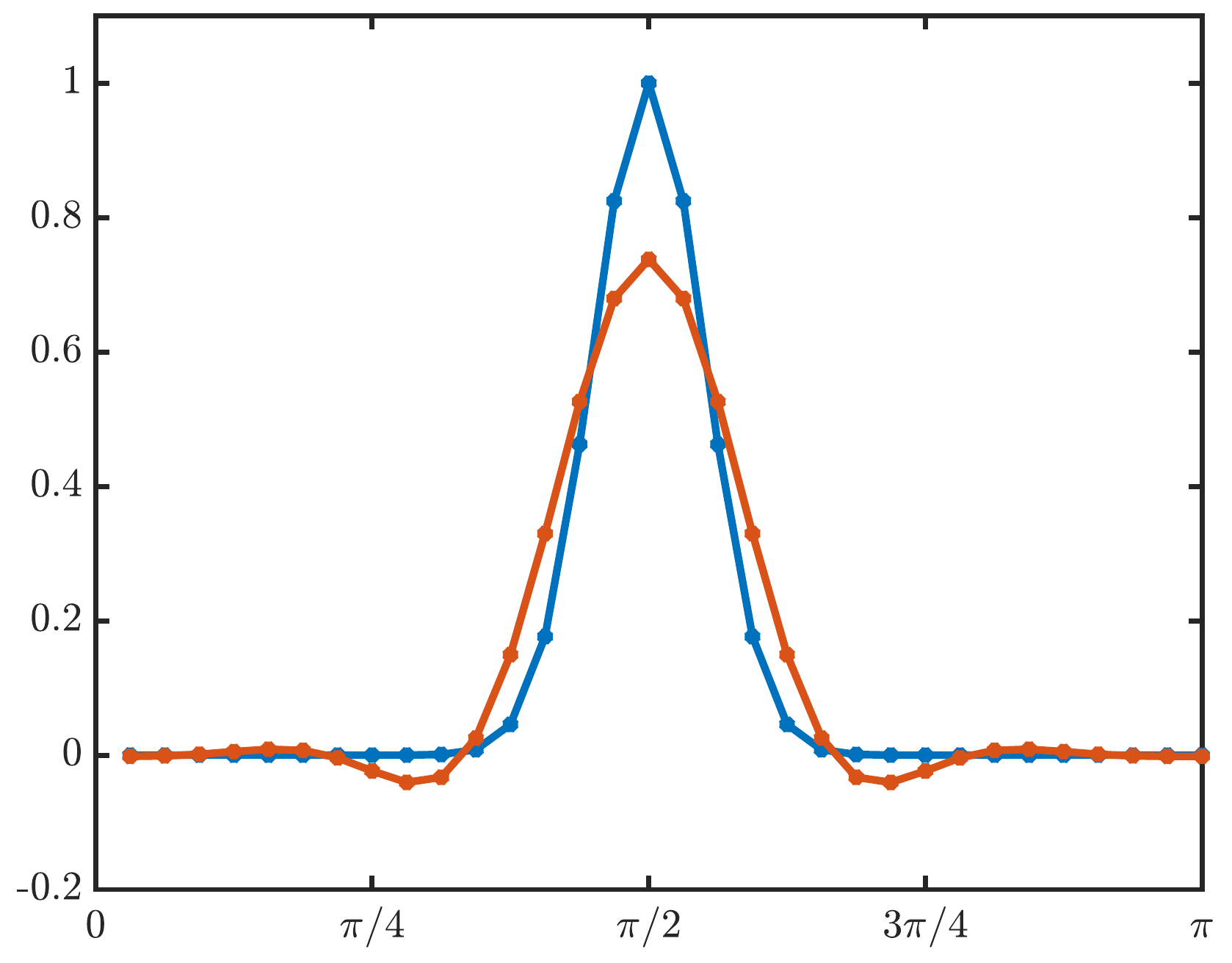}}
\subfigure[$n=300$.]{\includegraphics[width=.32\textwidth]{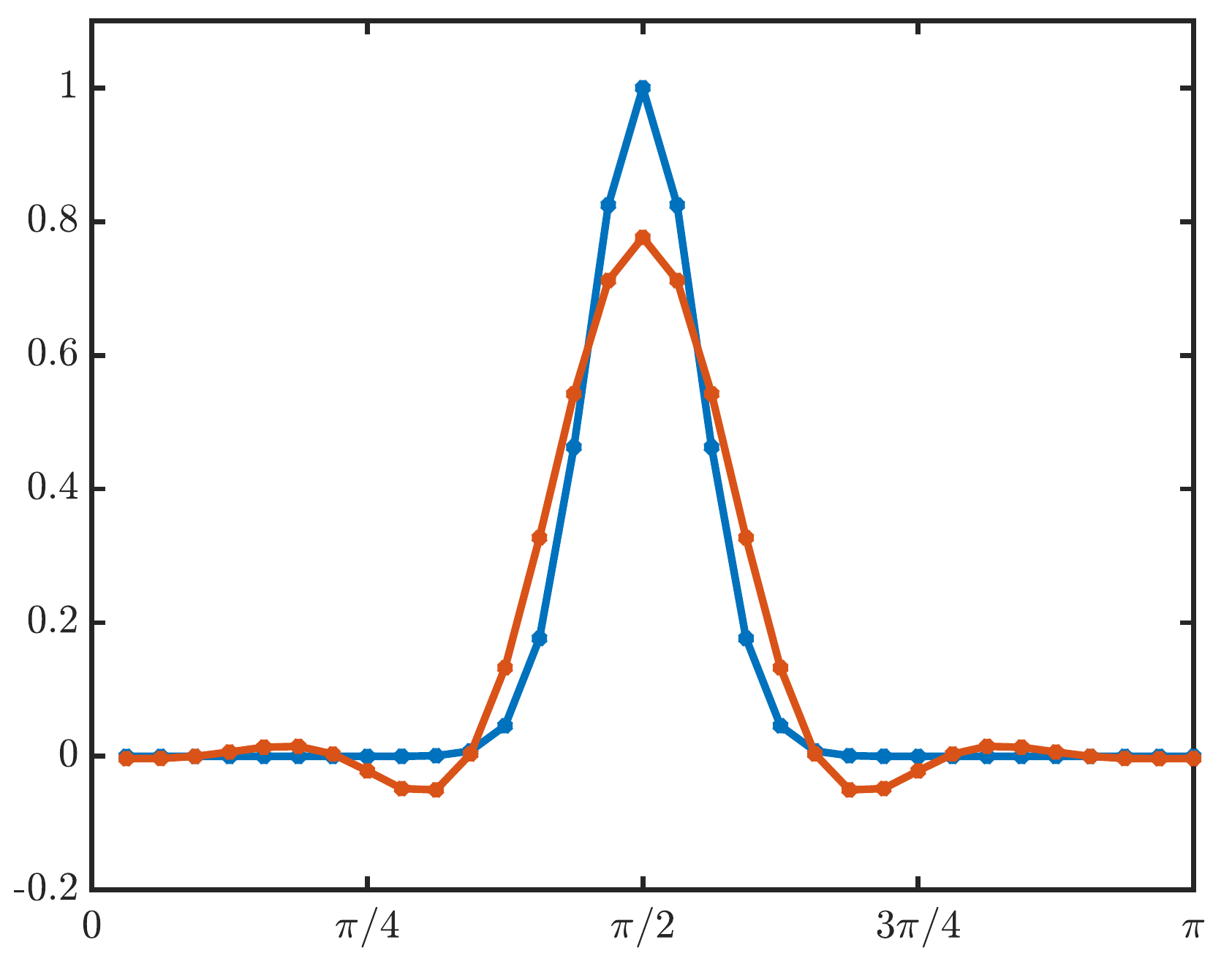}}
\subfigure[$n=400$.]{\includegraphics[width=.32\textwidth]{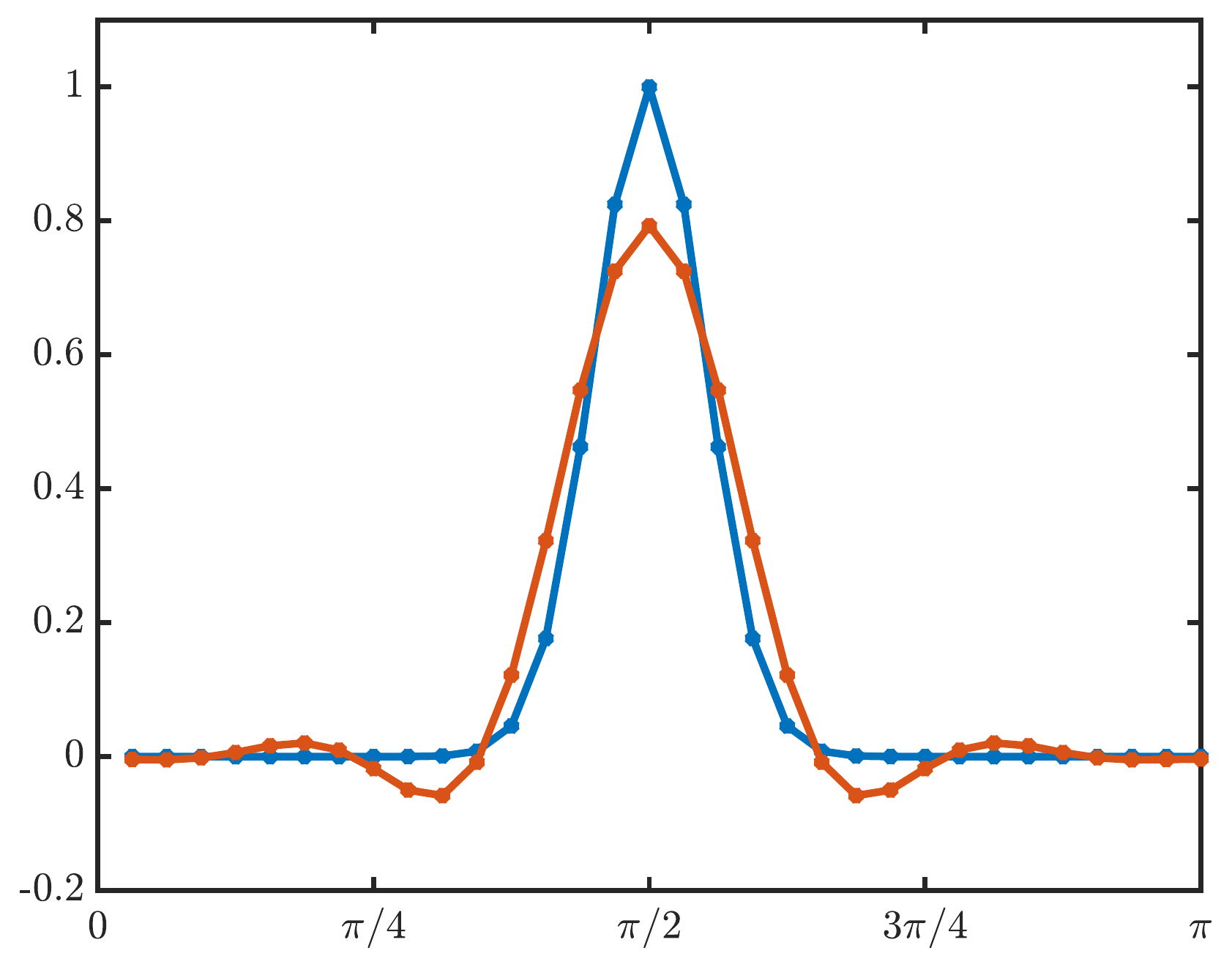}}
\caption{Time evolution of $\E_j^n$ at layer $j=10$ (orange) for $n\in\left\{0,20,40,60,80,100,200,300,400\right\}$ with fixed input $\E_0$ at layer $j=0$ (blue). The input $\E_0$ is a Gaussian centered at $\vartheta=\pi/2$.}
  \label{fig:TimeEvolJ10dirac}
\end{figure}

In a second example simulation (Figure~\ref{fig:TimeEvolJ10dirac}), we highlight the dynamics of the different modes in a situation where the input is a narrow Gaussian profile (close to a Dirac function), with a superposition of various Fourier modes. As expected from the different values of the propagation speed $c_0$ (Figure~\ref{fig:Speed}), we see that the mode associated with the first Fourier component is the first to reach layer $j=10$, later followed by successive modes associated with later Fourier components. In other words, this hierarchically higher layer $j=10$ first receives information about the coarse spatial structure of the input signal, and then gradually about finer and finer spatial details. 

\subsection{Summary}
In this section, we saw that the results obtained initially (The Identity Case) with the amplification function can be extended to more realistic situations with forward and backward connection matrices, for instance implementing (residual) convolutions or orientation processing. When we consider neural \textit{assemblies} capturing the principal components of the connection matrices, we see that each assembly can be treated independently in terms of stability and signal propagation speed and direction. The exact behavior of the system will depend on the actual connection matrices (and thus on the function that they implement in the neural network), but the important point is that our generic framework can always be applied in practice. In some example cases (ring model of orientations), we saw that only a few assemblies support signal propagation (implying that the system acts a filter on its inputs), and these assemblies propagate information at different speeds (implementing a coarse-to-fine analysis). In other cases (e.g. Figure~\ref{fig:WaveSpeedGam}), we have even seen that distinct assemblies can simultaneously propagate information in opposite directions, with one assembly supporting feedforward propagation while another entails feedback propagation.\\
We have extended our equations to the continuous limit in time, and found that the amplification factor function can give rise to qualitatively different stability regions compared to the discrete model. This served as a cautionary note for situations where the discrete implementation must be chosen; in that case, using smaller time steps will be preferable, because it makes such discrepancies less likely.\\
Finally, we also showed that it is possible to consider fully continuous versions of our dynamic system, where not only time but also network depth and neural layer width are treated as continuous variables. This gives rise to diffusion equations, whose stability can also be characterized as a function of hyperparameter values.\\
In the following, we address possible extensions of the model to more sophisticated and more biologically plausible neural architectures, taking into account the significant communication delays between layers.

\section{Extension of the model: taking into account transmission delays}

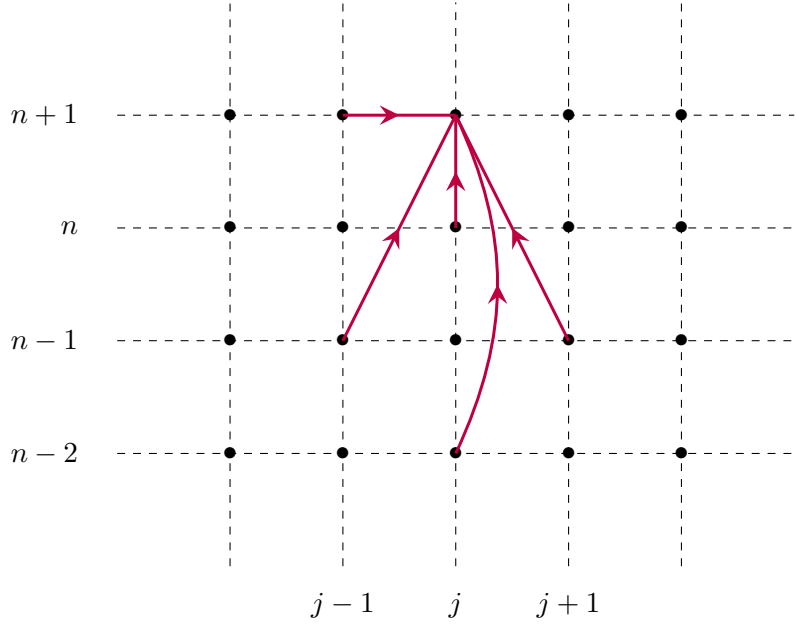
\begin{figure}
\begin{center}
\begin{tikzpicture}[scale=1.5]
\coordinate (A0) at (-1,0);
    \coordinate (A1) at (0,0);
    \coordinate (A2) at (1,0);
    \coordinate (A3) at (2,0);
    \coordinate (A4) at (3,0);
    \coordinate (A5) at (4,0);
 \coordinate (A6) at (5,0);

\coordinate (B0) at (-1,1);
   \coordinate (B1) at (0,1);
    \coordinate (B2) at (1,1);
    \coordinate (B3) at (2,1);
    \coordinate (B4) at (3,1);
    \coordinate (B5) at (4,1);
        \coordinate (B6) at (5,1);
    
        \coordinate (C2) at (1,-3.15);
    \coordinate (C3) at (2,-3.15);
 \coordinate (C4) at (3,-3.15);

    \coordinate (D1) at (0,2);
    \coordinate (D2) at (1,2);
    \coordinate (D3) at (2,2);
    \coordinate (D4) at (3,2);
    \coordinate (D5) at (4,2);

 \coordinate (F0) at (-1,-2);
    \coordinate (F1) at (0,-2);
    \coordinate (F2) at (1,-2);
    \coordinate (F3) at (2,-2);
    \coordinate (F4) at (3,-2);
    \coordinate (F5) at (4,-2);
  \coordinate (F6) at (5,-2);

\coordinate (E0) at (-1,-1);
    \coordinate (E1) at (0,-1);
    \coordinate (E2) at (1,-1);
    \coordinate (E3) at (2,-1);
    \coordinate (E4) at (3,-1);
    \coordinate (E5) at (4,-1);
       \coordinate (E6) at (5,-1);

   \coordinate (G1) at (0,-3);
    \coordinate (G2) at (1,-3);
    \coordinate (G3) at (2,-3);
    \coordinate (G4) at (3,-3);
    \coordinate (G5) at (4,-3);

% \coordinate (D1) at (1.5,1);

\node[left] at (B0) {$n+1\quad$};
\node[left] at (A0) {$n\quad$};
\node[left] at (E0) {$n-1\quad$};
\node[left] at (F0) {$n-2\quad$};

    \node[below] at (C2) {$j-1$};
    \node[below] at (C3) {$j$};
    \node[below] at (C4) {$j+1$};
    
  %  \node[above] at (D1) {$\W^f$};

    \node at (A1) {$\bullet$};
    \node at (A2) {$\bullet$};
    \node at (A3) {$\bullet$};
    \node at (A4) {$\bullet$};
    \node at (A5) {$\bullet$};
    
    \node at (B1) {$\bullet$};
    \node at (B2) {$\bullet$};
    \node at (B3) {$\bullet$};
    \node at (B4) {$\bullet$};
    \node at (B5) {$\bullet$};
    
    \node at (E1) {$\bullet$};
    \node at (E2) {$\bullet$};
    \node at (E3) {$\bullet$};
    \node at (E4) {$\bullet$};
    \node at (E5) {$\bullet$};

   \node at (F1) {$\bullet$};
    \node at (F2) {$\bullet$};
    \node at (F3) {$\bullet$};
    \node at (F4) {$\bullet$};
    \node at (F5) {$\bullet$};

    \draw[dashed] (A0) -- (A6);
    \draw[dashed] (B0) -- (B3);
    \draw[dashed] (B3) -- (B6);
     \draw[dashed] (G1) -- (D1);
 \draw[dashed] (G2) -- (D2);
    \draw[dashed] (G3) -- (E3);
  \draw[dashed] (G4) -- (D4);
   \draw[dashed] (G5) -- (D5);
\draw[dashed] (B3) -- (D3);
\draw[dashed] (E3) -- (A3);
\draw[dashed] (E0) -- (E6);
\draw[dashed] (F0) -- (F6);
\draw[line width=0.4mm,color=purple,directed] (F3) to [bend right=25] (B3) ;

\draw[line width=0.4mm,color=purple,directed] (B2) -- (B3) ;
\draw[line width=0.4mm,color=purple,directed] (E2) -- (B3) ;
\draw[line width=0.4mm,color=purple,directed] (A3) -- (B3) ;
\draw[line width=0.4mm,color=purple,directed] (E4) -- (B3) ;

\end{tikzpicture}
  \end{center}
\caption{Illustration of the network structure of model \eqref{modeldelay} for $k=1$ where the red arrows indicate the contributions leading to the update of $\E_j^{n+1}$.}
\label{fig:modeldelay}
\end{figure}

Deep feedforward neural networks typically implement \textit{instantaneous} updates, as we did in Eq~\eqref{model} with our feedforward term $\E_j^{n+1}=\beta \W^f \E_{j-1}^{n+1}+...$. Similarly, artificial recurrent neural networks sequentially update their activity from \textit{one time step to the next}, as we did with the other terms in our equation \eqref{model} (memory term, feedforward and feedback prediction error correction terms): $\E_j^{n+1}= ...+ \alpha (\W^b) ^\mathbf{t}\E_{j-1}^{n}+(1-\beta-\lambda)\E_j^{n}-\alpha (\W^b )^\mathbf{t} \W^b \E_j^{n} + \lambda  \W^b \E_{j+1}^{n}$. However, in the brain there are significant transmission delays whenever neural signals travel from one area to another. These delays could modify the system's dynamics and its stability properties. Therefore, in this section we modify model \eqref{model} by assuming that it takes $k$ time steps to receive information from a neighboring site in the feedback/feedforward dynamics, namely we consider the following recurrence equation
\bqq
\E_j^{n+1}-\beta \W^f \E_{j-1}^{n+1}=\alpha (\W^b) ^\mathbf{t}\E_{j-1}^{n-k}+(1-\beta-\lambda)\E_j^{n}-\alpha (\W^b )^\mathbf{t} \W^b \E_j^{n-2k} + \lambda  \W^b \E_{j+1}^{n-k},
\label{modeldelay}
\eqq
where $k\geq1$ is some given fixed integer (see Figure~\ref{fig:modeldelay} for an illustration with $k=1$), and we refer to \cite{Pang} for the justification of the derivation of the model. (Note in particular that we did not modify the \textit{instantaneous} nature of our feedforward updating term $\E_j^{n+1}=\beta \W^f \E_{j-1}^{n+1}+...$. This is because, as motivated in~\cite{choski21, Pang}, we aim for the feedforward part of the system to be compatible with state-of-the-art deep convolutional neural networks, and merely wish to investigate how adding recurrent dynamics can modify its properties.) We may already notice that when $k=0$, we recover our initial model \eqref{model}. In what follows, for the mathematical analysis, we restrict ourselves to the identity case $\W^f=\W^n=\mathbf{I}_d$ and when the model is set on $\Z$. Indeed, our intention is to briefly explain what could be the main new propagation properties that would emerge by including transmission delays. Thus, we consider
\bqq
e_j^{n+1}-\beta e_{j-1}^{n+1}=\alpha e_{j-1}^{n-k}+(1-\beta-\lambda)e_j^{n} -\alpha e_j^{n-2k} + \lambda  e_{j+1}^{n-k}, \quad j\in\Z.
\label{modelZdelay}
\eqq
%Of course, we could consider more general situations, but it is already enough to consider this setting to get an intuition about the key differences between the initial model \eqref{model}. 
Let us also note that the system \eqref{modelZdelay} depends on a ``history'' of $2k+1$ time steps; thus one needs to impose $2k+1$ initial conditions:
\bqs
e_j^m=h_j^m,\quad m=0,\cdots,2k, \quad j\in\Z,
\eqs
for $2k+1$ given sequences $(h_j^m)_{j\in\Z}$ with $m=0,\cdots,2k$.

To proceed in the analysis, we first introduce a new vector unknown capturing each layer's recent history:
\bqs
\mathbf{E}_j^n:=\left(\begin{array}{c}e_j^{n-2k}\\ \vdots \\ e_{j}^{n-2} \\ e_{j}^{n-1} \\ e_{j}^{n}\end{array}\right)\in\R^{2k+1}, \quad n\geq1, \quad j\in\Z,
\eqs
such that the above recurrence \eqref{modelZdelay} can then be rewritten as
\bqq
\mathbf{E}_j^{n+1}-\beta Q_{-1}\mathbf{E}_{j-1}^{n+1}=\alpha Q_1 \mathbf{E}_{j-1}^{n}+Q_0\mathbf{E}_j^{n}+\lambda Q_1 \mathbf{E}_{j+1}^{n},\quad n\geq1, \quad j\in\Z,
\label{modelZdelayextended}
\eqq
where the matrices $Q_1,Q_0,Q_{-1}\in\mathscr{M}_{2k+1}(\R)$ are defined as follows
\bqs
Q_{0}=\left(
\begin{matrix}
0 & 1 & 0 &  \cdots & \cdots & 0  \\
\vdots & \ddots & \ddots &  \ddots & \ddots & \vdots \\
\vdots & \ddots & \ddots &  \ddots & \ddots & \vdots \\
\vdots & \ddots & \ddots & \ddots & \ddots & 0 \\
0 & \ddots & \ddots & \ddots & 0 & 1 \\
-\alpha & 0 & \cdots & \cdots & 0 & 1-\beta-\lambda
\end{matrix}
\right),
\eqs
and $Q_{\pm1}$ have a single nonzero element on their last row:
\bqs
\left(Q_{-1}\right)_{2k+1,2k+1}=1, \quad \left(Q_{1}\right)_{2k+1,k+1}=1.
\eqs

\subsection{Mathematical study of the recurrence equation \eqref{modelZdelayextended}}

We now postulate an Ansatz of the form $\rho^ne^{\mbi \theta j}\mathbf{E}$ for some non zero vector $\mathbf{E}\in\C^{2k+1}$, and obtain
\bqs
%\left(\begin{matrix} \rho & -1 & 0 \\ 0 & \rho & -1 \\  \alpha & -\alpha  e^{-\mbi \theta}-\lambda e^{\mbi \theta } & \rho-(1-\beta-\lambda)-\beta  e^{-\mbi \theta} \end{matrix}\right)\mathbf{E}
\underbrace{\left( \rho \left[\mathbf{I}_{2k+1}-\beta e^{-\mbi \theta} Q_{-1}\right] -Q_0 -(\alpha e^{-\mbi \theta}+\lambda e^{\mbi \theta})Q_1 \right)}_{:=\mathcal{A}_k(\rho,\theta)}\mathbf{E}=\left(\begin{array}{c} 0\\ \vdots \\ 0\end{array}\right)
\eqs
which is equivalent to
\bqs
\det\left( \rho \left[\mathbf{I}_{2k+1}-\beta e^{-\mbi \theta} Q_{-1}\right] -Q_0 -(\alpha e^{-\mbi \theta}+\lambda e^{\mbi \theta})Q_1 \right)=0,
%\det\left(\begin{matrix} \rho & -1 & 0 \\ 0 & \rho & -1 \\  \alpha & -\alpha  e^{-\mbi \theta}-\lambda e^{\mbi \theta } & \rho-(1-\beta-\lambda)-\beta  e^{-\mbi \theta} \end{matrix}\right)=0,
\eqs
that is
\bqq
(1-\beta e^{-\mbi \theta})\rho^{2k+1}-\rho^{2k}\left( 1-\beta-\lambda\right)-\rho^k\left(\alpha  e^{-\mbi \theta}+\lambda e^{\mbi \theta }\right)+\alpha=0.
\label{eqDRdelay}
\eqq
The above system has $2k+1$ roots in the complex plane that we denote $\rho_m(\theta)$ for $m=1,\cdots 2k+1$. We remark at $\theta=0$, $\rho=1$ is always a root of the equation since in this case \eqref{eqDRdelay} reduces to
\bqq
(1-\beta)\rho^{2k+1}-\rho^{2k}\left( 1-\beta-\lambda\right)-\rho^k\left(\alpha  +\lambda \right)+\alpha=0.
\label{eqDRdelay0}
\eqq 
By convention, we assume that $\rho_1(0)=1$. We further note that $\mathbf{E}_1=(1,\cdots,1)^{\bf t}$ is the associated eigenvector. 
% while  $\rho_2(0)=-\frac{\lambda}{2}-\frac{\sqrt{\lambda^2+4\alpha}}{2}$ and $\rho_3(0)=-\frac{\lambda}{2}+\frac{\sqrt{\lambda^2+4\alpha}}{2}$. 
 As usual, we can perform a Taylor expansion of $\rho_1$ near $\theta=0$ and we obtain that
\bqs
\rho_1(\theta)=\exp\left(-\mbi \frac{\alpha+\beta-\lambda}{1-\beta+k(\lambda-\alpha)}\theta+\mathcal{O}(|\theta|^2)\right), \text{ as } \theta \rightarrow 0,
\eqs
%where
%\bqs
%\sigma_0=\frac{3\alpha^2\beta + 3\alpha\beta^2 - 2\alpha\beta\lambda - \beta^2\lambda - \beta\lambda^2 + 5\alpha^2 + 6\alpha\beta - 6\alpha\lambda + \beta^2 - 6\beta\lambda + \lambda^2 - \alpha - \beta - \lambda}{2(1+\lambda-\alpha)^3}.
%\eqs
so that the associated wave speed is this time given by
\bqs
c_0^k=\frac{\alpha+\beta-\lambda}{1-\beta+k(\lambda-\alpha)},
\eqs
and depends explicitly on the delay $k$. We readily conclude that:
\begin{itemize}
\item When $\alpha<\lambda$, then $c_0^k$ is well defined for all values of $k$. Furthermore, the amplitude of the wave speed $k\mapsto |c_0^k|$ decreases as $k$ increases with $|c_0^k|\rightarrow0$ as $k\rightarrow+\infty$. That is, the activity waves may go forward or backward (depending on the hyperparameter values), but the transmission delay always slows down their propagation.
\item When $\alpha=\lambda$, then $c_0^k=\frac{\beta}{1-\beta}>0$ is independent of the delay $k$. This is compatible with our implementation choice, where the initial feedforward propagation term (controlled by $\beta$) is not affected by transmission delays.
\item When $\lambda<\alpha$, then $c_0^k$ is well defined whenever $k\neq \frac{1-\beta}{\alpha-\lambda}>0$. Furthermore,  the wave speed $c_0^k>0$ for $1\leq k < \frac{1-\beta}{\alpha-\lambda}$ and increases with the delay $k$ on that interval. That is, in this parameter range neural activity waves propagate forward and, perhaps counterintuively, accelerate when the transmission delay increases. On the other hand $c_0^k<0$ for $k>\frac{1-\beta}{\alpha-\lambda}$ and $k\mapsto |c_0^k|$ decreases as $k$ increases on that domain with $|c_0^k|\rightarrow0$ as $k\rightarrow+\infty$. In this parameter range, waves propagate backward, and decelerate when the transmission delay increases.
\end{itemize}

Coming back to \eqref{eqDRdelay0}, we can look for other potential roots lying on the unit disk, i.e., marginally stable solutions. That is we look for $\omega\in(0,2\pi)$ such that $\rho=e^{\mbi \omega}$. We obtain a system of two equations
\bqq
\left\{
\begin{split}
(1-\beta)\cos((2k+1)\omega)-(1-\beta-\lambda)\cos(2k\omega)-(\alpha+\lambda)\cos(k\omega)+\alpha&=0,\\
(1-\beta)\sin((2k+1)\omega)-(1-\beta-\lambda)\sin(2k\omega)-(\alpha+\lambda)\sin(k\omega)&=0.
\end{split}
\right.
\label{systemdelay}
\eqq

\begin{figure}[t!]
  \centering
\subfigure[]{\includegraphics[width=.215\textwidth]{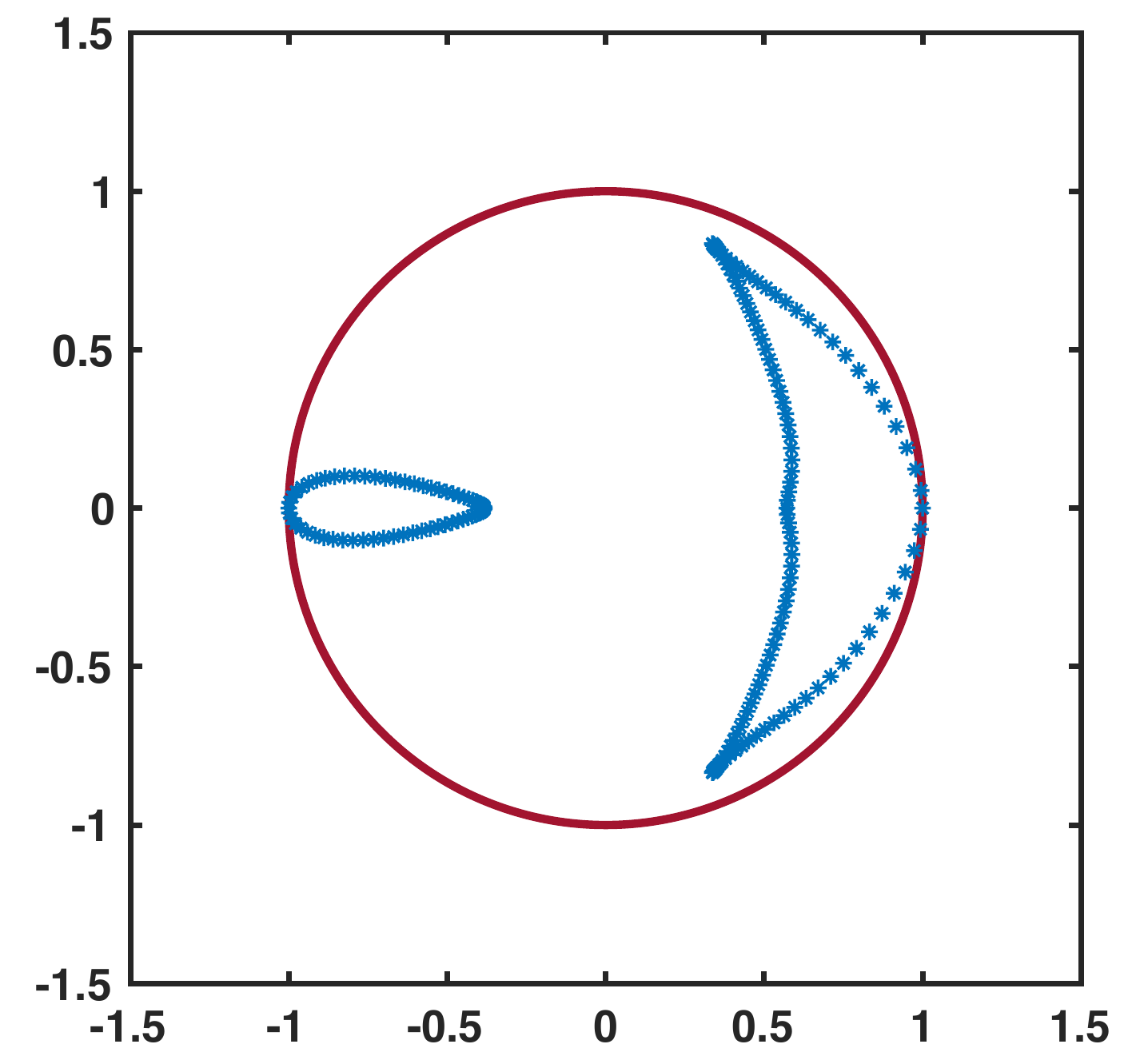}}
\subfigure[]{\includegraphics[width=.26\textwidth]{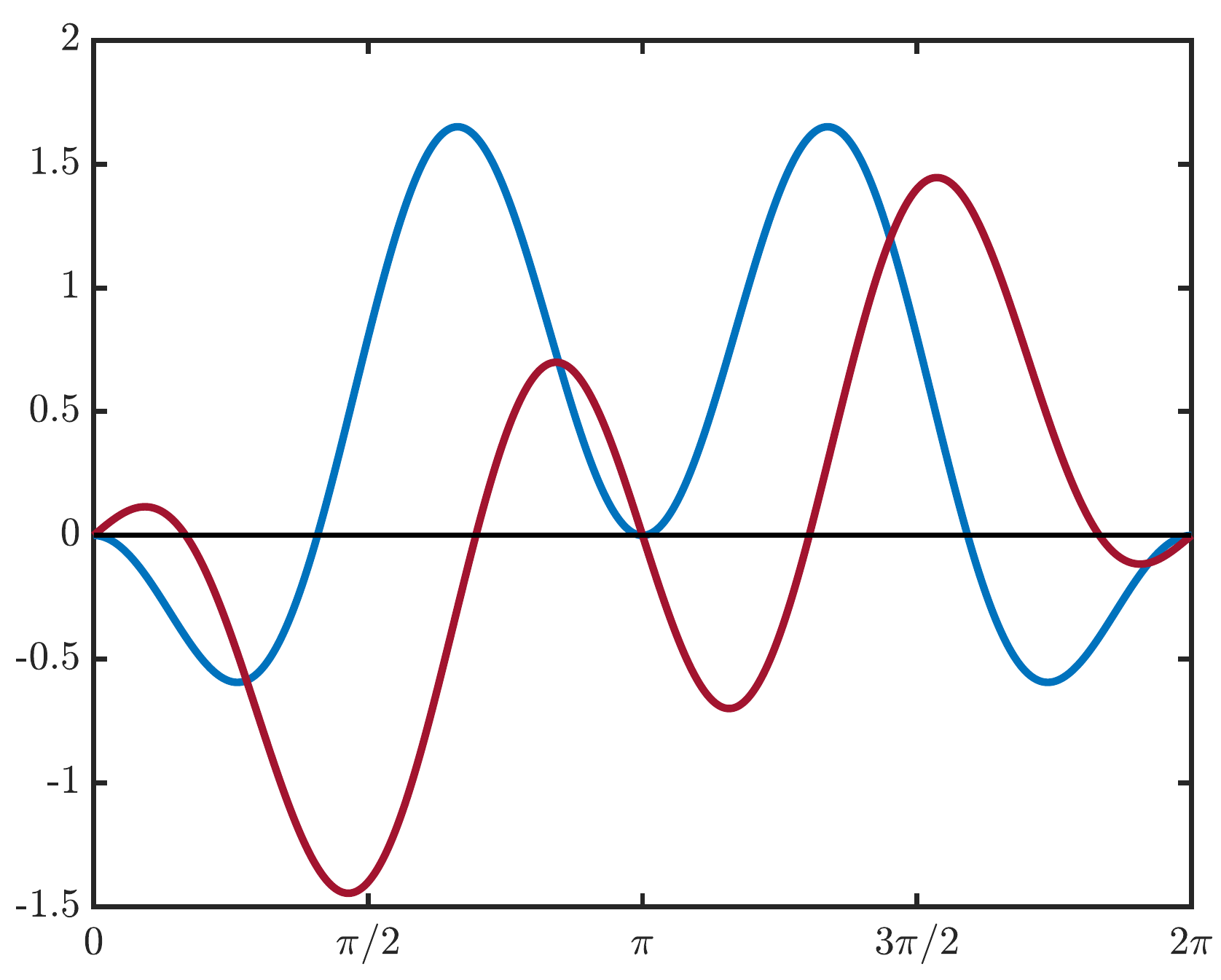}}
\subfigure[]{\includegraphics[width=.21\textwidth]{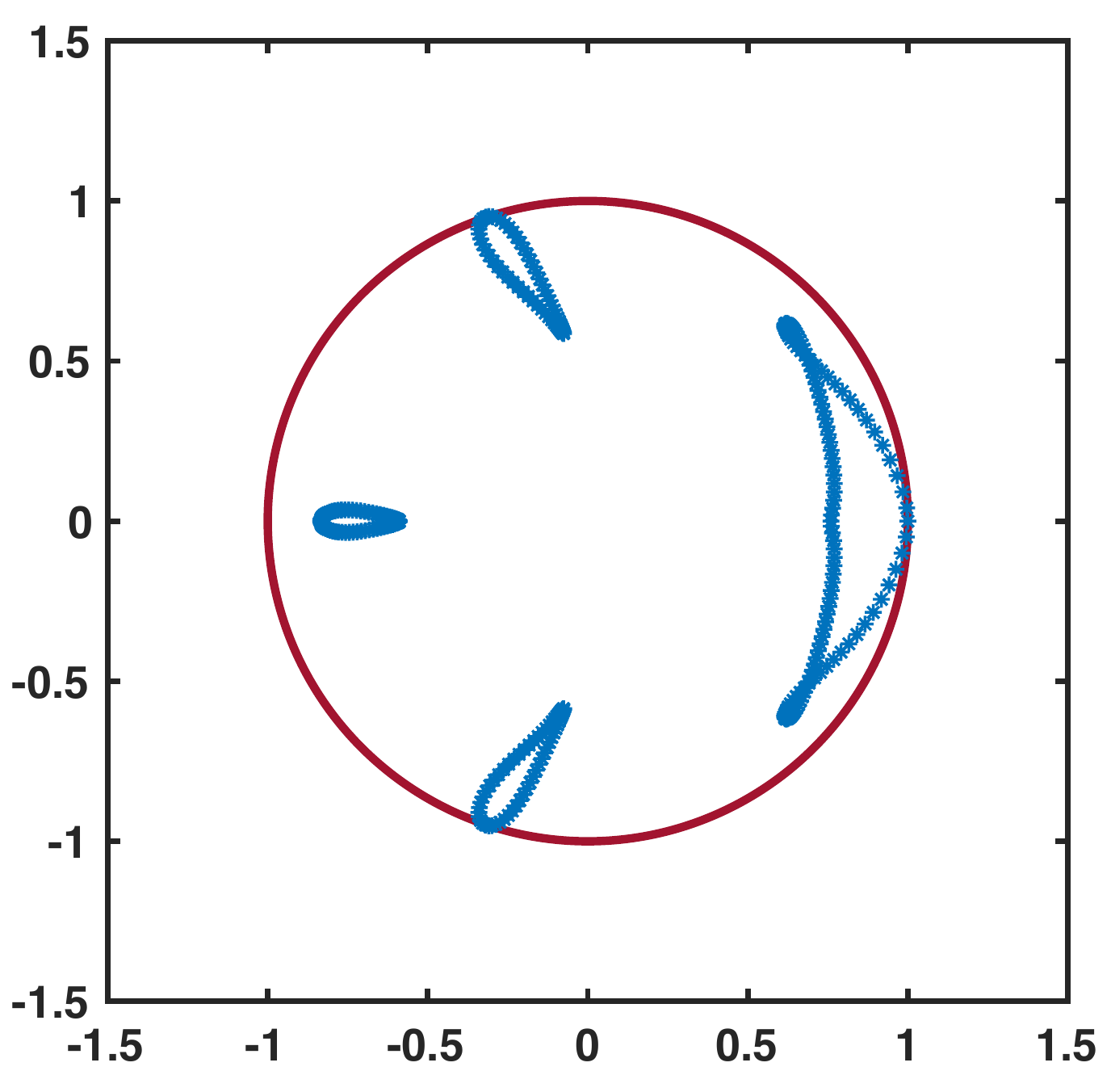}}
\subfigure[]{\includegraphics[width=.265\textwidth]{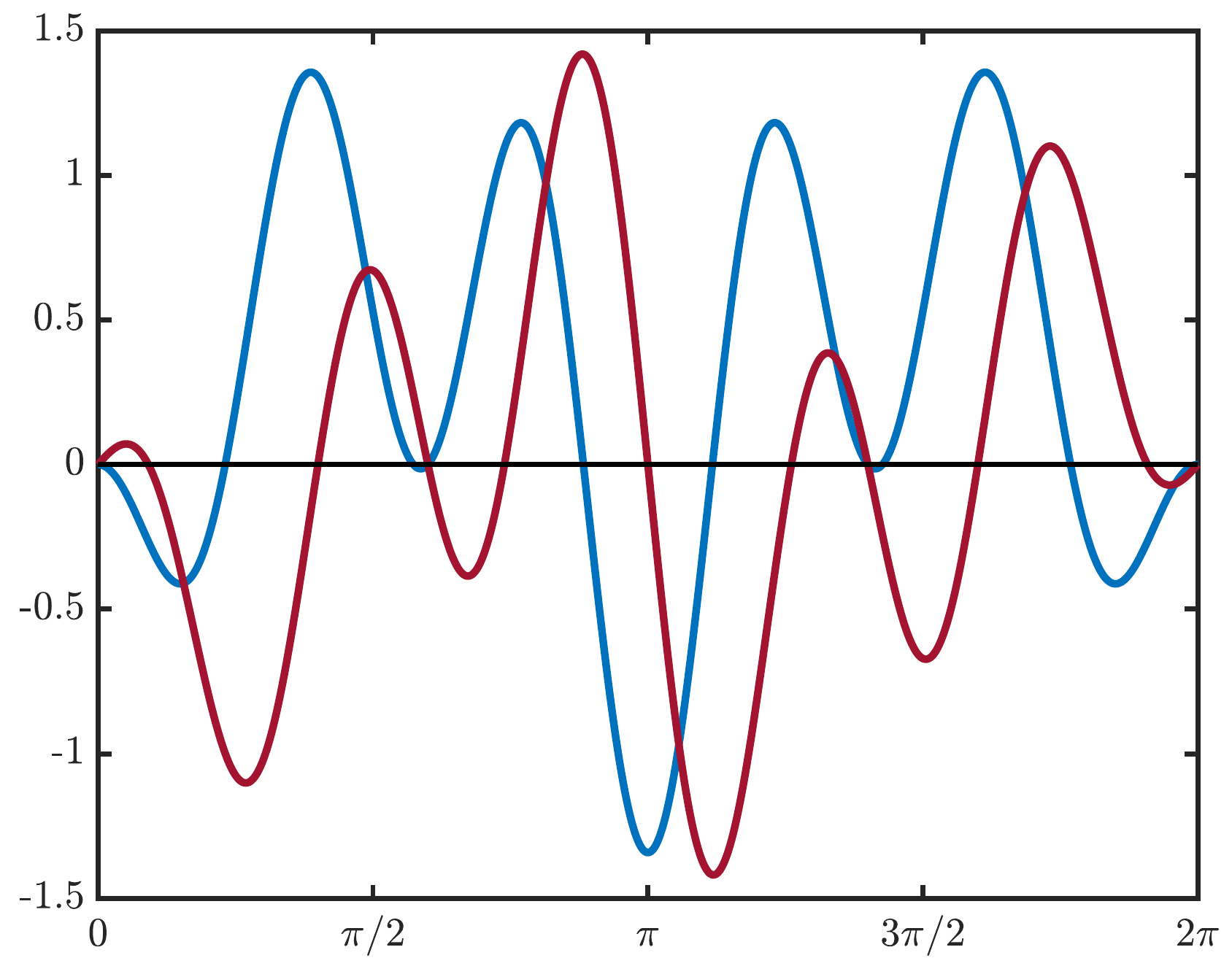}}
\caption{Spectral configurations in the case $k=1$ (a) and $k=2$ (c) with tangency points associated to $\theta=0$ in \eqref{eqDRdelay0}. In (b)-(d), we plot the left-hand side of the equations defining system \eqref{systemdelay} in the case $k=1$  and $k=2$ respectively where the first component is in blue and the second component in dark red.  For $k=1$, we have a solution at $\omega=0$ and $\omega=\pi$ which can be seen in panel (a) with the tangency points at $\pm1$. For $k=2$, we have three solutions $\omega=0$ and $\omega\sim 1.885$ and $\omega\sim 2\pi-1.885$ which can be seen in panel (c) with the tangency points at $1$ and $e^{\pm \mbi 1.885}$. Parameters are set to $(\alpha,\beta,\lambda)=(0.4,0.3,0.3)$ in (a)-(b) and $(\alpha,\beta,\lambda)=(0.3,0.3292,0.3)$.}
  \label{fig:spectrumdelay}
\end{figure}

\paragraph{Case $k=1$.} When $k=1$,  coming back to \eqref{eqDRdelay0}, we see that the two other roots are real and given by $-\frac{\lambda}{2(1-\beta)}\pm\frac{\sqrt{\lambda^2+4\alpha(1-\beta)}}{2(1-\beta)}$, such that when $\alpha+\beta+\lambda=1$ the negative root is precisely $-1$ such that $\omega=\pi$ is a solution which we assume, without loss of generality, to be the second root, that is $\rho_2(0)=-1$ whenever $\alpha+\beta+\lambda=1$. In this specific case, the associated eigenvector is $\mathbf{E}_{-1}=(1,-1,1)^{\bf t}$. Recall that $\mathbf{E}$ reflects the \textit{history} of activity across the $2k+1=3$ preceding time steps. In this case, the eigenvector $\mathbf{E}_{-1}$ is a rapid alternation of activity, i.e. an oscillation. We refer to Figure~\ref{fig:spectrumdelay}(a) for an illustration of the spectral configuration in that case. We can perform a Taylor expansion of $\rho_2$ near $\theta=0$ and we obtain that
\bqs
\rho_2(\theta)=-\exp\left(-\mbi \frac{\alpha+\beta-\lambda}{5-5\beta-\alpha-3\lambda}\theta+\mathcal{O}(|\theta|^2)\right), \text{ as } \theta \rightarrow 0,
\eqs
which provides an associated wave speed $\widetilde{c}_0$ given by
\bqs
\widetilde{c}_0=\frac{\alpha+\beta-\lambda}{5-5\beta-\alpha-3\lambda}.
\eqs

As a consequence of the above analysis, if $\mathbf{G}_j^n$ denotes the fundamental solution of \eqref{modelZdelayextended} starting from a Dirac delta mass centered at $j=0$ along the direction $\mathbf{E}\in\R^3$, then we have the following representation for  $\mathbf{G}_j^n$:
\begin{itemize}
\item If $\alpha+\beta+\lambda\neq1$, then
\bqs
\mathbf{G}_j^n\approx \frac{1}{\sqrt{4\pi \sigma_0^k n}}\exp\left( -\frac{|j-c_0^kn|^2}{4\sigma_0^k n}\right) \left\langle (0,0,1)^{\bf t}, \pi_1(\mathbf{E})\right\rangle_{\R^3},
\eqs
where $\pi_1$ is the spectral projection of $\R^3$ along the direction $\mathbf{E}_1$ and $\left\langle \cdot, \cdot\right\rangle_{\R^3}$ is the usual scalar product. Here $\sigma_0^k$ is some positive constant that can be computed explicitly by getting the higher order expansion of  $\rho_1(\theta)$.
\item If $\alpha+\beta+\lambda=1$, then
\begin{align*}
\mathbf{G}_j^n\approx \frac{1}{\sqrt{4\pi \sigma_0^k n}}\exp\left( -\frac{|j-c_0^kn|^2}{4\sigma_0^k n}\right)& \left\langle (0,0,1)^{\bf t}, \pi_1(\mathbf{E})\right\rangle_{\R^3}\\
&+\frac{(-1)^n}{\sqrt{4\pi \widetilde{\sigma_0} n}}\exp\left( -\frac{|j-\widetilde{c}_0n|^2}{4\widetilde{\sigma_0} n}\right) \left\langle (0,0,1)^{\bf t}, \pi_{-1}(\mathbf{E})\right\rangle_{\R^3},
\end{align*}
where $\pi_{-1}$ is the spectral projection of $\R^3$ along the direction $\mathbf{E}_{-1}$. Here $\widetilde{\sigma}_0$ is some positive constant that can be computed explicitly by getting the higher order expansion of  $\rho_2(\theta)$.
\end{itemize}

\begin{figure}[t!]
  \centering
\subfigure[$\mathbf{E}=\mathbf{E}_1=(1,1,1)^{\bf t}$.]{\includegraphics[width=.32\textwidth]{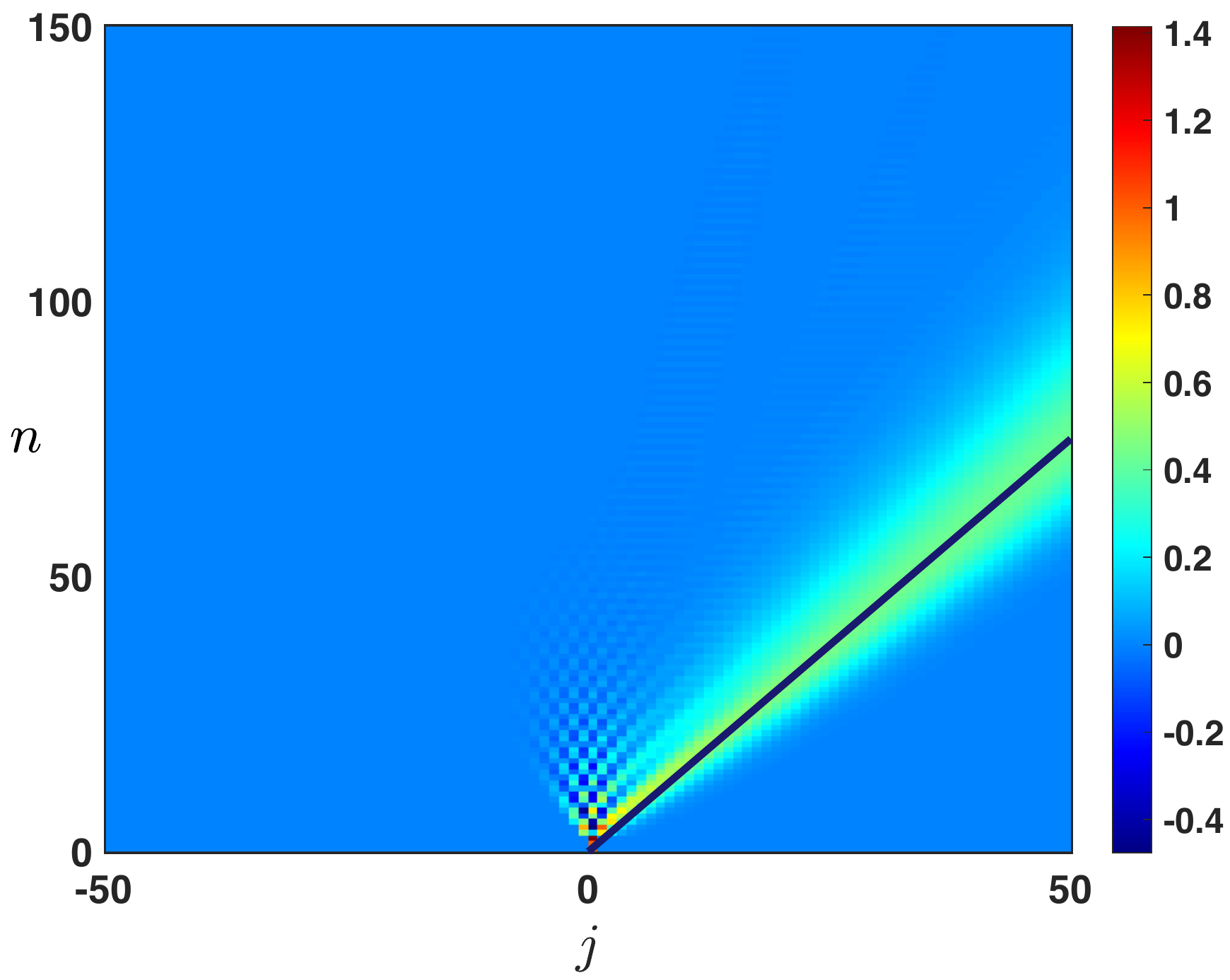}}
\subfigure[$\mathbf{E}=\mathbf{E}_{-1}=(1,-1,1)^{\bf t}$.]{\includegraphics[width=.32\textwidth]{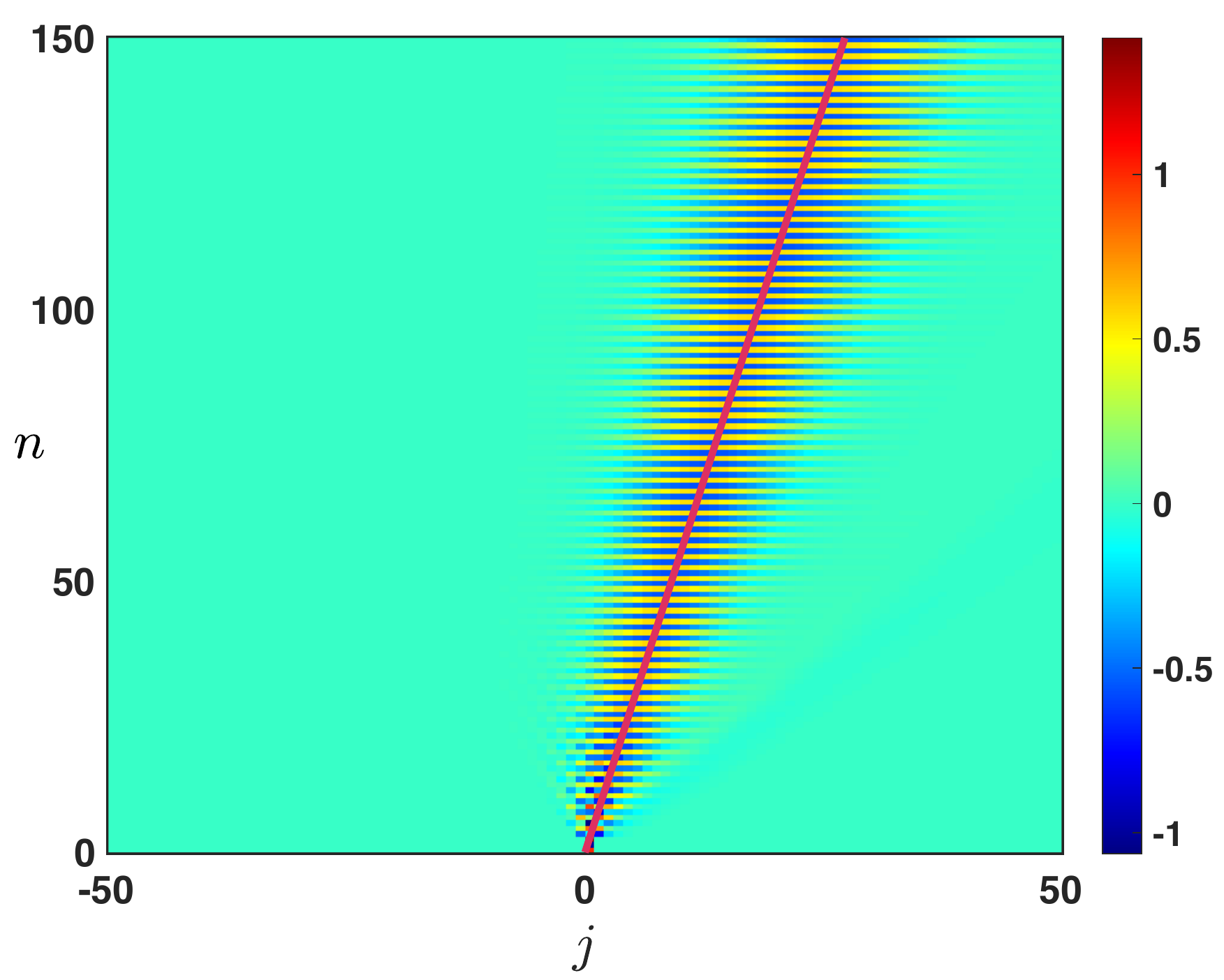}}
\subfigure[$\mathbf{E}=\frac{1}{2}\mathbf{E}_1+\frac{1}{2}\mathbf{E}_{-1}=(1,0,1)^{\bf t}$.]{\includegraphics[width=.32\textwidth]{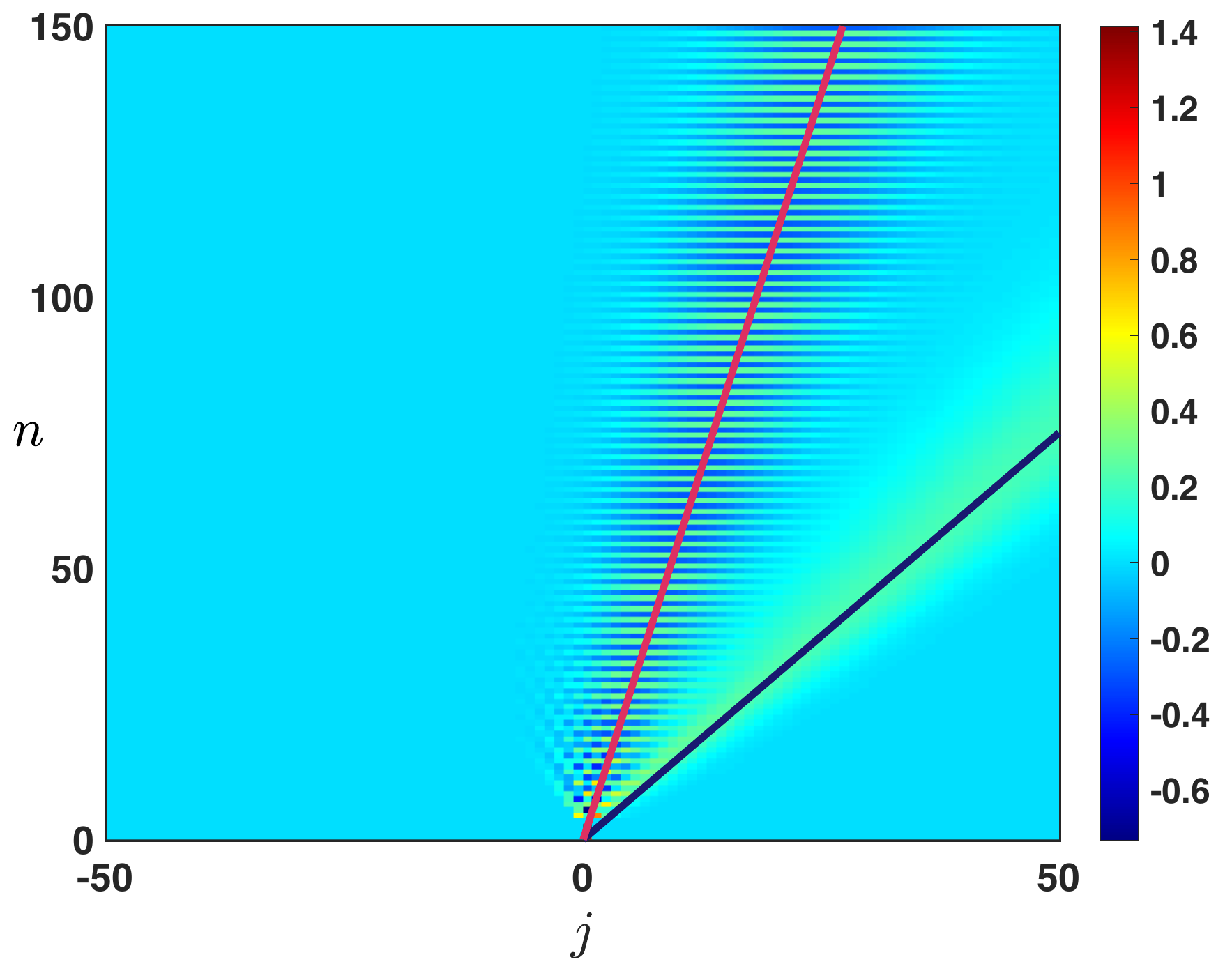}}
\caption{Space-time plots of the last component of the rescaled fundamental solution $\mathbf{E}_j^n$ of \eqref{modelZdelayextended} starting from a Dirac delta mass centered at $j=0$ along different directions $\mathbf{E}$ when $k=1$ and $\alpha+\beta+\lambda=1$. (a) When $\mathbf{E}=\mathbf{E}_1$  (constant history of activity) is the eigenvector associated to  $\mathcal{A}_1(1,0)$ we observe propagation at wave speed $c_0^1$. (b) When $\mathbf{E}=\mathbf{E}_{-1}$  (oscillating history of activity) is the eigenvector associated to  $\mathcal{A}_1(-1,0)$ we observe propagation  of an oscillatory wave at wave speed $\widetilde{c}_0$. (c) When $\mathbf{E}$ is a linear combination of $\mathbf{E}_{1}$ and $\mathbf{E}_{-1}$, we observe two propagating waves (one of them oscillating) at wave speed $c_0^1$ and $\widetilde{c}_0$ respectively. Parameter values are set to $(\alpha,\beta,\lambda)=(0.4,0.3,0.3)$.}
  \label{fig:spectrumdelayk1}
\end{figure}

In Figure~\ref{fig:spectrumdelayk1}, we illustrate the previous results in the case where $\alpha+\beta+\lambda=1$. In panel (a), we have set $\mathbf{E}=\mathbf{E}_1$ (a constant history of activity over the previous 3 time steps), such that $ \pi_1(\mathbf{E}_1)=\mathbf{E}_1$ and $ \pi_{-1}(\mathbf{E}_{1})=0_{\R^3}$ so that we only observe a Gaussian profile propagating at speed $c_0^k$. On the other hand in panel (b),  we have set $\mathbf{E}=\mathbf{E}_{-1}$ (an oscillating history of activity over the previous 3 time steps), such that $ \pi_1(\mathbf{E}_{-1})=0_{\R^3}$ and $ \pi_{-1}(\mathbf{E}_{-1})=\mathbf{E}_{-1}$ so that we only observe an oscillating (in time) Gaussian wave profile propagating at speed $\widetilde{c}_0$.  Note that in this case, the period of the oscillation is necessarily equal to $2k$, i.e. twice the transmission delay between layers. Finally in panel (c), we observe a super-position of the two Gaussian profiles propagating at speed $c_0^1$ and $\widetilde{c}_0$.

\begin{figure}[t!]
  \centering
\subfigure[$k=1$.]{\includegraphics[width=.3\textwidth]{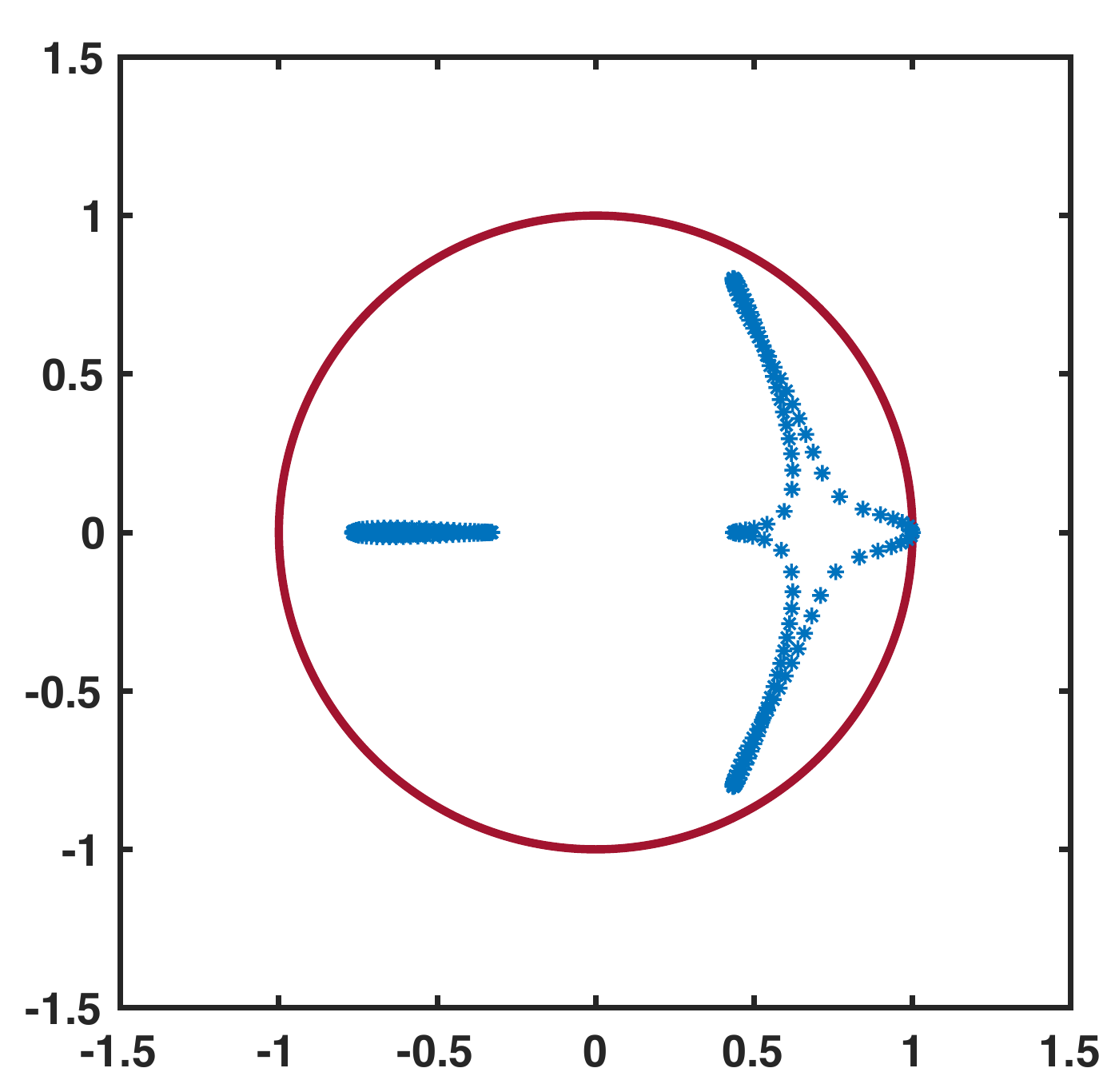}}
\subfigure[$k=3$.]{\includegraphics[width=.3\textwidth]{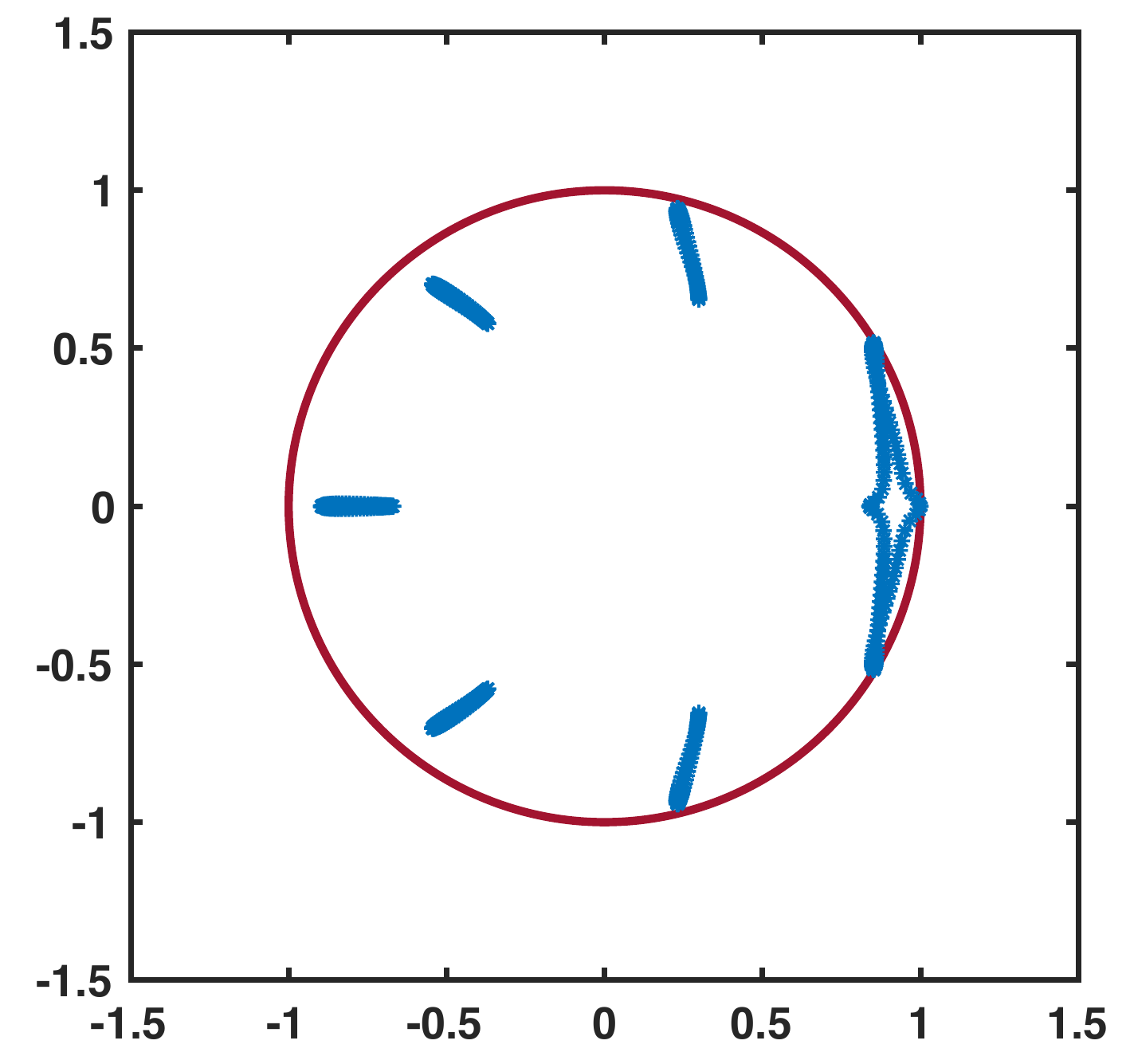}}
\subfigure[$k=5$.]{\includegraphics[width=.3\textwidth]{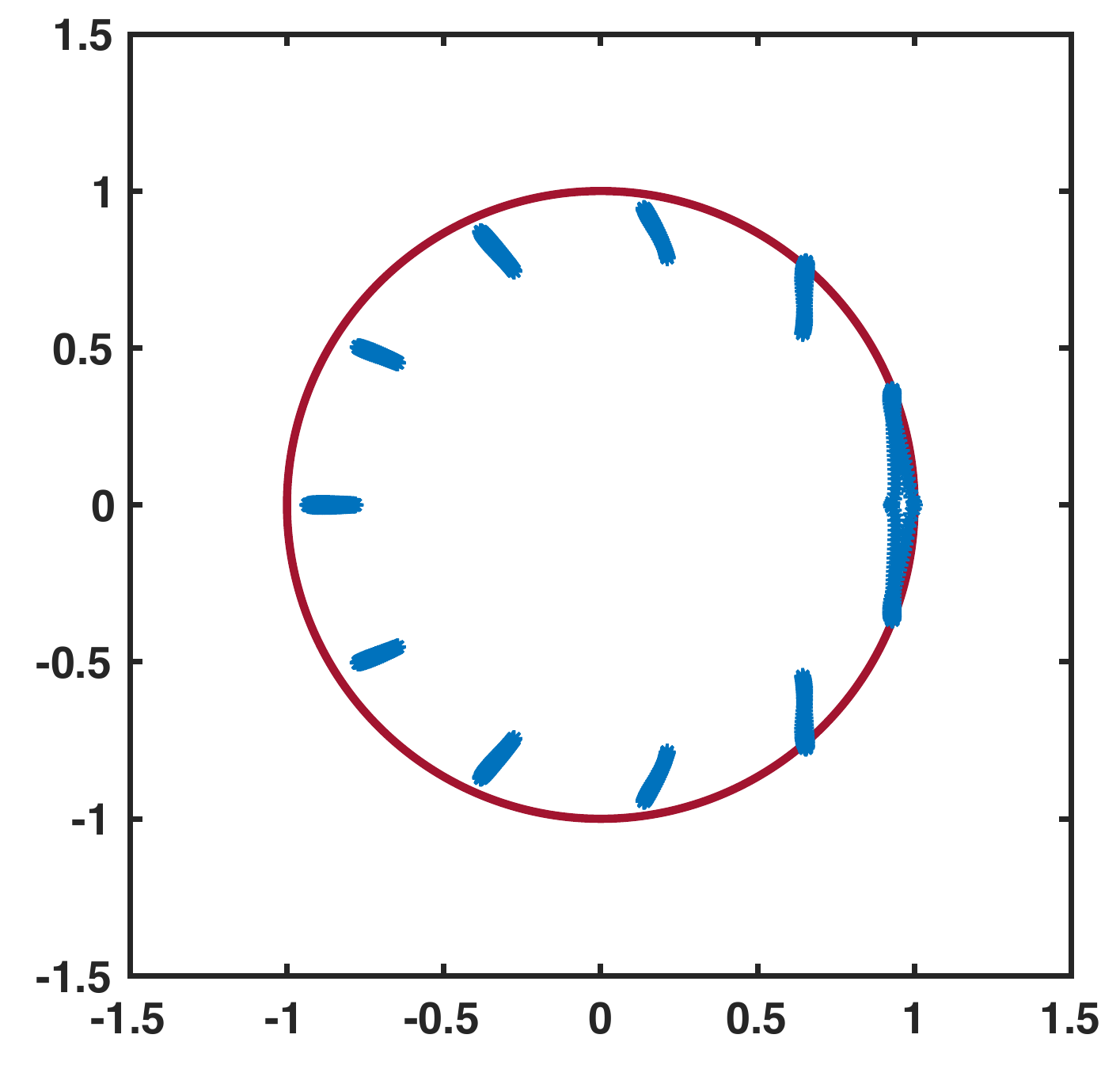}}
\caption{Destabilization of the spectrum by increasing the delay $k$ while keeping fixed the hyper-parameters to $(\alpha,\beta,\lambda)=(0.3,0.1,0.3)$.}
  \label{fig:spectrumdelaydest}
\end{figure}

\paragraph{Case $k\geq2$.} Studying the above system \eqref{systemdelay} in full generality is a very difficult task. We refer to Figure~\ref{fig:spectrumdelay}(c)-(d) for an illustration in the case $k=2$ with three tangency points associated to $\theta=0$ lying on the unit circle. Increasing the delay $k$ while keeping fixed the other hyper-parameters $(\alpha,\beta,\lambda)$ will generically tend to destabilize the spectrum (as shown in Figure~\ref{fig:spectrumdelaydest}).

\subsection{Continuous in time interpretation }

As done before, we now re-examine our model (with transmission delays) in the time-continuous limit. First, we recall our notations for the scaled parameters
\bqs
\widetilde{\beta}:=\frac{\beta}{\Delta t}, \quad \widetilde{\lambda}:=\frac{\lambda}{\Delta t}, \text{ and } \widetilde{\alpha}:=\frac{\alpha}{\Delta t},
\eqs
where $\Delta t>0$ is some time step. Next we introduce the following rescaled time delay (representing the transmission time for neural signals between adjacent areas)
\bqs
\tau := k \Delta t.
\eqs
Identifying $e_j^n$ as the approximation of some continuous fonction $\mathbf{e}_j(t_n)$ at $t_n=n\Delta t$, we readily derive a delayed version of \eqref{latticeODE}, namely
\bqs
\frac{\md }{\md t}{\bf e}_j(t)=\widetilde{\beta}{\bf e}_{j-1}(t)-(\widetilde{\beta}+\widetilde{\lambda}){\bf e}_j(t)+\widetilde{\alpha}{\bf e}_{j-1}(t-\tau)+\widetilde{\lambda}{\bf e}_{j+1}(t-\tau)-\widetilde{\alpha}{\bf e}_{j}(t-2\tau), \quad t>0, \quad j\in\Z.
\eqs

In what follows, we first investigate the case of homogeneous oscillations, which are now possible because of the presence of time delays into the equation. Then, we turn our attention to oscillatory traveling waves.

\subsubsection{Homogeneous oscillations} One key difference of the above delayed equation compared to \eqref{latticeODE} is that spatially homogeneous solutions (i.e., solutions ${\bf e}_j(t)$ that are independent of the layer $j$) may now have a non trivial dynamics, such as a broadly synchronized oscillation resembling brain rhythmic activity. Indeed, looking for solutions which are independent of $j$, we get the delayed ordinary differential equation
\bqs
\frac{\md }{\md t}{\bf e}(t)=-\widetilde{\lambda}{\bf e}(t)+(\widetilde{\alpha}+\widetilde{\lambda}){\bf e}(t-\tau)-\widetilde{\alpha}{\bf e}(t-2\tau), \quad t>0.
\eqs
Looking for pure oscillatory exponential solutions ${\bf e}(t)=e^{\mbi \omega t}$ for some $\omega\in\R$ we obtain
\bqs
\mbi \omega=-\widetilde{\lambda}+(\widetilde{\alpha}+\widetilde{\lambda})e^{-\mbi\tau\omega}-\widetilde{\alpha}e^{-2\mbi\tau\omega}.
\eqs

This leads to the system of equations
\bqs
\left\{
\begin{split}
0&=-\widetilde{\lambda}+(\widetilde{\alpha}+\widetilde{\lambda})\cos(\tau\omega)-\widetilde{\alpha}\cos(2\tau\omega),\\
\omega&=-(\widetilde{\alpha}+\widetilde{\lambda})\sin(\tau\omega)+\widetilde{\alpha}\sin(2\tau\omega).
\end{split}
\right.
\label{systtauomega}
\eqs
Introducing $\varrho=\widetilde{\lambda}/\widetilde{\alpha}>0$, we observe that the above system writes instead
\bqq
\left\{
\begin{split}
0&=-\varrho+(1+\varrho)\cos(\tau\omega)-\cos(2\tau\omega),\\
\omega&=\widetilde{\alpha} \left(-(1+\varrho)\sin(\tau\omega)+\sin(2\tau\omega)\right).
\end{split}
\right.
\label{systtauomega}
\eqq

\begin{figure}[t!]
  \centering
\subfigure[]{\includegraphics[width=.4\textwidth]{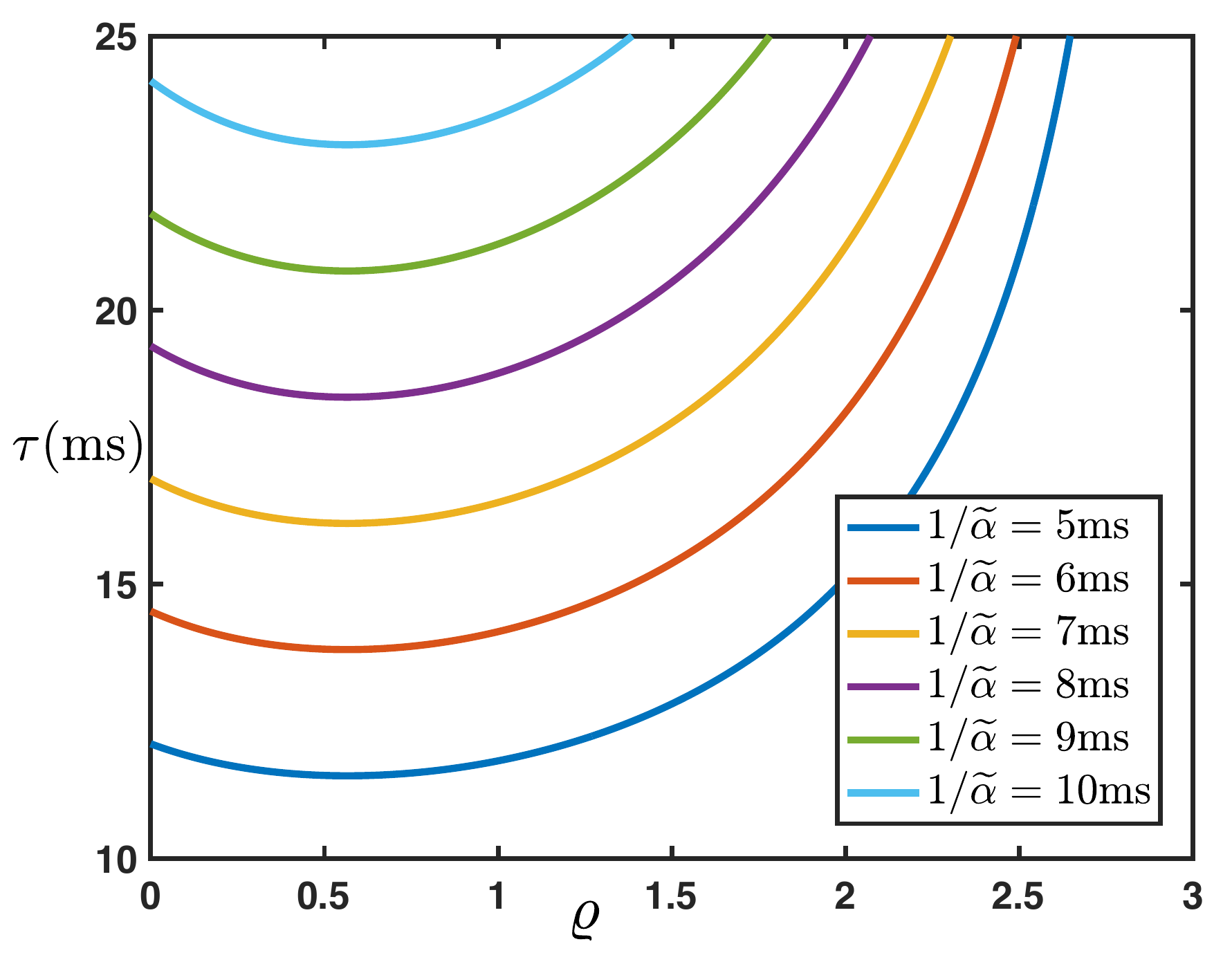}} \hspace{1cm}
\subfigure[]{\includegraphics[width=.4\textwidth]{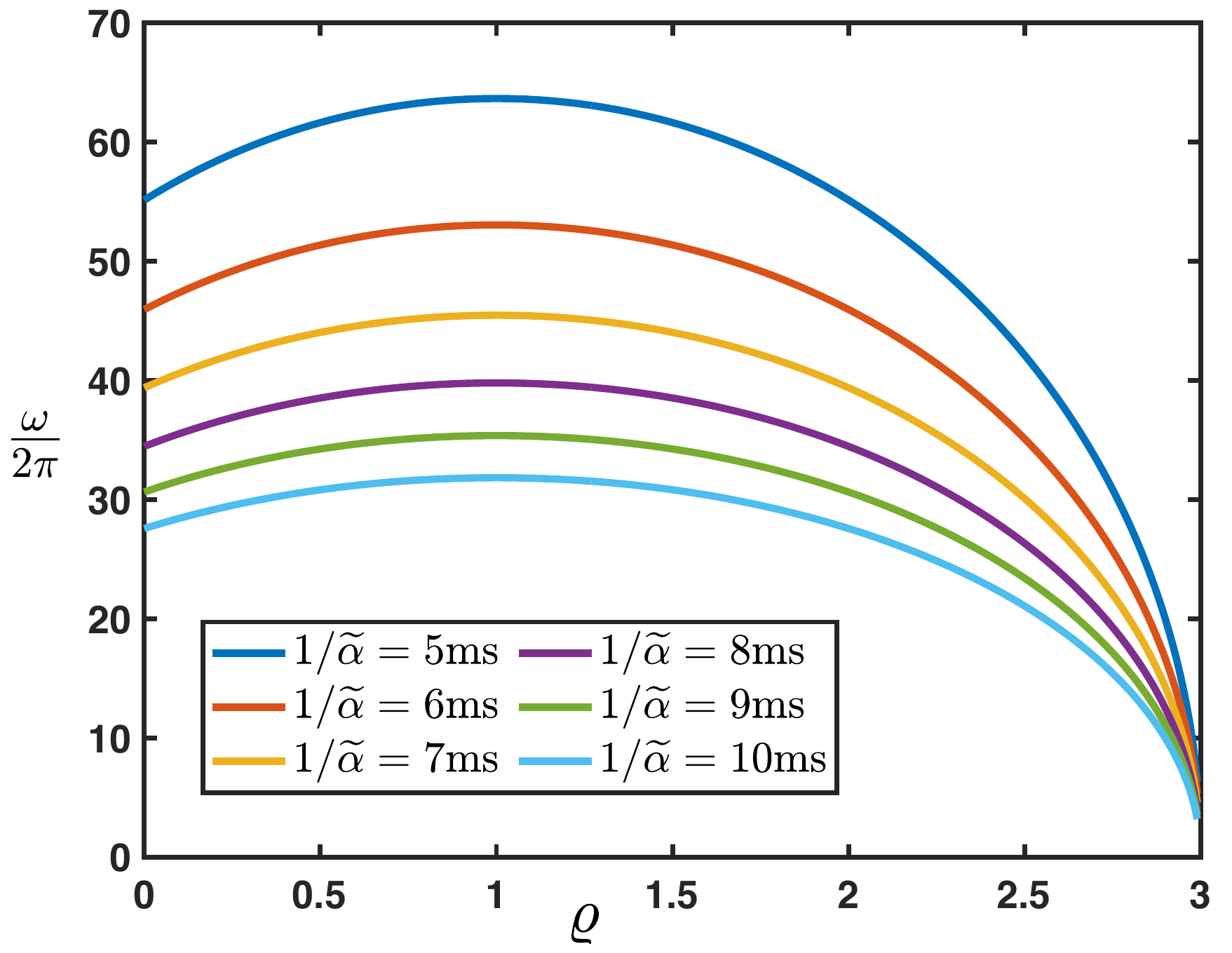}}
\caption{(a) Representation of the (minimal) time delay $\tau$ expressed in milliseconds as a function of the parameter $\varrho$ for various values of $1/\widetilde{\alpha}$ ranging from $5ms$ to $10ms$. We observe that for values of $\varrho$ in the range $(1/2,1)$ the corresponding time delay $\tau$ takes values between $12ms$ to $23ms$. (b) Representation of the frequency $\omega/2\pi$ (in Hertz) as a function of the parameter $\varrho$ for various values of $1/\widetilde{\alpha}$ ranging from $5ms$ to $10ms$. We observe that for values of $\varrho$ in the range $(1/2,1)$ the corresponding frequency $\omega/2\pi$ takes values between $30Hz$ to $60Hz$.}
  \label{fig:delaycontinuous}
\end{figure}

Using trigonometry identities, the first equation can be factorized as
\bqs
0=(1-\cos(\tau\omega))(\varrho-1-2\cos(\tau\omega)).
\eqs
We distinguish several cases. If $\varrho>3$, then the above equation has solutions if and only if $\tau\omega =2k\pi$ for $k\in\Z$. Inspecting the second equation, we see that necessarily $k=0$ and $\omega=0$ is the only possible solution. When $\varrho=3$, we notice that the equation reduces to $0=(1-\cos(\tau\omega))^2$, and the solutions are again given by $\tau\omega =2k\pi$ for $k\in\Z$, which yields $\omega=0$ because of the second equation.  Now, if $\varrho\in(0,3)$, we deduce that either $\tau\omega =2k\pi$ for $k\in\Z$ or 
\bqs
\cos(\tau\omega)=\frac{\varrho-1}{2}.
\eqs
In the first case, we recover that $\omega=0$. Assuming now that $\omega\neq0$, i.e., a true oscillation with non-zero frequency, we derive that
\bqs
\tau\omega=\pm \mathrm{arccos}\left(\frac{\varrho-1}{2}\right)+2k\pi, \quad k\in\Z.
\eqs
Injecting the above relation into the right-hand side of the second equation yields that
\bqs
\omega=\widetilde{\alpha} \left(-(1+\varrho)\sin(\tau\omega)+\sin(2\tau\omega)\right)=\mp \widetilde{\alpha} \sqrt{(1+\varrho)(3-\varrho)},
\eqs
and thus necessarily 
\bqs
(\tau,\omega) = \left(\frac{-\mathrm{arccos}\left(\frac{\varrho-1}{2}\right)+2k\pi}{ \widetilde{\alpha} \sqrt{(1+\varrho)(3-\varrho)}},\pm  \widetilde{\alpha} \sqrt{(1+\varrho)(3-\varrho)} \right), \quad k\in\Z.
\eqs
We recover the fact that the system \eqref{systtauomega} is invariant by $\omega\mapsto -\omega$. Since $\mathrm{arccos}\left(\frac{\varrho-1}{2}\right)\in[0,\pi]$, we deduce that the smallest positive $\tau$ is always achieved at $k=1$. We computed for several values of $\widetilde{\alpha}$ the corresponding values of $\tau$ and $\omega$ (for $k=1$) as a function of $\varrho$, which are presented in Figure~\ref{fig:delaycontinuous}(a)-(b). We observe that for values of $\varrho$ in the range $(1/2,1)$ the corresponding time delay $\tau$ takes values between $12ms$ to $23ms$ for values of $1/\widetilde{\alpha}$ ranging from $5ms$ to $10ms$. Correspondingly, in the same range of values for $\varrho$, the frequency $\omega/2\pi$ takes values between $30Hz$ to $60Hz$. 

This tells us that, when the time delay $\tau$ is chosen to be around $10-20ms$, compatible with biologically plausible values for communication delays between adjacent cortical areas, and when hyperparameters $\widetilde{\alpha}$ and $\widetilde{\lambda}$ are suitably chosen ($\widetilde{\alpha}$ in particular must be strong enough to allow rapid feed-forward error correction updates, i.e. around $1/\widetilde{\alpha}<8ms$, while $\widetilde{\lambda}$ can be chosen more liberally, as long as it stays $<3\widetilde{\alpha}$), then the network produces globally synchronized oscillations, comparable to experimentally observed brain rhythms in the $\gamma$-band regime (30-60Hz). In this context, it is interesting to note that theoretical and neuroscientific considerations have suggested that error correction in predictive coding systems is likely to be accompanied by oscillatory neural activity around this same $\gamma$-frequency regime~\cite{bastos2012}.

\subsubsection{Oscillatory traveling waves}

However, experimental and computational studies have also suggested that oscillatory signatures of predictive coding could be found at lower frequencies, in the so-called $\alpha$-band regime, around 7-15Hz. Furthermore, these oscillations are typically not homogeneous over space, as assumed in the previous section, but behave as forward- or backward-travelling waves with systematic phase shifts between layers~\cite{AVR19}. To explore this idea further, we
now investigate the possibility of having traveling wave solutions of the form
\bqs
\mathbf{e}_j(t)=e^{\mbi(\omega t +j\theta)}, \quad t>0, \quad j\in\Z,
\eqs
for some $\omega\in\R$ (representing the wave's temporal frequency) and $\theta\in[0,2\pi)$ (representing the wave's spatial frequency, i.e. its phase shift across layers), and we are especially interested in deriving conditions under which one can ensure that $\theta\neq0$ (since otherwise, we would be again facing the homogeneous oscillation case). We only focus on the case $\widetilde{\beta}=0$ (as postulated, e.g. in Rao and Ballard's work~\cite{RB99}) and leave the case  $\widetilde{\beta}>0$ for future investigations. As a consequence, the equation reduces to 
\bqs
\frac{\md }{\md t}{\bf e}_j(t)=-\widetilde{\lambda}{\bf e}_j(t)+\widetilde{\alpha}{\bf e}_{j-1}(t-\tau)+\widetilde{\lambda}{\bf e}_{j+1}(t-\tau)-\widetilde{\alpha}{\bf e}_{j}(t-2\tau), \quad t>0, \quad j\in\Z.
\eqs

Plugging the ansatz ${\bf e}_j(t)=e^{\mbi\left( \omega t+j \theta\right)}$, we obtain:
\bqs
\mbi \omega = \widetilde{\alpha} \left( e^{-\mbi(\omega\tau+\theta)}-e^{-2\mbi \omega \tau} \right) +\widetilde{\lambda} \left( e^{-\mbi(\omega\tau-\theta)}-1 \right).
\eqs
Taking real and imaginary parts, we obtain the system
\bqs
\left\{\begin{split}
0&=\widetilde{\alpha} \left(\cos(\omega\tau+\theta)-\cos(2\omega \tau) \right)+\widetilde{\lambda} \left( \cos(\omega\tau-\theta)-1 \right),\\
\omega&=-\widetilde{\alpha} \left(\sin(\omega\tau+\theta)-\sin(2\omega \tau) \right)-\widetilde{\lambda} \sin(\omega\tau-\theta).
\end{split}\right.\eqs
Once again, we introduce $\varrho:=\frac{\widetilde{\lambda}}{\widetilde{\alpha}}\geq0$ where we implicitly assumed that we always work in the regime $\widetilde{\alpha}>0$. Then, we note that the right-hand side of the first equation of the above system can be factored as
\bqs
\widetilde{\alpha} \left(\cos(\omega\tau+\theta)-\cos(2\omega \tau) \right)+\widetilde{\lambda} \left( \cos(\omega\tau-\theta)-1 \right)=-2\widetilde{\alpha}\sin\left(\frac{\theta-\omega\tau}{2}\right)\left( \varrho\sin\left(\frac{\theta-\omega\tau}{2}\right)+\sin\left(\frac{\theta+3\omega\tau}{2}\right) \right).
\eqs
As a consequence, either $\sin\left(\frac{\theta-\omega\tau}{2}\right)=0$, that is $\omega\tau=\theta+2k\pi$ for $k\in\Z$, which then leads, from the second equation, to $\omega=0$ and $\theta=0$ since we restrict $\theta\in[0,2\pi)$, or $\sin\left(\frac{\theta-\omega\tau}{2}\right)\neq0$. In the latter case, assuming that $\omega\tau\neq\theta+2k\pi$ for $k\in\Z$, we get that
\bqs
0=\varrho\sin\left(\frac{\theta-\omega\tau}{2}\right)+\sin\left(\frac{\theta+3\omega\tau}{2}\right).
\eqs
We will now study several cases.

\begin{figure}[t!]
  \centering
\subfigure[]{\includegraphics[width=.4\textwidth]{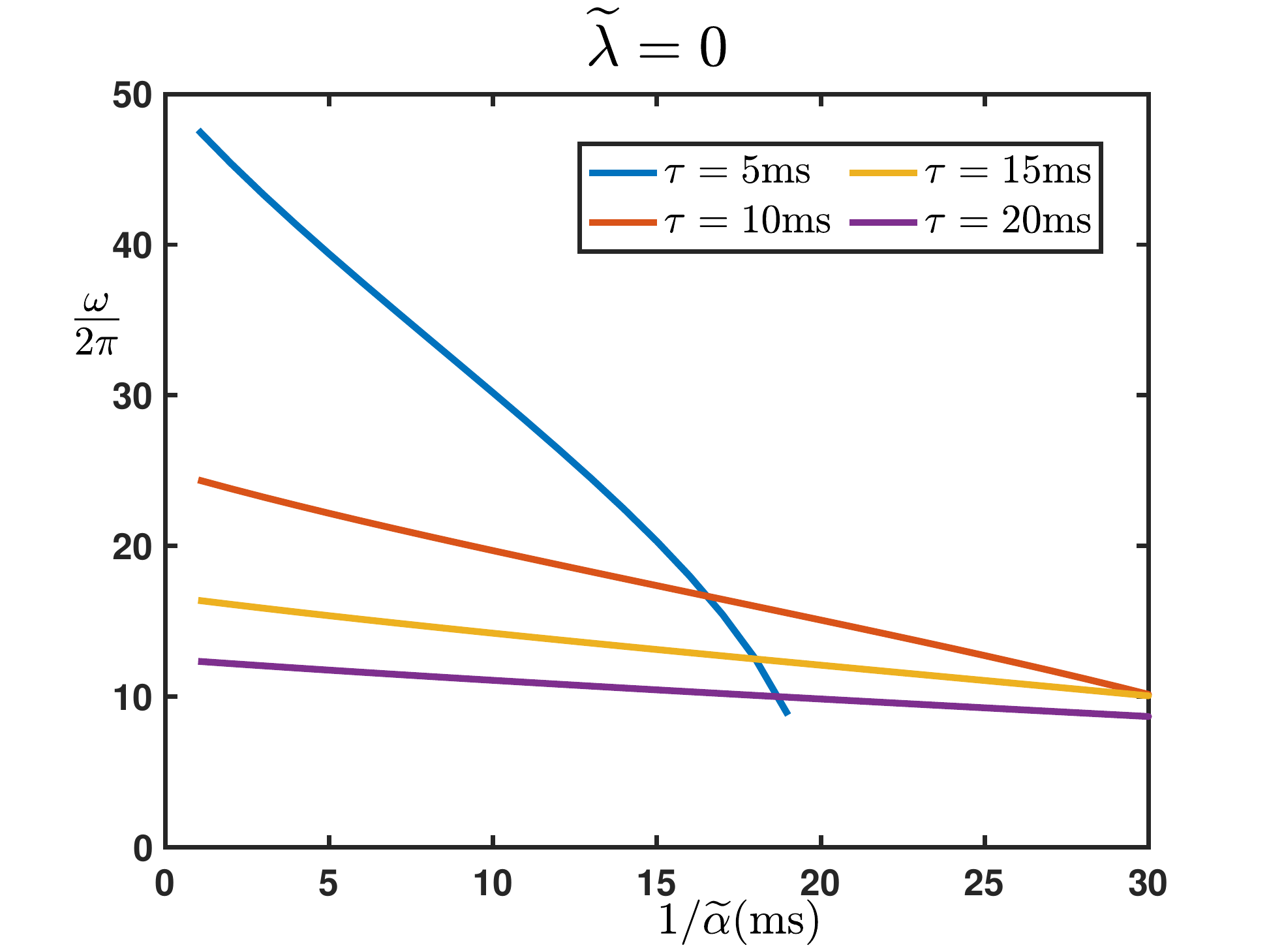}} \hspace{1cm}
\subfigure[]{\includegraphics[width=.4\textwidth]{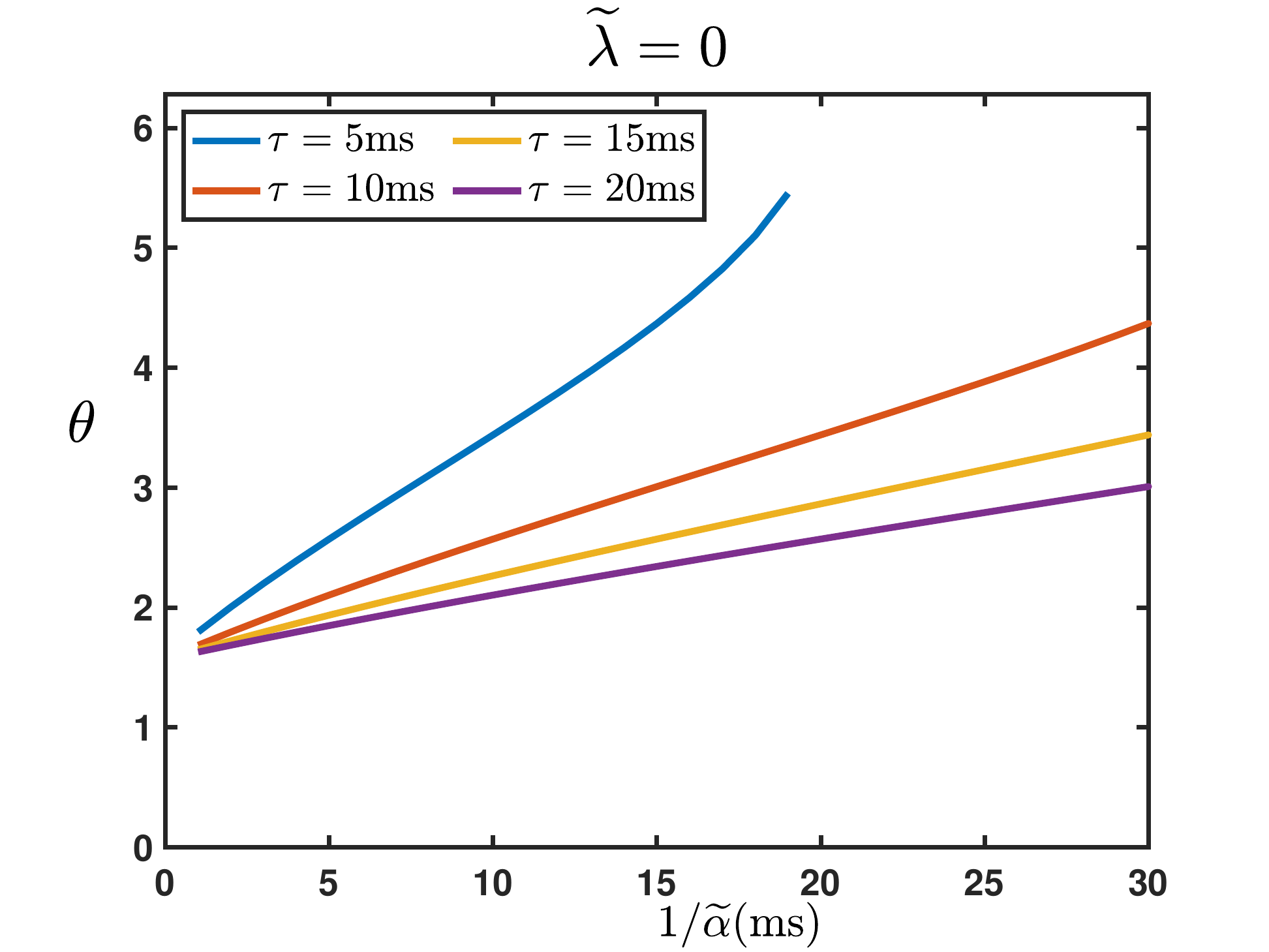}}
\caption{Representation of the temporal frequency $\omega/2\pi$ (in Hz) and the spatial frequency $\theta\in[0,2\pi)$ (panel (a) and (b) respectively) in the case $\widetilde{\lambda}=0$ as a function of $1/\widetilde{\alpha}$ (in ms) for several values of the time delay $\tau$.}
  \label{fig:varrho0}
\end{figure}

\paragraph{Case $\varrho=0$.} (In other words, this case implies $\widetilde{\lambda}=0$, that is, a system with no feedback error correction.) \\
From $\sin\left(\frac{\theta+3\omega\tau}{2}\right)=0$, we deduce that $\theta=-3\omega\tau+2k\pi$ for some $k\in\Z$, and reporting into the second equation of the system, we end up with
\bqs
\omega=2\widetilde{\alpha} \sin(2\omega \tau).
\eqs
We always have the trivial solution $\omega=0$ with $\theta=0$. In fact, when $4\widetilde{\alpha}\tau\leq1$, $\omega=0$ is the only solution of the above equation. On the other hand, when $4\widetilde{\alpha}\tau>1$, there can be multiple non trivial solutions. At least, for each $(\widetilde{\alpha},\tau)$ such that  $4\widetilde{\alpha}\tau>1$ there always exist a unique $\omega_c(\widetilde{\alpha},\tau)\in\left(0, \frac{\pi}{2\tau}\right)$ solution of the above equation. This gives a corresponding $\theta_c^k=-3\omega_c(\widetilde{\alpha},\tau)\tau+2k\pi$ with $k\in\Z$, and retaining the corresponding value of $\theta$ in the interval $[0,2\pi)$, we have $\theta_c=-3\omega_c(\widetilde{\alpha},\tau)\tau+2\pi$. We refer to Figure~\ref{fig:varrho0} for an illustration of the solutions $(\omega,\theta)$ for several values of the parameters.

Interestingly, we see that for biologically plausible values of the time delay $\tau$ between $10ms$ and $20ms$, the observed oscillation frequency is lower than in the previous case, and now compatible with the $\alpha$-frequency regime (between $10Hz$ and $20Hz$). Furthermore, the phase shift between layers $\theta$ varies roughly between 2 and 4 radians. As phase shifts below $\pi$ or above $\pi$ radians indicate respectively backward- or forward-travelling waves, we see that the exact value of the parameters $\tau$ and $\widetilde{\alpha}$ critically determines the propagation direction of the travelling waves: stronger feedforward error correction (lower values of $1/\widetilde{\alpha}$) and longer communication delays $\tau$ will tend to favor backward-travelling waves; and vice-versa, weaker feedforward error correction (higher values of $1/\widetilde{\alpha}$) and shorter communication delays $\tau$ will favor forward-travelling waves.

\paragraph{Case $\varrho=1$.} Now we assume that $\widetilde{\lambda}\neq0$, that is, the system now includes feedback error correction. At first, we consider the simpler case when $\varrho=1$, that is when $\widetilde{\alpha}=\widetilde{\lambda}$, where the equation can also be solved easily. Indeed, we have either
\bqs
\frac{\omega\tau-\theta}{2}=\frac{\theta+3\omega\tau}{2}+2k\pi, \quad k\in\Z,
\eqs
or 
\bqs
\frac{\omega\tau-\theta}{2}=\pi-\frac{\theta+3\omega\tau}{2}+2k\pi, \quad k\in\Z.
\eqs
This equivalent to
\bqs
\theta=-\omega\tau+2k\pi, \quad k\in\Z,
\eqs
or
\bqs
\omega\tau=\frac{\pi}{2}+k\pi, \quad k\in\Z.
\eqs
Let assume first that $\theta=-\omega\tau+2k\pi$ for some $k\in\Z$, then the second equation of the system gives $\omega=0$ since $\widetilde{\alpha}=\widetilde{\lambda}$ when $\varrho=1$, and thus we end up with $\theta=0$. Now, if $\omega\tau=\frac{\pi}{2}+k\pi$ for some $k\in\Z$, the second equation leads to
\bqs
\omega=-2\widetilde{\alpha}\cos(\theta+k\pi),
\eqs
from which we deduce that necessarily we must have
\bqs
\frac{(2k+1)\pi}{-4\widetilde{\alpha}\tau}=\cos(\theta+k\pi), \quad k\in\Z.
\eqs
We first remark that if $4\widetilde{\alpha}\tau<\pi$, then the above equation has no solution. On the other hand if $4\widetilde{\alpha}\tau\geq\pi$, one can obtain solutions to the above equation. Indeed, let us denote 
\bqs
p:=\left\lfloor \frac{4\widetilde{\alpha}\tau}{\pi} \right\rfloor \geq 1
\eqs
 the integer part of $ \frac{4\widetilde{\alpha}\tau}{\pi}$. Then, for each $k\in\Z$ such that $|2k+1|\leq p$, we have
 \bqs
 \theta=\pm \mathrm{arcos}\left( (-1)^{k+1} \frac{(2k+1)\pi}{4\widetilde{\alpha}\tau}\right)+2m\pi, \quad m\in\Z,
 \eqs
with corresponding $\omega$ given by
\bqs
\omega=\frac{\pi}{2\tau}+\frac{k\pi}{\tau}.
\eqs
That is, the set of solutions is given by
\bqs
(\omega,\theta)=\left(\frac{\pi}{2\tau}+\frac{k\pi}{\tau},\pm \mathrm{arcos}\left( (-1)^{k+1} \frac{(2k+1)\pi}{4\widetilde{\alpha}\tau}\right)+2m\pi\right),
\eqs
for each $k\in\Z$ such that $|2k+1|\leq \left\lfloor \frac{4\widetilde{\alpha}\tau}{\pi} \right\rfloor$ and $m\in\Z$. If we only retain the smallest $\omega$ and the corresponding value of $\theta\in[0,2\pi)$, we must take $k=0$, and we have two solutions 
\bqs
(\omega,\theta)=\left(\frac{\pi}{2\tau}, \mathrm{arcos}\left(-\frac{\pi}{4\widetilde{\alpha}\tau}\right)\right), \text{ and } (\omega,\theta)=\left(\frac{\pi}{2\tau}, -\mathrm{arcos}\left(-\frac{\pi}{4\widetilde{\alpha}\tau}\right)+2\pi \right) \quad \text{ if } 4\widetilde{\alpha}\tau\geq\pi.
\eqs
We note that in this case the temporal frequency only depends on the time delay $\tau$ since it is given by $\frac{\omega}{2\pi}=\frac{1}{4\tau}$ and ranges from 12.5Hz ($\alpha$-band regime) to 25Hz  ($\beta$-band regime) for values of $\tau$ between $10$ms to $20$ms (as long as $\widetilde{\alpha}$ is fixed such that $4\widetilde{\alpha}\tau\geq\pi$ is verified). The corresponding spatial frequencies are $\mathrm{arcos}\left(-\frac{\pi}{4\widetilde{\alpha}\tau}\right)\in(\pi/2,\pi)$ and $-\mathrm{arcos}\left(-\frac{\pi}{4\widetilde{\alpha}\tau}\right)+2\pi\in(3\pi/2,2\pi)$. 

In summary, when the feedforward and feedback error correction strengths are matched (that is when $\widetilde{\alpha}=\widetilde{\lambda}$) and sufficiently high (such that $4\widetilde{\alpha}\tau\geq\pi$), then the system will show two simultaneous travelling waves at the same frequency in the $\alpha$-band or $\beta$-band regime, but travelling in opposite directions, one as a feedforward wave and the other as a feedback wave.

\paragraph{Case $\varrho>1$.} Here, the feedback error correction $\lambda$ is stronger than the feedforward $\alpha$. In this case, we remark that
\bqs\varrho\sin\left(\frac{\theta-\omega\tau}{2}\right)+\sin\left(\frac{\theta+3\omega\tau}{2}\right)\\
=\left(\varrho+\cos(2\omega\tau)\right)\sin\left(\frac{\theta-\omega\tau}{2}\right)+\cos\left(\frac{\theta-\omega\tau}{2}\right)\sin(2\omega\tau).
\eqs
Since $\varrho>1$ and $\omega\tau\neq\theta+2k\pi$ for $k\in\Z$, we have $\left(\varrho+\cos(2\omega\tau)\right)\sin\left(\frac{\theta-\omega\tau}{2}\right)\neq0$, and thus $\cos\left(\frac{\theta-\omega\tau}{2}\right)\sin(2\omega\tau)\neq0$ otherwise we would reach a contradiction since we try to solve 
\bqs
\varrho\sin\left(\frac{\theta-\omega\tau}{2}\right)+\sin\left(\frac{\theta+3\omega\tau}{2}\right)=0.
\eqs
As a consequence, $\cos\left(\frac{\theta-\omega\tau}{2}\right)\neq0$ and we can rewrite the above equation as
\bqs
\tan\left(\frac{\theta-\omega\tau}{2}\right)=-\frac{\sin(2\omega\tau)}{\varrho+\cos(2\omega\tau)},
\eqs
so that
\bqs
\theta=\omega\tau-2\mathrm{arctan}\left(\frac{\sin(2\omega\tau)}{\varrho+\cos(2\omega\tau)}\right)+2k\pi, \quad k\in\Z.
\eqs
Injecting this expression for $\theta$ into the second equation, we find, after simplification, that
\bqs
\omega=\widetilde{\alpha}\frac{2\sin(2\omega\tau)\left(1-\varrho^2\right)}{2\varrho \cos(2\omega\tau)+\varrho^2+1}.
\eqs
We first remark that $\omega=0$ is always a solution, giving $\theta=0$. Now, inspecting the right-hand of the above expression, we get that
\bqs
\frac{2\widetilde{\alpha}\left(1-\varrho^2\right)}{2\varrho \cos(2\omega\tau)+\varrho^2+1}<0, \text{ for all } \omega\in\R.
\eqs
As a consequence, we look for the negative minima of the function $\omega\mapsto \frac{\sin(2\omega\tau)}{2\varrho \cos(2\omega\tau)+\varrho^2+1}$ which are given by $\omega_0=\frac{\pi}{2\tau}+\frac{1}{2\tau}\mathrm{arcos}\left(\frac{2\varrho}{1+\varrho^2}\right)+\frac{k\pi}{\tau}$ for $k\in\Z$, at such minima, one gets that
\bqs
\frac{\sin(2\omega_0\tau)}{2\varrho \cos(2\omega_0\tau)+\varrho^2+1}=\frac{1}{1-\varrho^2}.
\eqs
This implies that if $4\widetilde{\alpha}\tau<\pi+\mathrm{arcos}\left(\frac{2\varrho}{1+\varrho^2}\right)$, then there is no other solution than $(\omega,\theta)=(0,0)$. As a consequence, one needs to assume  $4\widetilde{\alpha}\tau\geq \pi+\mathrm{arcos}\left(\frac{2\varrho}{1+\varrho^2}\right)$ to ensure the existence of at least one non trivial solution. We remark that this condition is consistent with our condition $4\widetilde{\alpha}\tau\geq \pi$ derived in the case $\varrho=1$.

\paragraph{Case $0<\varrho<1$.} We start once again from the equation
\bqs
0=\left(\varrho+\cos(2\omega\tau)\right)\sin\left(\frac{\theta-\omega\tau}{2}\right)+\cos\left(\frac{\theta-\omega\tau}{2}\right)\sin(2\omega\tau).
\eqs
This time, it is possible that $\varrho+\cos(2\omega\tau)=0$, which gives necessarily that
\bqs
\omega \tau=\pm\frac{1}{2}\mathrm{arcos}(-\varrho)+k\pi, \quad k\in\Z.
\eqs
But if $\varrho+\cos(2\omega\tau)=0$, then one has $0=\cos\left(\frac{\theta-\omega\tau}{2}\right)\sin(2\omega\tau)$. 

Let us first assume that it is $0=\cos\left(\frac{\theta-\omega\tau}{2}\right)$, such that $\theta=\omega \tau+(2k+1)\pi$ for $k\in\Z$. Now, looking at the second equation, we find that
\bqs
\omega = 2\widetilde{\alpha}\sin(2\omega\tau)=\pm2\widetilde{\alpha}\sqrt{1-\varrho^2},
\eqs
which implies that it is possible only if
\bqs
\tau = \frac{\mathrm{arcos}(-\varrho)+k\pi}{4\widetilde{\alpha}\sqrt{1-\varrho^2}}, \quad k\geq0.
\eqs
As a conclusion, if $\tau$ and $0<\varrho<1$ satisfy $\tau = \frac{\mathrm{arcos}(-\varrho)+k\pi}{4\widetilde{\alpha}\sqrt{1-\varrho^2}}$ for some positive integer $k\geq0$, then
\bqs
( \omega,\theta)=\left(2\widetilde{\alpha}\sqrt{1-\varrho^2},\frac{1}{2}\mathrm{arcos}(-\varrho)-\pi\right), \quad \text{ and } \quad ( \omega,\theta)=\left(-2\widetilde{\alpha}\sqrt{1-\varrho^2},-\frac{1}{2}\mathrm{arcos}(-\varrho)+\pi\right),
\eqs
are corresponding solutions of the problem.

Next, let us assume that it is $\sin(2\omega\tau)=0$, implying that $2\omega\tau=k\pi$ for $k\in\Z$. Now we readily remark that since $0<\varrho<1$, we have $\mathrm{arcos}(-\varrho)\in(\pi/2,\pi)$. As a consequence, we should have
\bqs
\pm\mathrm{arcos}(-\varrho)+2k\pi = p\pi, \quad k,p\in\Z,
\eqs
this is impossible and thus $\sin(2\omega\tau)\neq0$ and we are back to the case treated before.

We now assume that $\varrho+\cos(2\omega\tau)\neq0$. In that case, we can proceed as in the case $\varrho>1$ and obtain that
\bqs
\theta=\omega\tau-2\mathrm{arctan}\left(\frac{\sin(2\omega\tau)}{\varrho+\cos(2\omega\tau)}\right)+2k\pi, \quad k\in\Z,
\eqs
which gives
\bqs
\omega=\widetilde{\alpha}\frac{2\sin(2\omega\tau)\left(1-\varrho^2\right)}{2\varrho \cos(2\omega\tau)+\varrho^2+1}.
\eqs
Once again, $(\omega,\theta)=(0,0)$ is always a solution. What changes in this case is that now \bqs
\frac{2\widetilde{\alpha}\left(1-\varrho^2\right)}{2\varrho \cos(2\omega\tau)+\varrho^2+1}>0, \text{ for all } \omega\in\R.
\eqs
This time, one needs to look at the positive maxima of the map $\omega\mapsto \frac{\sin(2\omega\tau)}{2\varrho \cos(2\omega\tau)+\varrho^2+1}$ which are given by $\omega_0=\frac{\pi}{2\tau}-\frac{1}{2\tau}\mathrm{arcos}\left(\frac{2\varrho}{1+\varrho^2}\right)+\frac{k\pi}{\tau}$ for $k\in\Z$, at such maxima, one gets that
\bqs
\frac{\sin(2\omega_0\tau)}{2\varrho \cos(2\omega_0\tau)+\varrho^2+1}=\frac{1}{1-\varrho^2}.
\eqs
As a consequence, if $4\widetilde{\alpha}\tau<\pi-\mathrm{arcos}\left(\frac{2\varrho}{1+\varrho^2}\right)$, then there is no other solution than $(\omega,\theta)=(0,0)$. To obtain at least one non trivial positive solution, one needs to impose that $4\widetilde{\alpha}\tau\geq\pi-\mathrm{arcos}\left(\frac{2\varrho}{1+\varrho^2}\right)$. Once again, this condition is consistent with the condition $4\widetilde{\alpha}\tau\geq\pi$ derived in the case $\varrho=1$. We can also derive a second simple condition which ensures the existence of a non trivial solution by looking at the behavior near $\omega\sim0$ where we have
\bqs
\widetilde{\alpha}\frac{2\sin(2\omega\tau)\left(1-\varrho^2\right)}{2\varrho \cos(2\omega\tau)+\varrho^2+1}\sim \frac{4\widetilde{\alpha}\tau\left(1-\varrho\right)}{1+\varrho}\omega.
\eqs
Thus if 
\bqs
4\widetilde{\alpha}\tau>\frac{1+\varrho}{1-\varrho},
\eqs
then there exists at least one positive solution $\omega\in(0,\pi/2\tau)$ to the above equation (and also one negative solution in $(-\pi/2\tau,0)$ by symmetry). Note that the condition $4\widetilde{\alpha}\tau>\frac{1+\varrho}{1-\varrho}$ is consistent with the condition $4\widetilde{\alpha}\tau>1$ derived in the case $\varrho=0$.

\begin{figure}[t!]
  \centering
\subfigure[$\omega/2\pi$ as a function of $\varrho$.]{\includegraphics[width=.3\textwidth]{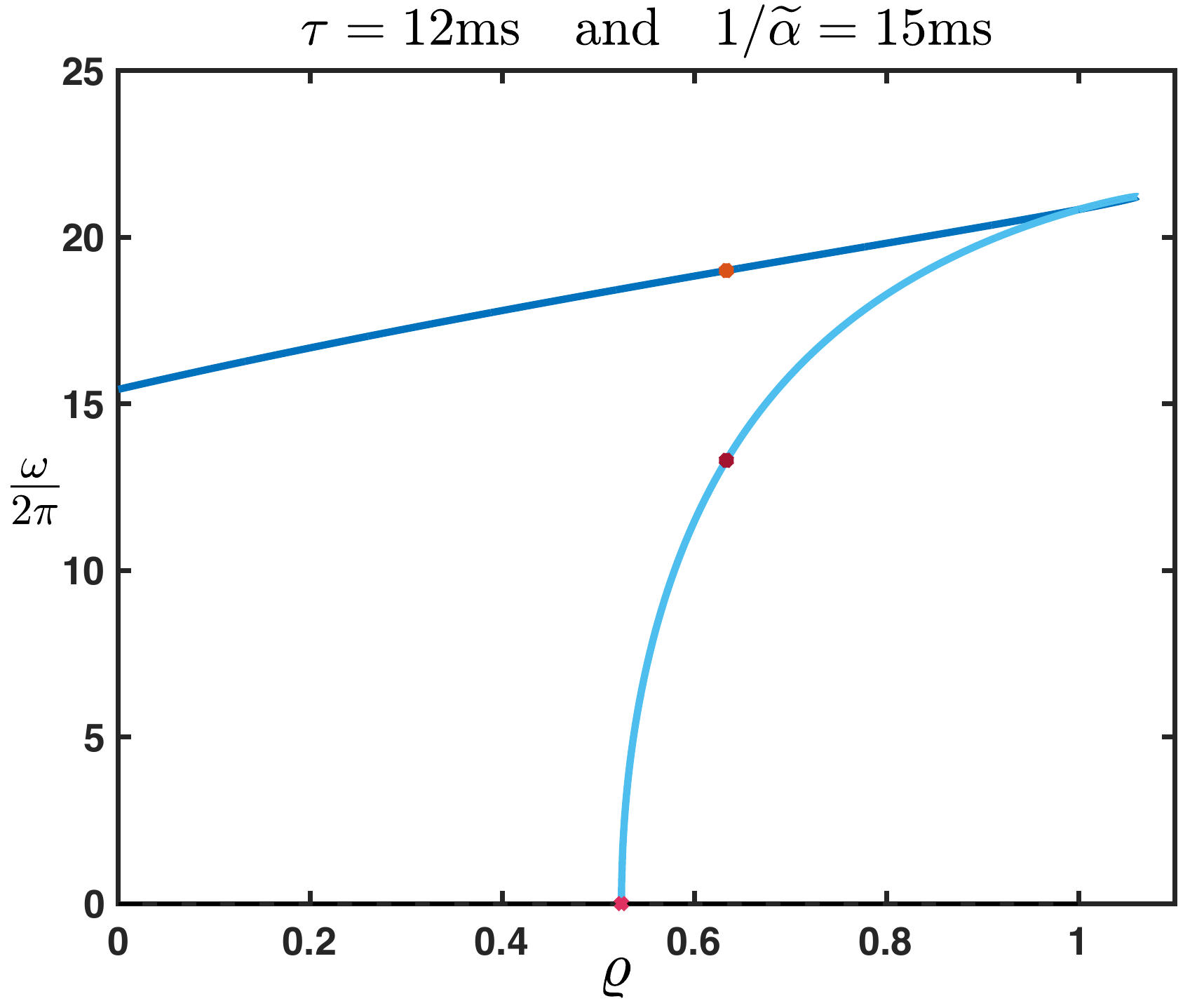}} 
\subfigure[Zoom of panel (a) near $\varrho\sim1$.]{\includegraphics[width=.3\textwidth]{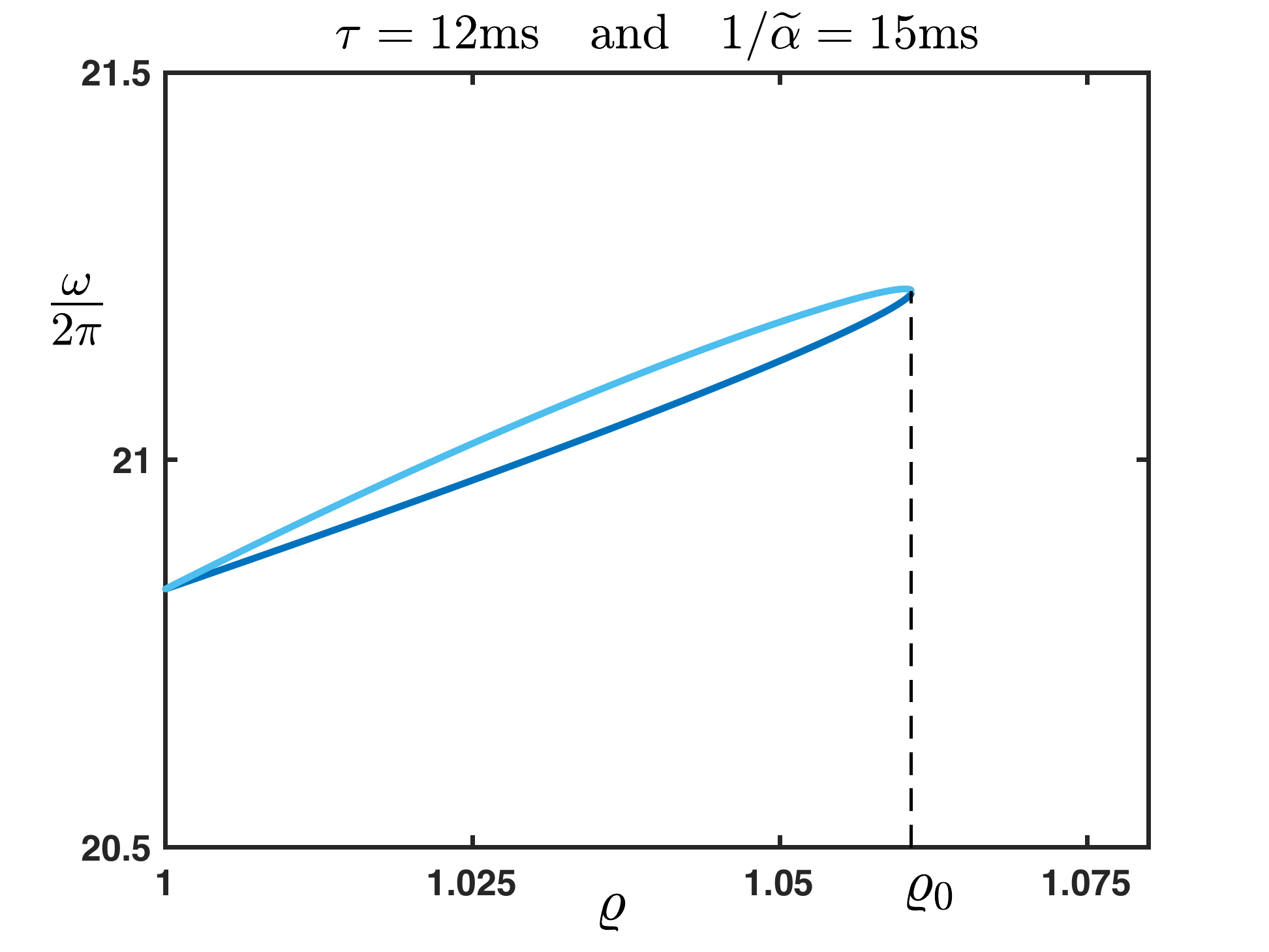}} 
\subfigure[$\theta$ as a function of $\varrho$.]{\includegraphics[width=.3\textwidth]{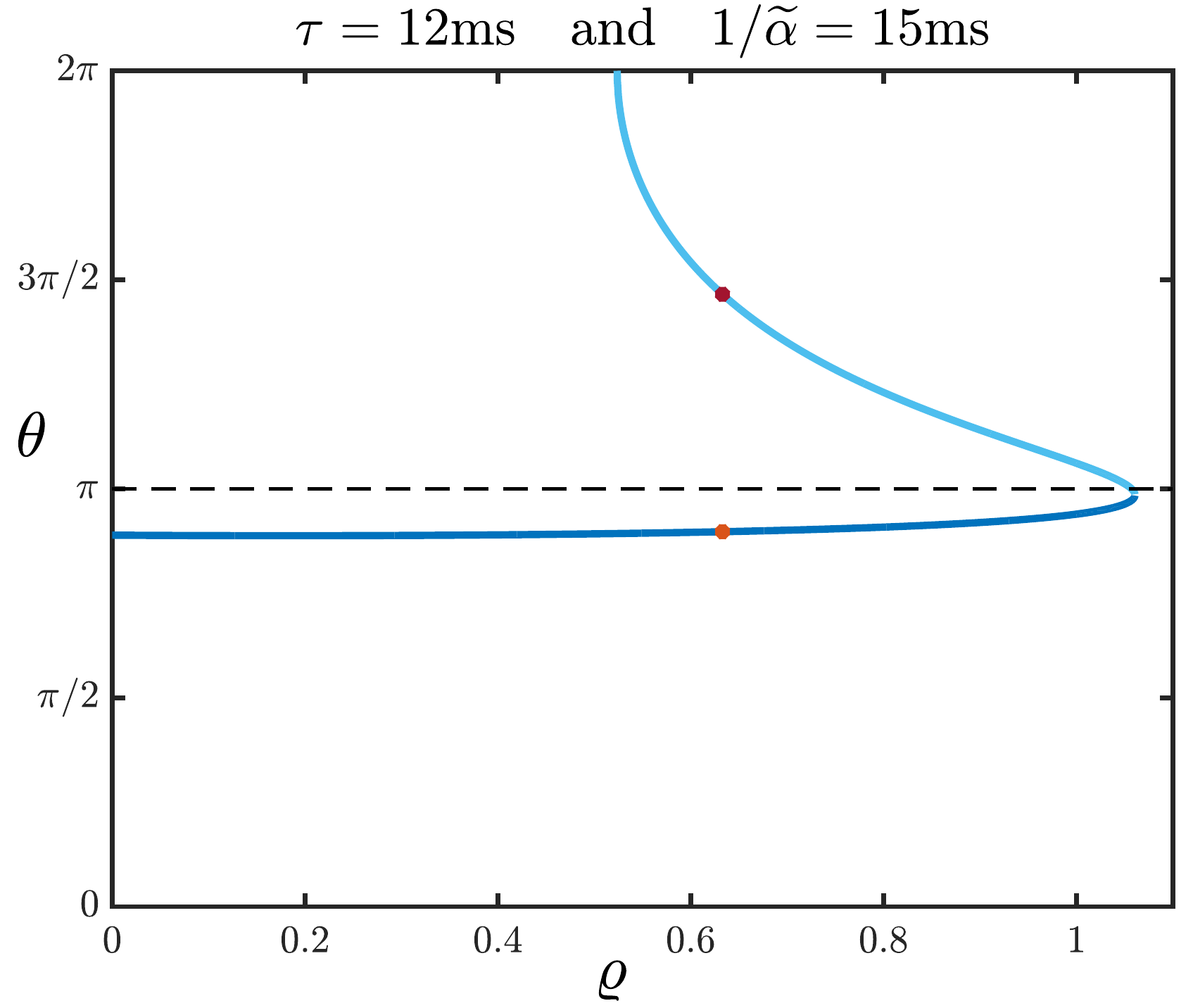}}
\caption{Representation of the temporal frequency $\omega/2\pi$ (in Hz) and the spatial frequency $\theta\in[0,2\pi)$ (panel (a) and (c) respectively) as a function of $\varrho$ for fixed values of the time delay $\tau=12$ms and $1/\widetilde{\alpha}=15$ms. Panel (b) represents a zoom of panel (a) near $\varrho\sim1$ where the two branches terminate.}
  \label{fig:varrhoneq0}
\end{figure}

\paragraph{Examples.}
To illustrate the different scenarios and their possible interpretations in terms of brain oscillations, we take here two distinct examples corresponding to the situations described above.
%the first for $\varrho<1$ (and thus $\widetilde{\lambda}<\widetilde{\alpha}$), the second extending to $\varrho>1$ (and thus $\widetilde{\lambda}>\widetilde{\alpha}$).}
We report in Figure~\ref{fig:varrhoneq0} the non trivial branches of solutions corresponding to positive values of $\omega$, as a function of $\varrho$ for fixed values of the time delay $\tau=12$ms and $1/\widetilde{\alpha}=15$ms. These values are biologically plausible and correspond to the values used in \cite{AVR19}. Upon denoting 
\bqs
\varrho_c:=\frac{1-4\widetilde{\alpha}\tau}{1+4\widetilde{\alpha}\tau}\in(0,1),
\eqs
for all $\varrho\in[0,\varrho_c)$, we get the existence of a unique branch of solution (blue curve) for the temporal frequency $\omega/2\pi$. A second branch (light blue curve) of solutions emerges precisely at $\varrho=\varrho_c$. These two branches cross at $\varrho=1$ where $\omega/2\pi=\frac{1}{4\tau}$ and terminate at a value of $\varrho=\varrho_0\sim1.06$ (see Figure~\ref{fig:varrhoneq0}(b)). The branch of solutions which exists for all values of $\varrho\in[0,\varrho_0]$ has an associated spatial frequency which is almost constant and whose value is around $\sim 2.82\in(0,\pi)$. On the other hand, the branch of solutions which only exists for values of $\varrho\in(\varrho_c,\varrho_0]$ has an associated spatial frequency which lies in $(\pi,2\pi)$. Let us remark that at $\varrho=1$, the spatial frequencies of the two solutions are different and symmetric with respect to $\pi$. Furthermore, at $\varrho=\varrho_0\sim1.06$ where the two branches collide the associated spatial frequency is $\theta\sim \pi$. Let us finally note that for $\varrho\in[1,\varrho_0]$, the spatial frequencies of the two branches are almost identical, although the secondary branch is slightly above the primary one.

\begin{figure}[t!]
  \centering
\subfigure[$(\omega,\theta)\sim (0.12,2.82)$]{\includegraphics[width=.4\textwidth]{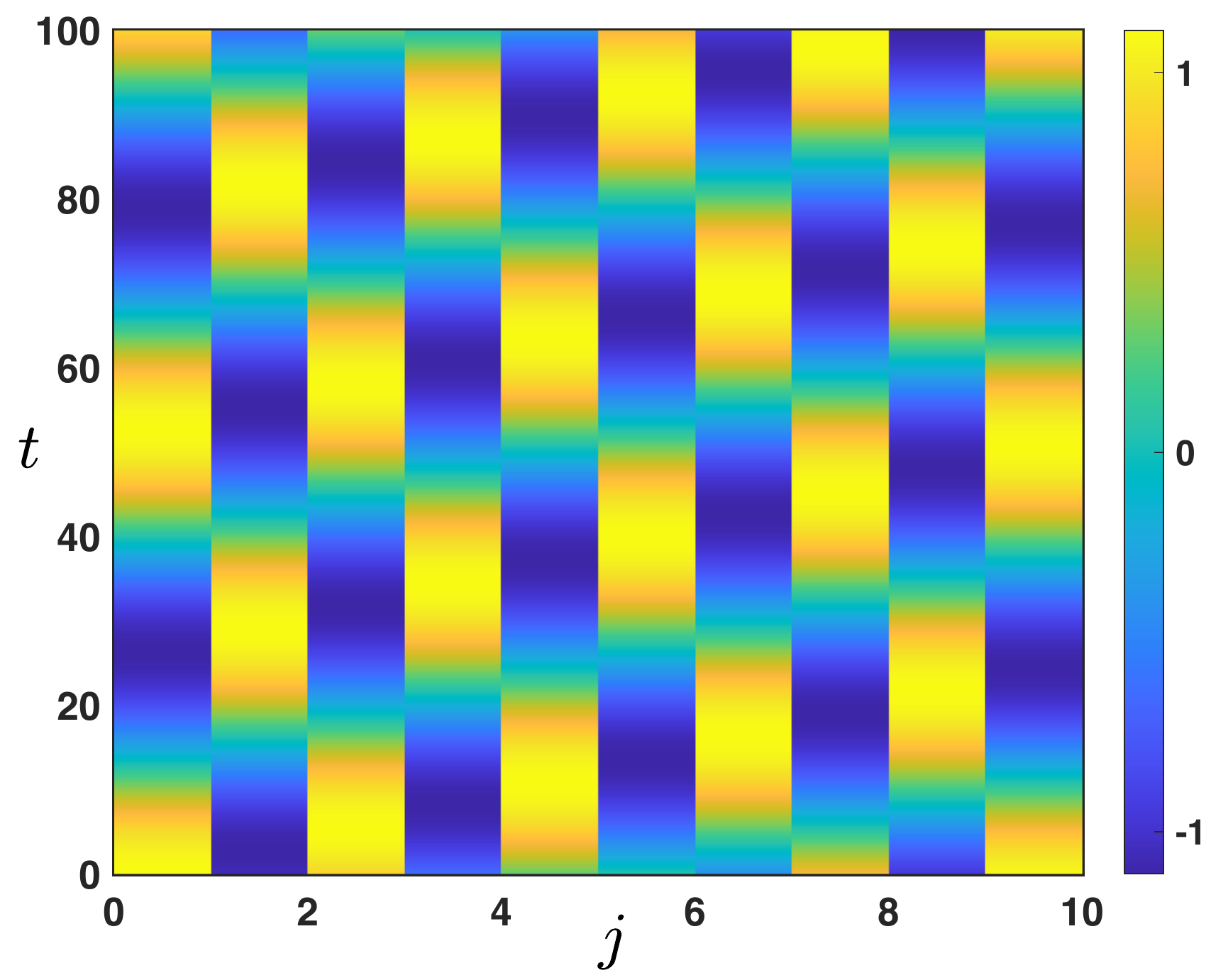}} \hspace{1cm}
\subfigure[$(\omega,\theta)\sim (0.08,4.60)$]{\includegraphics[width=.4\textwidth]{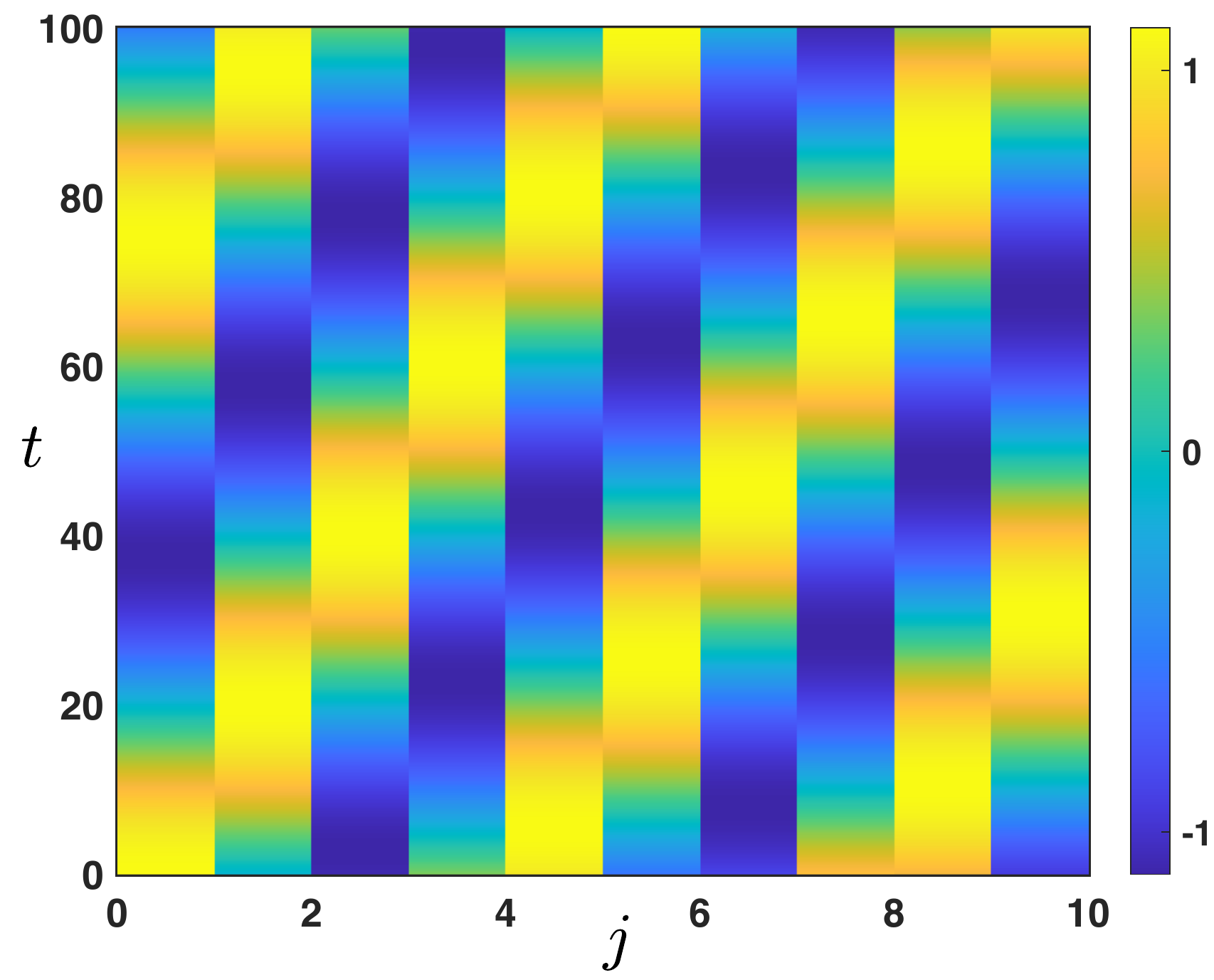}}
\caption{Space-time plot of $\cos(\omega t+\theta j)$ for values of $(\omega,\theta)$ which correspond to the orange and dark red points of Figure~\ref{fig:varrhoneq0} lying respectively on the blue and light blue curves, time $t$ is in ms. In (a), the temporal frequency is $\omega/2\pi\sim19$Hz while in (b) it is $\omega/2\pi\sim 13.3$Hz. In (a), since $\theta\in(0,\pi)$, we observe a backward propagation of the wave while in (b) we have a forward propagation since $\theta\in(\pi,2\pi)$.}
  \label{fig:STbranch}
\end{figure}

Correspondingly we illustrate in Figure~\ref{fig:STbranch}, the space-time plot of two points along the two different branches which correspond to the orange and dark red points in Figure~\ref{fig:varrhoneq0}. The corresponding values are $(\omega,\theta)\sim (0.12,2.82)$ and $(\omega,\theta)\sim (0.08,4.60)$ and associated to the same value of $\varrho\sim0.633$. In the first panel of Figure~\ref{fig:STbranch}(a), which corresponds to the point on the branch of solution defined for all $\varrho\in[0,\varrho_0]$, since the corresponding value of the spatial frequency is $\theta\in(0,\pi)$, we observe an apparent backward propagation, while in the second panel of Figure~\ref{fig:STbranch}(b), we observe a forward propagation. This corresponds to the point on the lower branch of the solutions defined for values of $\varrho\in(\varrho_c,\varrho_0]$ with associated spatial frequency $\theta\in(\pi,2\pi)$. From a biological point of view, this indicates that the more interesting range of the parameters is the one with $\varrho\in(\varrho_c,\varrho_0]$ and the corresponding branch of solutions which emerges at $\varrho=\varrho_c$ from the trivial solution $(\omega,\theta)\sim (0,0)$ since in this case we obtain an oscillatory traveling wave with forward propagation into the network.

\begin{figure}[t!]
  \centering
\subfigure[]{\includegraphics[width=.4\textwidth]{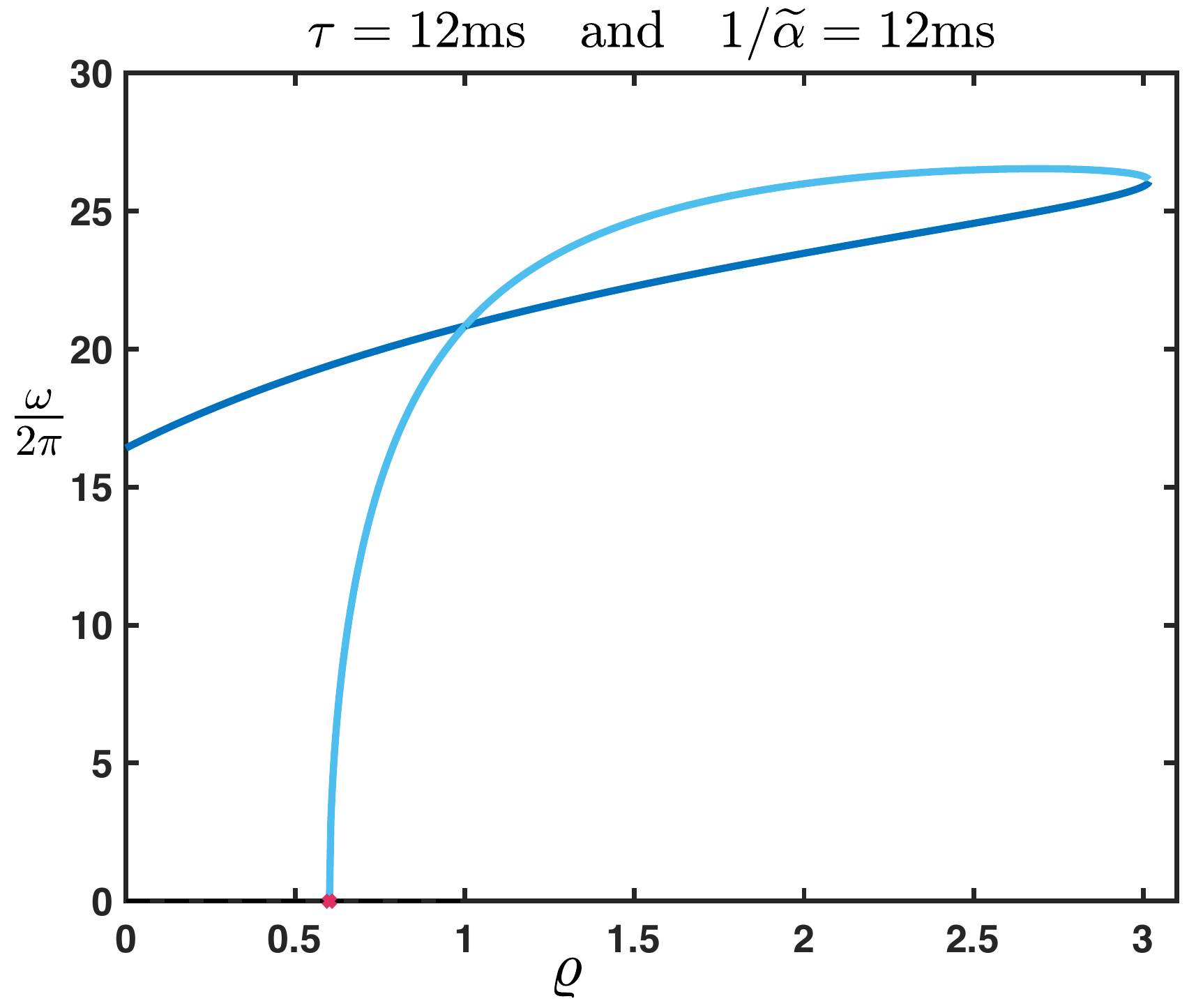}} \hspace{1cm}
\subfigure[]{\includegraphics[width=.4\textwidth]{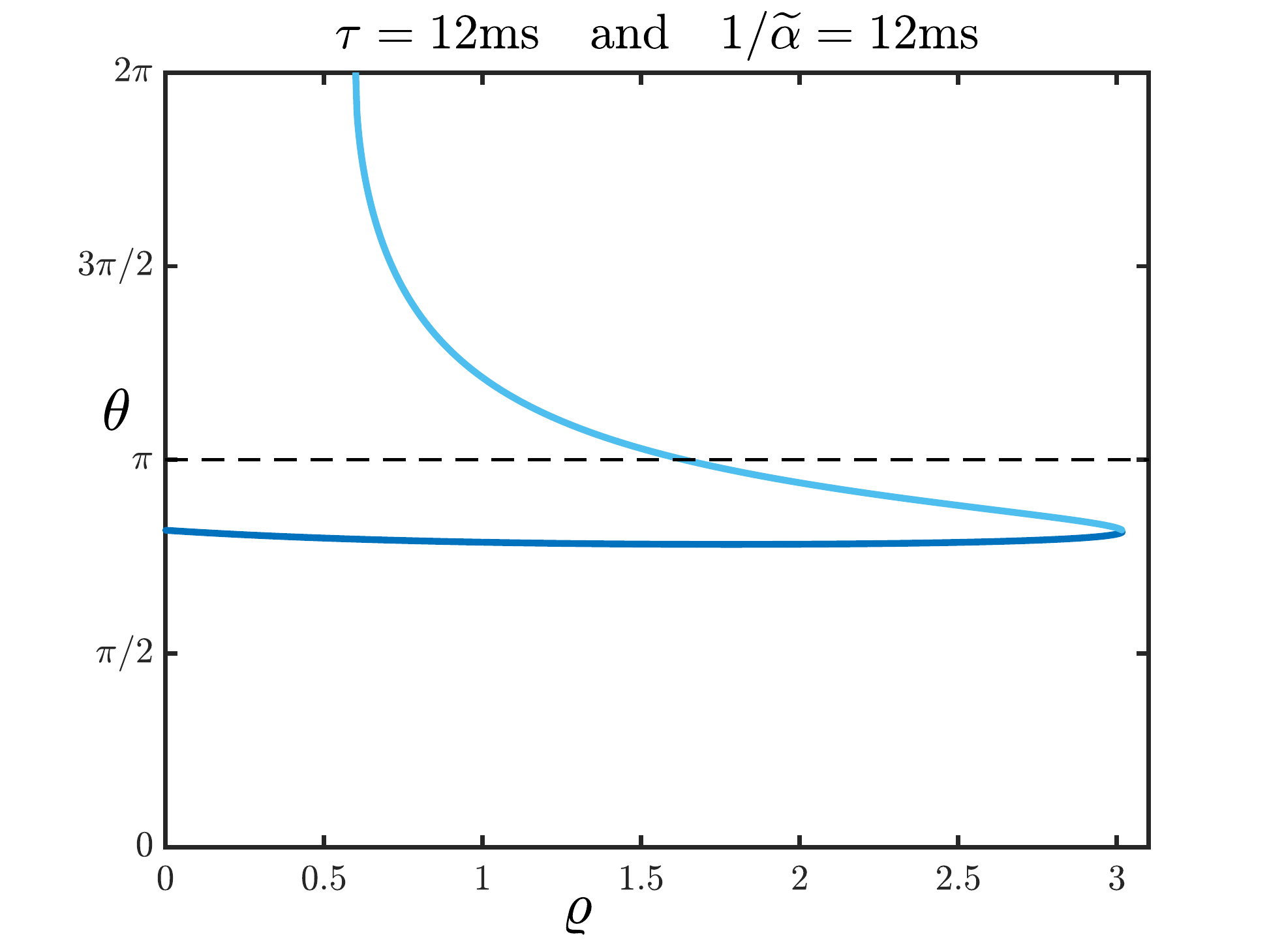}}
\caption{Representation of the temporal frequency $\omega/2\pi$ (in Hz) and the spatial frequency $\theta\in[0,2\pi)$ (panel (a) and (b) respectively) as a function of $\varrho\in[0,1]$ for fixed values of the time delay $\tau=12$ms and $1/\widetilde{\alpha}=12$ms.}
  \label{fig:varrhoneq012}
\end{figure}

In Figure~\ref{fig:varrhoneq012}, we show the global structure of the branches for a second example, with fixed values of the time delay $\tau=12$ms and $1/\widetilde{\alpha}=12$ms, which are still biologically relevant values. We observe that the two branches terminate at a value of $\varrho=\varrho_0\sim 3.03$ with a crossing at $\varrho=1$. For $\varrho\in[1,\varrho_0]$, the primary branch (blue curve) has a temporal frequency below the secondary branch (light blue curve), the difference in frequencies is almost $5$Hz for values of $\varrho\sim2$. Even more interestingly, we see that the corresponding spatial frequencies along the secondary branch are decreasing from $2\pi$ to a final value below $\pi$ at $\varrho_0$ indicating that by increasing the value of $\varrho$ we can reverse the direction of propagation from forward to backward oscillatory traveling waves. The transition occurs for $\varrho\sim1.65$, that is for values of $1/\widetilde{\lambda}\sim7-8$ms.
%that is for values of $1/\widetilde{\lambda}\sim20$ms. 
It is further noticed that the associated temporal frequencies in the backward regime are around $25$Hz ($\beta$-frequency regime) much higher than for forward traveling waves whose temporal frequencies range from $0$Hz to $20$Hz (and include the $\alpha$-frequency regime).

\paragraph{Summary}
In this section, we saw that including temporal delays in the time-continuous version of the system produces non-trivial dynamics that can be characterized analytically. Contrary to the discrete version of the system, which can only be analyzed for a few discrete time delays $k=1,2,...$, the continuous version is informative for a wide range of biologically plausible delays, and the resulting frequencies are very diverse. In particular, we observed homogenous synchronized oscillations in the gamma band ($30-60Hz$) that emerged when the feed-forward error correction term $\widetilde{\alpha}$ was strong enough (roughly, with $1/\widetilde{\alpha} < 8 ms$). But we also found situations in which the oscillatory activity was not homogenous, but propagated as a travelling wave through the network. With biologically plausible values for the various parameters, the waves could propagate forward in the alpha-band ($7-15Hz$) frequency range, and when the feedback error correction term $\widetilde{\lambda}$ was strong enough (e.g. $1/\widetilde{\lambda}<8ms$ while $1/\widetilde{\alpha}=12ms$), they started moving backward at a faster frequency in the beta-band ($15-30Hz$). Altogether, this pattern of results is compatible with various (sometimes conflicting) observations from the Neuroscience literature~\cite{AVR19, bastos2012}, and informs us about the conditions in which the corresponding dynamic behaviors might emerge in the brain.

\section{Discussion}
\subsection{Contributions}
We proposed a mathematical framework to explore the properties and stability of neural network models of the visual system comprising a hierarchy of visual processing areas (or ``layers''), mutually connected according to the principles of predictive coding. Using a discrete model, as is typically done in the recent deep learning literature, we introduced the amplification factor function, which serves to characterize the interesting (i.e., ``marginally stable'') regions as a function of the model hyperparameters. When considered on an infinite domain, 
we showed that the response of our linear neural network to a Dirac delta initialization presents a universal behavior given by a Gaussian profile with fixed variance and which spreads at a given speed. Both speed and variance could be explicitly characterized in terms of the model hyperparameters. This universal Gaussian profile was then the key to understand the long-time dynamics of the linear neural network set on a semi-infinite domain with a fixed constant source term at the left boundary of the network.

At first, we ignored the influence of neuronal selectivity and used feed-forward and feedback connection matrices set to the identity matrix. When $\beta=0$ (no feedforward update after the network initialization), we observed that hyperparameters $\alpha$ and $\lambda$ compete for forward and backward propagation, respectively. When $\beta>0$, the constant feedforward input makes things more complex, with $\lambda$ (feedback error correction) now competing with $\beta+\alpha$ (feedforward drive and feedforward error correction). In the special case when $\alpha+\lambda=1$, a second (but spurious) mode of propagation with rapidly alternating activity can emerge, whose direction is determined by the competition between $\alpha$ and $\beta+\lambda$. 

Next, to evaluate the influence of a more complex and functionally relevant connectivity matrix, we defined \textit{neural assemblies} reflecting the eigenvectors of the matrix. Each of these neural assemblies can be analyzed separately, and its behavior depends on the corresponding eigenvalue (in addition to the hyperparameters $\alpha$, $\beta$ and $\lambda$, as explained above). Different assemblies can simultaneously support different dynamics, so that some may propagate information forward, others may not propagate at all (acting as a filter on the inputs), while yet others might propagate backward (e.g. carrying ``priors'' set by preceding activations). We again saw a number of cases where ``fringe'' or spurious behavior arose, e.g. rapid alternations in activity, and understood that this could be caused by the discrete nature of our model, when the time steps defining the model's temporal resolution are too coarse.

The time-continuous version of the model helped us overcome this issue, and characterize dynamics in the limit of infinitely small time steps. The amplification factor function is still crucial in this situation, but it produces more robust results, without fringe behavior or spurious oscillations. In particular, the analysis of stability and propagation direction/speed was greatly simplified in this continuous case. 

The same time-continuous model also allowed us to investigate the inclusion of biologically plausible communication delays between layers. In this case, we demonstrated the emergence of genuine oscillatory dynamics and travelling waves in various frequency bands compatible with neuroscientific observations (alpha-band from 7 to 15Hz, beta-band from 15 to 30Hz and gamma-band from 30 to 60Hz).

Finally, we considered fully continuous versions of the model, not only in time but also in space, both across network depth (across neuronal layers) and width (across neurons in the same layer). This mathematical abstraction revealed that our model could be understood as a transport equation, and that it produced diffusion dynamics. 

\subsection{Biological interpretations}

The mathematical framework that we proposed naturally lends itself to interpretation in biological terms. The model's hyperparameters reflect the strength of feedforward and feedback signalling in the brain. These are determined not only by axonal density and synaptic strength (that vary slowly throughout development and learning), but can also be gated by other brain regions and control systems, e.g. through the influence of neurotransmitters, and thus vary much more dynamically. For instance, the feedforward drive $\beta$ could be more active to capture sensory information immediately after each eye movement, and decrease over time until the next eye movement~\cite{knoell2011}; similarly, feedback error correction $\lambda$ could dominate over the feedforward error correction $\alpha$ for one given second (e.g. because top-down attention drives expectation signals) and decrease in the next second (e.g. because unexpected sensory inputs have been detected)~\cite{tschantz2022}. In this dynamic context, it is fundamental to be able to characterize the dependence of the system's behavior on the exact hyperparameter values.
Fortunately, our framework reveals that when the hyperparameters vary, the stability of the system, and its ability to propagate signals and maintain activity, change in predictable ways. Some hyperparameter combinations would not support signal propagation at all; others would render the system unstable, e.g. because of runaway excitation. Under the (reasonable) assumption that the brain behaves as a predictive coding system, our equations inform us about the plausible parameter regimes for the brain. 

Using our time-continuous model, we found that predictive coding dynamics associated with inter-areal communication delays result in oscillatory activity. This finding resonates with both experimental observations and neuroscientific theories~\cite{bastos2012,AVR19}. 

Bastos and colleagues~\cite{bastos2012,bastos2015} suggested that feedforward error correction could be accompanied by gamma-band oscillations; this suggestion was verified in our model, with synchronized gamma rhythms appearing when the corresponding hyperparameter $\widetilde{\alpha}$ was strong enough (and with a frequency that monotonically increased from 30 to 60Hz when the value of $1/\widetilde{\alpha}$ decreased from 10ms to 5ms). However, considering that the communication delay $\tau$ between two adjacent brain regions is a fixed property of the system (a reasonable first approximation), our analysis shows that this oscillatory mode will only happen for a narrow range and a very precise combination of hyperparameter values $\widetilde{\alpha}$ and $\widetilde{\lambda}$ (see Figure~\ref{fig:delaycontinuous}). This could explain why gamma-band oscillations are not always found during electrophysiological recording experiments~\cite{ray2015}.

By relaxing the phase delay between brain areas, our equations also revealed the potential emergence of oscillatory travelling waves across brain regions, similar to those observed in human EEG experiments~\cite{AVR19,pang2020, alamia2020, alamia2023}. Again, for a fixed communication delay $\tau$, these waves may only happen for specific values and combinations of the hyperparameters $\widetilde{\alpha}$ and $\widetilde{\lambda}$. In certain regimes (see e.g. Figure~\ref{fig:varrhoneq012} with $1/\widetilde{\alpha}=1/\widetilde{\lambda}=12ms$), two waves might coexist at the same frequency, but going in opposite directions. This matches experimental reports of co-occurring feedforward and feedback waves in the brain~\cite{alamia2020, alamia2023}. Upon increasing the feedback strength $\widetilde{\lambda}$, we saw that an initial alpha-band (7-15Hz) feed-forward wave could accelerate (towards the beta-band, 15-30Hz) and eventually reverse its direction, producing a feedback wave. Similar reversal phenomena have also been reported for oscillatory waves in the human brain~\cite{pang2020, alamia2020, alamia2023}.

\subsection{Limitations and future extensions}

``All models are wrong, but some are useful''~\cite{box1979}. 
Our model, like all mathematical models, is based on simplifications, approximations and assumptions, and can only be valid under those assumptions. Some (if not all) of these assumptions are questionable, and future work will need to determine the robustness of the model, or its potential modifications, when relaxing these assumptions. 

Even tough we assumed that the brain follows the general principles of predictive coding~\cite{RB99}, our system's hyperparameters can in fact be modulated to accommodate many variants of this framework~\cite{Wen18, choski21, heeger2017, tschantz2022}.
One other important assumption that we made was to simplify the connectivity matrices between neuronal layers---which determines the selectivity of each neuron, and thus the functionality of the entire system. Even when we moved past the ``identity'' assumption, the connection matrices that we adopted were constrained to be symmetric, and most importantly, were assumed to be similar from one layer to the next. This made our equations tractable, but it constitutes a clear restriction, and a departure from biological plausibility that will need to be addressed in future extensions. Another important limitation that we wish to relax in future works is the fact that we have considered a linear model although real biological networks or deep neural networks are intrinsically nonlinear. Going beyond the linear analysis that we have presented here would need the development of new theoretical techniques which constitutes a major open problem to be addressed in forthcoming works.

\begin{figure}
\begin{center}
\begin{tikzpicture}[scale=1.5]
\coordinate (A0) at (-1,0);
    \coordinate (A1) at (0,0);
    \coordinate (A2) at (1,0);
    \coordinate (A3) at (2,0);
    \coordinate (A4) at (3,0);
    \coordinate (A5) at (4,0);
 \coordinate (A6) at (5,0);

\coordinate (B0) at (-1,1);
   \coordinate (B1) at (0,1);
    \coordinate (B2) at (1,1);
    \coordinate (B3) at (2,1);
    \coordinate (B4) at (3,1);
    \coordinate (B5) at (4,1);
        \coordinate (B6) at (5,1);
  
        \coordinate (C1) at (0,-1.15);  
        \coordinate (C2) at (1,-1.15);
    \coordinate (C3) at (2,-1.15);
 \coordinate (C4) at (3,-1.15);
  \coordinate (C5) at (4,-1.15);

    \coordinate (D1) at (0,2);
    \coordinate (D2) at (1,2);
    \coordinate (D3) at (2,2);
    \coordinate (D4) at (3,2);
    \coordinate (D5) at (4,2);

    \coordinate (E1) at (0,-1);
    \coordinate (E2) at (1,-1);
    \coordinate (E3) at (2,-1);
    \coordinate (E4) at (3,-1);
    \coordinate (E5) at (4,-1);

% \coordinate (D1) at (1.5,1);

\node[left] at (B0) {$n+1\quad$};
\node[left] at (A0) {$n\quad$};

    \node[below] at (C1) {$j-2$};
    \node[below] at (C2) {$j-1$};
    \node[below] at (C3) {$j$};
    \node[below] at (C4) {$j+1$};
    \node[below] at (C5) {$j+2$};
    
  %  \node[above] at (D1) {$\W^f$};

\draw[line width=0.4mm,color=purple,directed] (B2) -- (B3) ;
\draw[line width=0.4mm,color=purple,directed] (A3) -- (B3) ;
\draw[line width=0.4mm,color=purple,directed] (A2) -- (B3) ;
\draw[line width=0.4mm,color=purple,directed] (A4) -- (B3) ;

\draw[line width=0.4mm,color=cyan,directed] (A1) -- (B3) ;
\draw[line width=0.4mm,color=cyan,directed] (A5) -- (B3) ;

\draw[line width=0.4mm,color=cyan,directed] (B1) to [bend left=45] (B3) ;

    \node at (A1) {$\bullet$};
    \node at (A2) {$\bullet$};
    \node at (A3) {$\bullet$};
    \node at (A4) {$\bullet$};
    \node at (A5) {$\bullet$};
    
    \node at (B1) {$\bullet$};
    \node at (B2) {$\bullet$};
    \node at (B3) {$\bullet$};
    \node at (B4) {$\bullet$};
    \node at (B5) {$\bullet$};

    \draw[dashed] (A0) -- (A6);
    \draw[dashed] (B0) -- (B2);
    \draw[dashed] (B3) -- (B6);
     \draw[dashed] (E1) -- (D1);
 \draw[dashed] (E2) -- (D2);
  \draw[dashed] (E4) -- (D4);
   \draw[dashed] (E5) -- (D5);
\draw[dashed] (B3) -- (D3);
\draw[dashed] (E3) -- (A3);

\end{tikzpicture}
  \end{center}
\caption{Illustration of the network structure of model \eqref{modelZ2layer} where blue arrows indicate the new long-range interactions coming from layer $j\pm2$.}
\label{fig:higherorder}
\end{figure}
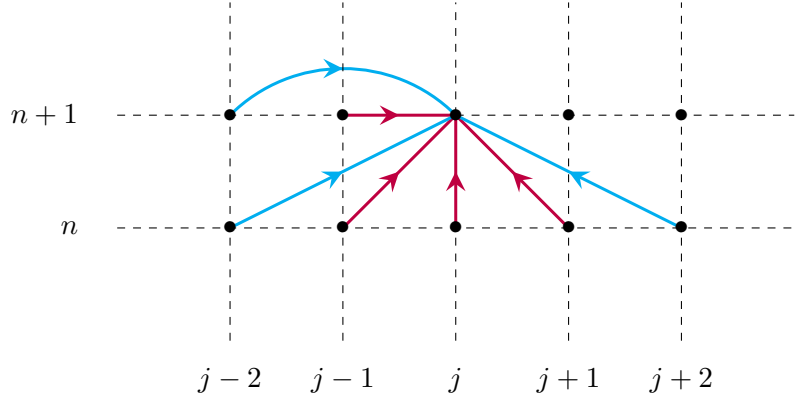  

Aside from exploring richer patterns of connectivity between adjacent layers, another natural extension of the model could be to incorporate long-range interactions, beyond the immediately adjacent layers. 
For instance, one could explore a simple second-order layer model (illustrated in Figure~\ref{fig:higherorder}), whose scalar version reads:
\bqq
e_j^{n+1}-\beta_1e_{j-1}^{n+1}-\beta_2e_{j-2}^{n+1}=\alpha_1e_{j-1}^{n}+\alpha_2e_{j-2}^{n}+(1-\beta^*-\lambda^*- \alpha^* )e_j^{n} + \lambda_1  e_{j+1}^{n}+ \lambda_2  e_{j+2}^{n}, \quad j\in\Z,
\label{modelZ2layer}
\eqq
where we have set $\beta^*:=\beta_1+\beta_2$, $\alpha^*:=\alpha_1+\alpha_2$ and $\lambda^*:=\lambda_1+\lambda_2$. Once again the fate of such a system \eqref{modelZ2layer} would be dictated by the amplification factor function
\bqs
\rho(\theta)=\frac{\alpha_1 e^{-\mbi\theta}+\alpha_2e^{-2\mbi\theta}+1-\beta^*-\lambda^*- \alpha^*  + \lambda_1  e^{\mbi\theta}+ \lambda_2e^{2\mbi\theta}}{1-\beta_1e^{-\mbi\theta}-\beta_2e^{-2\mbi\theta}}, \quad \theta\in[-\pi,\pi].
\eqs

This as well as higher-order interaction models, possibly including ``hub'' regions like the thalamus that would be mutually interconnected with all layers in the hierarchy~\cite{hwang2017}, are promising directions for follow-up studies.

\subsection{Conclusion}

The mathematical framework proposed here, guided by both computational considerations and neuroscientific inspiration, can be of use to both fields. In machine learning, the framework may serve to provide guarantees about the stability of a predictive coding system given its chosen hyperparameters, or to choose a valid range for these hyperparameters. For neuroscientists, our equations can be used directly to understand biological vision and to make predictions about biological behavior in various situations compatible with predictive coding.
But this general mathematical framework (a number of hierarchically connected layers with source terms, boundary conditions, feedforward and feedback connectivity matrices, analyzed via its amplification factor function) may also be adapted to fit other models of biological perception and cognition beyond predictive coding. We hope that the various derivations made in the present work can serve as a template for future applications in this direction. And more generally, that this study may be helpful to the larger computational neuroscience community.

\section*{Acknowledgements} 
G.F. acknowledges support from the ANR via the projects: Indyana under grant agreement ANR-21-CE40-0008, ChaMaNe under grant agreement ANR-19-CE40-0024  and an ANITI (Artificial and Natural Intelligence Toulouse Institute) Research Chair. R.V. is supported by ANR OSCI-DEEP (ANR-19-NEUC-004) and an ANITI (Artificial and Natural Intelligence Toulouse Institute) Research Chair (ANR-19-PI3A-0004).

\end{document}